\documentclass[binding=0.6cm]{sapthesis}
\usepackage{microtype}
\usepackage{hyperref}
\usepackage{graphicx}% Include figure files
\usepackage{dcolumn}% Align table columns on decimal point
\usepackage{bm}% bold math
\usepackage{lipsum}
\usepackage{cleveref}
\usepackage{amsmath,color}
\usepackage{tabularx}
\usepackage{comment}
\usepackage{multicol}
\usepackage{appendix}
\usepackage{caption}
\usepackage{subcaption}

\hypersetup{pdftitle={Enrico Ventura's PhD Thesis},pdfauthor={Enrico Ventura}}
\title{Demolition and Reinforcement of Memories in \\Spin-Glass-like Neural Networks}
\author{Enrico Ventura}
\IDnumber{1699775}
\course{Scuola di Dottorato "Vito Volterra" and École Doctorale Physique en Île de France (EDPIF)}
\courseorganizer{Joint International PhD in Physics (Tesi in cotutela/Cotutelle de thèse) \\between Università di Roma
La Sapienza and École Normale Supérieure - PSL}
\cycle{XXXVI}
\AcademicYear{2022-2023}
\advisor{Prof. Giancarlo Ruocco \\ Prof. Francesco Zamponi}
\examdate{December $1^{st}$, 2023}
\examiner[]{Prof. Raffaella Burioni, Università di Parma.} \examiner[]{Prof. Carlo Lucibello, Università Bocconi.}
\examiner[]{Dr. Beatriz Seoane, Université Paris-Saclay.}
\examiner[]{Dr. Silvia Bartolucci, University College London.}
\examiner[]{Prof. Giancarlo Ruocco, Università La Sapienza.}
\examiner[]{Prof. Francesco Zamponi, École Normale Supérieure-PSL.}
\authoremail{enrico.ventura@uniroma1.it}
\copyyear{2023}
\thesistype{PhD thesis}
\begin{document}
\frontmatter
\maketitle
%\dedication{}
\begin{abstract}
Statistical mechanics has made significant contributions to the study of biological neural systems by modeling them as recurrent networks of interconnected units with adjustable interactions. Several algorithms have been proposed to optimize the neural connections to enable network tasks such as information storage (i.e. associative memory) and learning probability distributions from data (i.e. generative modeling). Among these methods, the Unlearning algorithm, aligned with emerging theories of synaptic plasticity, was introduced by John Hopfield and collaborators. \\
The primary objective of this thesis is to understand the effectiveness of Unlearning in both associative memory models and generative models. Initially, we demonstrate that the Unlearning algorithm can be simplified to a linear perceptron model which learns from noisy examples featuring specific internal correlations. The selection of structured training data enables an associative memory model to retrieve concepts as attractors of a neural dynamics with considerable basins of attraction. 
Subsequently, a novel regularization technique for Boltzmann Machines is presented, proving to outperform previously developed methods in learning hidden probability distributions from data-sets. The Unlearning rule is derived from this new regularized algorithm and is showed to be comparable, in terms of inferential performance, to traditional Boltzmann-Machine learning.
\end{abstract}
\tableofcontents
\mainmatter
\chapter{\label{sec:chap1} Introduction: modeling intelligence with Recurrent Neural Networks}

%\section{Modeling intelligence with Recurrent Neural Networks}
Understanding the functioning of the human brain has been a central topic of investigation since a long time \cite{kendal_principles_2013, nicholls_neuron_2011, amit_modeling_nodate}. 
Among the main themes that have attracted the attention of scientists there are: the capability of the brain to store information and retrieve it from external stimulation (i.e. \textit{associative memory}), its ability at separating stimuli into different categories (i.e. \textit{classification}) and understanding the underlying structure that is necessary to generate new coherent information (i.e. \textit{generation}). 
All these tasks belong to the broader concept of intelligence, which is responsible for adaptation and survival in animals \cite{legg_collection_2007}. \\Early studies on the physiology of the brain \cite{kendal_principles_2013, nicholls_neuron_2011, amit_modeling_nodate} pointed out its particular composition: an ensemble of neuronal cells linked by wires (i.e. the \textit{synaptic apparatus}) reciprocally exchanging electric signals at deferred times (i.e. the \textit{postsynaptic potentials}). 
The underlying network structure is generally sparse, asymmetric in the connections, and it can contain closed loops. We call this type of system \textit{recurrent} neural network \cite{amit_modeling_nodate}.  
Given the experimental observations and the study of the performance of the brain, one might be encouraged to consider recurrent neural networks as a successful starting point to describe intelligence. \\  
No wonder that mathematicians, simultaneously with the development of neuroscience, discovered that some optimization problems could be mapped into graphical representations sharing important similarities with real neural networks. In this case stimuli experienced by the system were replaced by data-points belonging to a data-set.  
The most emblematic example is Rosenblatt’s perceptron and its generalizations \cite{rosemblatt_perceptron_1958, marvin_minsky_seymour_papert_perceptrons_1969}. The perceptron problem consists in separating clusters of data-points into different classes depending on reciprocal similarities: this operation is called \textit{classification} and the perceptron is a \textit{classifier}. One of the contact points with biology is the way artificial and real neurons work: the neuronal spike, i.e. the abrupt emission of a postsynaptic potential by a neuron, is triggered when the sum of all signals received by its neighbours overcomes a threshold in tension; at the same way, the perceptron assigns a class to an input vector depending on whether the full incoming field to an output unit overcomes a fixed value or not.  
Another aspect involves what neuroscientists refer to as \textit{synaptic plasticity} of the network \cite{kendal_principles_2013, nicholls_neuron_2011}: synapses tend to modify their conductivity in time, in response of both external and internal endogenous stimuli. Such modifications are associated to the act of learning to accomplish a particular task \cite{stampanoni_bassi_synaptic_2019}. 
\\At this point, it came natural to statistical physicists, who are specialists in applying probability and statistics for the study of interacting systems, to start contributing to the theory of neural networks.
At first, it was a simplification effort \cite{little_analytic_1978}: real systems of neurons were reduced to simple recurrent networks, where a number of binary variables, representing the two possible \textit{active/silent} states of the neuron, are mutually coupled by pairwise interactions.
These models strongly resembled what people called \textit{spin glasses} \cite{mezard_spin_1986}: the competition among the different strength of the interactions implies a dynamic multistability of the neural activity and a complex variety of possible equilibrium configurations. Most likely, a boost in the study of these very ancestral types of \textit{complex systems} \cite{anderson_more_1972}, taking place across the 70s, encouraged John Hopfield, with his pioneering work about Hebbian networks \cite{hopfield_neural_1982} published in 1982, to set a bridge between the statistical mechanics of disordered systems and neuro-physiology. According to these models, the more correlated neurons are, while reacting to external stimuli, the stronger is the synapse that connects them.  
Three years later, Daniel Amit, Hanoch Gutfreund and Haim Sompolinsky applied the physics of spin glasses to compute the thermodynamics and the critical capacity of a plausible memory model \cite{amit_storing_1985}. 
Further progresses on this line were made until the most recent years, giving an important boost in the field of theoretical neuroscience and the study of associative memory.
\\In parallel with John Hopfield, Geoffrey Hinton and Terrence Sejnowski (a former Hopfield's doctoral student) proposed another spin glass-like model that encoded the statistics of the stimuli in its parameters (i.e. the interactions and the external fields)\cite{hinton_optimal_1983, hinton_analyzing_1983}.
From sampling the typical configurations of the model one could create a new data-set being perfectly indistinguishable from the learnt one. 
This system, known as Boltzmann-Machine, was the first example of neural network-based generative model, and also a notable case of physics connecting the biological learning with the artificial one.
\\Nevertheless, the early most important success in connecting artificial with biological networks, was made by Elizabeth Gardner, in her late works from 1988-1989 \cite{gardner_space_1988, gardner_optimal_1989}, where she mapped the Rosemblatt's perceptron into a recurrent neural network, computed its critical capacity and advanced a training algorithm for the synaptic strength. 
The type of computation that she proposed is based on optimizing the interactions rather than fixing them a-priori. Once again, this optimization mechanism reflects the synaptic plasticity of a natural neural system.               
\\ \\ This thesis is centered on the Unlearning algorithm, a training procedure for recurrent neural networks introduced by John Hopfield and collaborators in 1983 to enhance the associativity of a memory model \cite{hopfield_unlearning_1983}. The original idea, from which the algorithm is generated, consists in pruning the system from spurious states, i.e. local attractors of the dynamics responsible for disturbing memory retrieval \cite{amit_modeling_nodate, gardner_structure_1986}. This removal of the spurious attractors is performed by iterating an anti-Hebbian learning rule over the synapses: the more neurons are correlated across spurious states, the the more their connection will be weakened by the algorithm.   
Since the moment that this technique was proposed, interesting similarities between the Unlearning procedure and neuroscience emerged, in particular concerning the way sleep is supposed to affect the functionality of the brain \cite{crick_function_1983}. Even Hinton himself noticed that a mechanism similar to the Unlearning rule contributed to the training of a Boltzmann Machine \cite{hinton_unsupervised_1999}.  
\\The goal of this work is thus threefold:
\begin{enumerate}
    \item Showing that the Unlearning rule naturally emerges from a perceptron algorithm regularized through noise injection, and that it reaches an optimal associative memory performance. Unlearning appears to be a valuable unsupervised alternative to the training of a maximally stable perceptron, i.e. a Support Vector Machine. These results are published into \cite{benedetti_supervised_2022, benedetti2}.
    \item Proposing a new type of regularization for Boltzmann Machines, that can be generalized to the Unlearning rule. We also investigate the inferential power of the standard Unlearning procedure, and the importance of the initialization of the parameters in a Boltzmann Machine. These results are contained into \cite{inprep}.  
    \item Using the Unlearning algorithm as a connection between two learning frameworks of artificial intelligence: associative memory (with its formal mapping to classification problems) and generative modeling, specifically Boltzmann Machine learning. In particular, we want to show a formal equivalence between three learning algorithms: Unlearning, Support Vector Machines, and Boltzmann Machines. These results are contained into \cite{inprep_single}.  
\end{enumerate}
This thesis is placed at the interface between statistical mechanics, theoretical neuroscience and artificial intelligence, and it provides some useful insights for the unification of these three fields of knowledge. 
\\ \\The manuscript is structured as follows: 
\begin{itemize}
    \item Chapter 1: an introduction to various learning algorithms associated to different tasks performed by neural networks. A distinction is made between three fundamental types of modeling: associative memory, classification and the generative one. 
    \item Chapter 2: a simple noise-injection training algorithm for recurrent neural networks, named Training-with-noise, is studied in the case of structured noise. While injecting a maximal amount of random noise would be deleterious in the standard scenario, including internal dependencies among the features of the noisy training data significantly improves the associativity power of the network. An analytical recipe for the best noise structure is derived and tested numerically. We display the emergence of the Unlearning rule from the Training-with-noise algorithm in presence of structured noise and study the cases of different types of training data-sets.
    \item Chapter 3: a new regularization for Boltzmann Machine learning is proposed. A particular limit of the regularization, that recovers a thermally averaged Hebbian Unlearning, is studied and the equivalence of the Unlearning rule with a two-steps Boltzmann Machine is displayed. Eventually, we show a formal equivalence between Boltzmann Machines (i.e. generative models), Support Vector Machines and the Hebbian Unlearning algorithm (i.e. associative memory models). 
    \item Chapter 4: we discuss the results and give some future perspectives of the research. 
\end{itemize}
Each chapter starts with an introduction explaining the research goals and ends with a summary listing the main points discussed in the text. Each section of the thesis ends with a \textit{checkpoint} paragraph briefly summarizing the main results of the analysis.\\ \\

For what concerns this chapter, its structure will be the following. 
The idea of associative memory modeling is introduced in \cref{sec:ann} together with the description of a recurrent neural network prototype that will be utilized in our analysis. 
The main observables representing the quality of the memory retrieval are also presented in detail. Three learning rules regarding associative memory are then described: Hebbian Learning, Hebbian Unlearning (HU), Linear perceptrons and Support Vector Machines (SVMs).
\\Furthermore, \cref{sec:class} defines what is meant for classification in statistical learning and discusses the differences and analogies between classifiers and associative memory models.
\\Eventually, \cref{sec:data_gen} treats the generative modeling approach and gives an example of a celebrated energy-based model, namely the Boltzmann Machine (BM).

%\section{Learning algorithms for different tasks}
\newpage
\section{Associative Memory Task \label{sec:ann}}

\begin{figure}[ht!]
\centering
\includegraphics[width=0.55\textwidth]{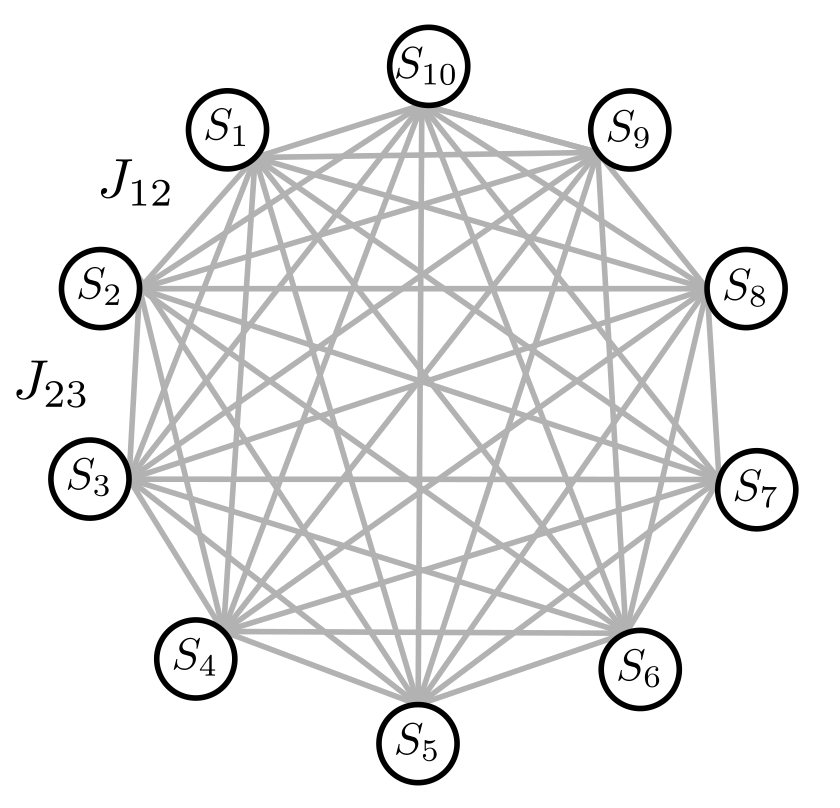}
    \caption{Graphical representation of a simple fully connected recurrent neural network of $N = 10$ neurons with symmetric couplings $J_{ij} = J_{ji}$. Autapses $J_{ii}$ are absent.}
\label{fig:memo_graph}           
\end{figure}

With the term \textit{associative memory} we define the capability of the neural system of recalling a given concept (i.e. a \textit{memory}) when a corrupted version of it is displayed as an input \cite{amit_modeling_nodate, hertz_introduction_1991}. 
%This type of task implies that a number of memories are generated a-priori and then labelled. 
Associative memory is extensively studied by neuroscience, in the terms of the capability of precise regions of the brain (e.g. prefrontal cortex, hippocampus) to perform \textit{long term} and \textit{short term} storage of information \cite{nicholls_neuron_2011, gerstner_neuronal_2014}. 
\\We will limit ourselves to the main class of simple models implemented by physicists i.e. a network of $N$ Ising variables $S_i$ mutually interacting through the couplings $J_{ij}$ with $J_{ii} = 0$. Fig. \ref{fig:memo_graph} depicts an example of fully connected recurrent neural network with symmetric couplings, which is similar to the ones employed in our further analysis. 
The memory retrieval operation implies an evolution of the network in time, i.e. a rule for the neural dynamics. An emblematic rule is
\begin{equation}
    \label{eq:dynamics}
    S_i(t+1) = \text{sign}\left(\sum_{j=1}^N J_{ij}S_j(t)\right), \hspace{0.7cm} i = 1,..,N
\end{equation}
which can be run either in parallel (i.e. \textit{synchronously}) or in series (i.e. \textit{asynchronously} in a random order) over the $i$ indices \cite{peretto}. We will mainly concentrate on asynchronous dynamics, in which case equation  (\ref{eq:dynamics}) can only converge to fixed points, when they exist \cite{amit_modeling_nodate}. This kind of network can be used as an associative memory device, namely for reconstructing a number $p$ of configurations $\{\vec{\xi}^{\mu}\}_{\mu = 1}^p$ called \textit{memories}, when the dynamics is initialized on configurations similar enough to them. Memories have binary entries $\xi^{\mu}_i = \pm 1$, $\mu\,\in [1,...,p]$. In this work, we will concentrate on Rademecher random memories, generated with a probability 
 \begin{equation}
 \label{eq:rade}
 P(\xi^{\mu}_i= +1) = P(\xi^{\mu}_i= -1) = 1/2\hspace{0.5cm}\forall i,\mu
 \end{equation}
With an appropriate choice of the couplings, the model can store an extensive number of memories $p = \alpha N$, where $\alpha$ is called \textit{load} of the network. This category of models, that aim at retrieving memories as stable fixed points of the dynamics, are called \textit{attractor neural networks}.\\We generally want to benchmark the neural network performance, specifically in terms of the dynamic stability achieved by the memory vectors and the ability of the system to retrieve them from blurry examples. Thus, some useful definitions and observables are now introduced for this purpose.  
\\ \\We define $\textit{perfect-retrieval}$ as the capability to perfectly retrieve each memory when the dynamics is initialized on the memory itself. It is here convenient to define a quantity, called \textit{stability}, defined as
\begin{equation}
    \label{eq:stab}
    \small
    \Delta_i^{\mu} = \frac{\xi_i^{\mu}}{\sqrt{N}\sigma_i}\sum_{j = 1}J_{ij}\xi_j^{\mu}, \qquad \sigma_i = \sqrt{\sum_{j=1}^N J_{ij}^2/N}.
\end{equation}
We call $n_{SAT}$ the fraction of $\xi_i^{\mu}$ units that satisfy the following inequality  
\begin{equation}
    \label{eq:perf_ret}
    \Delta_i^{\mu} > 0,  
\end{equation}
i.e. that are stable according to one step of the dynamics (\ref{eq:dynamics}). 
Perfect-retrieval is reached when $n_{SAT} = 1$.
It is thus important to remark the following implication of statements 
\begin{equation}
\label{eq:const1}
    n_{SAT} = 1 \hspace{0.2cm}\Longleftrightarrow \hspace{0.2cm} \Delta_i^{\mu} > 0 \hspace{0.5cm}\forall i,\mu.
\end{equation}
The label SAT derives from the nomenclature used in celebrated optimization problems \cite{altarelli2009review, marvin_minsky_seymour_papert_perceptrons_1969} computing the limits of satisfiability of the condition in \cref{eq:const1}. The phase of the model where \cref{eq:const1} is satisfied was called SAT phase, as opposed to the UNSAT phase. \\\\
We furthermore define $\textit{robustness}$ as the capability to retrieve the memory, or a configuration that is strongly related to it, by initializing the dynamics on a noise-corrupted version of the memory. This property of the neural network is related to the size of the basins of attraction to which the memories belong, and does not imply $n_{SAT}=1$. A good measure of the performance in this sense is the \textit{retrieval map}
\begin{equation}
\label{eq:m}
m_f(m_0, J) := \overline{\Big\langle\frac{1}{N}\sum_{i = 1}^N \xi_i^\mu S_i^{\mu}(\infty)\Big\rangle}.
\end{equation}
Here, $\vec{S}^{\mu}(\infty)$ is the stable fixed point reached by the network having couplings $J$, when it exists, when the dynamics is initialized on a configuration $\vec{S}^\mu(0)$ having overlap $m_0$ with a given memory $\vec{\xi}^\mu$. The symbol $\overline{\hspace{0.1cm}\cdot\hspace{0.1cm}}$ denotes the average over different realizations of the memories and $\langle \cdot \rangle$ the average over different realizations of $\vec{S}^\mu(0)$. In the perfect-retrieval regime, one obtains $m_f = 1$ when $m_0 = 1$. The analytical computation of the retrieval map might be challenging for some networks. Hence one can introduce another indicative observable for the robustness, i.e. the \textit{one-step retrieval map}  $m_1(m_0)$ \cite{kepler_domains_1988}, defined by applying a single step of \textit{synchronous} dynamics (\ref{eq:dynamics}):
\begin{equation}
\label{eq:onestep_ret_map}
    m_1(m_0, J) := \overline{\frac{1}{N}\sum_{i = 1}^N  \Big\langle \xi_i^\mu \;\text{sign}\Big(\sum_{j=1}^N J_{ij}S_j^\mu(0)\Big)\Big\rangle}\;,
\end{equation}
\\\\We now provide a list of notable learning prescriptions in attractor neural networks that will be useful for the rest of the dissertation.  
\subsection{Hebbian Learning}

Hebb's (or Hebbian) learning prescription \cite{hopfield_neural_1982, hebb_organization_1949} consists in building up the connections between the neurons as an empirical covariance of the memories, i.e. 
\begin{equation}
    \label{eq:hop}
    J_{ij}^{H} = \frac{1}{N}\sum_{\mu = 1}^p\xi_i^{\mu}\xi_j^{\mu}.
\end{equation}
This rudimentary yet effective rule allows to retrieve memories up to a critical capacity $\alpha_c^H = 0.138$. \cite{amit_storing_1985}. Notably, when $\alpha < \alpha_c^H$ memories are not perfectly recalled, but only reproduced with a small number of errors. In this phase, named \textit{retrieval} phase, memories show some robustness since they belong to large basins of attraction. On the other hand, when $\alpha > \alpha_c^H$ the statistical interference between the random memories impedes the dynamics to retrieve them, shifting the system into an oblivion regime. Notably, the landscape of attractors given by this rule is rugged and disseminated with \textit{spurious states}, i.e. stable fixed points of the dynamics barely overlapped with the original memories \cite{gardner_structure_1986}. \\An important notion to keep in mind for the rest of this work is that, whenever the network interactions are symmetric, as in the current case, the Lyapunov function of the dynamics, that is minimized by the attractor states, coincides with the \textit{energy} function of the model. Consequently, attractors are local minima of the energy landscape. For recurrent neural networks, the energy can be defined as
\begin{equation}
    \label{eq:ene}
    E[\vec{S}|J] = -\sum_{i,j>i}S_i J_{ij}S_j,
\end{equation}
equivalently to Ising spin systems in statistical mechanics \cite{amit_modeling_nodate, hertz_introduction_1991}. 

\subsection{Hebbian Unlearning \label{sec:UNL}}

Inspired by the brain functioning during REM sleep \cite{crick_function_1983}, the Hebbian Unlearning algorithm (HU) \cite{hopfield_unlearning_1983,crick_function_1983,van_hemmen_increasing_1990,taylor_unlearning_1992,benedetti_supervised_2022, wimbauer_universality_1994, Klein} is a training procedure leading to perfect-retrieval and good robustness in a symmetric neural network. This paragraph contains an introduction to the algorithm as well as a description of the gained performance in terms of perfect-retrieval. The robustness capability of the algorithm will be adressed further in the work. \\\\Training starts by initializing the connectivity matrix according to the Hebb's rule \cref{eq:hop} (i.e. $J^{(0)} = J^H$). Then, the following procedure is iterated at each time step $d$:
\begin{enumerate}
    \item Initialize the network on a random neural state. 
    \item Run the asynchronous dynamics     (\ref{eq:dynamics}) until convergence to a stable fixed point $\vec{S}^{*}$. 
    \item Update couplings according to:
    \begin{equation}
        \label{eq:unl_rule}
        \delta J_{ij}^{(d)} = -\frac{\lambda}{N}S_i^{*} S_j^{*}\hspace{1cm}J_{ii} = 0 \hspace{0.5cm}\forall i.
    \end{equation}
\end{enumerate}
This algorithm was first introduced to prune the landscape of attractors from proliferating spurious states, i.e. fixed points of (\ref{eq:dynamics}) not coinciding with the memories \cite{gardner_structure_1986, amit_modeling_nodate}. Such spurious states are only weakly correlated with the memories. Even though this pruning action leads to the full stabilization of the memories, the exact mechanism behind this effect is not completely understood.\\
HU is an \textit{unsupervised} algorithm, in the sense that it does not need to be provided explicitly with the memories $\{\vec{\xi}^\mu\}_{\mu = 1}^p$, and only exploits the information encoded in Hebbian initialization (see \cref{eq:hebbs}). 
The total number of iterations $D$ is a parameter of the algorithm, and must be chosen as to maximize the recognition performance at a given load $\alpha$. 
We report here the analysis of the perfect-retrieval properties of the resulting network, with an estimate of the critical capacity and the amount of iterations for an effective early-stopping of the algorithm, contained in \cite{benedetti_supervised_2022}.\\
The performance of the algorithm has been studied in terms of the stabilities $\Delta_i^\mu$. This approach has already been attempted~\cite{Horas}, but the following analysis pushes it further and reveals new unexpected features.  
\begin{figure}[ht!]
\centering
\includegraphics[width=.7\textwidth]{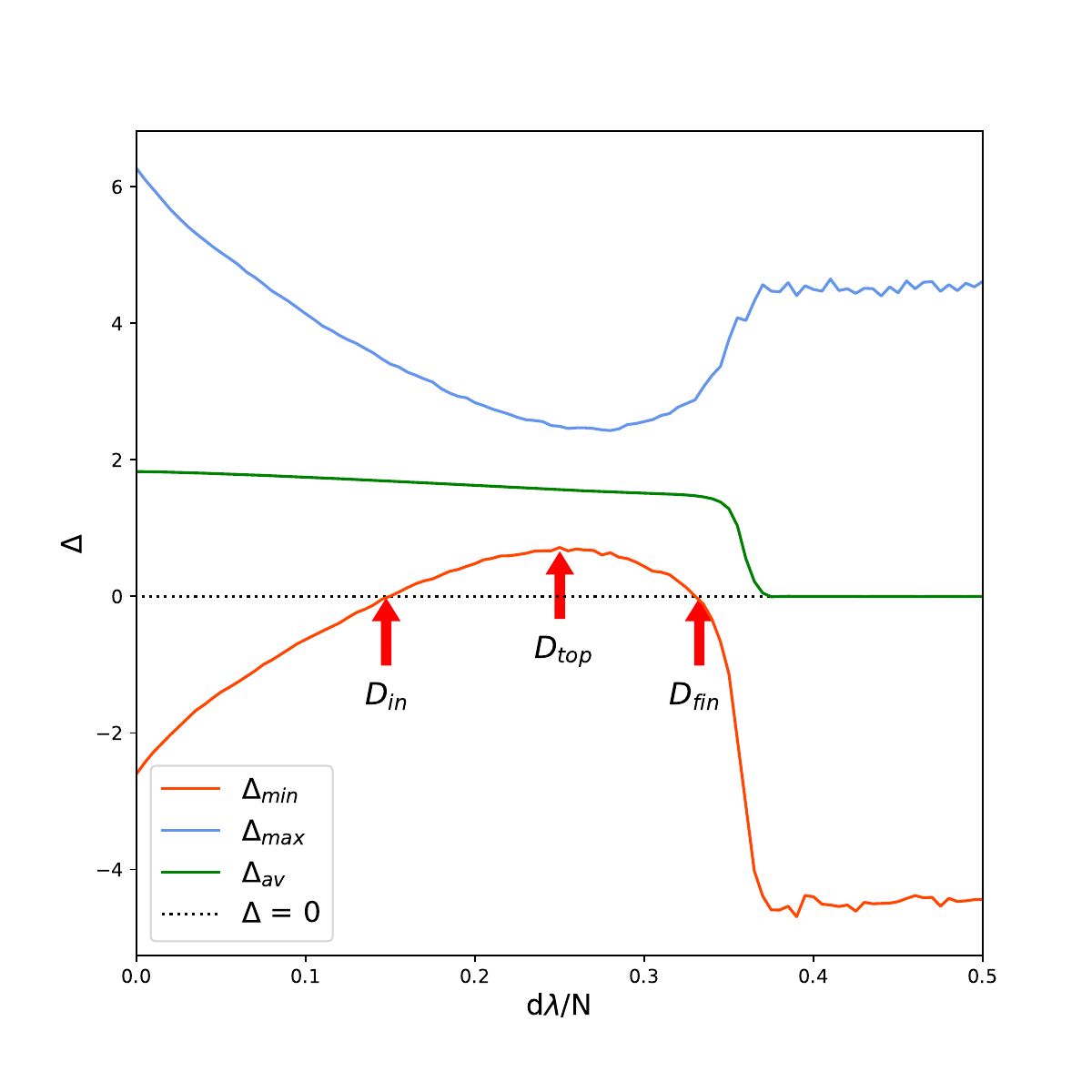}
    \caption{Values of the minimal stability $\Delta_{min}$ (orange), maximal stability $\Delta_{max}$ (blue) and average stability $\Delta_{av}$ (green) computed during Hebbian unlearning, averaged over $50$ realizations of the memories at $N = 800$, $\alpha = 0.3$, $\lambda = 10^{-2}$. The black dotted line represents the zero-stability threshold to be overcome by $\Delta_{min}$ to have all memories perfectly recalled. We denote the corresponding value of the number of iterations $D$ by $D_{in}$. $D_{top}$ is for the point where the algorithm reaches the maximum value of  $\Delta_{min}$, while $D_{fin}$ is the end of the perfect perfect-retrieval regime of the network. We used red arrows to point at $D_{in}$, $D_{top}$ and $D_{fin}$.}
\label{fig:stability_HU}           
\end{figure}
Fig.~\ref{fig:stability_HU} shows the typical behavior of the minimum stability $\Delta_{min}$, the average one $\Delta_{av}$ and the maximum stability $\Delta_{max}$ as the Unlearning procedure unfolds. The horizontal axis represents the number of steps $d$ performed by the algorithm, rescaled by a factor $\lambda/N$. 
%The reason for this choice will become clear later. 
Focusing on $\Delta_{min}$, we can see a non-monotonic behavior: the minimal stability grows to positive values, peaks at some value $d=D_{top}$ and then decreases back to negative values. Between $D_{in}$ and $D_{fin}$ every stability is positive or, equivalently, every memory is a fixed point for the dynamics. As we increase $\alpha$, the interval $[D_{in}, D_{fin}]$ shrinks, and the height of the peak at $D_{top}$ lowers, until we reach a critical load $\alpha_c$, above which $\Delta_{min}$ never goes above zero. Collecting data for networks of size $N=300,\,400,\,500,\,600,\,800$ and different values of $\alpha$ and $\epsilon$ it is possible to extrapolate the position of $D_{in},\,D_{top}$ and $D_{fin}$ as a function of $\alpha,\,\lambda$ and $N$, as well as the critical capacity $\alpha_c$. By fitting the data with respect to the model parameters one can find that the number of iterations is in every case linear in $N$ and $1/\lambda$. Moreover, $D_{top}$ also depends linearly on $\alpha$. At the critical capacity,
$$\alpha_c^{HU}=0.589\pm0.003 \ ,$$
and the value of $\Delta_{min}(D_{top})$ approaches zero. 
The value of the critical capacity $\alpha_c^{HU}$ as well as the linear dependence of $D_{in}$, $D_{top}$, $D_{fin}$ on $N/\lambda$ are consistent with evidence provided by past literature~\cite{taylor_unlearning_1992, van_hemmen_increasing_1990, Horas}. \\
Because the $\Delta_{min}(D)$ curve is quadratic around $D_{top}$, as illustrated in
\cref{fig:stability_HU},
$D_{in}$ and $D_{fin}$ both tend to $D_{top}$ at the critical capacity  with a critical exponent $1/2$. 
The resulting scaling relations are: 
\begin{equation}
    \label{eq:top}
    D_{top}(\lambda,\alpha,N)=\frac{N}{\lambda}(a\cdot\alpha+b) \ ,
\end{equation}
\begin{equation}
    \label{eq:in}
    D_{in}(\lambda,\alpha,N)=D_{top}-\frac{N}{\lambda}(c\cdot\alpha+d)^{1/2} \ ,
\end{equation}
\begin{equation}
    \label{eq:fin}
    D_{fin}(\lambda,\alpha,N)=D_{top}+\frac{N}{\lambda}(e\cdot\alpha+f)^{1/2} \ ,
\end{equation}
with
$$ a=1.02\pm0.02 \ , \hspace{0.5cm} b=-0.05\pm0.01 \ ,$$
$$ c=-0.039\pm0.003 \ ,\hspace{0.5cm} d=0.023\pm0.002 \ ,$$
$$ e=-0.022\pm0.001 \ ,\hspace{0.5cm} f=0.013\pm0.001 \ .$$

All the statistical errors have been evaluated using the jackknife method \cite{jack}.\\

The study of the evolution of the basins of attraction during Unlearning is one of the main contribution of this manuscript, and it will be deepened in detail in the next Chapter. 

\subsection{Linear Perceptrons \& Support Vector Machines}
The linear perceptron algorithm \cite{gardner_space_1988, gardner_phase_1989, marvin_minsky_seymour_papert_perceptrons_1969}, which is one of the pillars of modern artificial intelligence, is an iterative procedure allowing to fully stabilize the memories and tune their robustness capabilities. 
Specifically, we are going to refer to \textit{linear perceptron} as the adaptation of the classical perceptron to recurrent neural networks, already introduced in \cite{gardner_space_1988, gardner_phase_1989, battista_capacity-resolution_2020, brunel_optimal_2004, brunel_is_2016}. 
In this case a $N$-dimensional input layer is fully connected by the connections $J$ to a $N$-dimensional output layer. This architecture can thus be mapped into a biologically inspired recurrent neural network. Further details about this mapping will be provided in the next paragraph about classification. 
\\Given the fully connected architecture, we want to find a set of couplings that satisfy the constraints 
\begin{equation}
\label{eq:const}
    \Delta_i^{\mu} > k , \qquad \forall \mu, i \ .
\end{equation}
with $k$ being a parameter called \textit{margin}. An elegant way to interpret this problem is to recast it in terms of a linear regression~\cite{marvin_minsky_seymour_papert_perceptrons_1969} in the $J$. In terms of the network dynamics the larger $k\geq 0$ is, the more robust memories will be under perturbation of the dynamics. Specifically, given a value of $\alpha$ all memories will be stable up to a maximum value of $k_{max}(\alpha)$. The solution of the problem such that $k = k_{max}(\alpha)$ can be proved to be unique, given one realization of the memories. The maximum capacity achievable by the network is $\alpha_c^{P} = 2$ such that $k_{max}(2) = 0$. Following previous work, we call SAT phase the region in the space $(\alpha, k)$ such that \cref{eq:const} is satisfied, while the rest of the phase space will be said to be UNSAT.\\
Inside these limits all constraints in (\ref{eq:const}) will be satisfied after a number of iterations of the following serial update for the couplings
\begin{equation}
    \small
    \label{eq:lp_update}
    \delta J_{ij}^{(d)} = +\lambda\sum_{\mu=1}^p\epsilon^\mu_i(d)\xi^\mu_i\xi^\mu_j, \hspace{0.3cm} J_{ii}=0,\hspace{0.3cm}\lambda >0
\end{equation}

    $$\epsilon^\mu_i(d) =\frac{1}{2}\left(1 + \text{sign}(k-\Delta^\mu_i(d))\right),$$
where $\lambda$ is the learning rate, considered to be small, $d$ is the descrete algorithm time on which the mask $\epsilon_i^{\mu}$ depends.  
One can also symmetrize equation (\ref{eq:lp_update}) to train symmetric couplings by redefining the mask $\epsilon_i^{\mu}$ as
\begin{equation}
    \label{eq:lp_sym_update}
    \epsilon_i^{\mu} \rightarrow \epsilon_{ij}^{\mu} = \frac{1}{2}(\epsilon^\mu_i+\epsilon^\mu_j),
\end{equation}
where the dependence on $d$ has been removed for clarity. The perceptron algorithm is \textit{supervised}, because it needs to be provided explicitly with the memories $\{\vec{\xi}^\mu\}_{\mu = 1}^p$ to update the couplings.
In the symmetric case, that will be shortened as SP for the rest of the manuscript, the function $k_{max}(\alpha)$ has been determined analytically \cite{gardner_phase_1989} for slightly diluted recurrent networks, i.e. networks with an average connectivity scaling as $\log N$. 
\begin{figure}[ht!]
\centering
\includegraphics[width=.7\textwidth]{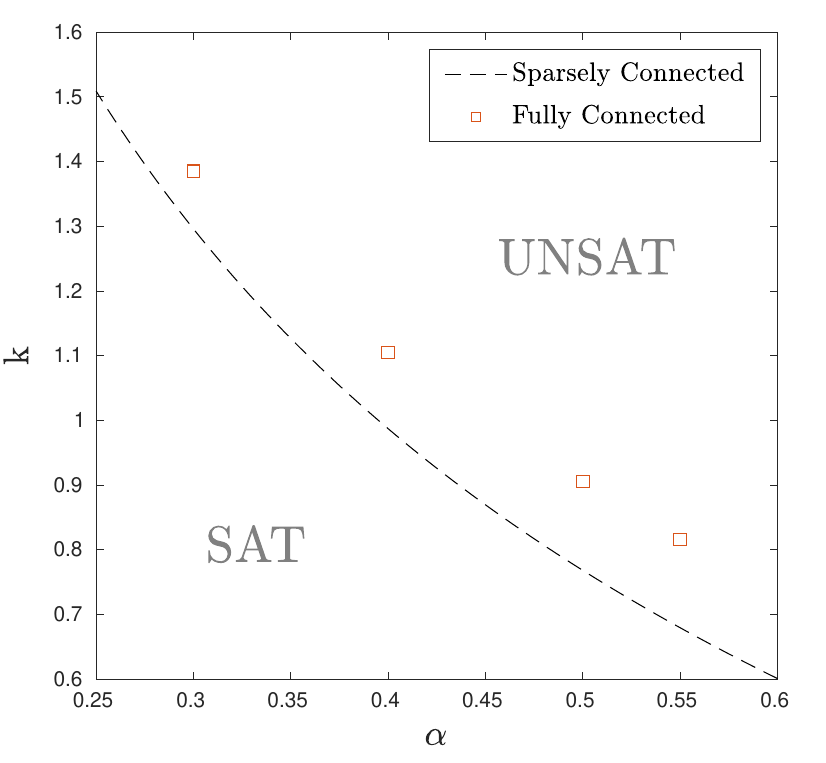}
    \caption{Phase diagram of the linear symmetric perceptron in the plane defined by the load parameter $\alpha$ and the margin $k$. The dashed line is the analytical result for $k_{max}(\alpha)$ obtained for slightly diluted networks~\cite{gardner_phase_1989}. Squares show numerical results for $k_{max}(\alpha)$ in a fully connected model at $\alpha \in \{0.3, 0.4, 0.5, 0.55\}$. Simulations have been run at different sizes of the network to measure the probability for the algorithm to converge before $10^3$ steps of the training (hence providing a lower bound to the actual value of stability). A standard finite size scaling analysis~\cite{barber1983domb,altarelli2009review} has been used to extrapolate the value of $k_{max}(\alpha)$ to the thermodynamic limit.}
\label{fig:sat_unsat}           
\end{figure}
Numerical results from the study of the algorithm defined in \cref{eq:lp_update} on networks that are both fully connected and fully symmetric suggest that, for the same degree of symmetry, $k_{max}$ at a given $\alpha$ is located slightly above the one predicted by \cite{gardner_phase_1989}. This finding, discussed in \cref{fig:sat_unsat}, suggests to reconsider previous interesting analyses \cite{theumann} and opens the road to further investigations of the critical capacity as a function of the network connectivity \cite{maurofabian}. 
It has been proved numerically that, given $\alpha$, the larger $k$, with $0 < k < k_{max}(\alpha)$, the wider the basins of attraction \cite{forrest_content-addressability_1988, benedetti_supervised_2022}. In line with previous literature \cite{battista_capacity-resolution_2020,scholkopf_learning_2018} we call a maximally stable perceptron, such that $k = k_{max}(\alpha)$, a Support Vector Machine (SVM). 
\subsection*{Checkpoint}
In this section we have seen that:
\begin{itemize}
    \item Associative memory, i.e. the ability of a system to store information and retrieve it when stimulated, can be modeled by recurrent neural networks. 
    \item The definition of memory performance used in this work is based on two properties: perfect-retrieval, i.e. the capability of the network to retrieve a memory vector with no errors; robustness, the capability to associate corrupted versions of a memory to the memory itself. 
    \item Learning algorithms can build the neural network from a specific realization of the memory in an iterative way: Hebbian learning does not reach perfect-retrieval, yet its retrieval phase extends up to $\alpha_c^H \simeq 0.14$; Hebbian Unlearning (HU) reaches perfect-retrieval up to $\alpha_c^{HU}\simeq 0.6$ and it is unsupervised; the linear perceptron gains perfect-retrieval up to $\alpha_c^P = 2$ and it is supervised. 
\end{itemize}
\newpage

\section{\label{sec:class} Classification Task}
In the classification problem we have a set of $N$ dimensional data-points $\{\vec{S}^{\mu}\}_{\mu=1}^M$ assigned to $C$ possible classes by an unknown function $\phi(\vec{S^{\mu}})$, i.e. 
\begin{equation}
    \label{eq:class}
    \vec{S}^{\mu} \xrightarrow{\phi} \xi^{\mu}\hspace{1.2cm}\text{with}\hspace{1.2cm}\xi^{\mu} \in \{\text{class 1},\text{class 2},...,\text{class C}\}, \hspace{0.3cm}\forall \mu.
\end{equation}
Generally speaking, each class is associated to a number, as data-points are also encoded in vector of numbers (i.e. the \textit{features}), allowing proper statistical calculations to be performed. 
The goal of a classifier, as a method to solve the classification problem, is to learn the class to which each data-point belongs and how to assign new unseen data to their most suitable classes. Literature usually refer to the act of learning as \textit{training} of the model: the data used for this purpose are named \textit{training-set} while the ones used to test the classification performance form the \textit{testing-set}.   
\\Let us consider the case of only $C = 2$ classes, that we translate into two possible values for $\xi^{\mu}$, i.e. $\xi^{\mu} \in \{-1,+1\}$, $\forall \mu$. 
This problem can be translated into a simple neural network problem, with nodes and couplings. Yet in this case there is not recurrency in the graph and the simplest classification method is called \textit{linear regression}. 
\begin{figure}[ht!]
\centering
\includegraphics[width=\textwidth]{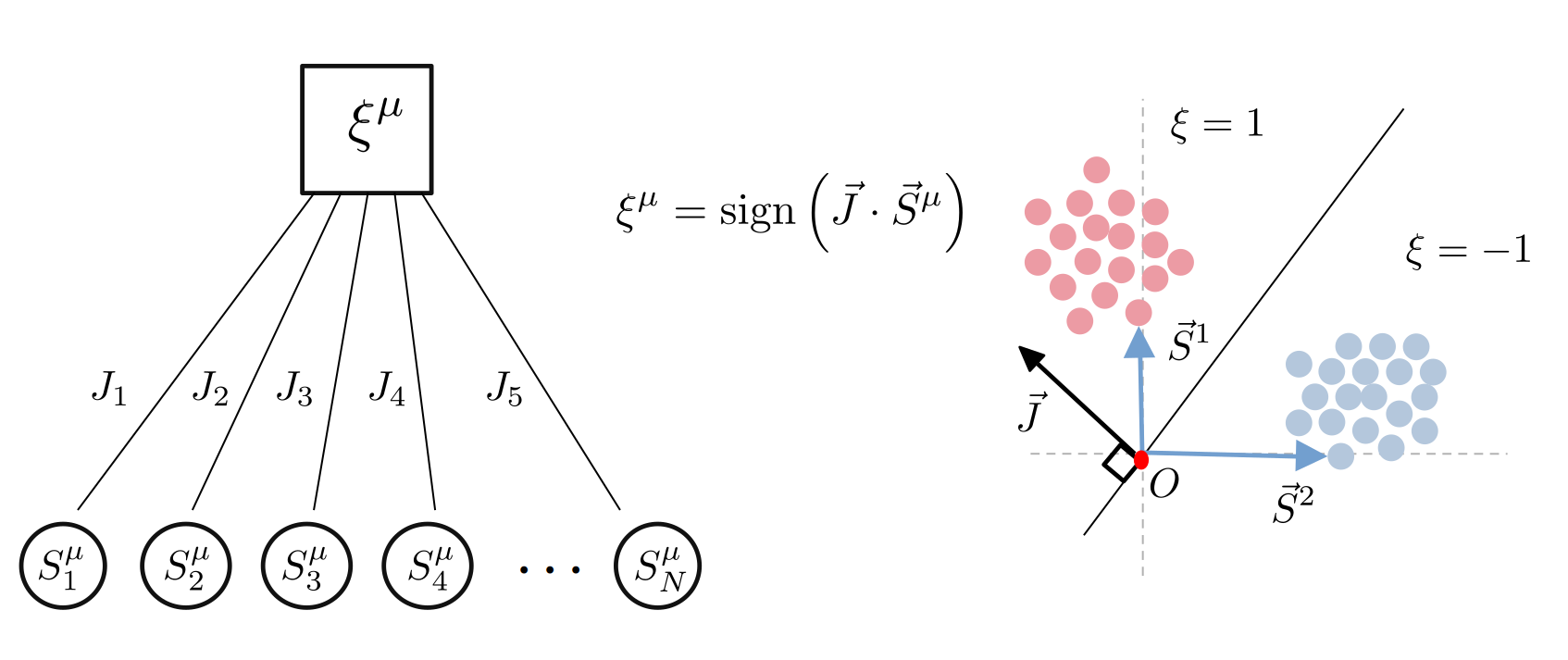}
    \caption{Graphical representation of a simple classification problem with two classes. Data-points are separated by a hyperplane in the space of the data features.}
\label{fig:lin_reg}           
\end{figure}
Fig. \ref{fig:lin_reg} provides a graphical representation of a classification problem translated into a neural network. Given the classes for each data-point, we can search for the realization of the parameters $\vec{J}$ such that 
\begin{equation}
    \label{eq:percept_class}
    \xi^{\mu} = \text{sign}\left(\vec{J}\cdot \vec{S}^{\mu} \right),\hspace{1.0cm}\forall \mu.
\end{equation}
It is now evident that the $N-1$ dimensional hyperplane orthogonal to $\vec{J}$ separates the two classes in the space of the features. 
In principle there is not one single solution to the problem (i.e. one realization of $\vec{J}$ and the hyperplane), and one can modify the classification rule to make the separation even more robust to new data to be showed to the system. In statistical learning the capability of a network to classify unseen data belonging to the testing-set is called \textit{generalization}. The absence of generalization is called \textit{overfitting}. A broader description of the concepts of overfitting and generalization will be provided further in the manuscript.\\
This model coincides with the \textit{linear perceptron} that we previously introduced in the context of associative memory. Though, as a difference with the previous case, now we have an input layer that converges into an output variable through a single vector $\vec{J}$. The practical method to learn $\vec{J}$ from the knowledge of the classes is analogous to the one introduced in \cref{eq:lp_update}, i.e.
\begin{equation}
    \small
    \label{eq:lp_class_update}
    \delta J_{i}^{(d)} = +\lambda\sum_{\mu=1}^M\epsilon^\mu_i(d) S^\mu_i\xi^\mu\hspace{0.5cm}\lambda > 0,
\end{equation}

    $$\epsilon^\mu_i(d) =\frac{1}{2}\left(1 - \text{sign}(\xi^{\mu}\vec{J}(d)\cdot \vec{S}^{\mu})\right),$$
where $\lambda$ is the learning rate, considered to be small, and $d$ is the discrete algorithm time, on which the mask $\epsilon_i^{\mu}$ depends. It can be proved that, after a suitable number of iterations, and up to a maximum amount of training data that is the double of the number of neurons, the matrix $J$ converges to one configuration that satisfies \cref{eq:percept_class} \cite{gardner_space_1988, cocco_statistical_2022}. 

\subsection{\label{sec:analogies} The formal analogies between associative memory and classification}
The strong similarity between the concepts of associative memory and classification now looks clear from the previous sections. Nevertheless, it is important to recognize the mutual differences as well as their points of contact. 
\\Associative memory is a dynamic process, because the retrieval of a memory from an example relies on the existence of a basin of attraction, which is a pure dynamic entity. 
Hence we might think to use basins of attraction as \textit{dynamic classes}, areas of influence of the memories to which new unseen neural configurations belong. In other words, we might treat each memory as a distinct dynamic class. 
If this is the case, each class would be no more encoded into one numerical variable, as it was for classification, but rather into a vector, having the same dimension of the neural configurations. 
\begin{figure*}[ht!]
\centering
\begin{subfigure}[b]{0.49\textwidth}
\centering
\includegraphics[width=\textwidth]{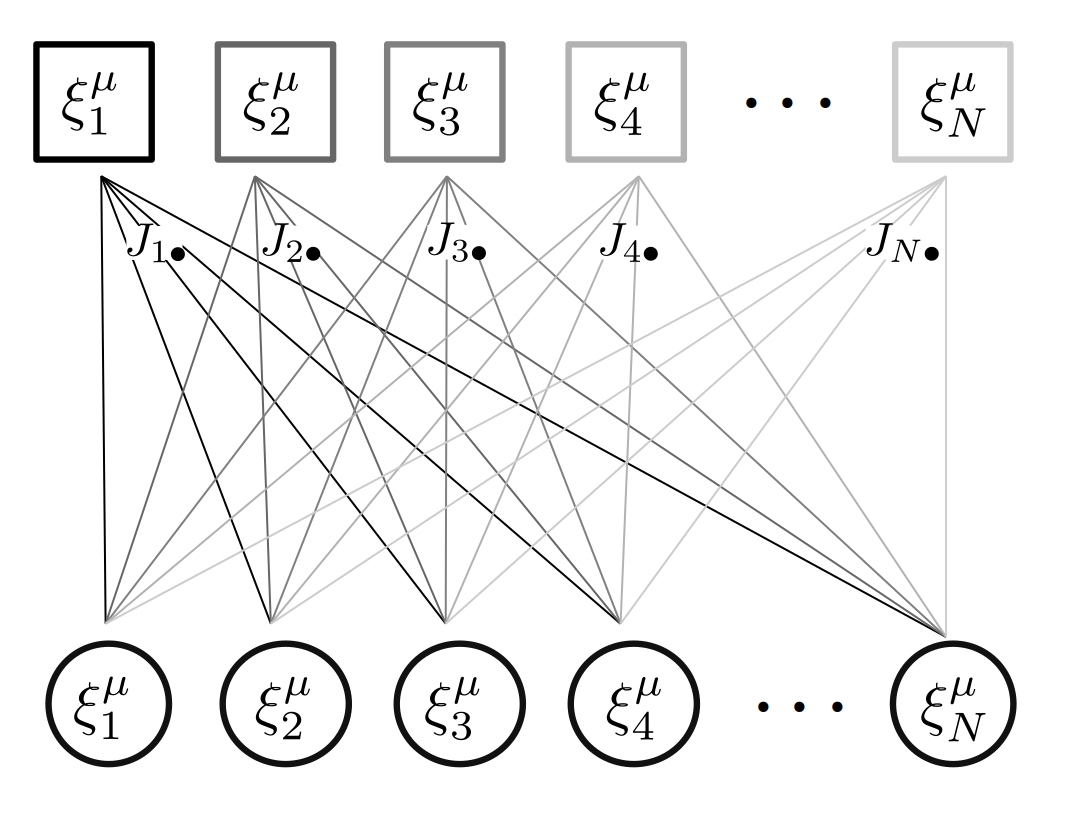}
\caption{Associative memory model}
\label{fig:memory_a} 
\end{subfigure}
\hfill
\begin{subfigure}[b]{0.43\textwidth}
\centering
\includegraphics[width=\textwidth]{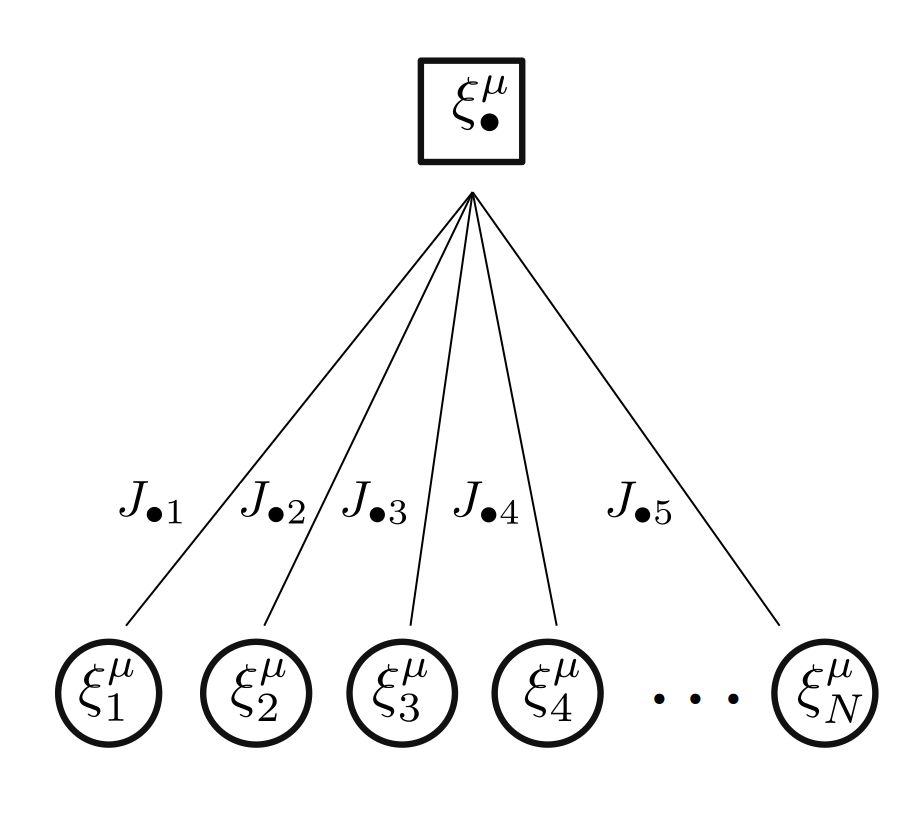}
\caption{Gardner's linear perceptron}
\label{fig:memory_b} 
\end{subfigure}
\hfill
    \caption{Graphical representations of associative memory models, in analogy with traditional classification problems.}          
\end{figure*}
Fig. \ref{fig:memory_a} graphically represents an associative memory model as a feed-forward neural network with a $N$ dimensional input layer and a $N$ dimensional output layer being fully connected by the elements of the couplings matrix $J$. The output is no more a single label but, instead, a collection of labels $\vec{\xi}^{\mu}$. This feed-forward representation can be mapped into a recurrent neural network, of the same kind as in \cref{fig:memo_graph}. 
\\Furthermore, an associative memory model is built by learning $J$ directly from the memories, i.e. the dynamic classes are learnt by means of the dynamic classes themselves. This means that the classification rule is no more the one expressed in \cref{eq:class} but it is instead given by 
\begin{equation}
    \label{eq:class_memo}
    \xi_i^{\mu} = \text{sign}\left(\vec{J_{i}}\cdot \vec{\xi}^{\mu}\right)\hspace{0.5cm}\forall i,\mu,
\end{equation}
where $\vec{J_{i}}$ is the $i$ line of the connectivity matrix of the recurrent neural network. 
This picture can be interpreted as a Cartesian product of $N$ classifiers where each class is a memory: perfect-retrieval is obtained when each memory is fitted by the intersection of the hyper-planes found by the rows of the connectivity matrix $J$, i.e. when $N$ parallel classification problems are correctly solved. In principle, if no assumptions about the structure of the couplings are made, such classification problems are independent, but this is not always the case, especially when the symmetry of the connections is constrained by the network model.  
\\\\By contrast with the associative memory picture explained above, classifiers are not dynamic machines. In fact, once the classes have been learned, new data to be classified are instantaneously associated to one or the other side of the hyperplane (or to one of the multiple half-planes if more than two classes are provided). Hence they are not classified by any dynamic process. 
\\We might propose two points of contact between the two types of problems.
Classifiers usually rely on a number of classes $C$ that is much smaller than the dimension of the data-points, i.e. $C \ll N$. 
On the other hand, associative memory models have been developed with the explicit aim of dealing with a number of memories $p = \mathcal{O}(N)$: this implies the emergence of spurious attractors and similar dynamic phenomena that, apparently, have nothing to share with classification problems. However, when $p \ll N$ memories are well separated with each other, their basins of attraction are large, and new unseen configurations are rapidly associated to one memory. 
In this manner we can relate the concepts of \textit{perfect-retrieval} in memory models and \textit{classification} as well as the ones of \textit{robustness} and \textit{generalization}. 
\\Another case where the two tasks overlap significantly is when we map a linear perceptron into a recurrent neural network, i.e. when we solve the perceptron problem à la Gardner \cite{gardner_space_1988, brunel_is_2016}. 
In this case we want to perfectly retrieve $p$ randomly generated memories. To do so, we build $N$ independent perceptrons: each of them receives one memory as an input and returns one of the entries of the same memory as an output. 
This picture is depicted, once again, in \cref{fig:memory_a}. 
However, if perceptrons are independent, and memories are generated by the same random process, we can treat the $N$ parallel problems as one single perceptron classification problem, represented in \cref{fig:memory_b}. 
It is not a case that, by construction, Gardner's perceptron algorithm \cite{gardner_space_1988, gardner_phase_1989, forrest_content-addressability_1988} forces memories to be retrieved in one step of the dynamics, when the network is initialized elsewhere in their basins of attraction (see \cref{eq:lp_update}). 
This problem is thus a clear connection between dynamic classes embodied by basins of attraction and standard, static classes used in classifiers. 
\subsection*{Checkpoint}
In this section we have seen that:
\begin{itemize}
    \item Classification can be performed by feed-forward neural networks. Classic perceptron models are examples of classifiers.
    \item Classification and Associative memory are different tasks since the former relies on one step of the network dynamics, the latter needs convergence into a fixed point. Associative memory resembles classifiers when the number of memories is small and dynamics easily converges onto attractors.  
\end{itemize}
\newpage

\section{\label{sec:data_gen} Data Generation Task}
By referring to \textit{generative modeling} we describe the capacity of specific neural networks to learn the probability distribution of a data-set and generate brand new data that exhibit maximum coherence with the same statistics \cite{jebara_machine_2012, cocco_statistical_2022}. Imagine to have a collection of $N$ variables in a vector $\vec{S} = (S_1,...,S_N)$. Data are realizations of such a vector grouped into a set $\{\vec{S}^{\mu}\}_{\mu = 1}^M$ and sampled from a joint distribution $P_{true}(\vec{S})$. Given $M$ data, we have access to the frequency of occurrence of a certain variable, i.e. the empirical distribution
\begin{equation}
    \label{eq:Pemp}
    P_{data}(\vec{S}) = \frac{1}{M}\sum_{\mu = 1}^{M}\delta\left(\vec{S}^{\mu} - \vec{S} \right). 
\end{equation}
The generative approach finds the model described by the joint distribution $P_{mod}(\vec{S}|\hat{\theta})$ where the parameters $\hat{\theta}$ are inferred from the training data such that $P_{mod}$ is the closest possible to $P_{data}$. 
However, $P_{data}$ might still differ from the ground-truth distribution $P_{true}$.
As a remedy, it is useful to reduce the number of degrees of freedom of the problem by designing the model depending on some hints that we have about $P_{true}$. For instance one might choose a \textit{graphical model}, where variables are sketched as the nodes of graph \cite{cocco_statistical_2022} with interactions to be inferred, rather than setting a prior distribution for the variables as a \textit{mixture of Gaussians} of unknown means and unit variances \cite{bishop_neural_1995}. \textit{Regularization} techniques are used for this purpose, i.e. to reduce the number of parameters, and thus allowing $P_{mod}$ to get closer to $P_{true}$.
\\ The generative approach is largely implemented across several disciplines such as computational neuroscience \cite{cocco_neuronal_2009, schneidman_weak_2006}, bio-informatics \cite{morcos_direct-coupling_2011}, animal behaviour \cite{chen_modelling_2023}, physical simulations \cite{sims_evolving_1994}, image and text synthesis \cite{brock_large_2019, kawthekar_evaluating_2017, sohl-dickstein_deep_2015}. 

\subsection{\label{sec:BML} Boltzmann Machine Learning}
We now describe a specific graphical model of generative neural networks which will be particularly relevant to the rest of our dissertation. It is inspired by the statistical mechanics at the equilibrium and it is called Boltzmann Machine (BM) \cite{hinton_boltzmann_2007, hinton_unsupervised_1999, jordan_graphical_2001, cocco_statistical_2022}. \\     
Consider a fully connected network of binary $N$ Ising variables $\vec{S} \in \{-1,+1\}^N$ with the following energy function
\begin{equation}
\label{eq:ene_bm}
    E[\vec{S}| J, \vec{h}] = -\sum_{i, j >i}^N S_i J_{ij} S_j - \sum_i^N h_i S_i,
\end{equation}
where $J_{ij}$ are symmetric couplings and $h_i$ are the fields acting on each neuron site. 
We know, from statistical mechanics, that such a system at equilibrium at a temperature $\beta^{-1}$ will obey the following joint probability density function
\begin{equation}
\label{eq:boltzpdf}
\small
    P_{mod}(\vec{S}|J, \vec{h}, \beta) = \frac{1}{Z_{\beta}} \exp{\left(-\beta E[\vec{S}| J, \vec{h}]\right)},\hspace{1.5cm}Z_{\beta} = \sum_{\vec{S}}\exp{\left(-\beta E[\vec{S}| J, \vec{h}]\right)},    
\end{equation}
which is the \textit{Gibbs-Boltzmann distribution}, where $\beta$ can be set to one without any effect in the training. Imagine a data-set $\{\vec{S}^{\mu}\}_{\mu = 1}^M$ satisfying the following empirical distribution
\begin{equation}
    \label{eq:Pemp_BML}
    P_{data}(\vec{S}) = \frac{1}{M}\sum_{\mu = 1}^M\prod_{i = 1}^N \delta_{S_i^{\mu}, S_i}.
\end{equation}
%with respect to the which one can measure the empirical mean $\langle S_i \rangle_{data}$ and covariance $\langle S_i S_j \rangle_{data}$.
Training a Boltzmann Machine means finding the parameters $J$ and $\vec{h}$ that minimize the distance between $P_{data}$ and $P_{mod}$. Thus it comes natural to impose the minimization of the Kullback-Leibler divergence between $P_{data}$ and $P_{mod}$, i.e.
\begin{equation}
\label{eq:kull}
    \mathcal{L}(J,\vec{h}) = -\sum_{\vec{S}}P_{data}(\vec{S})\log{\left( \frac{P_{mod}(\vec{S}|J, \vec{h})}{P_{data}(\vec{S})}\right)},
\end{equation}
which is equivalent to maximize the cross-entropy of $P_{mod}$ with respect to the empirical distribution $P_{data}$. Derivating \cref{eq:kull} with respect to $J_{ij}$ and $h_i$ we obtain the gradient of the Loss, i.e.
\begin{equation}
    \label{eq:grad_ij}
    \nabla_{ij}\mathcal{L} = \langle S_i S_j \rangle_{mod} - \langle S_i S_j \rangle_{data},
\end{equation}
\begin{equation}
    \label{eq:grad_i}
    \nabla_{i}\mathcal{L} = \langle S_i \rangle_{mod} - \langle S_i \rangle_{data},
\end{equation}
where $\langle \hspace{0.1cm}\cdot\hspace{0.1cm}\rangle_{data}$ and $\langle \hspace{0.1cm}\cdot\hspace{0.1cm}\rangle_{mod}$ are the averages over the respective probability distributions. Therefore, the parameters can be found by iterating the following gradient descent equations
\begin{equation}
\label{eq:Jup}
    \delta J_{ij} = -\lambda\nabla_{ij}\mathcal{L} = \lambda \left(\langle S_i S_j \rangle_{data} - \langle S_i S_j \rangle_{mod}\right),
\end{equation}
\begin{equation}
\label{eq:hup}
    \delta h_{i} = -\lambda\nabla_{i}\mathcal{L} = \lambda \left(\langle S_i \rangle_{data} - \langle S_i \rangle_{mod}\right),
\end{equation}
with $\lambda$ being a small positive learning rate. 
\\To go technical in the training algorithm, the mean and covariance over the data can be computed upstream, because they only depend on the training data-set; on the other hand, the moments of $P_{mod}$ must be sampled step-by-step during the process, because their exact calculation would involve a sum over all possible $\vec{S}$. 
Sampling can be performed by a sufficient number of Monte Carlo chains at the equilibrium at $\beta = 1$, implying an algorithm time that is long enough to ensure ergodicity for each chain. Usually the number of chains should be of the same order of magnitude of $M$, the number of training data-points, in order for the corrections to the empirical averages not to be sub-dominant with respect to the sampled ones.\\When the process converges to the fixed points of \cref{eq:Jup} and \cref{eq:hup} we obtain a condition of \textit{moment matching}, i.e. the first and the second moments of the two probability distributions coincide. In principle the training of a BM is a convex problem \cite{hinton_boltzmann_2007}, however there might be some initial conditions that push the parameters closer to their target configuration. A good choice is, for instance
\begin{equation}
    \label{eq:J0}
    J_{ij}^{(0)} = \langle S_i S_j \rangle_{data} - \langle S_i \rangle_{data} \langle S_j \rangle_{data},     
\end{equation}
and
\begin{equation}
    \label{eq:h0}
    h_i^{(0)} = \text{asinh}\left(\langle S_i \rangle_{data}\right).
\end{equation}
because \cref{eq:J0} and \cref{eq:h0} are generally close to the fixed point of equations (\ref{eq:Jup}) and (\ref{eq:hup}).\\
In order to measure the quality of the training of a BM one can measure whether the moment matching condition is met or not. The observable to measure in this case is the Pearson coefficient between the moments of the $P_{mod}$ and $P_{data}$ distributions at the end of the training. Let us collect the 2-point correlation matrices $\langle S_i S_j \rangle$ in a vector $\vec{c}$ where each entry runs over the indices $i, j > i$. Let us group the means $\langle S_i \rangle$ in a similar vector $\vec{\mu}$ where each entry runs over the index $i$. Then the Pearson coefficients will be defined as
\begin{equation}
    \label{eq:rho}
    \rho_J = \frac{\sum_{i}c^{mod}_i c^{data}_i}{\sqrt{\sum_i (c^{mod}_i)^2}\sqrt{\sum_i (c^{data}_i)^2}} \hspace{1cm}\rho_h = \frac{\sum_{i}\mu^{mod}_i \mu^{data}_i}{\sqrt{\sum_i (\mu^{mod}_i)^2}\sqrt{\sum_i (\mu^{data}_i)^2}}. 
\end{equation}
\subsection*{Checkpoint}
In this section we have seen that:
\begin{itemize}
    \item Generative models learn from data a joint probability distribution which can be used to sample new examples. Efficient models generate examples that are indistinguishable from the training data. 
    \item A good generative model does not learn the probability distribution of the visible data, but rather the ground-truth distribution that generated them. 
    \item Boltzmann-Machines are generative models that can be mapped into a recurrent neural network. The parameters of the model, i.e. couplings and fields, help a Gibbs-Boltzmann distribution to fit the statistics of the data. The training of the neural network is unsupervised. 
\end{itemize}
\newpage

\section{Summary \& Conclusions}
In this chapter we have seen that: 
\begin{itemize}
    \item An associative memory model is a recurrent neural network that stores information in form of binary configurations of the neurons, named memories. These models can be characterized in terms of two properties of the memories: perfect-retrieval and robustness. Hebbian networks are simple memory models that never reach the perfect-retrieval condition. Hebbian Unlearning (HU) is an unsupervised learning algorithm that reaches perfect-retrieval up to a critical value of the load $\alpha_c^{HU} \simeq 0.6$. Linear perceptrons, instead, reach both perfect-retrieval and robustness through a full supervised training procedure, also up to a critical capacity $\alpha_c^{P} = 2$ (when the symmetry of the couplings is not constrained a-priori). 
    \item There are some differences and formal analogies between the associative memory task, accomplished by recurrent neural networks, and the classification one, performed by feed-forward networks. The main difference is: memory models use a neural dynamics, to be iterated for several steps, to associate one given neural state to a memory, when the former state belongs to the basin of attraction of the latter; classifiers associate each data-point to a class in one single step of the neural dynamics. Associative memory tend to behave similarly to classifiers when the number of memories is subdominant with respect to the number of neurons. In this case one can relate perfect-retrieval to accomplished classification and robustness to the generalization properties of classifiers. The mapping of linear perceptrons into recurrent neural networks established by Elisabeth Gardner is a successful bridge between these two learning frameworks.
    \item Boltzmann Machines (BMs) are generative models that share the same architecture of associative memory models. As a conceptual difference, BMs learn the probability distribution of a data-set, without storing the data themselves. Once the distribution is learned, one can sample new examples that are statistically coherent to the training data. The way a BM can learn the unknown distribution of data can be improved by regularizing the Loss function of the problem.    
\end{itemize}
In the next section we will extensively analyze the associative memory performance of HU, highlighting its excellent robustness properties. We advance a theoretical explanation for the approaching of such wide basins of attraction by employing the theory behind a noise-injection based learning algorithm resembling a supervised linear perceptron. Eventually, the analytical argument is tested on different types of data-set, such as MNIST or spatially correlated Euclidean maps.    

\chapter{\label{sec:noise_inject} Unlearning as noise injection: approaching maximally stable Perceptrons}

An important challenge for modern artificial neural networks is to improve their performance by means of regularization techniques. One class of techniques relies on injecting noise in the training process, with the idea of teaching the machine how to better infer the hidden data structure from errors. Examples of these types of regularization are the drop-out \cite{labach_survey_2019, achille_information_2018, noh_regularizing_2017, drop} or data-augmentation procedures \cite{ shorten_survey_2019, zhao_maximum-entropy_2020}. The former consists in stochastically deactivating neurons across the network, the latter acts on the training data themselves, by performing transformations (e.g. translations, rotations and other symmetries) to increase the heterogeneity of the data-set. \\
Even if such procedures are practical and successful in the world of deep networks, there are only few examples of theory-based criteria for choosing how to engineer noise to inject in the training \cite{achille_information_2018, zhao_maximum-entropy_2020}. 
%where with \textit{type} of noise we refer to a particular internal structure of the features in each data-point, namely internal relations due to a specific way to generate the training-set.  
Our work considers a simple learning algorithm for attractor neural networks (see \cref{sec:ann}), employing noise-injection during training.
We are interested in this type of networks for multiple reasons: they are suitable to be modeled through the tools of statistical mechanics; 
they constitute a simplified version of modern neural networks, and yet they present a resemblance with biological neuronal systems. In the context of attractor neural networks, \textit{noise-injection} refers to a random alteration of the neuronal activity encoding the memories. 
The objective of this chapter is twofold: 
\begin{enumerate}
    \item To establish a theoretical criterion for generating the most effective noise to be utilized in training, thereby optimizing the performance of an associative memory model. Specifically, we demonstrate that generating optimal noise translates into producing training data-points whose features adhere to specific constraints, which we refer to as the \textit{structure} of the noise.
    \item To show that injecting a maximal amount of noise, subject to specific constraints, turns the learning process into an unsupervised procedure, which is faster and more biologically plausible. We will also delineate some important connections between training procedures with optimally structured noise and other consolidated learning prescriptions present in literature, such as the Hebbian Unlearning rule. Specifically we will re-interpret the Unlearning algorithm in terms of an effective noise-injection regularization for a Hebbian network.  
\end{enumerate}
The structure of the chapter will be the following. Gardner's original \textit{training-with-noise} algorithm is presented in section \ref{sec:twn} and the consistency of the numerical results with the theory developed by \cite{wong_optimally_1990, wong_neural_1993} is showed; 
we will deal in particular with the perfect-retrieval and robustness capabilities of the algorithm on fully connected networks. \\We then propose a derivation of the constraints that should be satisfied by maximally noisy training data for the same algorithm in order to mimic SVM learning. This will be treated in section \ref{sec:choice_of_training_conf} where we also show how relevant the initial conditions over the couplings are to sample the good training configurations. 
Specifically, we conclude that low saddles in an initially Hebbian-shaped energy landscape are the most effective data for the training-with-noise algorithm. 
\\Furthermore, section \ref{sec:U} focuses on the use of stable fixed points of the neural dynamics as training data, proving that the celebrated Unlearning rule is contained into the training-with-noise procedure when noise is maximal. Both the perfect-retrieval and robustness properties of the trained network are examined: consistently with the theoretical results, the system correctly reproduces a SVM for $\alpha$ up to $\alpha_c^{HU}\simeq 0.6$. Eventually, the original Unlearning routine is generalized by implementing a training-with-noise procedure with a moderate amount of noise, showing an improvement in the critical capacity. \\Section \ref{sec:corr} deals with the case of correlated memories, such as the pictures from a MNIST data-set and the paramagnetic configurations of a 2-dimensional Ising model. \\Section \ref{sec:cann} follows up with the evaluation of another type of data, i.e. spatial maps in a D-dimensional Euclidean space. The performance of the Unlearning algorithm is yet again compared to the one of a SVM in terms of robustness and diffusive dynamics in the real Euclidean space.\\To end up, a sampling procedure for optimal noisy training data in the case of maximal noise is proposed and implemented in section \ref{sec:samp}, proving to outperform both the training-with-noise and the HU learning procedures.

%\textit{The scientific results presented in this Chapter result from a collaboration established with Marco Benedetti and Enzo Marinari at University La Sapienza that resulted in the following papers \cite{benedetti_supervised_2022}, as well as the joint work with Simona Cocco and Rémi Monasson at LPENS which is still unpublished.} 

\newpage
\section{Training with noise \label{sec:twn}}
The concept of learning from noisy examples, introduced for the first time in \cite{le_cun_learning_1986}, is at the basis of a study performed by Gardner and co-workers \cite{gardner_training_1989}, a pioneering attempt to increase and control robustness through the introduction of noise during the training phase of recurrent neural networks.  Here, we report the algorithm and characterize, for the first time, its performance over fully connected neural networks. All the observables used for this purpose have been defined in \cref{sec:ann}.
\\ 
The training-with-noise (TWN) algorithm \cite{gardner_training_1989} consists in starting from any initial coupling matrix $J_{ij}^{(0)}$ with null entries on the diagonal, and updating recursively the couplings according to 
\begin{equation}
\label{eq:lwn}
    \delta J_{ij}^{(d)} = \frac{\lambda}{N}\epsilon_i^{\mu_d} \xi_i^{\mu_d} S_j^{\mu_d}, \hspace{1cm} \delta J_{ii}^{(d)} = 0\hspace{0.3cm}\forall i,
\end{equation}
where $\lambda$ is a small learning rate, $\mu_d \in [1,...,p]$ is a randomly chosen memory index and the mask $\epsilon_i^{\mu_d}$ is defined as
\begin{equation}
    \label{eq:eps}
    \epsilon_i^{\mu_d} = \frac{1}{2}\Big(1 - \text{sign}\Big(\xi_i^{\mu_d}\sum_{k=1}^N J_{ik}S_k^{\mu_d}\Big)\Big).
\end{equation}
In this setting, $\vec{S}^{\mu_d}$ is a noisy memory, generated according to a Bernoulli process 
\begin{equation}
    \label{eq:p_s}
    \small
    P(S_i^{\mu_d} = x) = \frac{(1+m_t)}{2}\delta(x - \xi_i^{\mu_d}) + \frac{(1-m_t)}{2}\delta(x + \xi_i^{\mu_d}).
\end{equation}
 The $\textit{training overlap}$ $m_t$ is a control parameter for the level of \textit{noise} injected during training, corresponding to the expected overlap between $\vec{S}^{\mu_d}$ and $\vec{\xi}^{\mu_d}$, i.e.
\begin{equation}
    \label{eq:mt}
    m_t = \frac{1}{N}\sum_{j=1}^N\xi_j^{\mu_d} S_j^{\mu_d} + O\Big(\frac{1}{\sqrt{N}}\Big).
\end{equation}
Each noisy configuration can be expressed in terms of a vector of \textit{noise units} $\vec{\chi}$, such that
\begin{equation}
    \label{eq:chi2}
    S_i^{\mu_d} = \chi_i^{\mu_d}\xi_i^{\mu_d}.
\end{equation}
In this setting, noise units are i.i.d variables, distributed according to
\begin{equation}
    P(\chi_i^{\mu_d} = x) = \frac{(1+m_t)}{2}\delta(x - 1) + \frac{(1-m_t)}{2}\delta(x + 1).
\end{equation}
The algorithm would converge when every configuration with overlap $m_t$ with a memory generates on each site a local field aligned with the memory itself. Let us define the function
\begin{equation}
    \label{eq:loss_ws}
    \mathcal{L}(m, J) = -\frac{1}{\alpha N^2}\sum_{i,\mu}^{N,p} \text{erf} \left(\frac{m\Delta_i^{\mu}}{\sqrt{2(1-m^2)}}\right).
\end{equation}
Calculations contained in \cref{app:m1} prove that $-\mathcal{L}(m = m_0,J)$ is equal to the one-step retrieval map for one realization of $J$ and averaged over all the configurations having an overlap $m_0$ with the memories, that we refer to as $m_1(m_0,J)$.
%Assuming that stabilities are self-averaging quantities, it can be proven that $-\mathcal{L}(m = m_0,J)$ tends to $m_1(m_0)$ when $N \rightarrow \infty$ (see \cref{app:m1}), where the definition of $m_1(m_0)$ is given by \cref{eq:onestep_ret_map}. 
Wong and Sherrington \cite{wong_optimally_1990, wong_neural_1993} propose an elegant analysis of a  network designed to optimize $\mathcal{L}(m, J)$, i.e. whose couplings $J_{WS}(m)$ correspond to the global minimum of $\mathcal{L}(m, J)$. Some of their findings, relevant to this work, are:
\begin{enumerate}
    \item For any $m_0$, the maximum value of $m_1(m_0, J_{WS}(m))$ with respect to $m$ is obtained when $m = m_0$. This result is important because it suggests that a network with $J = J_{WS}(m)$ tends to increase the influence of the memory over the surrounding configurations, possibly increasing the basins of attraction to which they belong. 
    \item When $m \rightarrow 1^-$, the minimization of the $\mathcal{L}(m,J)$ trains a linear perceptron with maximal stability, i.e. a SVM. This result translates into having $J_{WS}(m\rightarrow 1^{-}) = J_{SVM}$. This result is not trivial, since $m = 1$ would reproduce a linear perceptron with zero margin.
    \item When $m \rightarrow 0^+$, the  minimization of $\mathcal{L}(m,J)$ leads to a Hebbian connectivity matrix. This result translates into having $J_{WS}(m\rightarrow 0^+) \propto J^H$. On the other hand, the case $m = 0$ would be trivial, since no learning would be possible. 
\end{enumerate}
Now, we want to examine the TWN procedure defined
above in light of the results obtained by Wong and Sherrington. When the network is trained through TWN, the resulting coupling matrix depends on $m_t$, i.e. $J = J(m_t)$. It is crucial to stress the difference between the variables $m$ and $m_t$: the former is a parameter of eq.~(\ref{eq:loss_ws}), the latter is the level of noise used by the training algorithm (\ref{eq:lwn}). Eq. \ref{eq:loss_ws} is relevant to the TWN procedure, since \cref{eq:lwn} leads to a reduction of $\mathcal{L}(m, J(m_t))$, for any value of $m$ and $m_t$. In fact, considering a small variation of the stabilities induced by the algorithm update
\begin{equation*}
     \Delta_i^{\mu} \rightarrow \Delta_i^{\mu} + \delta \Delta_i^{\mu},
\end{equation*}
and performing a Taylor expansion of (\ref{eq:loss_ws}) at first order in $\mathcal{O}(N^{-1/2})$, one obtains (see Appendix \ref{sec:appA1})
\begin{equation}
    \label{eq:new_loss3}
    \mathcal{L}^{'} = \mathcal{L} + \sum_{i=1}^{N}
    \delta\mathcal{L}_{i}
\end{equation}
where 
\begin{equation}
\label{eq:delta_loss}
\small
    \delta\mathcal{L}_{i} = -\frac{\epsilon_i^{\mu_d}\lambda}{\alpha\sigma_i N^{5/2}}\frac{\sqrt{2}m \cdot m_t}{\sqrt{\pi(1-m^2)}}\exp{\left(-\frac{m^2 \Delta_i^{\mu_d^2}}{2(1-m^2)}\right)}.
\end{equation}
Hence, $\delta\mathcal{L}_{i}$ is strictly non-positive when $\frac{\lambda}{N}$ is small, so that the Taylor expansion is justified. 
\begin{figure}[ht!]
\centering
\includegraphics[width=0.7\linewidth]{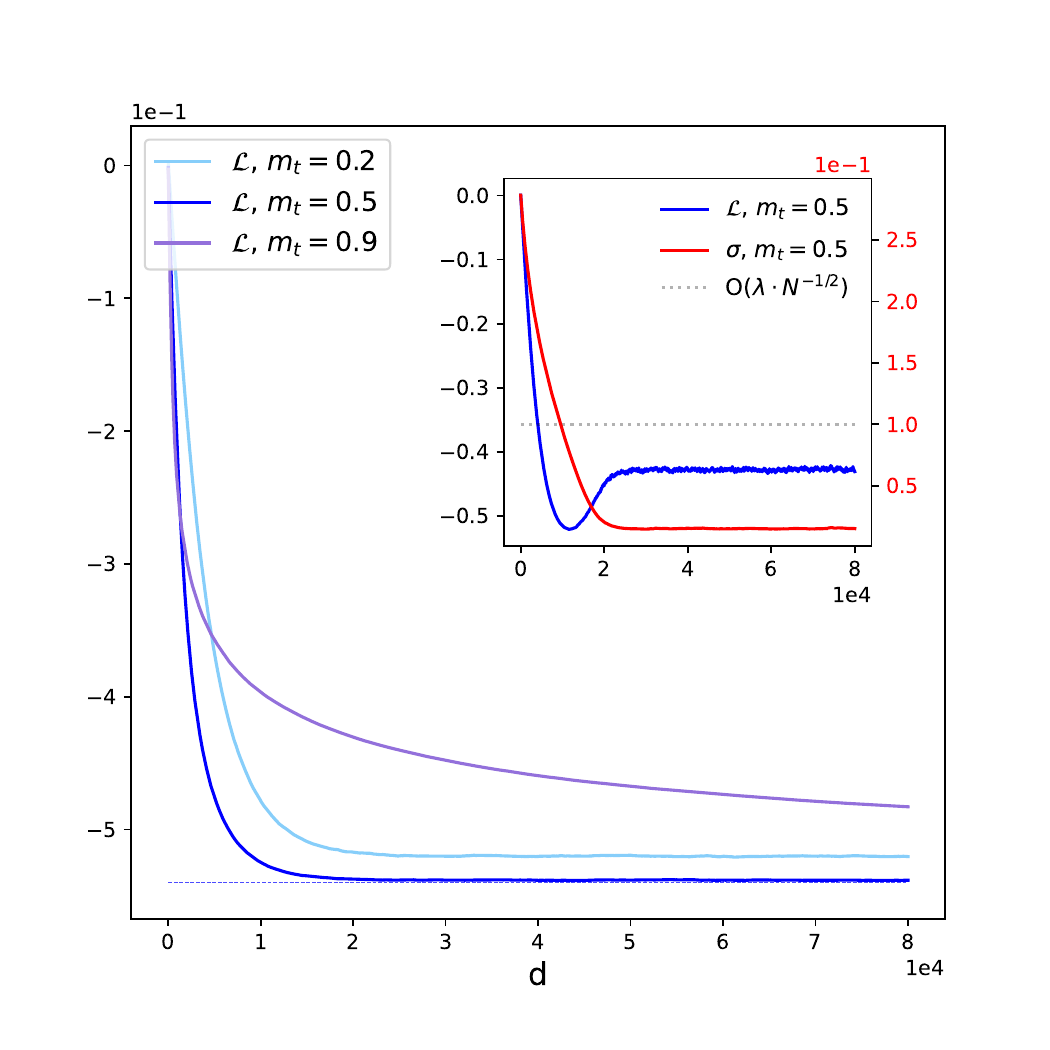}
\caption{The lines in the main plot report the function $\mathcal{L}(m = 0.5, J(m_t))$ for different training overlaps as functions of the number of algorithm steps $d$. The \textit{dotted} line represents the theoretical minimum value from \cite{wong_neural_1993}. The learning strength $\lambda$ has been rescaled by the standard deviation of the couplings as described in the text. The subplot reports the case $m_t = 0.5$ when the learning strength is not rescaled: $\mathcal{L}$ is in \textit{blue}, while a measure of the standard deviation of the couplings, defined as $\sigma = \frac{1}{N}\sum\sigma_i$, is reported in \textit{red}.  The value $\lambda\cdot N^{-1/2}$ of the standard deviation is also depicted in \textit{light gray} to properly signal the moment when equation (\ref{eq:delta_loss}) loses its validity. All measures are averaged over $5$ realizations of the couplings $J$. Choice of the parameters: $N = 100$, $\alpha = 0.3$, $\lambda = 1$, the initial couplings are Gaussian with unitary mean, zero variance and $J_{ii}^{(0)} = 0$ $\forall i$.}
\label{fig:loss}
\end{figure}
Moreover, we numerically find that iterating (\ref{eq:lwn}) with a given value of $m_t$ drives $\mathcal{L}(m, J(m_t))$ to its theoretical absolute minimum computed in \cite{wong_neural_1993}, as reported in \cref{fig:loss} for one choice of $N, \alpha$ and $m$. This means that the performance of the TWN algorithm can be completely described in the analytical framework of \cite{wong_neural_1993} and, that $\mathcal{L}(m, J(m_t))$ can be considered as the Loss function optimized by the TWN algorithm. 
As a technical comment, note that standard deviation along one row of the couplings matrix $\sigma_i$ (see \cref{eq:stab}) is a variable quantity over time, and numerics suggest that it is slowly decreasing. As a result, the expansion performed to determine the variation of $\mathcal{L}$ (see \cref{eq:new_loss2} in Appendix \ref{sec:appA1}) might not be justified after a certain number of steps, leading to a non-monotonic trend of the Loss function. The non-monotonic trend of $\mathcal{L}(m, J(m_t))$ due to this effect is showed in the inset of \cref{fig:loss}. However, this inconvenience can be overcome by rescaling the learning rate $\lambda$ into $\lambda_i = \lambda\cdot \sigma_i$ at each iteration, as also the curves in \cref{fig:loss} display.
%%%%%%%%%%%%%%
\subsection{\label{sec:3a}Perfect-retrieval}
The computations contained in \cite{
wong_neural_1993}, and resumed in \cref{app:rep_twn}, are now used to calculate $n_{SAT}$ as a function of $m_t$ and $\alpha$ in the TWN problem.
\begin{figure}[ht!]
\centering
\includegraphics[width=0.8\linewidth]{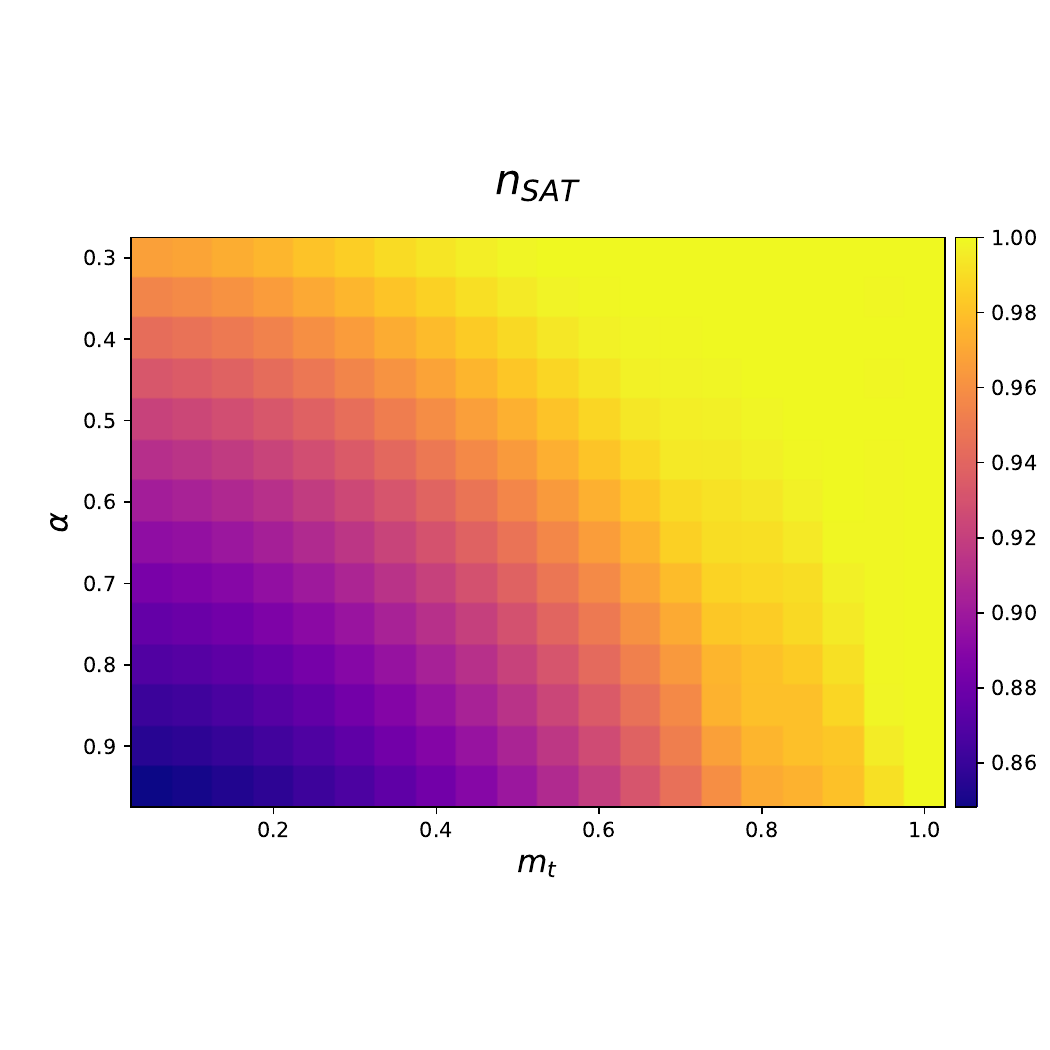}
\caption{$n_{SAT}$ as a function of $m_t$ and $\alpha$. Warmer shades of colour are associated to higher retrieval performances.}
\label{fig:nsat}
\end{figure}
The probability distribution function of the stabilities in the trained network (see equation (\ref{eq:rho1})) has always a tail in the negative values, implying that perfect-retrieval is never reached. The only exception to this statement is the trivial case of $m_t = 1^-$, where $n_{SAT} = 1$ for $\alpha \leq 2$. Nevertheless, the values of $n_{SAT}$ remain close to unity for relatively high values of $m_t$ and relatively low values of $\alpha$ (see \cref{fig:nsat}).

%%%%%%%%%%%%%%%%%%%%%%%%%%%%%%%%%%%%%%%%%%%%%%%%%%%%%%%%%%%%%%%%%%
\subsection{\label{sec:3b}Robustness}
\begin{figure}[ht!]
\centering
\includegraphics[width=0.75\linewidth]{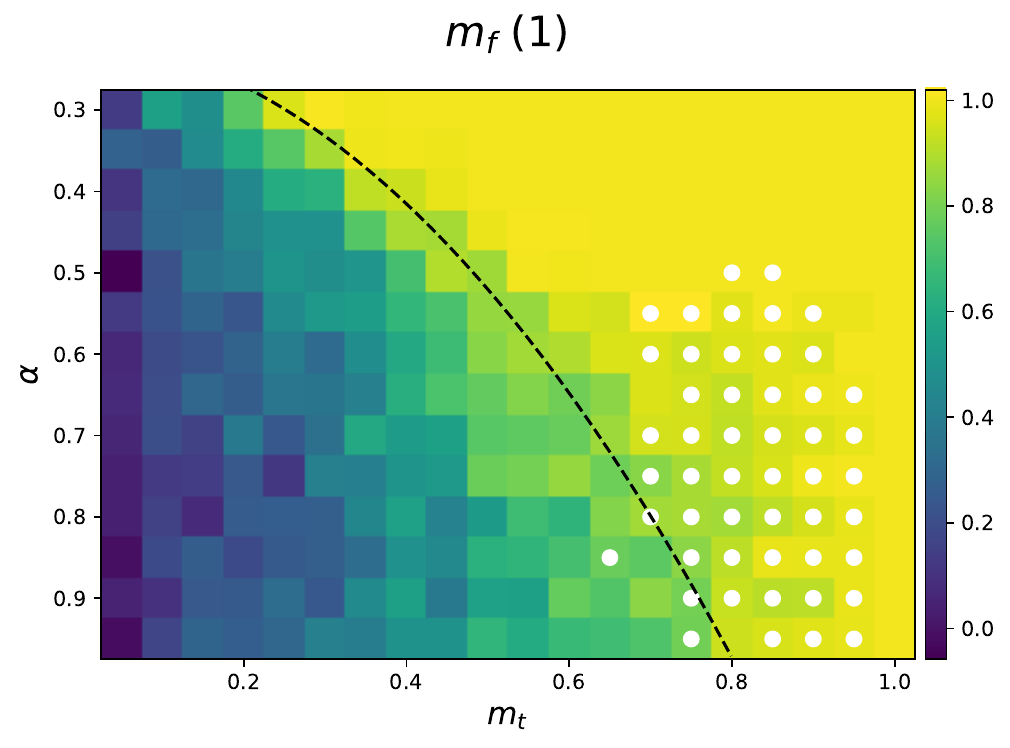}
\caption{$m_f(1)$ as a function of $m_t$ and $\alpha$. Warmer shades of color are associated to higher retrieval performances. The \textit{black dashed} line represents the boundary of the retrieval regime according to the criterion in \cref{app:gen_twn}, \textit{white} dots signal the points where basins of attraction to which memories belong are larger than those obtained from a SVM at $N = 200$.} 
\label{fig:mf1}
\end{figure}
The robustness properties of a network trained through TWN are now discussed. The color map in \cref{fig:mf1} reports the estimate of the retrieval map $m_f(m_0)$ at $m_0 = 1$ in the limit $N\rightarrow \infty$, i.e. a measure of the distance between a given memory and the closest attractor. Notice the emergent separation between two regions: one where $m_f(1)$ is mostly smaller than $0.5$, and memories are far from being at the bottom of the basin; another region where $m_f(1)$ is mostly close to unity, i.e. the memory is very close to the center of the basin. Such separation reminds the typical division between \textit{retrieval} and \textit{non retrieval} phases in fully connected neural networks \cite{amit_storing_1985}, differently from sparse neural networks \cite{wong_neural_1993,derrida_exactly_1987} where the possible topologies of the basins result more various yet harder to get measured by experiments. In \cref{app:gen_twn} we propose an empirical criterion to separate these two regions and so limit ourselves to the retrieval one. 
Consider the retrieval map $m_f(m_0)$ measured with respect to the attractor of the basin to which the memory belongs and not to the memory itself. Then one must have $m_f(1) = 1$.
Our criterion is based on assuming that $m_f(m_0)$ always develops a plateau starting in $m_0 = 1$ and ending in some $m_c < 1$ when $N \rightarrow \infty$. The behavior of the basin radius can be observed numerically as a function of $m_t$: when the plateau disappears (i.e. $m_c = 1$) then one can suppose that basins get shattered in the configurations space due to the interference with the other attractors. Given $\alpha$, this occurs at some value of $m_t$. The empirical transition line is reported in a dashed style in \cref{fig:mf1}.
Limiting ourselves to the retrieval region we employ a procedure also described in \cref{app:gen_twn} to compute the typical size of the basins of attraction. White dots in \cref{fig:mf1} signal the combinations of $(m_t, \alpha)$ where the basins of attraction found by TWN algorithm resulted larger than the ones obtained by a SVM at the same value of $\alpha$. We want to stress the importance of a comparison between the TWN and the corresponding SVM, since numerical investigations have shown the latter to achieve extremely large basins of attraction, presumably due to the maximization of the stabilities \cite{forrest_content-addressability_1988, benedetti_supervised_2022}. One can conclude that for most of the retrieval region the robustness performance is worse than the SVM, which maintains larger basins of attraction; on the other hand, at higher values of $\alpha$ the trained-with-noise network sacrifices its perfect-retrieval property to achieve a basin that appears wider than the SVM one. In conclusion, the TWN algorithm never outperforms the relative SVM without reducing its retrieval capabilities.   
\subsection*{Checkpoint}
In this section we have seen that:
\begin{itemize}
    \item The TWN algorithm can be interpreted as a linear perceptron that learns to align noisy versions of one memory to the memory itself, enhancing associativity. The control parameter of the algorithm is the training overlap $m_t$. 
    \item The TWN algorithm trains a network that solves the optimization problem studied by Wong and Sherrington. Hence, by tuning the training overlap one can interpolate between the Hebbian model (i.e. $m_t = 0^+$) and a SVM (i.e. $m_t = 1^-$). 
    \item The TWN never reaches perfect retrieval when $m_t < 1^-$. In all this region of the network configurations the TWN algorithm never outperforms the robustness performance of a SVM without moving each memory away from the closest attractor.  
\end{itemize}
\newpage

\section{\label{sec:choice_of_training_conf}Optimal noisy data-points}
As previously stated, SVMs are considered to be highly efficient associative memory models, due to their very good classification and generalization capabilities. Wong and Sherrington's analytical argument proves that minimization of $\mathcal{L}(m=1^-, J(m_t))$ trains a SVM. 
The TWN procedure proposed by Gardner and collaborators, relying on a Bernoulli process to generate noise, accomplishes this task only when $m_t = 1^-$ (see section \ref{sec:twn}). Injecting a larger amount of noise during training would deteriorate the performance: specifically, $m_t = 0^+$ trains a Hebbian neural network. Nevertheless, training a network with examples that are nearly uncorrelated with the memories can significantly speed up the sampling process, since such states can be generated in an unsupervised fashion, without knowledge of the memories (as seen for HU in section \ref{sec:UNL}). Such unsupervised processes are also considered more biologically plausible. \\ 
In this section, we show that it is possible to use maximally noisy configurations (i.e. $m_t = 0^+$) to train a network approaching the performance of a SVM, by means of the TWN algorithm. For this purpose, one must change the way noisy data are generated: they need to meet specific constraints which lead to internal dependencies among the features (i.e. a \textit{structure}). We derive a theoretical condition characterizing the optimal structure of noise, and show that specific configurations in the Hebbian energy landscape, including local minima, match well the theoretical requirements. \\
\\It will be helpful for our purposes to implement a symmetric version of rule (\ref{eq:lwn}), i.e.
\begin{equation}
    \label{eq:lwn2}
    \delta J_{ij}^{(d)} = \frac{\lambda}{N}\left(\epsilon_i^{\mu_d} \xi_i^{\mu_d} S_j^{\mu_d} + \epsilon_j^{\mu_d} \xi_j^{\mu_d} S_i^{\mu_d}\right).
\end{equation}
Equation (\ref{eq:lwn2}) can be rewritten explicitly making use of (\ref{eq:eps}), leading to
\begin{equation}
    \begin{aligned}
    \label{eq:lwsn}
    \delta J_{ij}^{(d)} = &\frac{\lambda}{2N}\big(\xi_i^{\mu_d}S_j^{\mu_d} + S_i^{\mu_d}\xi_j^{\mu_d}\big) + \\ -&\frac{\lambda}{2N}\big(S_i^{1,\mu_d}S_j^{\mu_d} + S_i^{\mu_d}S_j^{1,\mu_d}\big)
\end{aligned}
\end{equation}
where $S_i^{1,\mu_d} = \text{sign}\left(\sum_{k=1}^N J_{ik} S_k^{\mu_d}\right)$.
The total update to the coupling at time $D$ can be decomposed as a sum of two contributions
\begin{equation}
\label{eq:decomp}
    \Delta J_{ij}(D) = \Delta J_{ij}^N(D) + \Delta J_{ij}^U(D).
\end{equation}
The first term on right-hand side, which will be referred to as \textit{noise} contribution, is expressed in terms of noise units as
\begin{equation}
    \label{eq:Npart}
    \Delta J_{ij}^N(D) = \frac{\lambda}{2N}\sum_{d = 1}^D \xi_i^{\mu_d}\xi_j^{\mu_d}\chi_j^{\mu_d} + \frac{\lambda}{2N}\sum_{d = 1}^D\xi_j^{\mu_d}\xi_i^{\mu_d}\chi_i^{\mu_d},
\end{equation}
while the second term, which will be referred to as \textit{unlearning} contribution, is given by
\begin{equation}
    \label{eq:Upart}
    \Delta J_{ij}^U(D) = -\frac{\lambda}{2N}\sum_{d = 1}^D \left(S_i^{1,\mu_d}S_j^{\mu_d} + S_i^{\mu_d}S_j^{1,\mu_d}\right).
\end{equation} 
In the maximal noise case $m_t=0^+$, $\Delta J_{ij}^N(D)$ averages to zero over the process because its variance is $\lambda/N$ when the number of steps $D$ is proportional to $N/\lambda$, leading to 
\begin{equation}
    \Delta J_{ij}^N(D) = 0^+ + \mathcal{O}\left(\sqrt{\frac{\lambda}{N}}\right).
 \end{equation}

\subsection{Characterizing the good training configurations}%Trademark of virtuous training configurations}
\label{sec:trademark_virtuous} 
The variation of $\mathcal{L}(m, J)$ can be expressed (see Appendix \ref{sec:appA2}) as $$\delta \mathcal{L} = \delta\mathcal{L}_{N} + \delta \mathcal{L}_{U}.$$ 
As detailed in Appendix \ref{sec:appA2}, in the case of maximal noise (i.e. $m_t = 0^+$) $\delta\mathcal{L}_{N}$ is negligible, and the only relevant contribution is
\begin{equation}
    \label{eq:deltaL_lwsn}
    \small
    \delta\mathcal{L}_{U} \propto \frac{m}{\sqrt{2\pi(1-m^2)}}\sum_{i,\mu}^{N,p}\omega_i^{\mu}\exp{\left(-\frac{m^2\Delta_i^{\mu^2}}{2(1-m^2)}\right)}, 
\end{equation}
where $\omega_i^{\mu}$ play the role of weights to the positive Gaussian terms and they are given by
\begin{equation}
    \label{eq:omegatilde}
    \omega_i^{\mu} =\frac{1}{2\sigma_i}\left( m_{\mu}\chi_i^{1,\mu} + m_{1,\mu}\chi_i^{\mu}\right),
\end{equation}
with 
\begin{equation}
    \chi_i^{\mu} = \xi_i^{\mu}S_i^{\mu_d}\hspace{1cm}\chi_i^{1,\mu} = \xi_i^{\mu}S_i^{1,\mu_d},
\end{equation}
and
\begin{equation}
    m_{\mu} = \frac{1}{N}\sum_{j = 1}^N S_j^{\mu_d}\xi_j^{\mu}\hspace{1cm}m_{1,\mu} = \frac{1}{N}\sum_{j = 1}^N S_j^{1,\mu_d}\xi_j^{\mu}.
\end{equation}
For the case of fixed points, we have $\omega_i^{\mu} = m_{\mu}\chi_i^{\mu}$. We know that minimization of $\mathcal{L}(m, J(m_t))$ trains a SVM when $m \rightarrow 1^-$. When $m\rightarrow 1^-$, the Gaussian terms contained in the sum become very peaked around $0$. Since we want $\delta \mathcal{L}$ to be negative, we need, for most of the pairs $i,\mu$,
\begin{equation}
    \label{eq:deltaL_lwsn3}
    \omega_i^{\mu} < 0 \hspace{0.2cm}\text{when}\hspace{0.2cm}|\Delta_i^\mu|< 0^+.
    %\delta\mathcal{L}_{U} \propto \rho(0)\omega(0).
\end{equation}
 \\The more negative $\omega_i^{\mu}$ is when $\Delta_i^\mu \sim 0$, the more powerful is its contribution to approach the SVM performances. 
 Selecting training data that satisfy equation (\ref{eq:deltaL_lwsn3}) amounts to imposing specific internal dependencies among the noise units $\vec{\chi}$, which are no more i.i.d. random variables, as it was in \cite{gardner_training_1989}. We refer to such dependencies as \textit{structure} of the 
 noise.
 One should also bear in mind that training is a dynamic process: to reduce $\mathcal{L}(m=1^-, J(m_t = 0^+))$  condition (\ref{eq:deltaL_lwsn3}) should hold during training.
\subsection{Position of good training configurations in the energy landscape}
%The weights $\omega_i^{\mu}$ have a precise geometric interpretation in the case of HU. Define $N\text{-dimensional}$ vectors     $\vec{\eta}_i^\mu=\xi_i^\mu\vec{\xi}^\mu$ and $\vec{\eta_i}^{\mu_d} = S_i^{\mu_d}\vec{S}^{\mu_d}$, where $\vec{S}^{\mu_d}$ is the training configuration. Then, one has $\omega_i^{\mu} = (\vec{\eta_i}^{\mu_d} \cdot \vec{\eta_i}^{\mu})/N$. On the other hand, stabilities and the HU learning rule can be written as
%\begin{align}
%    &\Delta_i^\mu=\frac{\vec{J_i} \cdot \vec{\eta}_i^\mu }{\sqrt{N}\sigma_i},\\
%    &\vec{J_i}^{(d+1)} = \vec{J_i}^{(d)} - \frac{\lambda}{N}\vec{\eta_i}^{\mu_d}\ .
%\end{align}
%It follows that, a negative value of $\omega_i^{\mu}$ increases the degree of alignment between $\vec{J_i} \text{ and } \vec{\eta}_i^\mu$, improving the stability (see also \cite{benedetti_supervised_2022}).
The HU algorithm is based on choosing training configurations on a dynamical basis, namely as fixed points of zero temperature dynamics in the energy landscape dictated by Hebb's learning rule.
On the other hand, traditional TWN relies on fully random states, i.e. very high states in the energy landscape. %In this section, we analyze the effectiveness of this choice, and of its finite temperature robustness, on the basis of the concepts introduced above. 
In this section, we generalize this dynamical approach by evaluating the performance of training configurations which lie at different altitudes in the energy landscape of a symmetric neural network, i.e. states that are not limited to be stable fixed points or random states.  
To do so, we sample training configurations 
%from a network initialized with the Hebbian rule 
by means of a Monte Carlo routine at temperature $T$. 
%relying on the numerical observation that associates configurations probed at a particular temperature with a stable fraction of unstable directions under the dynamics \cite{aspelmeier_free_2006}. %This is also supported by previous studies on spin glasses \cite{aspelmeier_free_2006}. 
Temperature acts as a control parameter: when $T=0$ training configurations are stable fixed points of \cref{eq:dynamics}, as in standard HU. Higher values of $T$ progressively reduce the structure of noise in training configurations, and in the limit $T\to\infty$, training configurations are the same as in the TWN algorithm. The Monte Carlo of our choice is of the Kawasaki kind \cite{newman_monte_1999}, to ensure that all training configurations are at the prescribed overlap $m_t = 0^+$. We are going to use this technique to probe the states across two types of landscapes: the one resulting from a Hebbian initialization and the one resulting from a SK model \cite{sherrington_solvable_1975}. \\

\begin{figure*}
     \centering
     \begin{subfigure}[b]{0.245\textwidth}
         \centering
         \includegraphics[width=\textwidth]{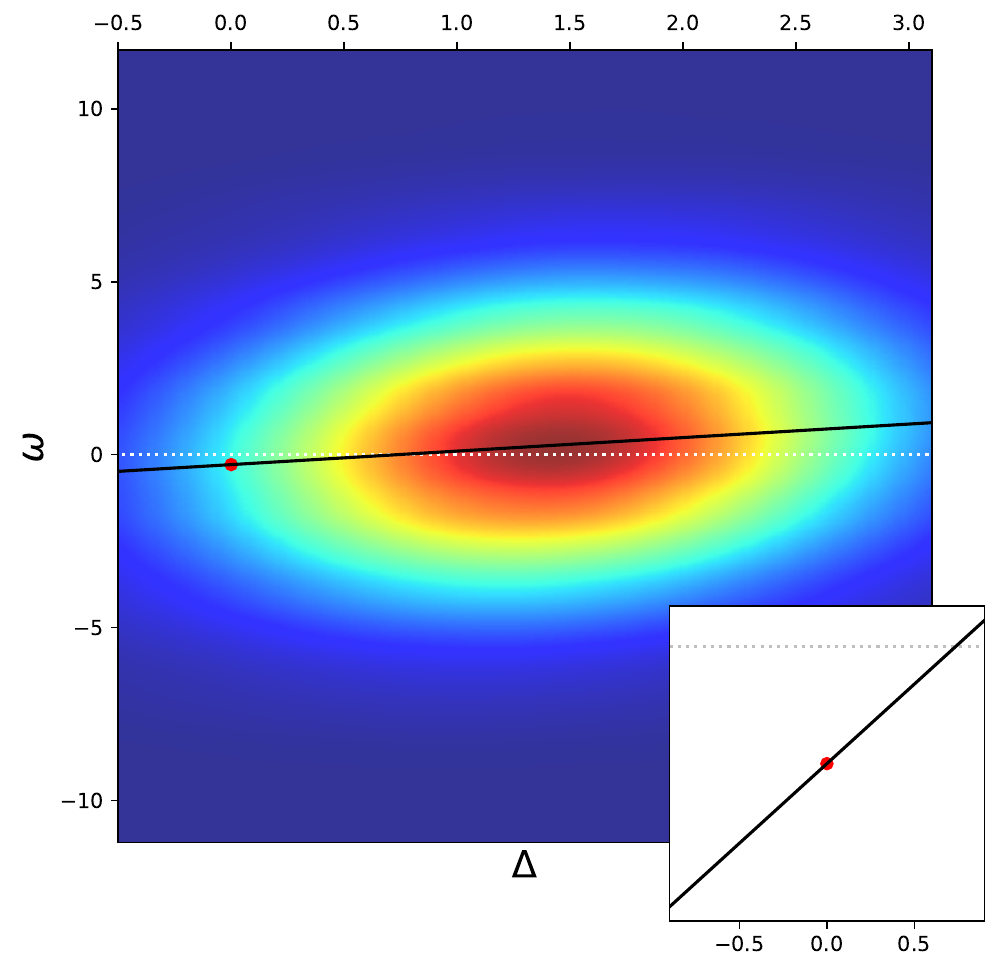}
         \caption{$T = 0$}
     \end{subfigure}
     \hfill
     \begin{subfigure}[b]{0.23\textwidth}
         \centering
         \includegraphics[width=\textwidth]{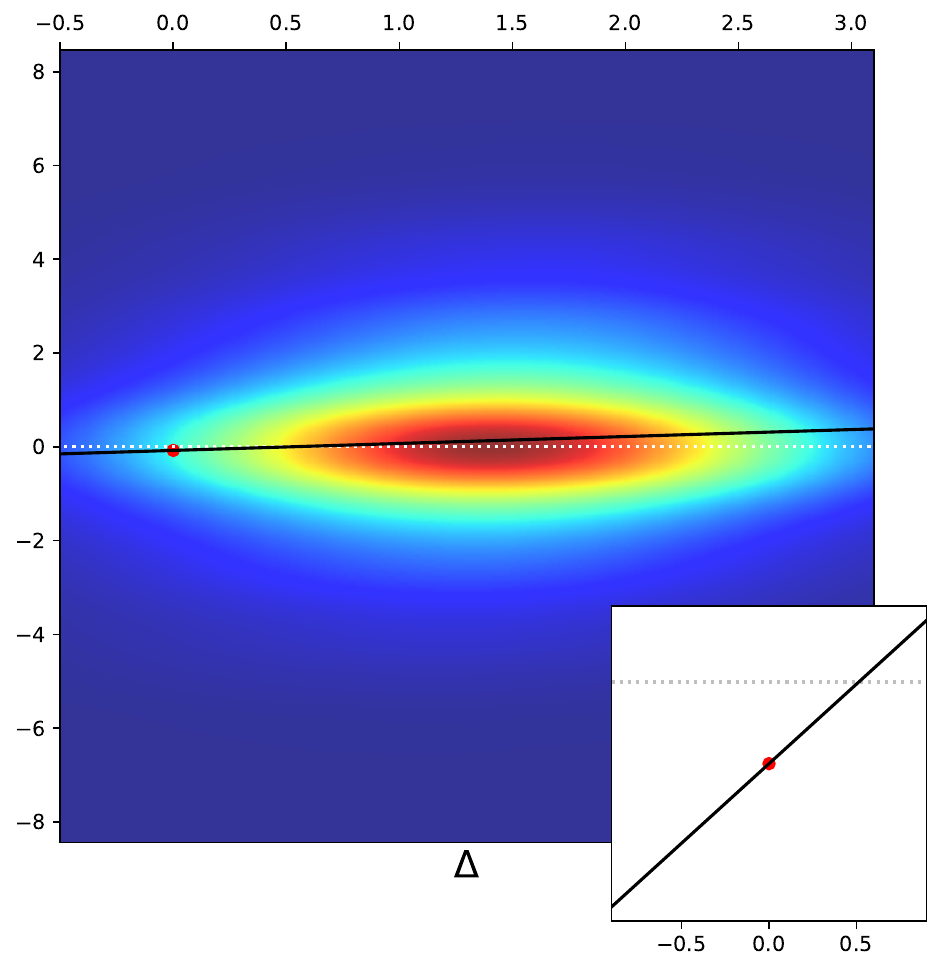}
         \caption{$T = 0.5$}
     \end{subfigure}
          \begin{subfigure}[b]{0.23\textwidth}
         \centering
         \includegraphics[width=\textwidth]{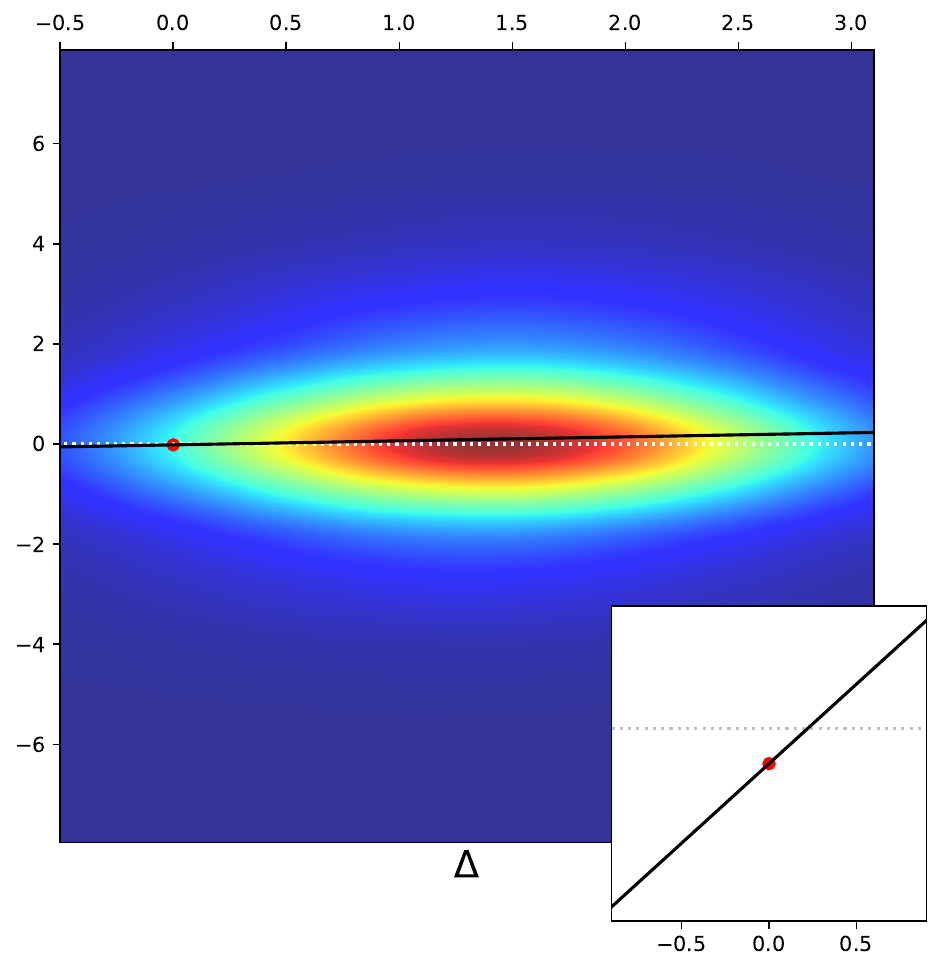}
         \caption{$T = 1$}
     \end{subfigure}
     \hfill
     \begin{subfigure}[b]{0.23\textwidth}
         \centering
         \includegraphics[width=\textwidth]{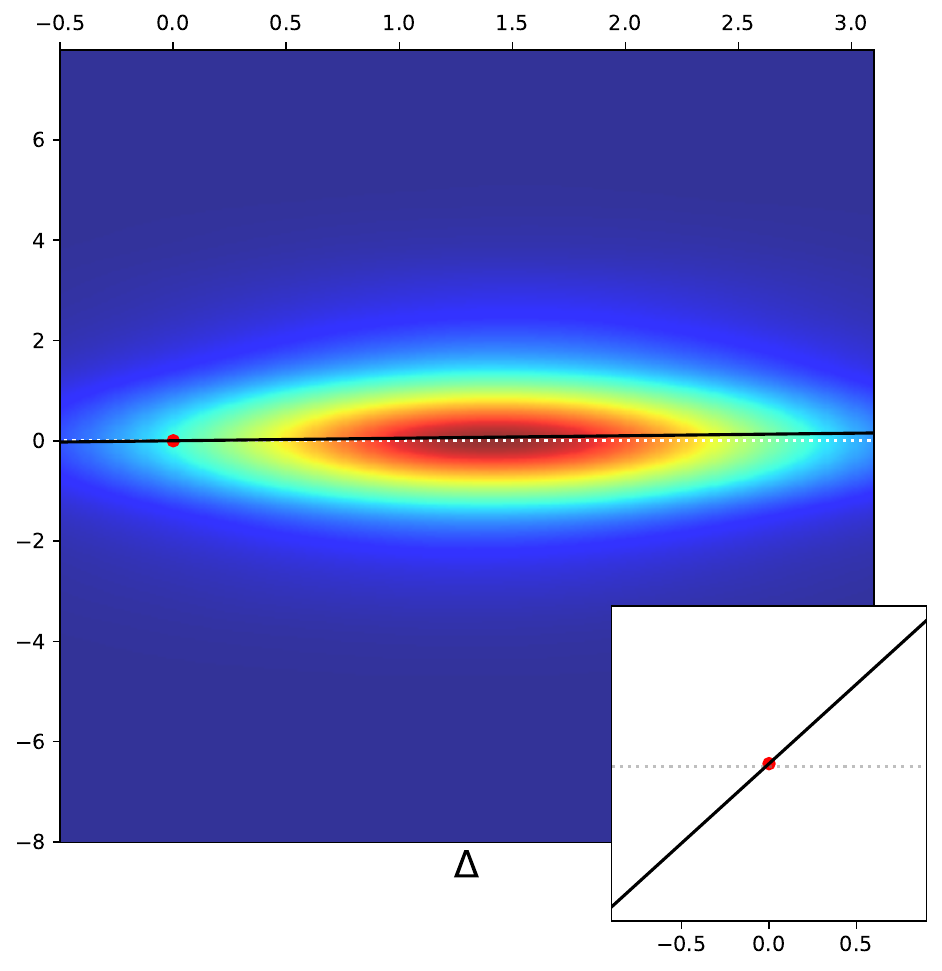}
         \caption{$T = 8$}
     \end{subfigure}
     \hfill
        \caption{Distribution of $\omega_i^{\mu}$ as a function of $\Delta_i^{\mu}$ for training configurations sampled with a Monte Carlo at temperature $T = 0$ i.e. stable fixed points only (a), $T = 0.5$ (b), $T = 1$ (c), $T = 8$ (d), on a Hebbian network. Warmer colors represent denser region of data points. The \textit{full black} line is the non-weighted best fit line for the points, the \textit{dotted white} line represents $\omega = 0$, the \textit{red dot} is the value of the best fit line associated with $\Delta = 0$. Sub-panels to each panel report a zoom of the line around $\Delta = 0$. Measures have been collected over $15$ samples of the network. Choice of the parameters: $N = 500$, $\alpha = 0.5$.}
        \label{fig:omegatilde}
\end{figure*}

\begin{figure*}
\begin{tabularx}{\linewidth}{*{3}{X}}
     %\centering
     \begin{subfigure}[b]{\linewidth}
         \centering
         \includegraphics[width=\textwidth]{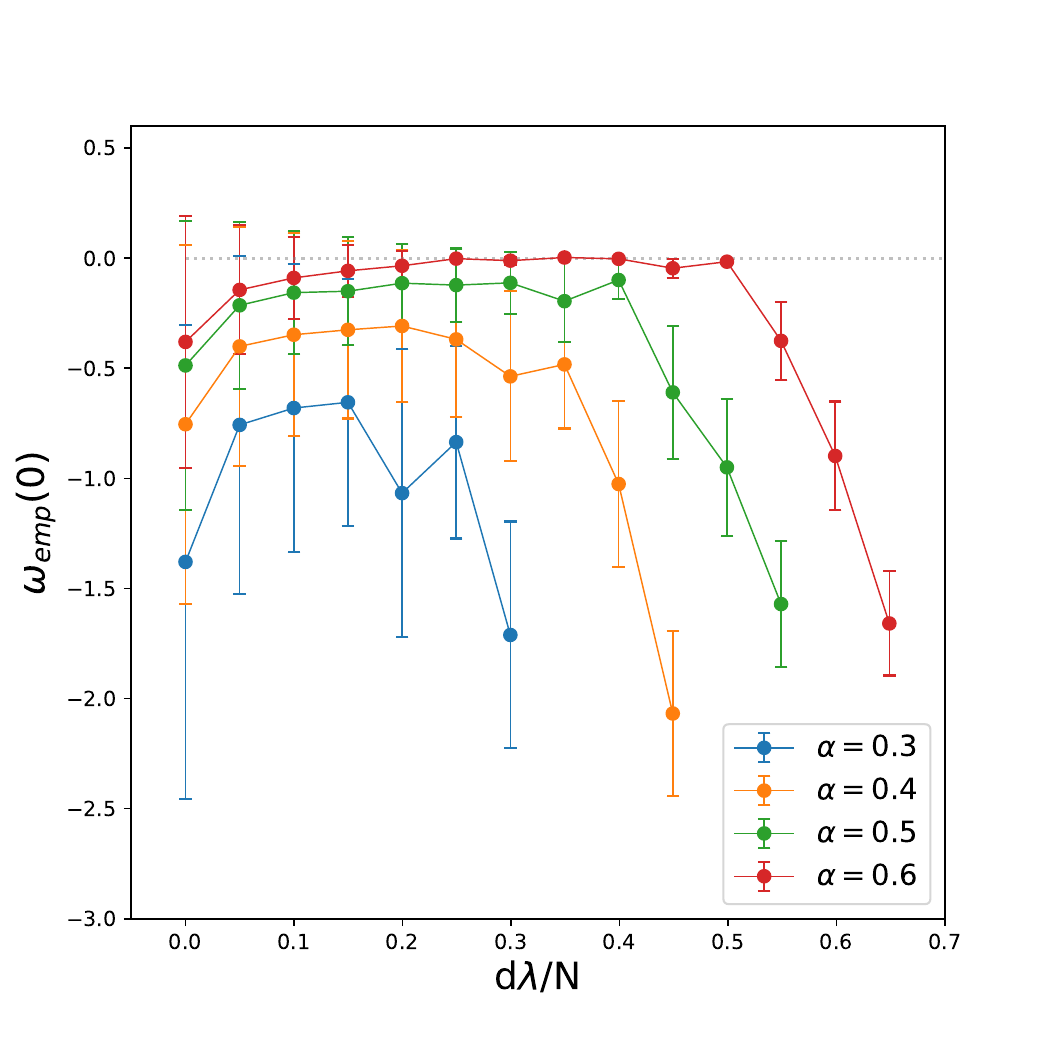}
         \caption{}
     \end{subfigure}
     &
     \hfill
     \begin{subfigure}[b]{\linewidth}
         \centering
         \includegraphics[width=\textwidth]{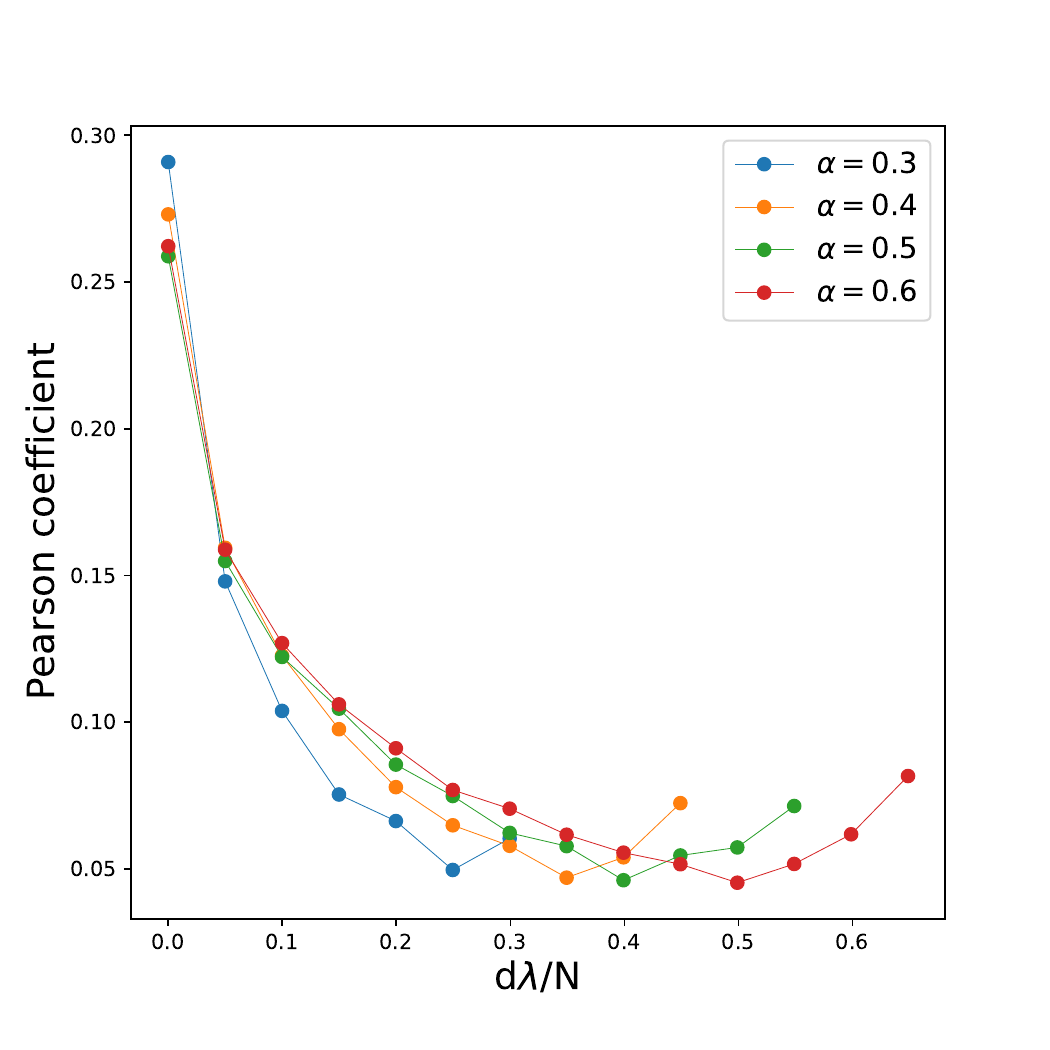}
         \caption{}
     \end{subfigure}
     &
     \hfill
     \begin{subfigure}[b]{\linewidth}
         \centering
         \includegraphics[width=\textwidth]{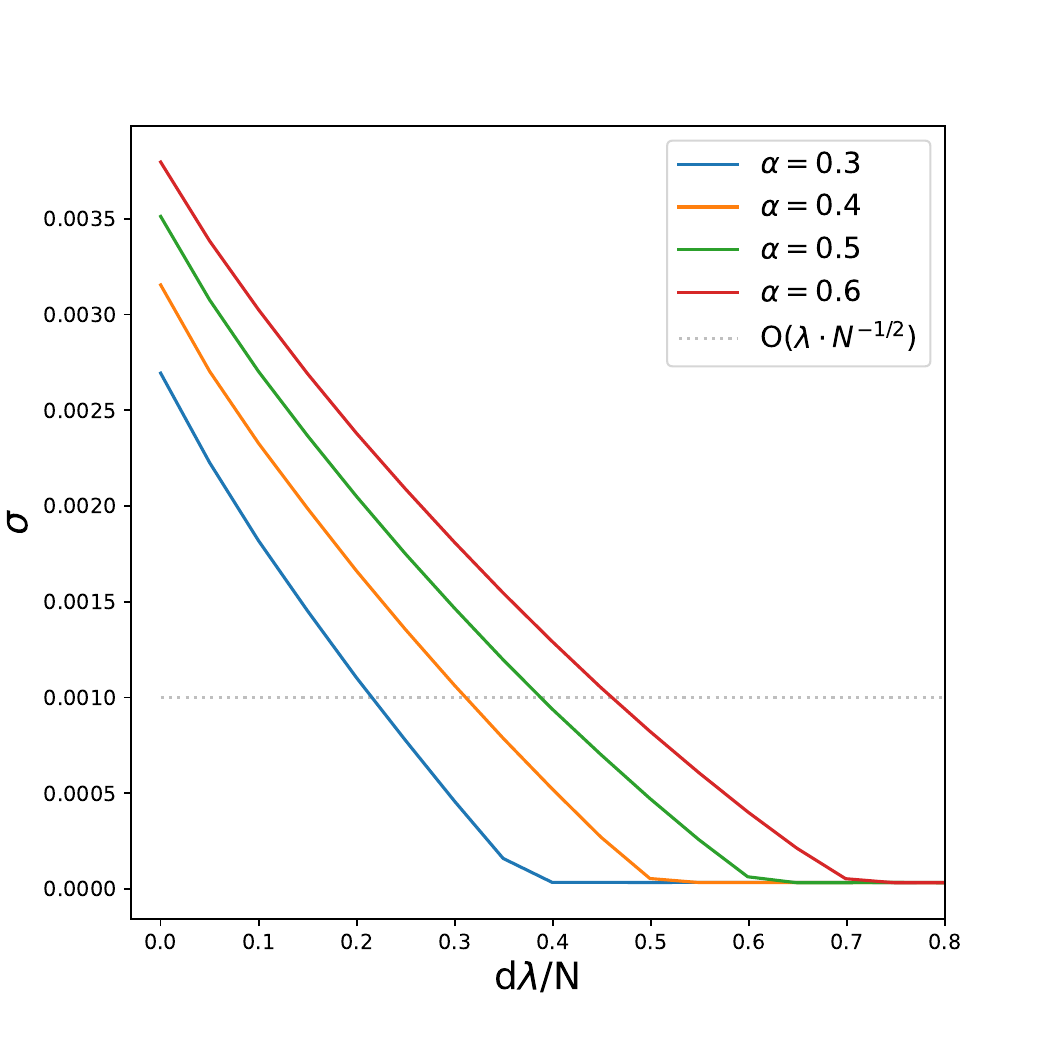}
         \caption{}
     \end{subfigure}
\end{tabularx}
\caption{The TWN algorithm is implemented by sampling stable fixed points of the network dynamics with $m_t = 0^+$. (a) The empirical measure of $\omega_i^{\mu}$ around $\Delta_i^\mu = 0$ for the case of stable fixed points as a function of the rescaled number of iterations of the learning algorithm. Errorbars are given by the standard deviations of the measures. (b) Pearson coefficient measured between $\omega_i^{\mu}$ and $\Delta_i^{\mu}$. (c) The standard deviation of the couplings during learning, defined as $\sigma = \frac{1}{N}\sum\sigma_i$. Points are averaged over $50$ samples and the choice of the parameters is: $N = 100$, $\lambda = 10^{-2}$.}
\label{fig:omega_dynamic}     
\end{figure*}

\begin{figure*}[ht!]
     \centering
     \begin{subfigure}[b]{0.24\textwidth}
         \centering
         \includegraphics[width=\textwidth]{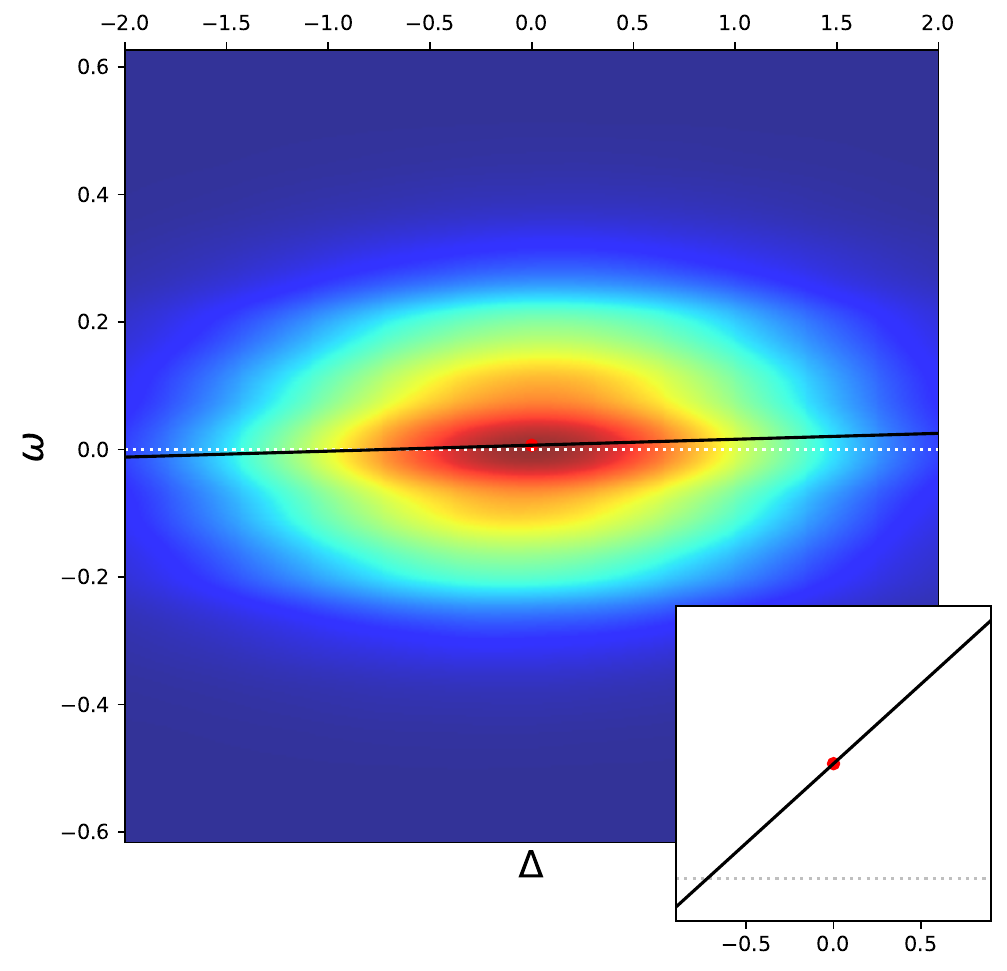}
         \caption{T = 0}
     \end{subfigure}
     \hfill
     \begin{subfigure}[b]{0.225\textwidth}
         \centering
         \includegraphics[width=\textwidth]{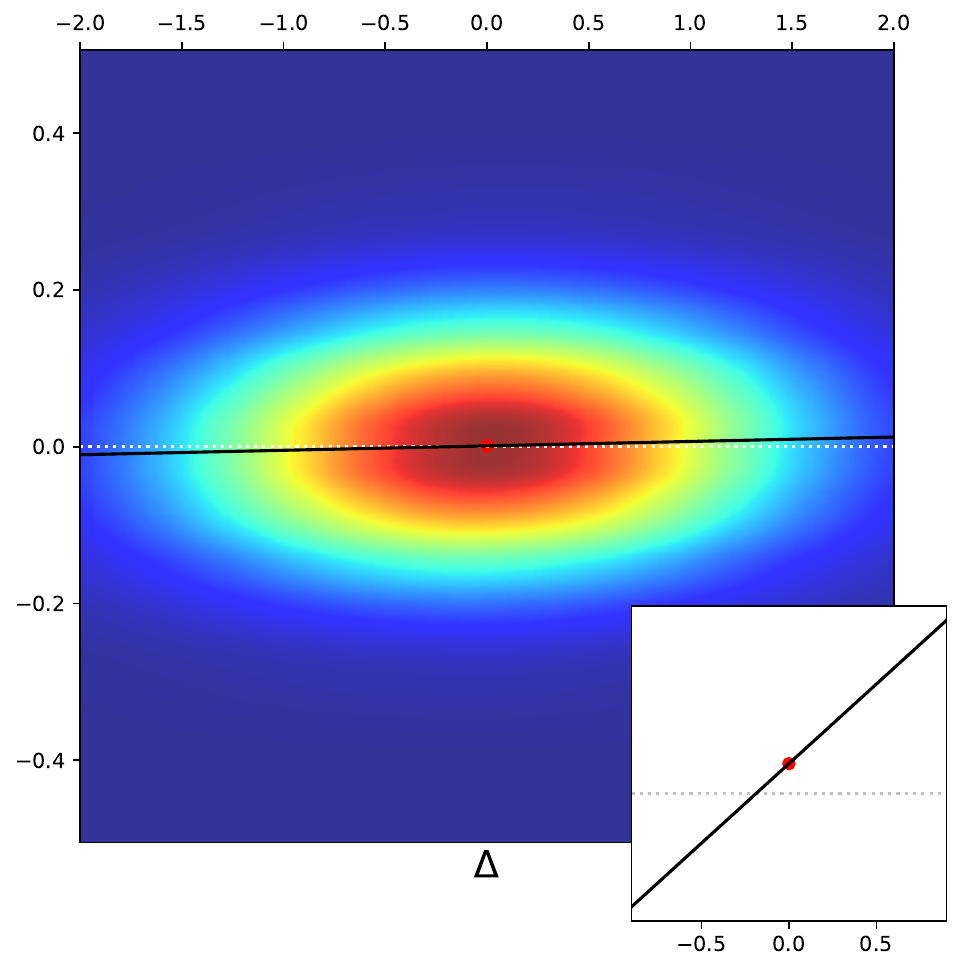}
         \caption{T = 8}
     \end{subfigure}
          \begin{subfigure}[b]{0.25\textwidth}
         \centering
         \includegraphics[width=\textwidth]{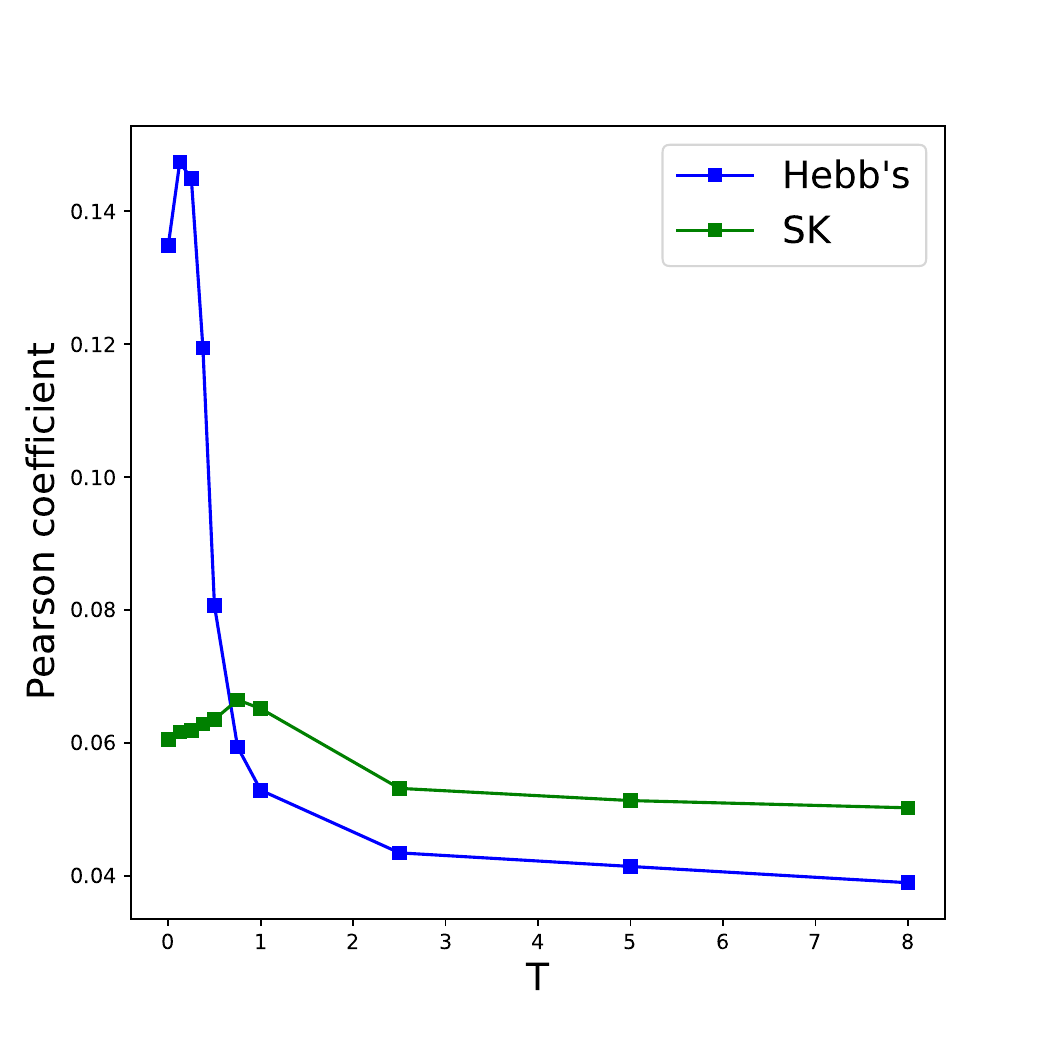}
         \caption{}
     \end{subfigure}
     \hfill
     \begin{subfigure}[b]{0.25\textwidth}
         \centering
         \includegraphics[width=\textwidth]{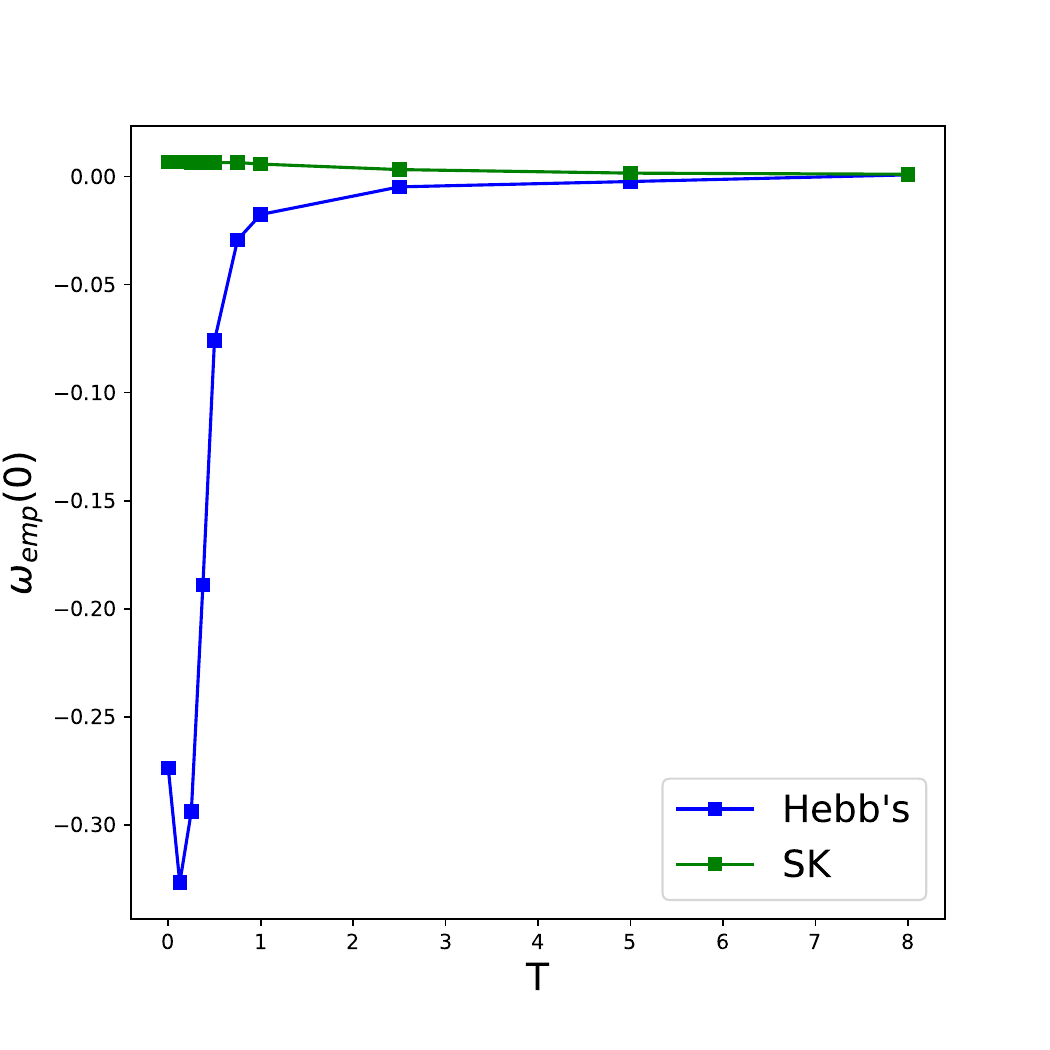}
         \caption{}
     \end{subfigure}
     
     \hfill
        \caption{(a), (b): Distribution of $\omega_i^{\mu}$ as a function of $\Delta_i^{\mu}$ for training configurations sampled with a Monte Carlo at temperature $T = 0$ i.e. stable fixed points only (a), and $T = 8$ (b) on a SK model. Warmer colors represent denser region of data points. The \textit{full black} line is the non-weighted best fit line for the points, the \textit{dotted white} line represents $\omega = 0$, the \textit{red spot} is the value of the best fit line associated with $\Delta = 0$. Sub-panels to each panel report a zoom of the line around $\Delta = 0$. (c), (d): Comparison between the Hebbian initialization and the Random one through evaluation of: the Pearson coefficient between $\omega_i^{\mu}$ and $\Delta_i^{\mu}$ (c) and the estimated value of $\omega_{emp}(0)$ from the dispersion plots (d). Measures have been collected over $15$ samples of the network. Choice of the parameters: $N = 500$, $\alpha = 0.5$.}
        \label{fig:omegatilde_SK}
\end{figure*}

Regarding the Hebbian initialization of the network, numerical results are reported in  
\cref{fig:omegatilde} for four different temperatures. Each panel shows the distribution of $(\omega_i^{\mu}, \Delta_i^\mu)$. 
Data points are collected over fifteen realizations of the network, then plotted and smoothed to create a density map. We are interested in the \textit{typical} behavior of $\omega_i^{\mu}$ when $\Delta_i^\mu \sim 0$, which can be estimated by a linear fit of the data. We consider the intercept of the best fit line as an \textit{indicator} $\omega_{emp}(0)$ of the typical value of $\omega_i^{\mu}$ around $\Delta_i^{\mu} = 0$.
We find that at lower temperatures the sampled configurations favor both perfect-retrieval and robustness because $\omega_{emp}(0)$ is more negative. As temperature becomes too high, $\omega_{emp}(0)$ gets closer to zero, suggesting low quality in terms of training performance. \\
%Results appear to be independent of the system size $N$, implying that finite size effects are not strong for this kind of measure. This is likely because both the mean of the distribution of the points along the $\omega$ direction and their dispersion, i.e. how far they span the y-axis, are decreasing as $\mathcal{O}(N^{-1/2})$.\\
%Another interesting aspect emerging from the analysis is that, by calling $a$ the line intercept and $b$ its angular coefficient, the position of the right extremum of the negative band $\frac{|a|}{b}$ is independent of the system size $N$, implying that finite size effects are not strong for this kind of measure. This is likely because both the mean of the distribution of the points along the $\omega$ direction and their dispersion, i.e. how far they span the y-axis, are decreasing as $\mathcal{O}(N^{-1/2})$.\\
One can also study how the distribution of $(\omega_i^{\mu}, \Delta_i^\mu)$ evolves during the training process. Fig. \ref{fig:omega_dynamic}a shows the value of $\omega_{emp}(0)$ at different time steps of the training process, for different values of $\alpha$, when configurations at $T = 0$ are given to the algorithm. We find that $\omega_{emp}(0)<0$ for $\alpha \leq 0.6$. The progressive increase of $\omega_{emp}(0)$ means that the structure of the fixed points is more effective in the starting Hebbian landscape compared to intermediate stages of training. 
In the last part of training, points reacquire more negative values, but this is not a reliable indication of good performance: as shown in \cref{fig:omega_dynamic}c, in this part of the process the standard deviation of the couplings $\sigma_i$ is comparable to $\mathcal{O}(\lambda\cdot N^{-1/2})$, and the expansion of the $\mathcal{L}$ in \cref{eq:deltaL_lwsn} is not valid. 
The last part of the training, where $\sigma_i \simeq 0$ $\forall i$, has been neglected from the plot. The experiment is thus consistent with the characterization of the HU algorithm presented in \cite{benedetti_supervised_2022}, which showed decreasing perfect-retrieval and robustness when increasing $\alpha$. This is confirmed by the study of the Pearson correlation coefficient between $\omega_i^{\mu}$ and the associated stabilities $\Delta_i^{\mu}$ (see \cref{fig:omega_dynamic},b). 
High values of the Pearson coefficient show a strong dependence of the structure of noise on the relative stabilities. For all $\alpha$, the Pearson coefficient is highest at $d = 0$, and progressively decreases during training, suggesting that the quality of the training configurations is deteriorating. The final increase in the coefficient is, again, due to the vanishing of the standard deviations $\sigma_i$ of the couplings, and does not indicate good performance.\\
For a comparison, one can study the distribution of $\omega_i^{\mu}$ in the case of a random initialization of the coupling matrix $J$. We chose the Sherrington-Kirkpatrick (SK) model \cite{sherrington_solvable_1975} as a case of study. Panels (a) and (b) in \cref{fig:omegatilde_SK} report the smoothed distribution of $(\omega_i^{\mu}, \Delta_i^\mu)$ showing a different scenario with respect to the Hebbian one. The distribution looks anisotropic, as in the Hebbian case, yet the stabilities are centered Gaussians, so $\omega_{emp}(0)$ is positive. 
In particular, things appear to improve when $T$ increases, in contrast with the previous case of study, though in accordance with the Hebbian limit of the TWN algorithm explained in section \ref{sec:twn}. Panel (c) displays more clearly the mutual dependence between $\omega_i^{\mu}$ and stabilities $\Delta_i^\mu$ by reporting the Pearson coefficient at the various $T$ between these two quantities: both Hebb's and SK show some mutual dependence, but the distribution in the Hebbian landscape is way more skewed. Furthermore, panel (d) shows $\omega_{emp}(0)$ in both cases. This measure is consistent with the indication coming from the Pearson coefficient: while for the Hebb's initialization $\omega_{emp}(0)$ remains significantly negative and reaches the lowest values at low temperatures, the random case shows the opposite trend, with the estimated $\omega_{emp}(0)$ staying generally close to zero.  
\subsection{The role of saddles}
\begin{figure}[ht!]
\centering
\includegraphics[width=\linewidth]{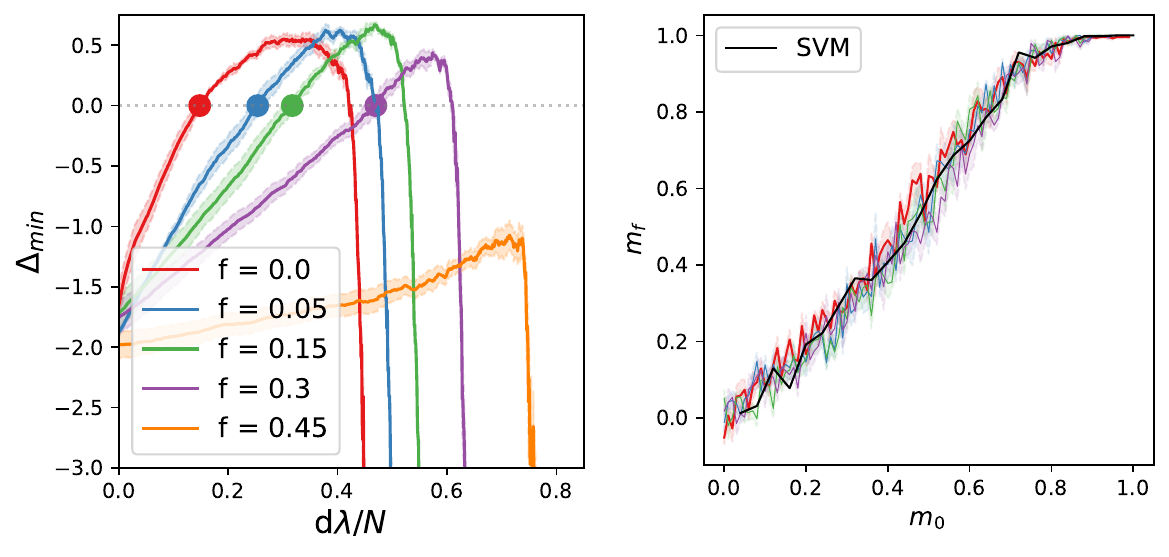}
\caption{Left: Minimum stability $\Delta_{\text{min}}$ as a function of the algorithm steps on a network trained with the symmetric TWN routine that learns saddles of various indices $f$. The initial matrix is assembled according to the Hebb's rule. Full dots report the amount of iterations needed to accomplish perfect-retrieval. Right: the retrieval map $m_f(m_0)$ as measured on the positions of the colored dots from the right panel, with the same color code being used. A comparison with a SVM trained with the same choice of the parameters is also presented through the \textit{dashed blue} line. All measures are averaged over $5$ samples with the shaded region indicating the experimental errors. The choice of the parameters is: $N = 100$, $\alpha = 0.35$, $\lambda = 10^{-3}$.}
\label{fig:stab_sads}
\end{figure}

Notice, from both panels (c), (d) of \cref{fig:omegatilde_SK}, the existence of an optimum which does not coincide with the stable fixed points of the dynamics (i.e. $T = 0$). As noted in previous studies on spin glasses \cite{aspelmeier_free_2006}, one can associate configurations probed by a Monte Carlo at finite temperatures with configurations which are typically saddles in the energy landscape with a given $\textit{saddle index}$. The $\textit{saddle index}$ is defined as the ratio between the number of unstable sites under the dynamics (see \cref{eq:dynamics}) and the total number of directions $N$. Stable fixed points have $f = 0$, while random configurations are expected to have $f = 1/2$. In order to check whether a particular $f$ is capturing relevant features of the virtuous training configurations, we sampled training data according to the requirement that their saddle fraction assumes a specific value $f$ and $m_t=0^+$. Saddles are then employed for training the network according to \cref{eq:lwn}. Sampling is performed by randomly initializing the network on a configuration having training overlap $m_t = 0^+$ with a reference memory, and performing a zero temperature dynamics on the landscape defined by the energy
\begin{equation}
    \label{eq:sad_energy}
    E(\vec{S}|f, J) = \frac{1}{2}\left(\frac{1}{N}\sum_{i=1}^N\Theta(S_i\sum_{j = 1}^N J_{ij}S_j) - f\right)^2,
\end{equation}
where $\Theta(x)$ is the Heaviside function. Yet again, the value of $m_t$ was maintained constant during the descent. 
The left panel in \cref{fig:stab_sads} shows how the minimum stability evolves during the training process while a symmetric TWN algorithm is initialized in the Hebbian matrix and learns saddles of different indices. For a network of $N = 100$ and $\alpha = 0.35$, we found that perfect-retrieval is reached until a certain value of $f$, suggesting that saddles belonging to this band are indeed good training data. The band of saddles that are suitable for learning is reduced when $\alpha$ increases until such states do not significantly satisfy \cref{eq:deltaL_lwsn3} anymore. Such limit capacity is located around the critical one for HU.  
\begin{comment}
Notice, in particular, that the range of saddles indexes that support good training performances are consistent with the ones spanned by the points in the sub-panel in \cref{fig:lwn_sampled}c.
\end{comment}
It should be stressed that the precise performance as a function of $f$ is quite sensitive to the sampling procedure. Simulated annealing routines \cite{s_optimization_1983} have also been employed to minimize (\ref{eq:sad_energy}), obtaining qualitatively similar results yet not coinciding with the ones reported in \cref{fig:stab_sads}. A qualitative study of the basins of attraction of the network has been performed and reported in the right panel in \cref{fig:stab_sads}. In particular, the retrieval map $m_f(m_0)$ has been measured relatively to the saddle indices $f$ at the first time they reached perfect-retrieval, in analogy to what has been measured in \cite{benedetti_supervised_2022}. The curves coincide quite well, suggesting that finite sized networks trained with different $f$ assume similar volumes of the basins of attraction when they are measured at the very first instant they reach perfect-retrieval. The plot also shows that the robustness performance is comparable with the one of a SVM trained with the same choice of the control parameters.
\subsection{Going unsupervised}
It is essential to note that, since the training overlap under consideration is close to $0$, stable fixed points and saddles in their proximity can be sampled in a full unsupervised fashion: when the dynamics (e.g. zero temperature Monte Carlo or gradient descent over (\ref{eq:sad_energy})) is initialized at random, which implies having an overlap $m_t = 0^+$ with some memory, it will typically conserve a small overlap with the same memory even at convergence \cite{kepler_domains_1988}. As a consequence, the argument presented above holds, and a supervised algorithm as TWN can be reduced to a more biologically plausible and faster unsupervised learning rule. 
This aspect will be deepened by the next section, where we show a particular scenario where TWN coincides with the HU rule, a fully unsupervised learning procedure \cite{hopfield_unlearning_1983}.
\subsection*{Checkpoint}
In this section we have seen that:
\begin{itemize}
    \item While the standard TWN employed random noise to train a SVM by tuning $m_t = 1^-$, one can approach a SVM using structured noise with $m_t = 0^+$. At each step of the TWN algorithm, the best training data are the ones satisfying condition (\ref{eq:deltaL_lwsn3}). Maximizing the noise is useful because it can be sampled in an unsupervised manner, i.e. by initializing the dynamics on a random network state and then descending the energy function.
    \item While the energy landscape of random networks contain few good training data, a Hebbian landscape contains many of these states in form of local minima and surrounding saddle points. A Hebbian initialization of the couplings is thus a good starting point for the TWN algorithm. 
\end{itemize}
\newpage
\section{\label{sec:U}Quasi-optimality of the Unlearning algorithm}
%Motivated by the results of the previous sections on the characterization of effective training configuration we select training configurations which are not only characterized by a fixed value of overlap with the memories $m_t$, but are also fixed points of the dynamics, after initializing the coupling according to Hebb's rule.

\begin{figure}[ht!]
\centering
\includegraphics[width=0.7\linewidth]{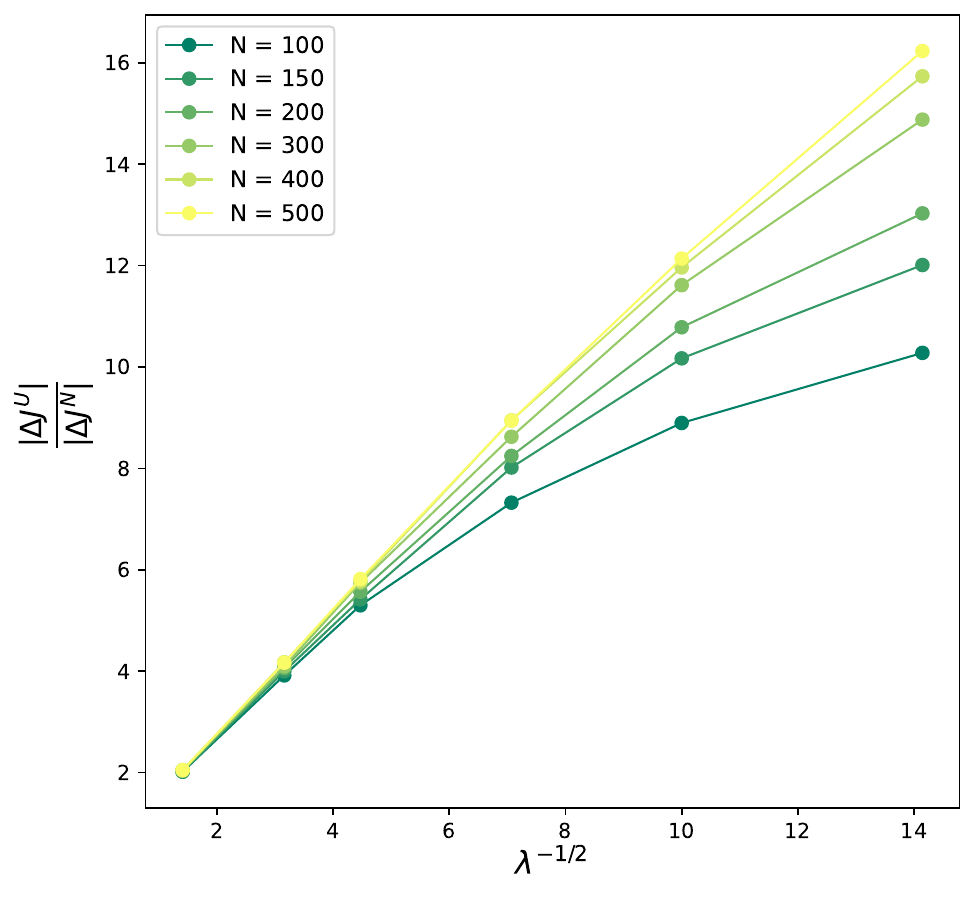}
\caption{Estimates of the ratio $|\Delta J^U|/|\Delta J^N|$ as a function of $\lambda^{-\frac{1}{2}}$ and $N$ for $\alpha = 0.5$. Measures are averaged over $5$ samples. Error bars are not indicated because smaller than the symbols.}
\label{fig:conv1}     
\end{figure}

\begin{figure}[ht!]
\centering
\includegraphics[width=0.7\linewidth]{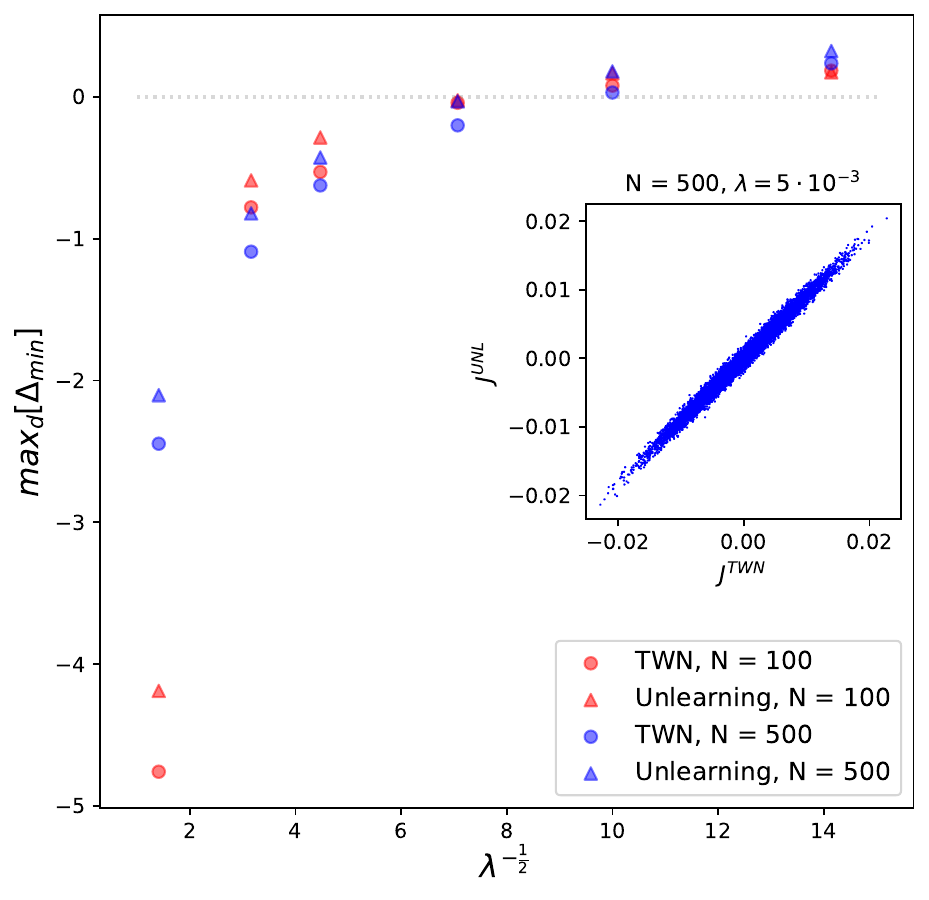}
\caption{The quantity $\text{max}_d\left(\Delta_{\text{min}}\right)$ as a function of $\lambda^{-1/2}$. Colors are: \textit{red} for $N = 100$ and \textit{blue} for $N = 500$. The \textit{gray} line represents the null value for the stability. Symbols are: \textit{circles} for the TWN rule, \textit{triangles} for the HU rule. In the subplot on the center right, the couplings obtained through the HU algorithm are plotted as a function of the ones resulting from the TWN, at the same amount of iterations, for one sample at $N = 500$ and $\lambda = 5\cdot 10^{-3}$. Measures are averaged over $50$ samples. The choice of the parameters is: $\alpha = 0.5$, $m_t = 0^+$ for TWN.}
\label{fig:conv_stab}
\end{figure}

%We firstly evaluate the maximal noise case $m_t = 0^+$, where the considerations of Sec. \ref{sec:trademark_virtuous} apply, and then move to $m_t > 0^+$. 
As shown in \cref{sec:choice_of_training_conf}, local minima of the Hebbian energy landscape are good training configurations, yet they are necessarily not the optimal ones (see \cref{fig:omegatilde_SK}). In this section, we analyze the performance of TWN when training data are minima in the energy landscape characterized by an overlap $m_t$ with the memories. Local minima in the energy landscape are convenient training data, since they can be easily and efficiently sampled by the asynchronous dynamics in \cref{eq:dynamics}. After initializing the couplings according to the Hebbian rule, we will study two different scenarios: we firstly evaluate training in the maximal noise case $m_t = 0^+$, when stable fixed points can be sampled in a unsupervised fashion, i.e. by the network dynamics when started on a fully random state. In this setting, where the considerations of section \ref{sec:trademark_virtuous} apply, we show that the TWN algorithm converges to the traditional HU rule in the small $\lambda/N$ limit. Then, we examine in detail training with finite overlap (i.e. $m_t > 0^+$), providing an estimate of the critical capacity reached by the neural network, and showing that results from the previous section about the effectiveness of stable fixed points can be generalized to this case. \\ \\
As mentioned in sec. \ref{sec:choice_of_training_conf}, in this case the only relevant contribution to the update rule is
\begin{equation}
    \label{eq:Upart_fp}
    \Delta J_{ij}^U(D) = -\frac{\lambda}{N}\sum_{d = 1}^T S_i^{\mu_d}S_j^{\mu_d},
\end{equation}
which is the classic HU update rule. 
As a result, when $\lambda/N \rightarrow 0$ the TWN algorithm and the HU algorithm will converge to the same updating rule for the couplings when stable fixed points of the dynamics are used in the training. The same  argument can be applied to the original asymmetric rule (\ref{eq:lwn}), however asymmetric networks may have no stable fixed points of (\ref{eq:dynamics}) that can be easily reached and employed in the learning. \\
We now perform a numerical test of the argument above, in the case of a symmetric connectivity matrix. 
At each step of the algorithm, the network is initialized with an initial overlap contained in $(0, N^{-1/2})$ with one memory $\vec{\xi}^{\mu_d}$. Then, asynchronous dynamics (\ref{eq:dynamics}) is run  until convergence, and the final overlap $m_t$ is measured. If $m_t\in (0, N^{-1/2})$, we use the sampled configuration for training, otherwise the process is repeated. Typically, an initial overlap equal to $0^+$ implies a similar order of magnitude for the final overlap, hence no reiteration is usually needed.\\
The algorithm (\ref{eq:lwn2}) is repeated for $D = \mathcal{O}(N/\lambda)$ steps.
%If the learning rate $\lambda/N$ is sufficiently small, there will exist a characteristic number of updates over which  the couplings will change very little, so that the properties of the network will be essentially unchanged. We will refer to this characteristic timescale as an \textit{epoch}. Rewriting equation (\ref{eq:Upart_fp}) in terms of epochs we obtain
%\begin{equation}
%\small
%    \Delta J_{ij}^U(T) = -\frac{\lambda}{N}\sum_{epochs}T_{epoch}\cdot\langle S_i S_j\rangle_{epoch} + \text{err}(N,\lambda),
%\end{equation}
%where $T_{epoch}$ is the duration of each epoch and $err$ is a correction that is supposed to vanish when $\lambda/N \rightarrow 0$. Since the order of magnitude of $\Delta J_{ij}^U(T)$ is supposed to be the same of $J_{ij}^{(0)}$ it is reasonable to conjecture that $T_{epoch} = \mathcal{O}\left(\frac{\lambda}{N}\right)$, the number of epochs is $\mathcal{O}(1)$ and 
%\begin{equation}
%\label{eq:ss}
%\langle S_i S_j \rangle_{epoch} = \mathcal{O}(J_{ij}^{(0)}).
%\end{equation}
The order of magnitude of $\Delta J_{ij}^U(D)$ is supposed to be the same of $J_{ij}^{(0)}$, in order to see significant modifications to the initial connectivity matrix.  
The network is initialized according to the Hebb's rule (\ref{eq:hop}), i.e. $J_{ij}^{(0)} = \mathcal{O}(N^{-1/2})$ which implies $\Delta J_{ij}^U(D) = \mathcal{O}(N^{-1/2})$ at leading order. The contributions U and N are compared by computing the norm of the relative $\Delta J$ matrix and evaluating the ratio $|\Delta J^U|/|\Delta J^N|$. From our previous considerations we expect $|\Delta J^U|/|\Delta J^N|$ to be linear in $\lambda^{-1/2}$ when corrections vanish. 
Results are reported in \cref{fig:conv1}: $|\Delta J^U|/|\Delta J^N|$ grows when $N$ increases and $\lambda$ decreases, according to the scaling relation predicted by our argument. In addition to this, curves are collapsing on the expected line when $\lambda \rightarrow 0$ and $N \rightarrow \infty$.\\ 
We also measured $\Delta_{\text{min}}$ at its maximum over the course of the algorithm (as described in \ref{sec:3a}). Results are reported in \cref{fig:conv_stab}. $\Delta_{\text{min}}$ produced by TWN and HU are found to coincide when $\lambda$ is sufficiently small. Moreover, the number of steps necessary to reach the maximum are the same for both algorithms, confirming that couplings are transforming at the same way. This last aspect is corroborated by the subplot in \cref{fig:conv_stab}, representing the set of $J_{ij}$ obtained with the traditional HU algorithm as a function of the one resulting from the TWN algorithm, for one realization of the network. The strong correlation is evident, as predicted from our pseudo-analytical arguments.  \\\\During the rest of the section we will drop the noisy part of the synaptic update and use the traditional rule $$\delta J_{ij} = -\frac{\epsilon}{N}S_i^* S_j^* $$ where we renamed $\lambda$ with $\epsilon$ not to confuse it with the learning rate used for the linear perceptron. Specifically the symmetric perceptron (SP) rule with a tunable margin $k$ will be applied (see \cref{eq:lp_update}, \cref{eq:lp_sym_update}). We also substituted the index $\mu_d$ with a star since, for the rest of this section, the HU procedure will be implemented in a fully unsupervised fashion. 
%\begin{figure}
%\centering
%\includegraphics[width=.45\textwidth]{Figures/deltaJ_vs_N_comparison_alpha0.5_lambda0.005.pdf}\hfill
%\includegraphics[width=.45\textwidth]{Figures/deltaJ_vs_lamb_comparison_alpha0.5_lambda0.005.pdf}\hfill
%end{figure}
\\

\subsection{Robustness performance and basins of attraction}

\begin{figure}[ht!]
\centering
\includegraphics[width=.7\textwidth]{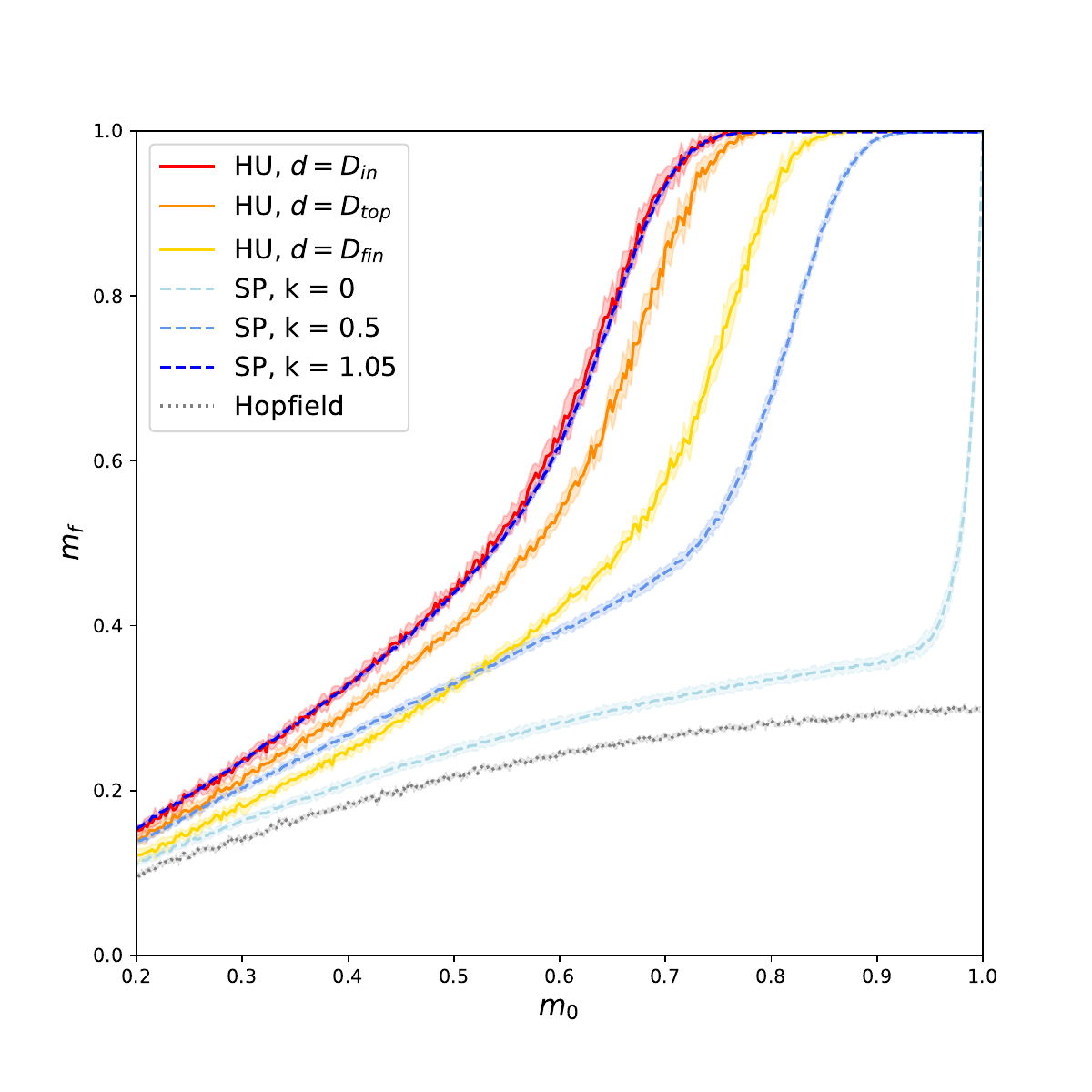}
    \caption{The average size of basins of attraction at $N = 800$, $\alpha = 0.4$ for both symmetric perceptron (SP) and Hebbian Unlearning (HU), averaged over several realizations of the disorder. The colored area around each curve represents the statistical errors. Continuous lines are for basins at $D_{in}, D_{top}$ and $D_{fin}$ for HU with $\epsilon = 10^{-2}$. Dashed lines are for three values of $k$, including the $k \simeq k_{max}$, used for SP with $\lambda = 1$. The gray dotted line represents the performance of the Hopfield model at the same value of $\alpha$. Notice that attraction basins of the SP and HU almost coincide when the two algorithms operate in their optimal regime, namely $d=D_{in}$ and $k\simeq k_{max}$.}
\label{fig:mf_mi}           
\end{figure}

We have compared the performance of SP and HU by measuring the shape of the basins of attraction around each memory. This is done by measuring the retrieval map $m_f(m_0)$. In \cref{fig:mf_mi} we plot $m_f$ as a function of $m_0$. Colored dashed curves refer to SP for different values of $k$, up to the highest $k$ that allows the algorithm to converge in $\mathcal{O}(10^3)$ iterations. This slight underestimations of the real $k_{max}(\alpha)$ bares very little consequences to our results. Related to this, we underline the importance of the choice of $\lambda$ at a given value of $N$. This is another crucial topic that is rarely discussed in the literature. Higher values of $\lambda$ imply larger learning steps, while smaller values are associated to a finer exploration of space of coupling matrices during training. It is observed that the algorithm, operating at $\lambda = \{1,10^{-1}, 10^{-2}\}$, converges to almost identical matrices already when $k$ is equal to the maximal stability for diluted networks~\cite{gardner_phase_1989} that, according to \cref{fig:sat_unsat}, is slightly lower than the actual $k_{max}$.
This suggests that the final state lies very closely to the unique optimal solution even when we are not exactly at $k_{max}$. Hence, no significant changes are expected in our numerical results when $k$ is pushed further towards its maximal value. On the other hand, when $\lambda$ assumes smaller values, i.e. $\lambda = \{10^{-3}, 10^{-4}, 10^{-5}\}$, basins are observed to be smaller in size and the volume of solutions is larger, indicating that the final state remains further from the maximal performance. In order to recover the numerical results obtained at a larger $\lambda$, one needs to progressively increase $k$ to values that are difficult to reach numerically. As a result, the choice of $\lambda = 1$ in this section seems to us well justified to reproduce the optimal performance of the symmetric perceptron at $k \simeq k_{max}$. \\
Consistently with the literature, we find that increasing the stability leads to an increase of $m_f$ at fixed $m_0$~\cite{forrest_content-addressability_1988}. In particular, when the stability is equal to zero, the memories, albeit being fixed point of the dynamics, have zero basin of attraction, as indicated by the very low values of $m_f$ for $m_0\neq 1$. The gray dashed line at the bottom of \cref{fig:mf_mi} refers to the Hopfield model without unlearning: since $\alpha>\alpha_c^H$ the model does not learn. The colored continuous lines refer to HU, for different amounts of unlearning. More specifically, we measured the performance of the model for the three values $d = D_{in},\,D_{top},$ and $D_{fin}$ defined in section \ref{sec:UNL}. It is clear how Unlearning improves the performance of the network, which is not maximized at $d=D_{top}$ as one could expect~\cite{Horas}, but at $d = D_{in}$, where the requirement for perfect-retrieval of the memories is satisfied with zero margin. 

\begin{figure}[ht!]
\centering
\includegraphics[width=.7\textwidth]{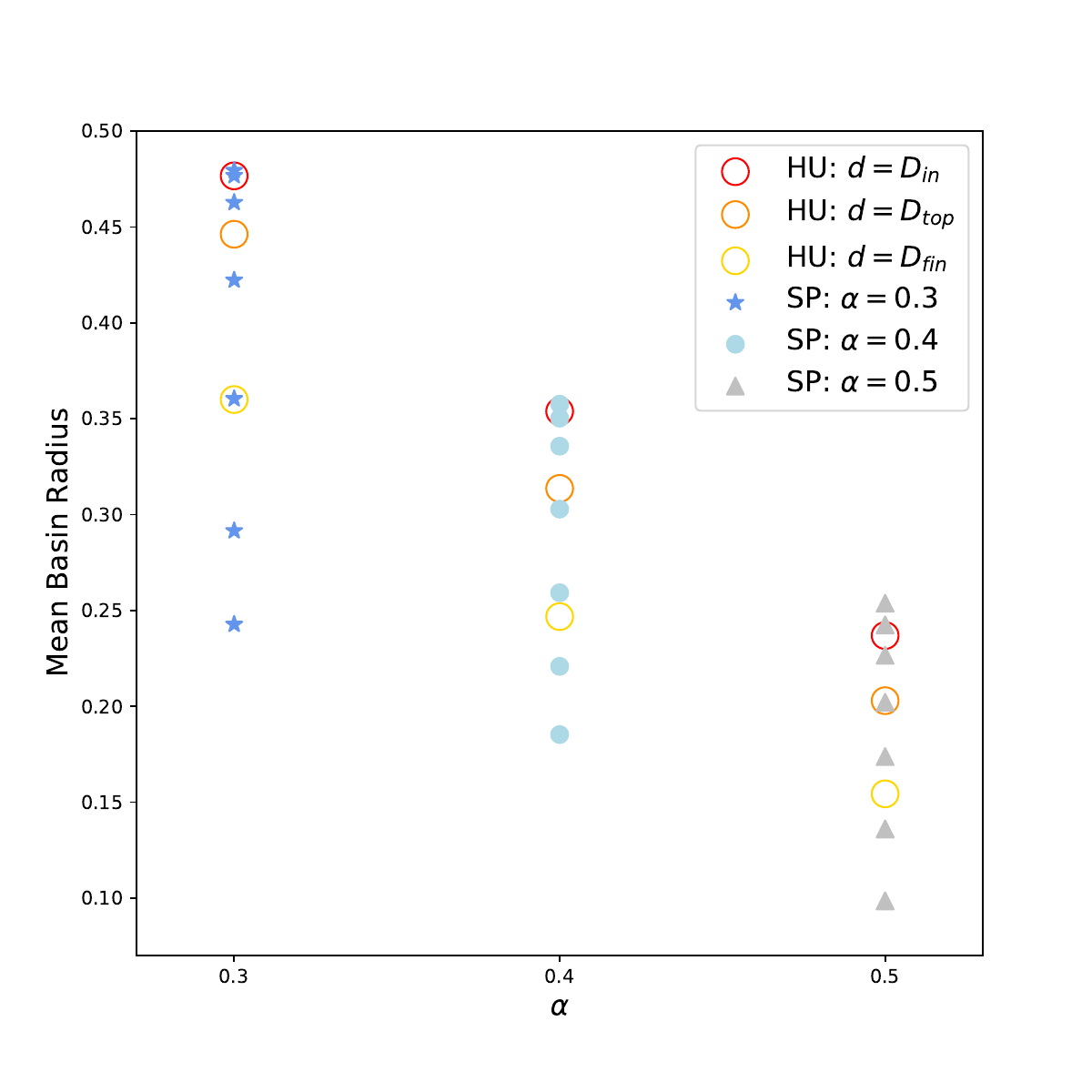}
    \caption{ Mean attraction basin radius for symmetric perceptron (SP) and Hebbian Unlearning (HU) measured as in~\cite{forrest_content-addressability_1988} and extrapolated to $N\to\infty$, for $\alpha = 0.3, 0.4, 0.5$ and $\lambda = 1$. Points for the SP correspond to the following values of $k$: $\alpha = 0.3\rightarrow k\in\{0.4, 0.5, 0.7, 0.9, 1.1, 1.296, 1.32\}$; $\alpha = 0.4\rightarrow k\in\{0.4, 0.5, 0.6, 0.75, 0.9, 0.988, 1.05\}$; $\alpha = 0.5\rightarrow k\in\{0.3, 0.4, 0.5, 0.6, 0.7, 0.768, 0.85\}$. Error bars are smaller than the symbols size. } 
\label{fig:basins_Ninf}           
\end{figure}

We also found that the performance of HU at $d=D_{in}$ and the one of the SP at $k\simeq k_{max}$ are indistinguishable within our numerical resolution. This is a remarkable fact, since the two algorithms have a radically different structure: the SP algorithm is supervised, i.e. it needs to have access at every step to all the memories that the network needs to memorize, while the HU is not, and only exploits the topology of the spurious states generated by Hebb's prescription. These findings are robust to change in the load $\alpha$ and to finite size effects, as illustrated in \cref{fig:basins_Ninf}. The mean basin radius at finite $N$ is defined as $1 - m_0$, selecting the value of $m_0$ below which more than $30\%$ of the memories are reconstructed with more then $5\%$ error. The dots represent our extrapolation of this quantity to the limit $N\to\infty$, for different values of $\alpha$. The lower dots relative to the SP correspond to $k<k_{max}$, and the value of the mean basin radius gets higher as $k$ is increased up to $k \simeq k_{max}$. Again, one can see that even in the thermodynamic limit, our simulations suggest that in their optimal regime the two algorithms perform essentially in the same way.

\subsection{Learning paths in the space of the interactions}
One way to visualize the solutions of the optimization problem, and the way these solutions are reached by means of the algorithm, is to exploit the space of interactions as conceived by Gardner~\cite{gardner_space_1988}. Consider a spherical surface in $N(N-1)/2 - 1$ dimensions where each point $\vec{J}$ is a vector composed by the off-diagonal elements $J_{j > i}$ of the connectivity matrix normalized by their standard deviation. These position vectors hence will be
\begin{equation}
    \label{eq:r}
    \vec{r} = \vec{J}/\sigma_J\ ,
\end{equation}
with
\begin{equation}
    \label{eq:sigma}
    \sigma_J = \sqrt{\frac{2}{N(N-1)}\sum_{i < j}^{1,N} J_{ij}^2}\ .
\end{equation}
For what concerns the SP, after fixing the value of $\alpha$ and a set of memories, one can imagine the sphere as composed by an UNSAT and a SAT region. These regions are connected sub-spaces of the original sphere, so that one can go from a matrix to another one in a continuous fashion. The SAT region contains the point relative to the unique solution at $k = k_{max}(\alpha)$.

We now define an \textit{overlap} parameter quantifying the covariance of two generic symmetric matrices $J_{ij}$ and $U_{ij}$
\begin{equation}
    \label{eq:q}
    q = \frac{2}{N(N-1)} \sum_{i < j}^{1,N}\overline{ \frac{J_{ij}U_{ij}}{\sigma_J \sigma_U}} \ ,
\end{equation}
where $\overline{\hspace{0.1cm}\cdot\hspace{0.1cm}}$ is the average over the disorder.

%\begin{multicols}
\begin{figure*}[ht!]
\centering
\begin{subfigure}[b]{0.49\textwidth}
\centering
\includegraphics[width=\textwidth]{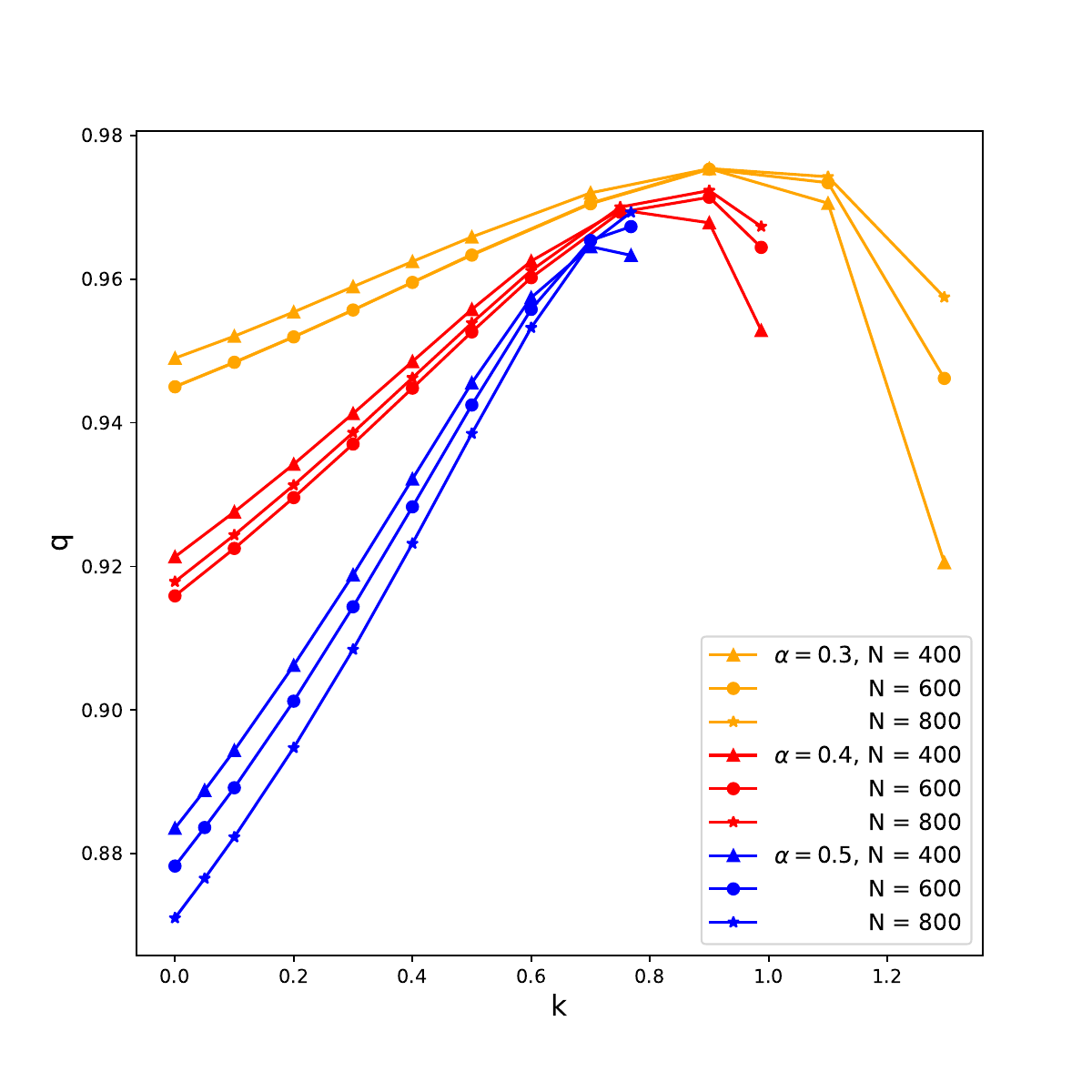}
\caption{}
\label{fig:overlap_a} 
\end{subfigure}
\hfill
\begin{subfigure}[b]{0.49\textwidth}
\centering
\includegraphics[width=\textwidth]{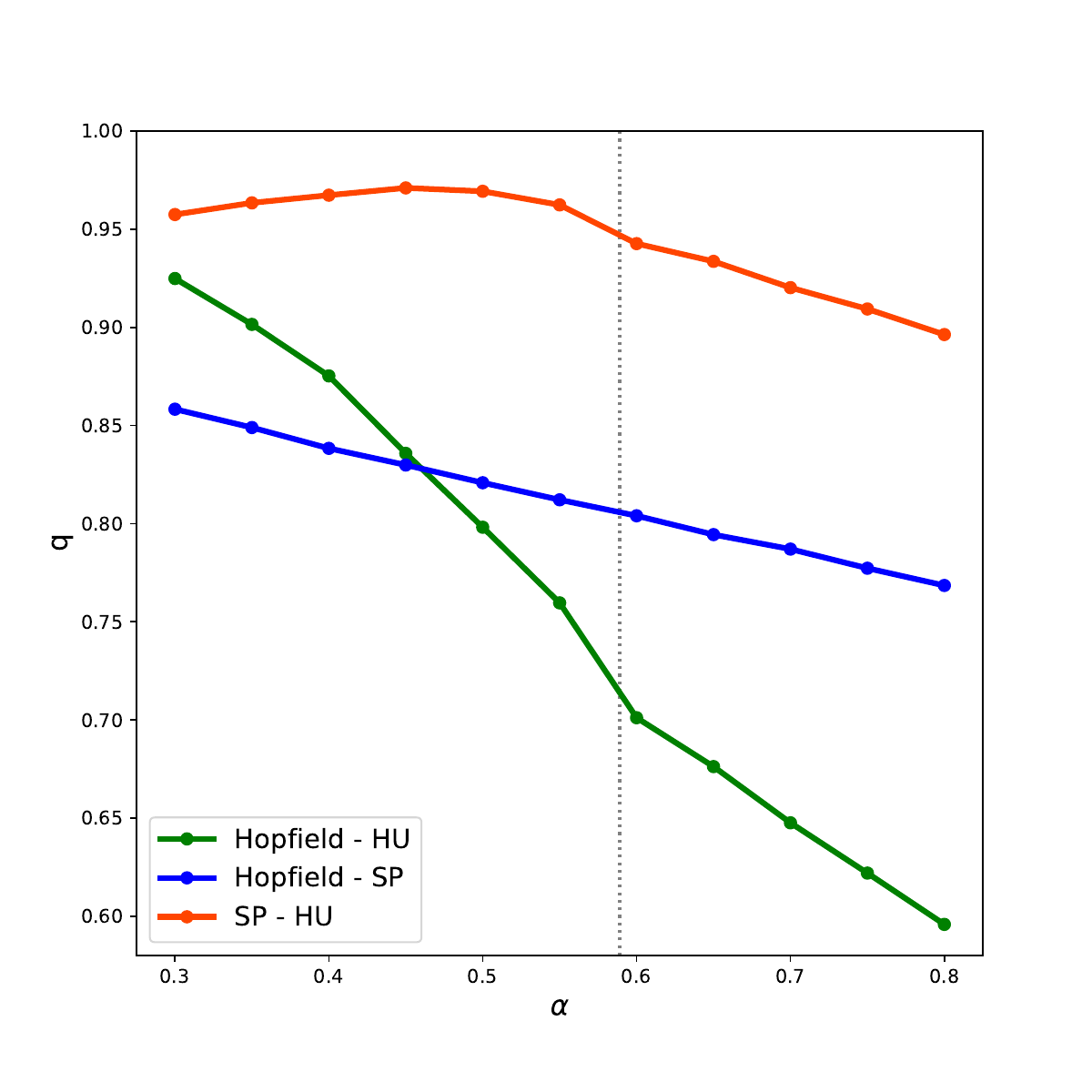}
\caption{}
\label{fig:overlap_b} 
\end{subfigure}
\hfill
    \caption{(a) Overlap $q$ between the final states of Hebbian Unlearning (HU) with $\epsilon = 10^{-2}$ at $d = D_{in}$ and symmetric perceptron (SP) with $\lambda = 1$ having reached a stability $k$. Measures are for different values of $N$ and $\alpha$ and points represent the mean computed over $5$ realizations of the disorder. Error bars are smaller than the data symbol. Values of $k$ range from $0$ to slightly below $k_{max}(\alpha)$. For each $\alpha$, $q$ peaks around $k_{max}(\alpha)$, indicating that the two algorithms converge to coupling matrices which are closest near to the value $k=k_{max}(\alpha)$. (b) Overlap $q$ as a function of $\alpha$ at $N = 800$. Points represent the mean of $10$ realizations of the disorder and error bars are  smaller than the data symbol. The orange symbols correspond to the overlap between the final states of SP with $\lambda = 1,\,k\simeq k_{max}(\alpha)$ and HU with $\epsilon = 10^{-2}$. For HU we chose $d = D_{in}$ for $\alpha<\alpha_c$, while for $\alpha>\alpha_c$ we chose $d=D_{top}$ ($\alpha_c$ is represented by the gray dotted line). The overlap between the initial Hebbian matrix and the final state of the HU (green) or SP (blue) with the same choice of the parameters is also shown. While in both algorithms the distance between initial and final matrix increases as $\alpha$ is increased, the distance between the final points remains small up to $\alpha_c$.}          
\end{figure*}
%\end{multicols} 

We first evaluate the final points where the two algorithms converge in the space of interactions. HU is stopped at $d = D_{in}$ as that is the relevant amount of iterations identified in section \ref{sec:UNL}. The SP is run at $\lambda = 1$. Fig. \ref{fig:overlap_a}, displays the overlap between the resulting matrices when the SP is performed at different values of $k$ before reaching $k_{max}(\alpha)$. The plot shows that $q$ increases with $k$, suggesting that HU pushes the system to the same region of solutions where the SP converges when $k$ is close to $k_{max}$. Finite size effects evidently appear near the abrupt transition from SAT to UNSAT, but the increase of $q$ with the size of the network suggests that the maximum overlap might be associated to the maximal stabilities when $N$ becomes large enough. 

The plot of $q$ as a function of $\alpha$, see \cref{fig:overlap_b}, shows how the distance between the final points and the initial Hebbian matrix increases when the number of memories becomes larger, while the distance between the two final points remains small and stable for $\alpha < \alpha_c^{HU}$. 
By comparing the final states of convergence we conclude that two networks, starting from the same initial matrix, end up in very similar configurations of the couplings $J_{ij}$. Now we analyze  the whole trajectory traced by the two algorithms in the space of interactions. 

\begin{figure*}[ht!]
\centering
\begin{subfigure}[b]{0.49\textwidth}
\centering
\includegraphics[width=\textwidth]{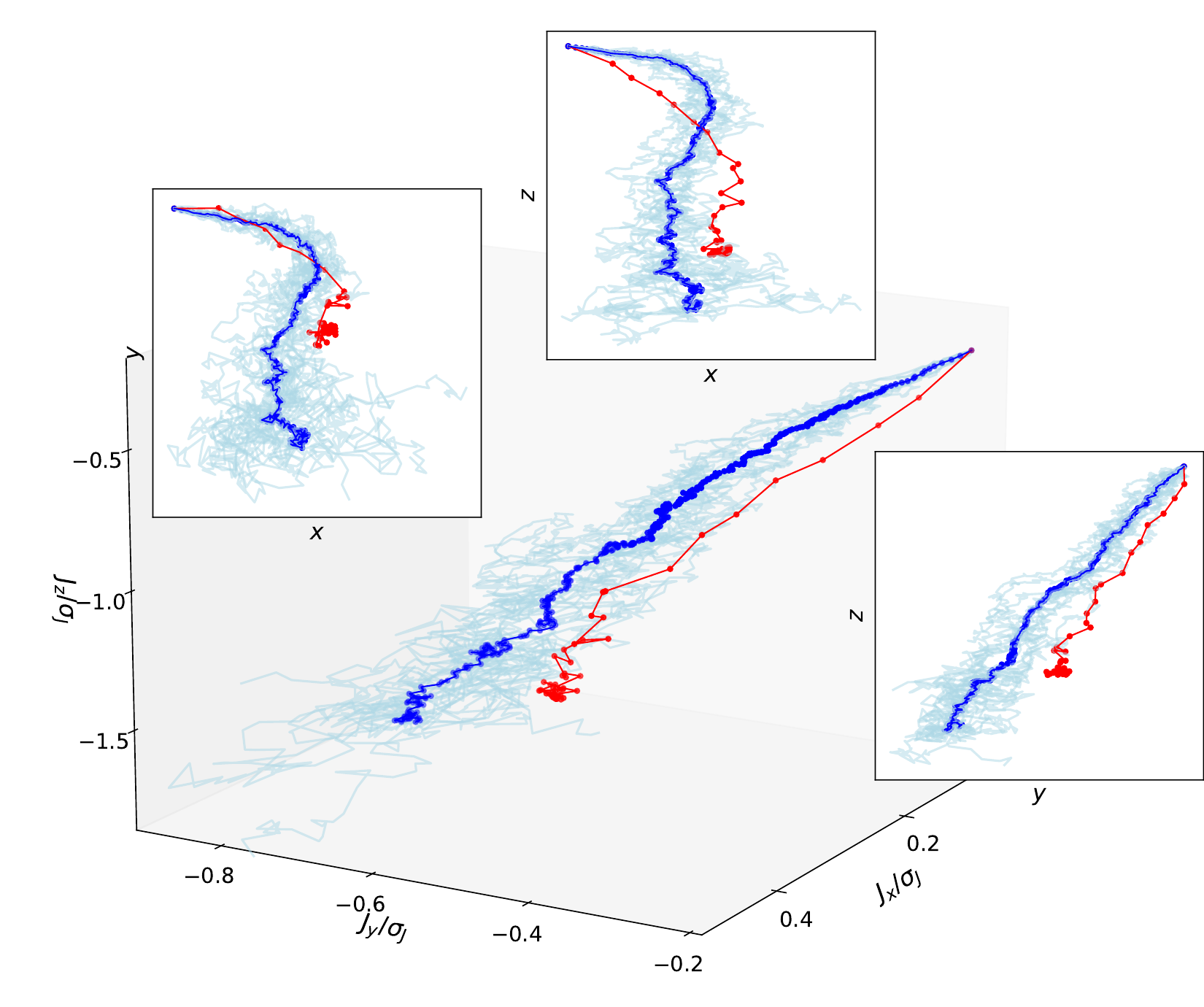}
\caption{}
\label{fig:traj_a} 
\end{subfigure}
\hfill
\begin{subfigure}[b]{0.45\textwidth}
\centering
\includegraphics[width=\textwidth]{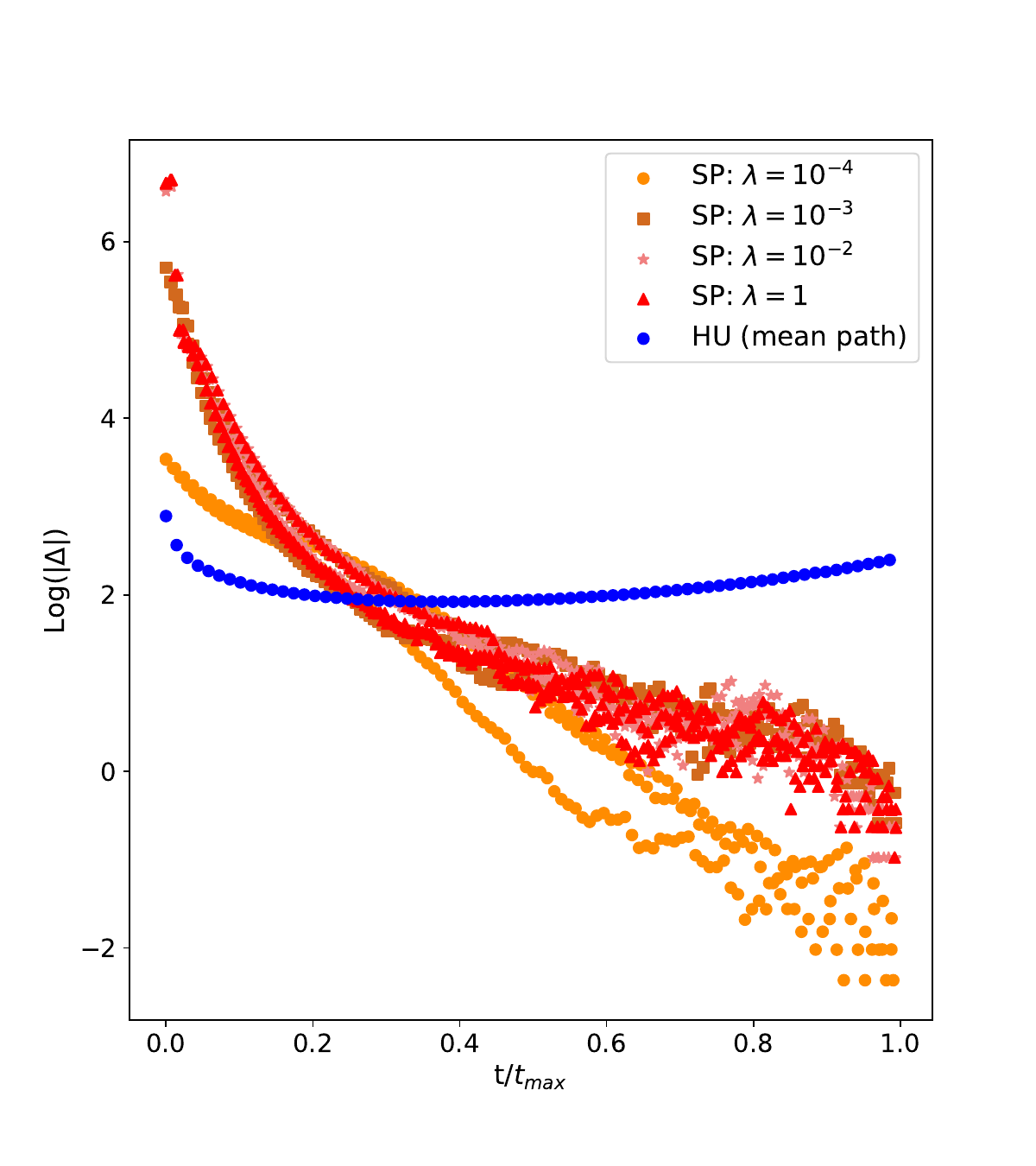}
\caption{}
\label{fig:traj_b} 
\end{subfigure}
\hfill
    \caption{(a) 3-dimensional projection of the trajectories followed by the system  in the space of interactions during the dynamics of Hebbian Unlearning (HU) and symmetric perceptron (SP). Numerical measurements have been taken for one sample at $N = 800$ and $\alpha = 0.55$, $\epsilon = 10^{-2}$, $\lambda = 10^{-4}$. $10$ trajectories of the HU are drawn in light blue, while the average Unlearning path is in blue. The path followed by the SP is depicted in red. Points represent different steps of the algorithms. HU has been resampled at regular intervals along the trajectory for simplicity of the data analysis. (b) Absolute value of the variation $\Delta \vec{J}$ in logarithmic scale as a function of the normalized time scale $t/t_{max}$ where $t_{max}$ is the maximum number of steps reached by the algorithm in a given sample. Numerical measurements are for one sample at $N = 800$ and $\alpha = 0.55$, $\epsilon = 10^{-2}$, $\lambda = 1, 10^{-2}, 10^{-4}$.  Three samples were simulated for the SP and one sample for the HU}          
\end{figure*}
%\end{multicols} 

We set $\alpha = 0.55$, so that the overlap between the initial and the final state is small enough, i.e. they are distant on the sphere, $N = 800$, $\lambda = 10^{-4}$ and $k$ close to $k_{max}$ in one single sample. The choice of a small value of the learning rate $\lambda$ allows to trace a continuous path in the space of the interactions. HU is run choosing $d = D_{in}$ for $10$ samples in total. Fig.~\ref{fig:traj_a} reports the projection of the resulting trajectories in the space of $J$ along three randomly chosen directions. The plot shows that the two algorithms explore the same region of the space of interactions, proceeding along a similar direction. We also observe that the convergence velocities of the two algorithms are very different. Indicating with $t$ the time steps for both processes, \cref{fig:traj_b} shows the logarithm of the absolute value of the \textit{variation} of vector $\vec{J}$, defined as 
\begin{equation}
    \label{eq:gradient}
    {\Delta}\vec{J}^{(t)} = \vec{J}^{(t+1)}/\sigma_{J}^{(t+1)} - \vec{J}^{(t)}/\sigma_{J}^{(t)} \ .
\end{equation}
The direction of this vector coincides with the one of the gradient followed by the algorithm in the space of interactions at a given time step.  

While the convergence speed of the HU does not significantly vary, the SP shows, at any scale of $\lambda$, an acceleration in time that resembles an exponential law. In other words, while HU explores the space of interactions nearly uniformly in speed, the SP takes about $15 \div 20$ time steps to reach a smaller condensed region where it gets confined until convergence.

%\begin{multicols}
\begin{figure*}[ht!]
\centering
\begin{subfigure}[b]{0.32\textwidth}
\centering
\includegraphics[width=\textwidth]{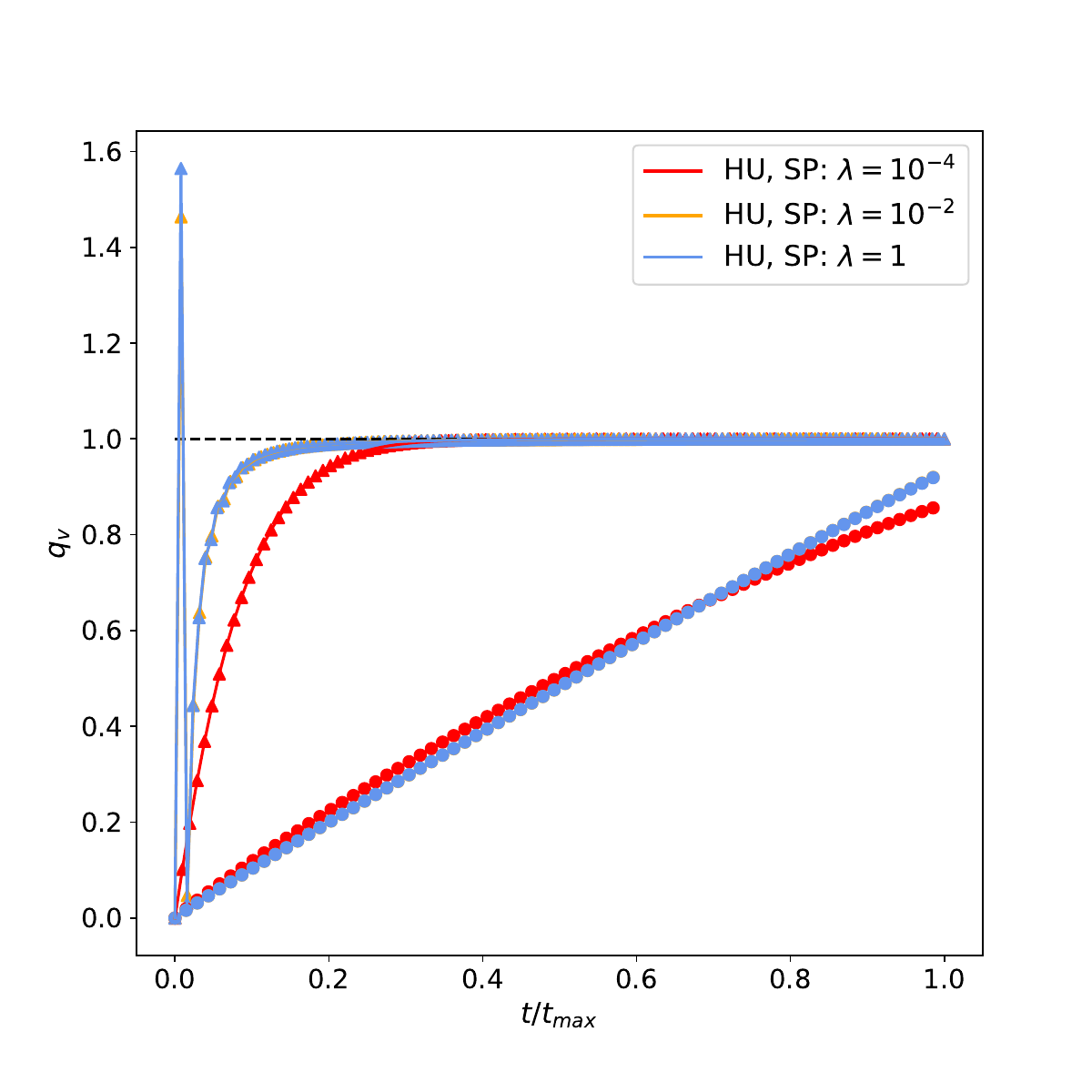}
\caption{}
\label{fig:gradients_a} 
\end{subfigure}
\hfill
\begin{subfigure}[b]{0.32\textwidth}
\centering
\includegraphics[width=\textwidth]{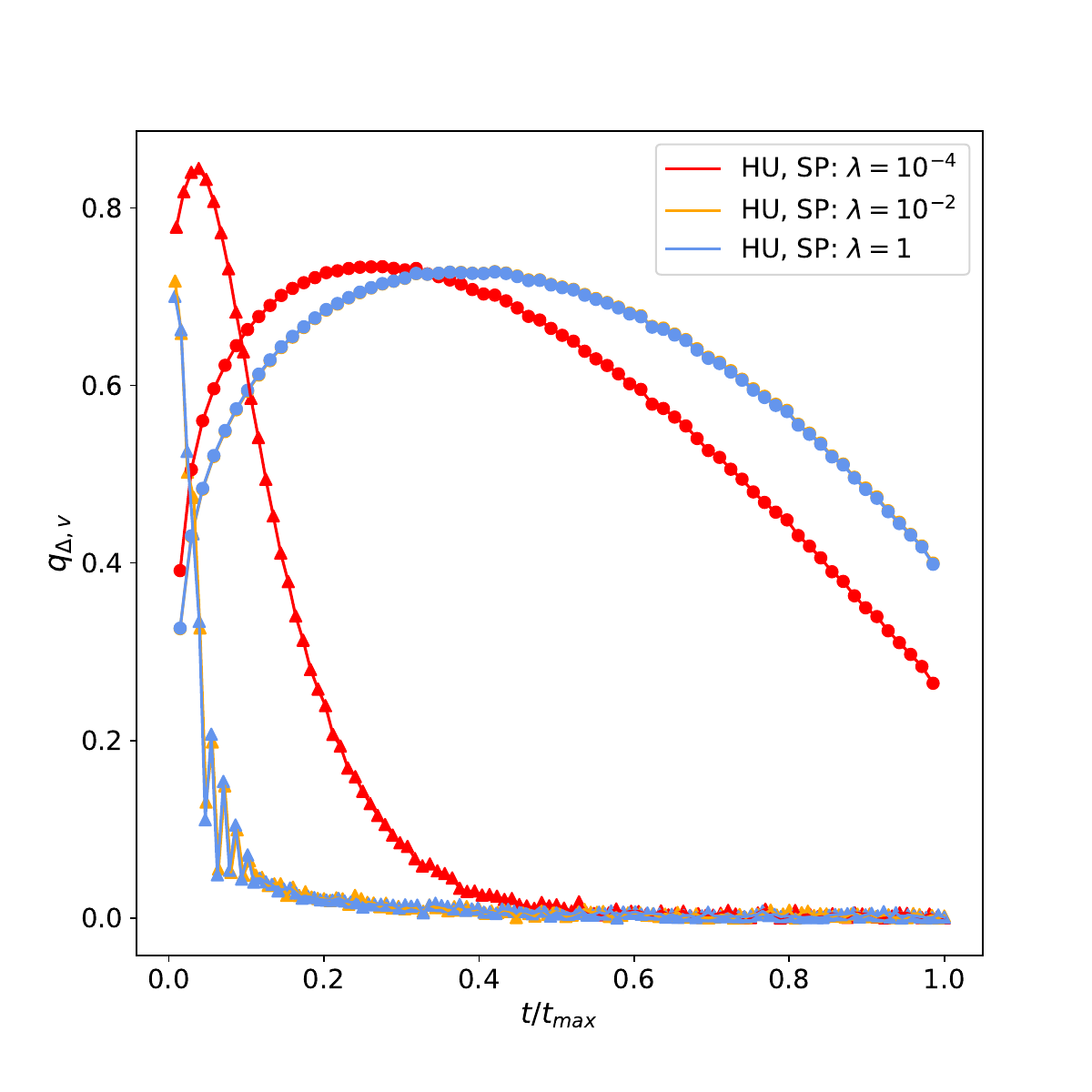}
\caption{}
\label{fig:gradients_b} 
\end{subfigure}
\hfill
\begin{subfigure}[b]{0.32\textwidth}
\centering
\includegraphics[width=\textwidth]{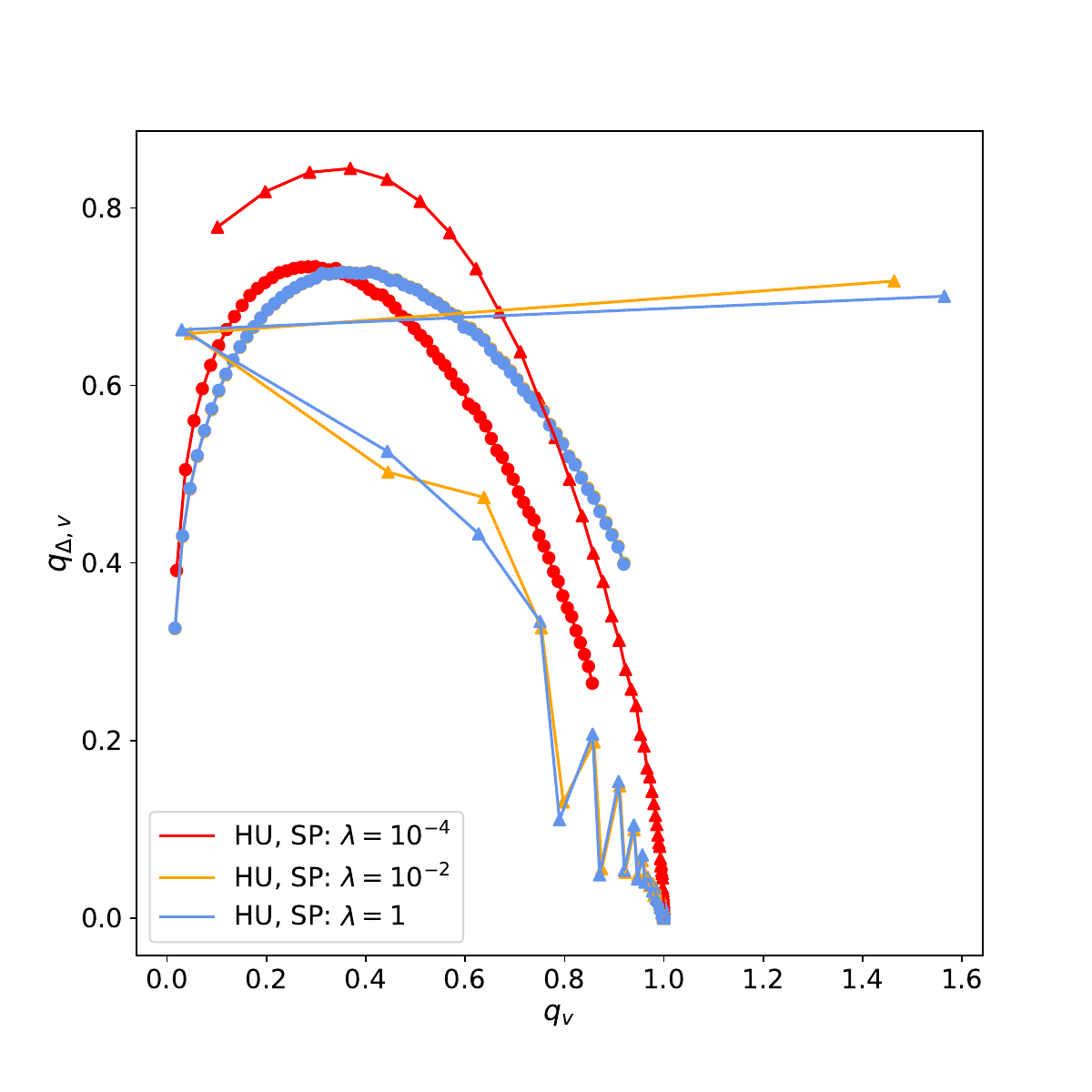}
\caption{}
\label{fig:gradients_c} 
\end{subfigure}
\hfill
    \caption{(a) $q_v$, angular distance of the trajectory from the reference direction $\hat{v}$,  as a function of time. (b) $q_{\Delta,v}$, projection of the variation along the direction $\hat{v}$, as a function of time. (c) $q_{\Delta,v}$ as a function of $q_v$. Numerical measurements are collected from one single sample at $N = 800$ and $\alpha = 0.55$ at $\epsilon = 10^{-2}$ for the Hebbian Unlearning (HU) (\textit{circles}) and different values of $\lambda$ for the symmetric perceptron (SP) (\textit{triangles}).}
\label{fig:gradients}           
\end{figure*}
%\end{multicols} 

The different speeds of the algorithms imply an inherent difficulty in comparing the trajectories \textit{point-by-point}. Our analysis will thus rely on defining a particular direction $\hat{v}$ in the space of interactions that we will use to compare the two trajectories and their gradients. Such a direction is defined by the line that connects the initial Hebbian matrix with the point of convergence of the SP,  
\begin{equation}
    \label{eq:v}
    \hat{v} = \frac{\vec{J}_{SP}^{(t_{max})}/\sigma_{SP}^{ (t_{max})} - \vec{J}^{(0)}/\sigma^{(0)}}{| \vec{J}_{SP}^{(t_{max})}/\sigma_{SP}^{ (t_{max})} - \vec{J}^{(0)}/\sigma^{(0)}|} \ .
\end{equation}
We can now define two time-dependent observables that can help us in the analysis of the trajectories:
\begin{equation}
    \label{eq:h}
    q_{v}(t) =  \frac{\vec{J}^{(t)}/\sigma^{(t)} - \vec{J}^{(0)}/\sigma^{(0)}}{|\vec{J}^{(t_{max})}/\sigma^{(t_{max})} - \vec{J}^{(0)}/\sigma^{(0)}|} \cdot \hat{v} \ ,
\end{equation}
which is a measure of the angular distance of any point of the trajectory from the line traced by the direction $\hat{v}$ at time $t$. The smaller this quantity is, the more evidently the trajectory is diverging from $\hat{v}$. One can also introduce
\begin{equation}
    \label{eq:q_grad}
    q_{\Delta, v}(t) =  \frac{\vec{J}^{(t+1)}/\sigma^{(t+1)} - \vec{J}^{(t)}/\sigma^{(t)}}{|\vec{J}^{(t+1)}/\sigma^{(t+1)} - \vec{J}^{(t)}/\sigma^{(t)} |} \cdot \hat{v} \ ,
\end{equation}
that is, the projection of the variation of $\vec{J}$ at the step $t$ along the direction $\hat{v}$. The larger this quantity is, the more aligned to $\hat{v}$ the trajectory is.

Fig.~\ref{fig:gradients_a} represents the values of $q_v$ during the same trajectory that is depicted in 
fig.~\ref{fig:traj_a}. 
One can see that $q_v(0)$ is small for both the HU and the SP. This means  that they both start in the wrong direction: this is particularly reasonable for the HU, which involves a random picking of the initialization state. However, while the SP rapidly reaches $q_v = 1$ because of its high initial acceleration at all the considered values of $\lambda$, in the HU algorithm $q_v$ is increasing at lower rate. An initial overshooting of the SP at high values of $\lambda$ is signaled by an anomalously high value of $q_v$ at the second step of the process.

The directions followed by the two algorithms with respect to $\hat{v}$ are shown in \cref{fig:gradients_b}. The SP at $\lambda = 10^{-4}$ has a peak in $q_{\Delta,v}$ in the first part of the trajectory, signaling a high degree of alignment between the gradient and $\hat{v}$. Later on, the SP rapidly converges towards the condensed region, where gradients lose their polarization with $\hat{v}$, but the convergence point has already been reached. When higher values of $\lambda$ are used, no relevant polarization is measured. The HU shows a similar behavior: the trajectory starts along a direction that is barely aligned with $\hat{v}$ but a consistently high degree of alignment is obtained after more or less half of the iterations. Eventually, the HU also converges towards the final state losing the alignment with $\hat{v}$. Fig.~\ref{fig:gradients_c} displays the direction of the variation as a function of the distance from $\hat{v}$. Three different behaviors of the trajectory can be thus recognized for both algorithms. First the trajectory moves away from $\hat{v}$ after a bad start, in a second phase it aligns to $\hat{v}$, and in a third phase the matrix plunges towards the convergence state.

\subsection{A geometric interpretation of the effective weights}

\begin{figure}[ht!]
\centering
\includegraphics[width=.7\textwidth]{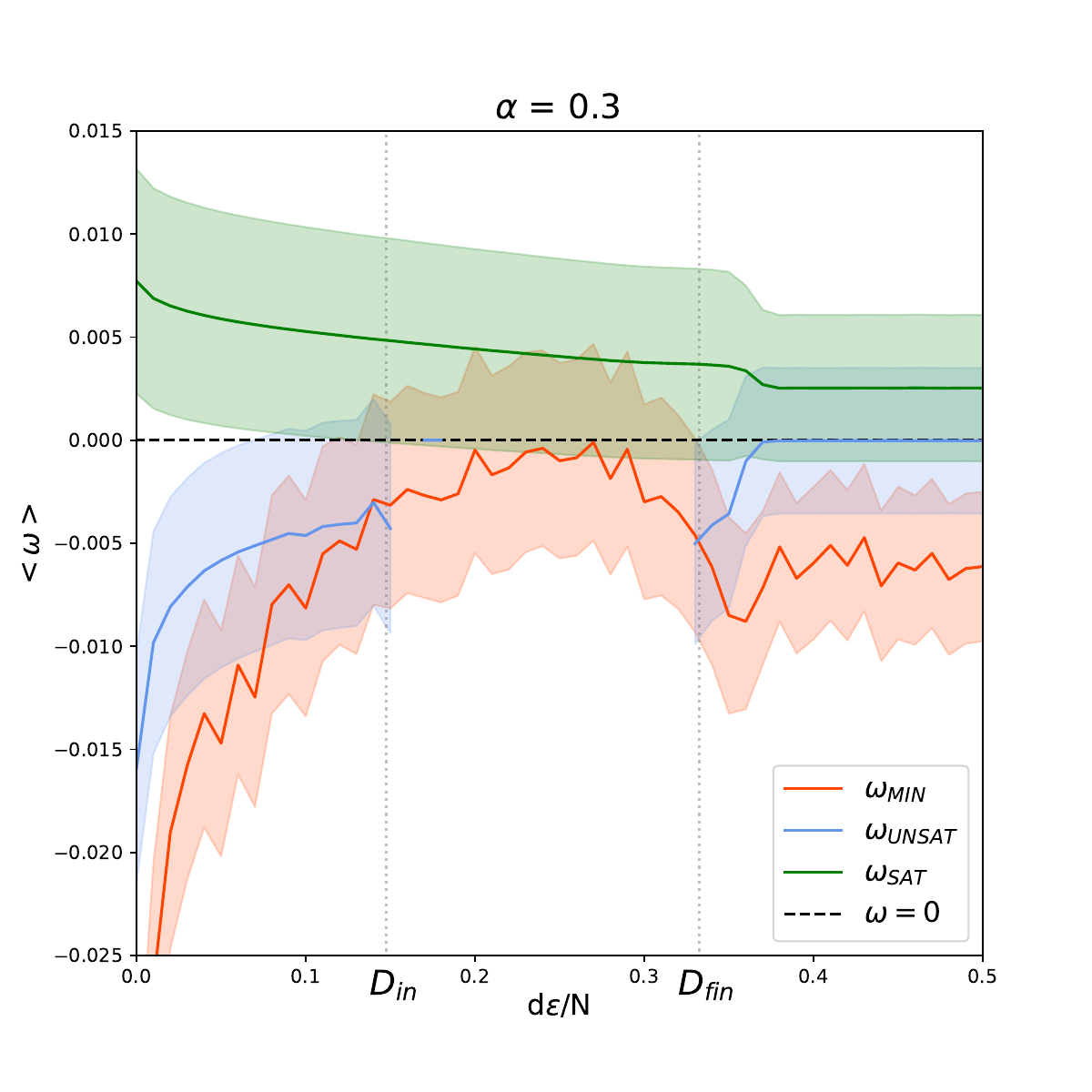}
    \caption{Mean values assumed by the perceptron overlap $\omega$ for SAT, UNSAT and $min$ patterns during the Hebbian Unlearning process. Measurements are performed at $N = 800$, $\alpha = 0.3$, $\epsilon = 10^{-2}$ over $200$ spurious states and several realizations of the disorder. Error bars are indicated by the shaded region.}
\label{fig:percept_overlap}           
\end{figure}
Considering our analysis of the optimal structure of noise and the robustness properties of a network trained via HU, we provide for a physical interpretation of the effective weights $\omega_i^{\mu}$ that were so fundamental for the TWN algorithm (and so to HU when fixed points are employed) to reproduce the SVM performance.\\ Let us introduce the vector $\vec{J}_i$ as the collection of the elements contained in the $i^{th}$ row of the connectivity matrix, $\vec{\eta_i}^{\mu}$ as the \textit{memory pattern} defined as 
\begin{equation}
    \vec{\eta_i}^{\mu} = \xi_i^{\mu}\vec{\xi^{\mu}},
\end{equation}
and eventually the \textit{glassy pattern}, defined in analogy with the memory patterns as
\begin{equation}
    \label{eq:glassy_pattern}
    \vec{\eta_i}^* = S_i^*\vec{S}^*.
\end{equation} 
Let us now rewrite a \textit{positive} perceptron update, i.e. one where the mask $\epsilon_{ij}^{\mu}$ assumes a non-zero value, as
\begin{equation}
    \delta \vec{J}_i^{P,d} = +\frac{\lambda}{N}\vec{\eta_i}^{\mu}.
\end{equation}
We can also rewrite the HU update in the same vectorial fashion, i.e.
\begin{equation}
    \label{eq:vec_unlearning}
    \delta\vec{J_i}^{HU,d} = - \frac{\epsilon}{N}\vec{\eta_i}^*\ ,
\end{equation}
Hence, the effective weights $\omega_i^{\mu}$, for the case of data-points being fixed points of the dynamics, are
\begin{equation}
    \label{eq:percept_overlap}
    \omega_i^{\mu} =  \frac{1}{N}\vec{\eta_i}^* \cdot \vec{\eta_i}^{\mu}\ .
    %\omega_i^{\mu} =  \frac{1}{N}\overline{\langle\vec{\eta_i}^* \cdot \vec{\eta_i}^{\mu}\rangle}\ ,
\end{equation}
It comes natural to reinterpret $\omega_i^{\mu}$ as the projection of the synaptic update for HU on the SP positive one, i.e.
\begin{equation}
    \label{eq:percept_overlap2}
    \omega_i^{\mu} \propto -\delta\vec{J_i}^{SP} \cdot \delta\vec{J_i}^{HU}
\end{equation}
at each iteration $d$. This implies that the more negative $\omega_i^{\mu}$ the higher is its contribution to align HU with SP learning. 
We know that each pair $(i,\mu)$ is related to a given constraint of the associated optimization problem, with $\Delta_i^{\mu} \geq 0$ for SAT constraints, and $\Delta_i^{\mu} < 0$ for UNSAT ones. Fig.~\ref{fig:percept_overlap} shows $\overline{\langle\omega_i^{\mu}\rangle}$, with the brackets indicating the average over the spurious states in a given realization of the disorder, at $\alpha = 0.3$ and $N = 800$ for the three types of constraints: SAT, UNSAT and \textit{minimally satisfied}, i.e. the SAT constraints with the lowest measured stability. The fact that the perceptron overlap is negative for both UNSAT and \textit{min} constraints, but positive for SAT constraints, suggests that the distribution of the $\vec{\eta}_i^{\mu}$ looks anisotropic from the reference frame of the glassy patterns. This is certainly induced by the fact that glassy patterns $\vec{\eta_i}^*$ are SAT by definition, so they are more likely to be contained in the same half of hyperspace, defined by the orthogonal plane to $\vec{J}_i$, that contains SAT memory patterns. 
Consequently, since there is a minus sign on r.h.s. of \cref{eq:vec_unlearning}, HU is performing the same geometric transformation of the perceptron in order to align the $\vec{J}_i$ vectors to the memory patterns $\vec{\eta_i}^{\mu}$. By only exploiting the rugged shape of the Hebbian landscape out of the retrieval regime, the HU algorithm manages to accomplish this task in an quasi-optimal way. 
\subsection{Evolution of the Unlearning algorithm}
\begin{figure*}
\centering
\includegraphics[width=0.6\textwidth]{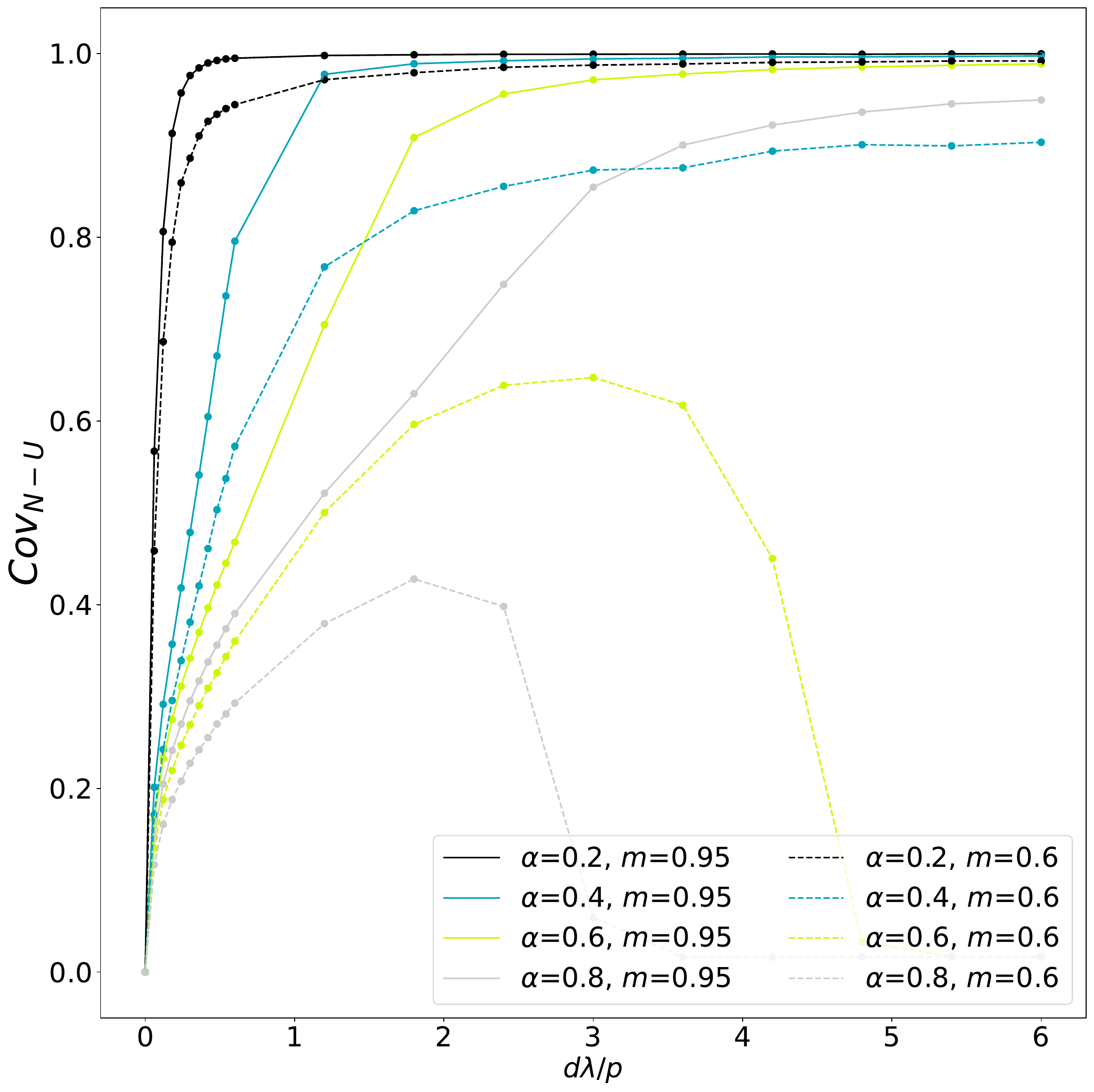}
   \caption{Correlation between \textit{noise} and \textit{unlearning} contributions to $\delta J$ as a function of the rescaled number of training steps $\frac{d\lambda}{p}$, for two values of the training parameter $m$, and different values of $\alpha$. When $\text{Cov}_{\text{N-U}}=1$, the algorithm stops modifying the coupling matrix. Results are averaged over $100$ samples. Choice of the parameters: $N=400$, $\lambda=10^{-2}$.}
\label{fig:ave_contribution_corr}      \end{figure*}

\begin{figure*}
\centering
    \includegraphics[width=0.6\textwidth]{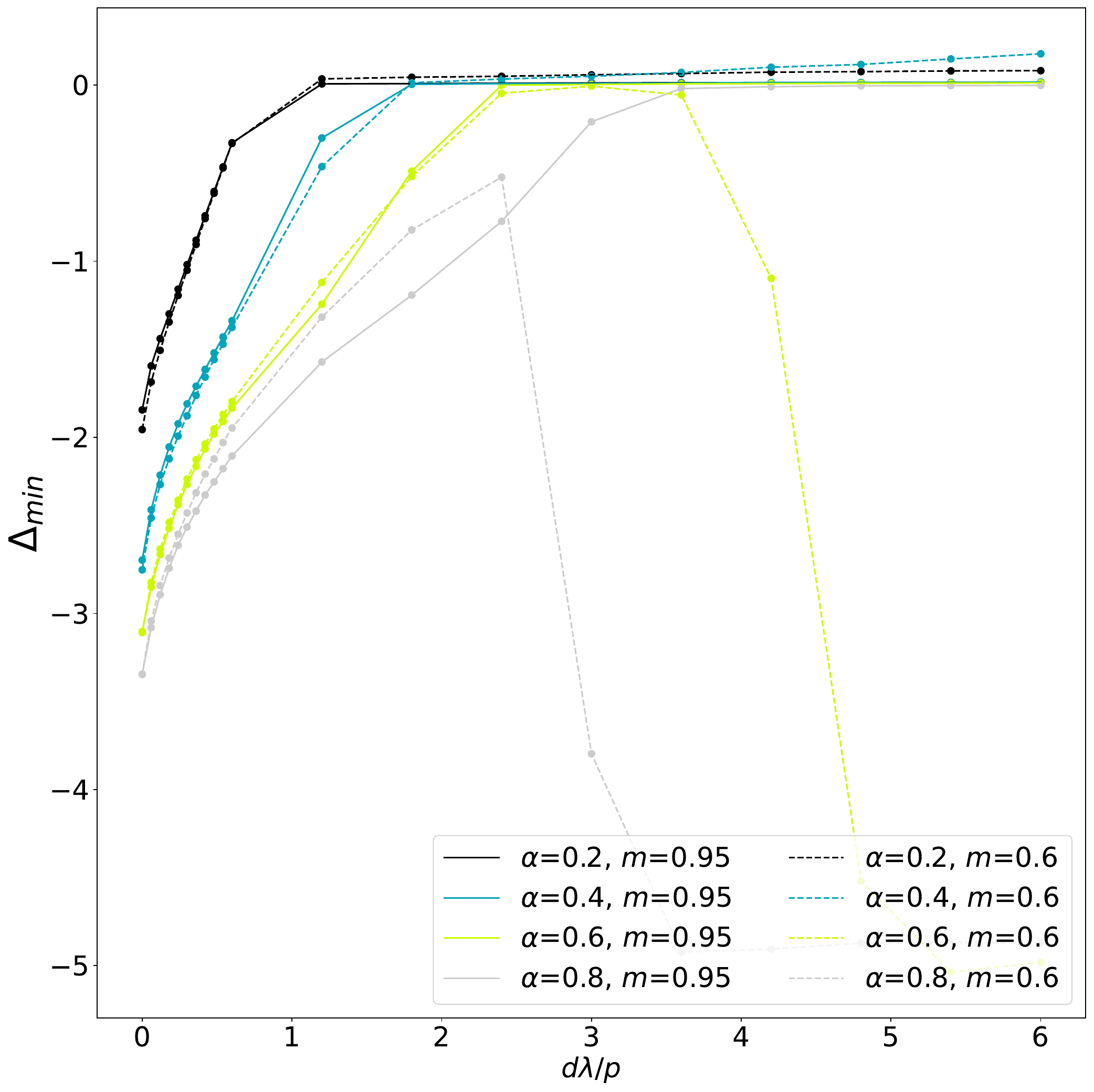}
   \caption{Minimum stability as a function of the rescaled number of training steps $\frac{d\lambda}{p}$, for two values of the training parameter $m$, and different values of $\alpha$. When $\Delta_{\text{min}}\geq 0$, each memory is a stable fixed point of the dynamics. Results are averaged over $100$ samples. Choice of the parameters: $N=400$, $\lambda=10^{-2}$.}
\label{fig:ave_stab_min}           
\end{figure*}

When training configurations are fixed points of the dynamics with $m_t>0^+$, the considerations of the previous section no longer apply. Such configurations can be generated by initializing the network at an overlap $m >0^+$ with a memory, and let it evolve according to (\ref{eq:dynamics}). The connectivity matrix is initialized according to Hebb's learning rule (\ref{eq:hop}). In this setting, if at some point during training $m$ happens to enter the basin of attraction of the memories, the fixed point reached by the dynamics will be the memory itself, and the noise contribution will cancel exactly the learning contribution, giving $\delta J=0$. On the other hand, for sufficiently high values of the load $\alpha$, at the start of the training procedure memories will have zero size basins of attraction and trajectories will drift away from the memories following the dynamics. In this scenario, the two contributions $\delta J^N$ and $\delta J^U$ decorrelate, and $\delta J^N$ will take again a similar role to what described in the previous section. The result is an algorithm which interpolates between HU when $d$ is small and a supervised algorithm when basins increase to a size close to $(1-m)$. In this regime, $\delta J^N$ acts as a breaking term, preventing the algorithm to further modify the coupling matrix $J$. A similar mechanism has been studied in \cite{poppel_dynamical_1987} where a supervised term was added to the standard HU update rule, leading to $\delta J_{ij}\propto-S^{\mu_d}_iS^{\mu_d}_j+ \xi^{\mu_d}_i\xi^{\mu_d}_j$. Notice that the term $\xi^{\mu_d}_i\xi^{\mu_d}_j$ is deterministically reproducing Hebb's learning rule, while in our study the HU is modified by a stochastic term, whose nature we have already discussed. Given a sufficiently small learning rate $\lambda$, there will exist a characteristic number of steps of the algorithm over which the coupling matrix does not change significantly. We will refer to this timescale as \textit{epoch}. Averaging the effect of training steps over an epoch, we get a snapshot of how the algorithm is affecting the couplings at a given point during training. This can be used to study the relation between $\delta J^N$ and $\delta J^U$, quantified by the connected correlation coefficient 
\begin{equation}
\small
    \text{Cov}_{\text{N-U}}:=\frac{2}{N(N-1)}\sum_{i,j>i}\big(\langle\delta J^N_{ij}\delta J^U_{ij}\rangle_{epoch} -\langle\delta J^N_{ij}\rangle_{epoch}\langle\delta J^U_{ij}\rangle_{epoch}\big),
\end{equation}
where the average is computed over an epoch. When this quantity equals one, there is no effective update of the couplings over an epoch. Results are presented in \cref{fig:ave_contribution_corr} for different values of $m$ and of $\alpha$, as a function of the number of training steps $d$. The number of iterations has been rescaled by a factor $p/\lambda$ for clarity of the plot. At any given $\alpha$, training with higher $m$ results in a faster increase of $\text{Cov}_{\text{N-U}}$, i.e. a faster convergence of the algorithm. If the value of $m$ is too low, the algorithm never manages to build a big enough basin of attraction, and never stops. As $\alpha$ is increases, higher and higher values of $m$ are required for the algorithm to stop, since the typical size of the basins of attraction shrinks. \\
The network performance can be benchmarked by tracking the evolution of the lowest stability $\Delta_{\text{min}}$ throughout the training procedure. Results are presented in \cref{fig:ave_stab_min} for different values of $\alpha$ and $m$. For sufficiently low values of $\alpha$ and sufficiently high values of $m$, $\Delta_{\text{min}}$ surpasses the zero. Once this condition is met, the value $\Delta_{\text{min}}$ becomes essentially constant, even if \text{Cov}$_{\text{N-U}}<1$, signaling that the update of the coupling matrix is still in progress. The result is a curve $\Delta_{\text{min}}(d)$ which barely surpasses zero. For each value of $m$ there exists a critical value $\alpha_c(m)$ beyond which no amount of steps is able to produce $\Delta_{\text{min}}>0$. Extrapolating empirical results to the $N\to\infty$ limit, one finds 
\begin{equation*}
    \alpha_c(m)=A\cdot(m)^B+C,
\end{equation*}
where
\begin{equation*}
    A=0.35\pm0.01, \quad B=6.9\pm0.5, \quad C=0.58\pm0.01
\end{equation*}
Consistently with what presented in the previous section, in the limit $m\to 0^+$ one finds the critical capacity of the Unlearning algorithm \cite{van_hemmen_increasing_1990, benedetti_supervised_2022}. The critical capacity increases up to a value $\alpha_c(1)=0.93\pm 0.01$ when $m$ reaches unity.

\begin{figure*}
\centering
\includegraphics[width=0.6\linewidth]{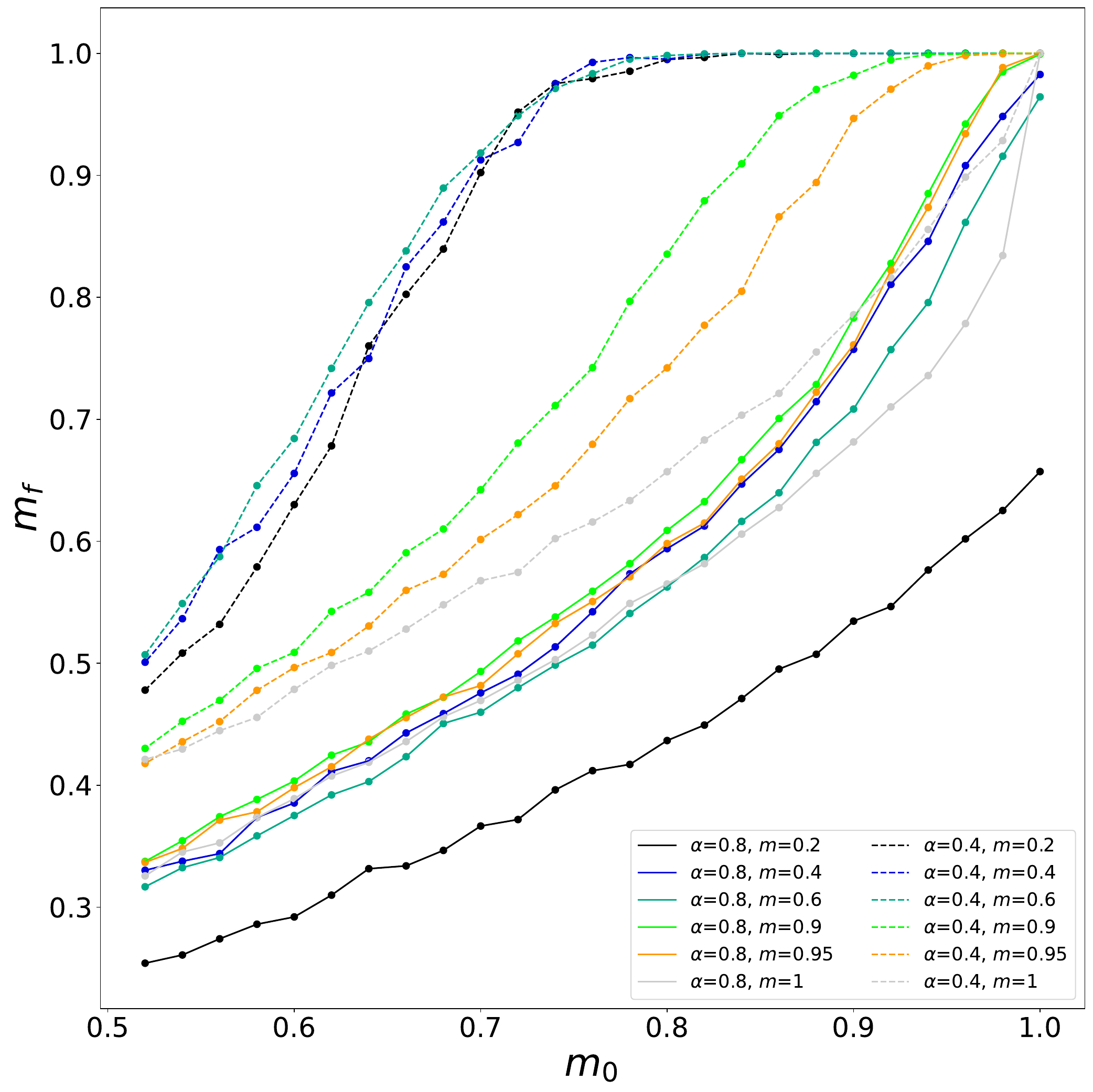}
\caption{Retrieval map $m_f(m_0)$ for two values of $\alpha$ and different values of the training parameter $m$. At $\alpha=0.4$, every $m$ leads to stable memories, i.e. $m_f(1)=1$. At $\alpha=0.8$, only the highest values of $m$ lead to stable memories, while for low values of $m$ one has $m_f(1)<1$. Results are averaged over $100$ samples. Choice of the parameters: $N=400$, $\lambda=10^{-2}$.}
\label{fig:mfmi}           
\end{figure*}

Regardless of whether $\Delta_{\text{min}}>0$ is obeyed, one can monitor the network performance as an associative memory device by measuring the retrieval map $m_f(m_0)$. We find that the best performance always corresponds to the number of training step maximizing $\Delta_{\text{min}}$, hence the curves $m_f(m_0)$ are all relative to this number of steps.
 Results are presented in \cref{fig:mfmi}. When perfect-retrieval is achieved (i.e. $\alpha < \alpha_c(m)$), lower values of $m$ increase the degree of robustness of the network (i.e. enlarge the basins of attraction), at the cost of a lower critical capacity.

\subsection*{Checkpoint}
In this section we have seen that:
\begin{itemize}
    \item The TWN algorithm formally converges to HU when training data are stable fixed points having an overlap $m_t = 0^+$ with the memories. 
    \item Consistently with the theory, the HU algorithm approaches the performance of a SVM (both in terms of perfect-retrieval and robustness) in a full unsupervised way: it is thus faster and more biologically relatable. This behaviour is conserved until a critical capacity $\alpha_c^{HU}\simeq 0.6$.  
    \item An explanation for the quality of the stable fixed points that can be sampled from a landscape initialized with the Hebbian rule can be found in the disposition of this states in the configuration space. When applying a particular transformation of the space of the network configurations (see \cref{eq:glassy_pattern}), stable fixed points appear to be distributed anisotropically around the memories.
    \item The TWN algorithm has be exploited to propose a supervised version of the HU algorithm that is fast (because it keeps sampling the stable fixed points of the landscape) and reaches a higher critical capacity, i.e. $\alpha_c \simeq 0.93$.
\end{itemize}
\newpage
\section{\label{sec:corr}The case of correlated memories}
\begin{figure*}[ht!]
\begin{tabularx}{\linewidth}{*{2}{X}}
     \centering
     \begin{subfigure}[b]{0.45\textwidth}
         \centering
         \includegraphics[width=\textwidth]{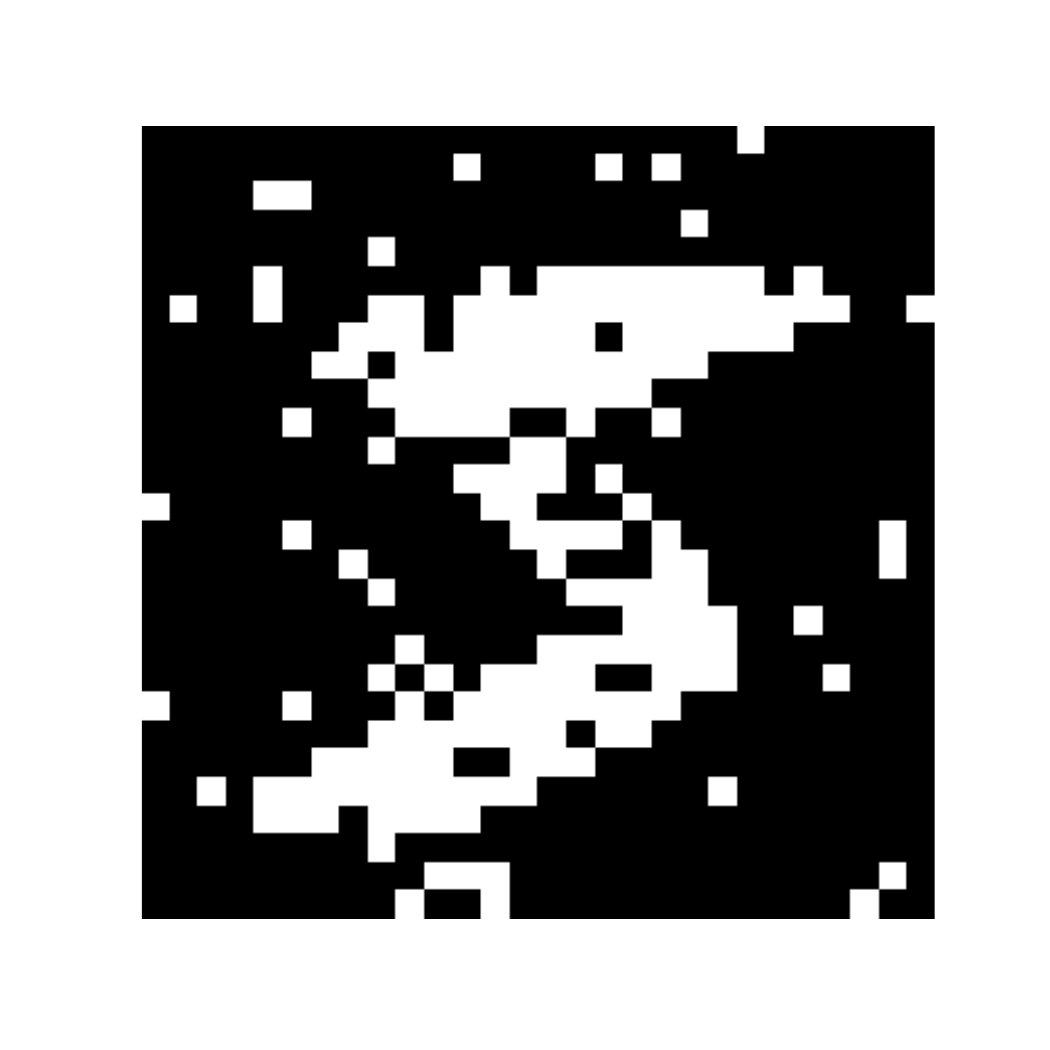}
         \caption{Data-point}
         \label{fig:mnist_a}
     \end{subfigure}
     \hfill
     \begin{subfigure}[b]{0.45\textwidth}
         \centering
         \includegraphics[width=
        \textwidth]{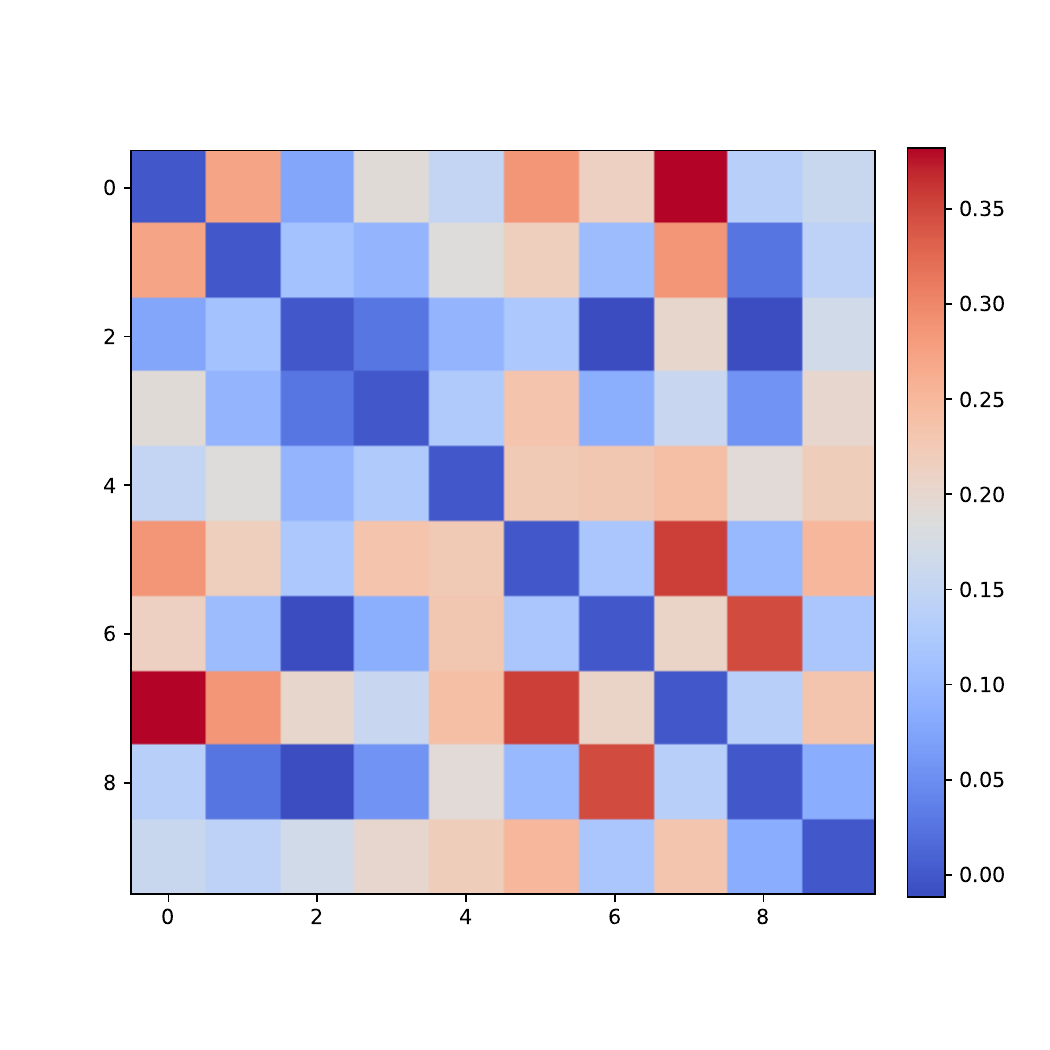}
         \caption{Correlation matrix}
         \label{fig:mnist_b}
     \end{subfigure}
     \hfill  
\end{tabularx}     
\caption{Representation of the disordered MNIST set. The data-point is represented in the matrix shape, while the correlation matrix is computed with $10$ examples and the unitary diagonal entries is set to $0$ to enhance the off-diagonal terms.}
\end{figure*}

\begin{figure*}[ht!]
\begin{tabularx}{\linewidth}{*{2}{X}}
     \centering
     \begin{subfigure}[b]{0.45\textwidth}
         \centering
         \includegraphics[width=\textwidth]{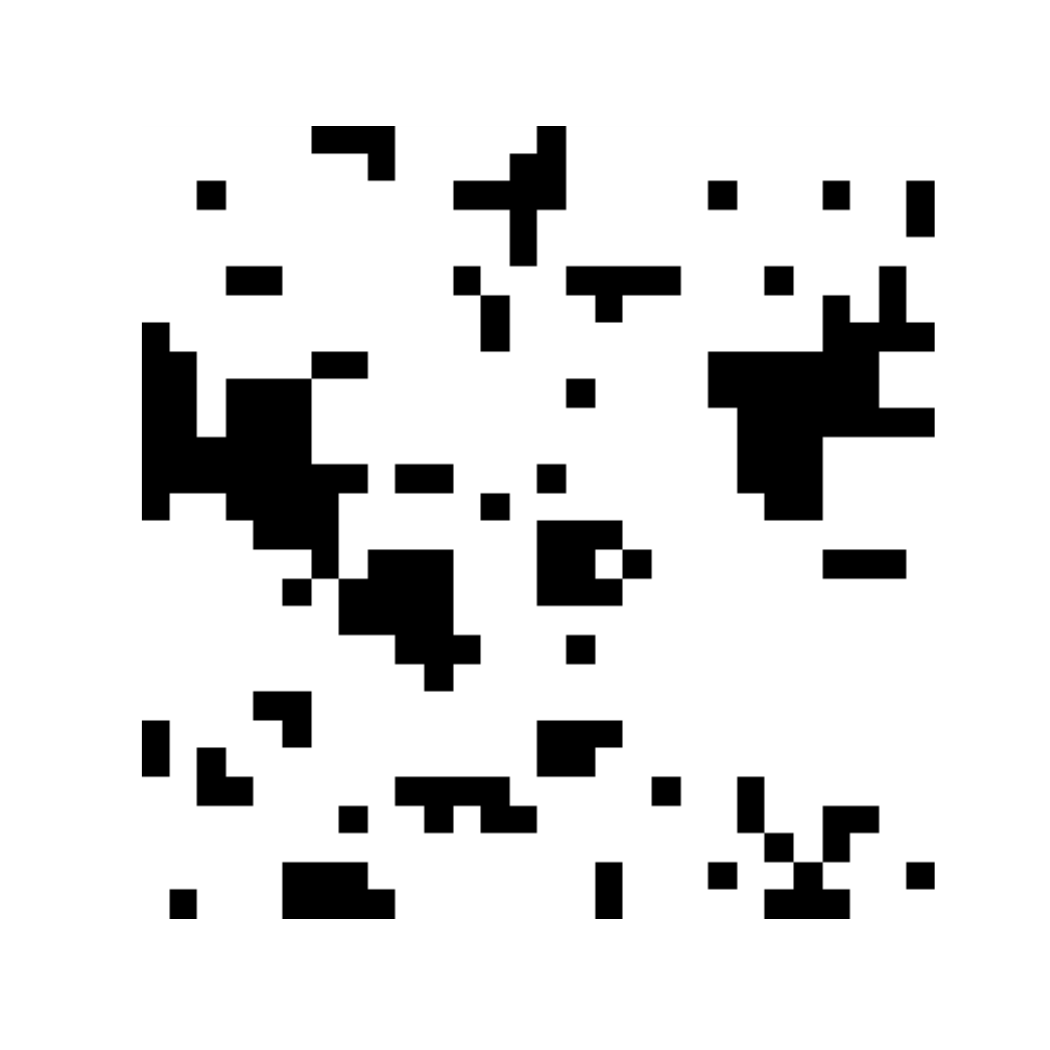}
         \caption{Data-point}
         \label{fig:ising_a}
     \end{subfigure}
     \hfill
     \begin{subfigure}[b]{0.45\textwidth}
         \centering
         \includegraphics[width=
        \textwidth]{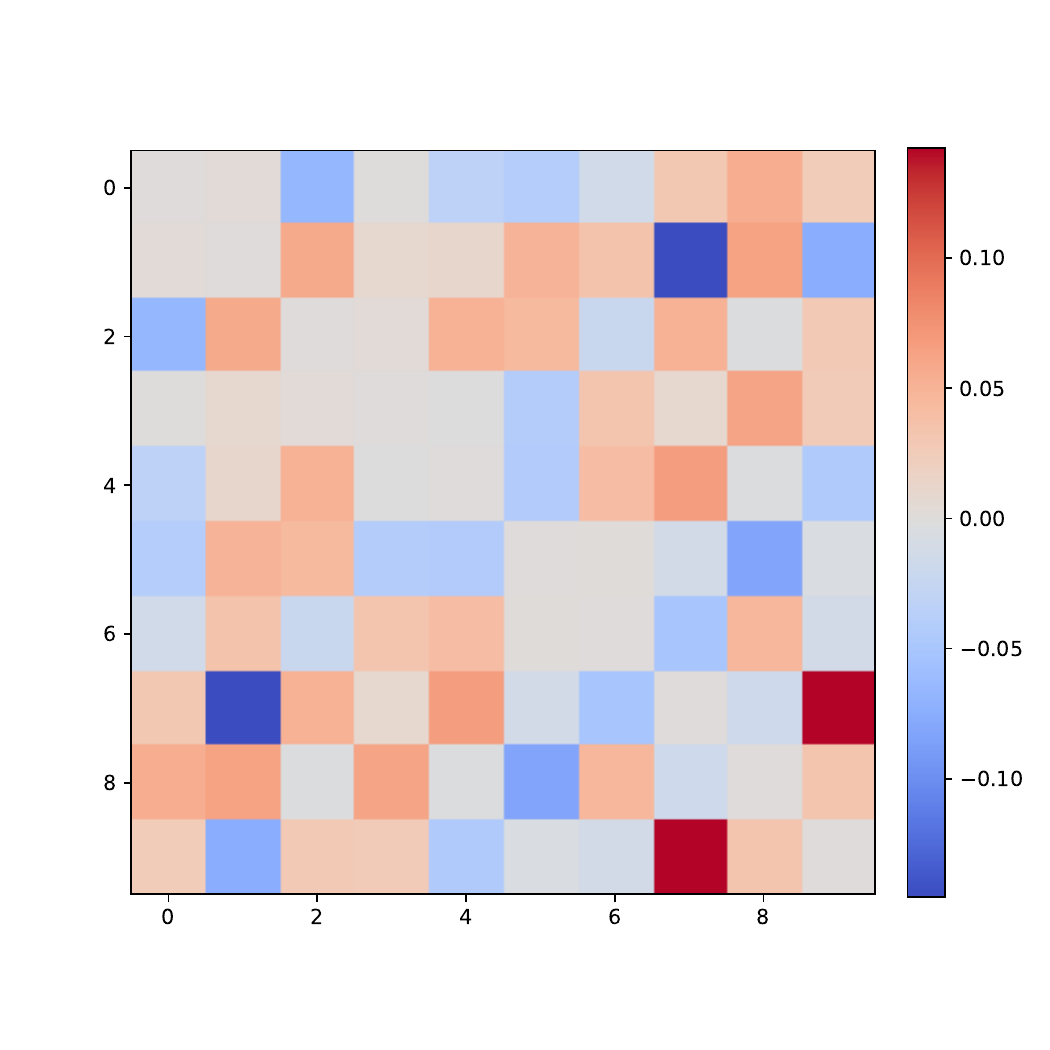}
         \caption{Correlation matrix}
         \label{fig:ising_b}
     \end{subfigure}
     \hfill  
\end{tabularx}     
\caption{Representation of the Ising set. The data-point is represented in the matrix shape, while the correlation matrix is computed with $10$ examples and the unitary diagonal entries is set to $0$ to enhance the off-diagonal terms.}
\end{figure*}

We now consider two sets of memories $\{\vec{\xi^{\mu}}\}_{\mu = 1}^{\alpha N}$ that contain different types of internal correlations and evaluate whether the new regularization technique we found in terms of noise structure can be applied in the case of correlated memories. \\ One set is composed by MNIST images representing handwritten digits, conveniently reshaped into one-dimensional arrays. Each image is a square matrix of $N = 28^2 = 784$ pixels assuming values between $0$ and $255$. Each pixel $\xi_i^{\mu}$ is then transformed into $-1$ if its original value is $0$ and $+1$ if its original value is larger than $0$. In addition to this we flip the entries of the vector with a probability $p = 0.1$ to inject some disorder in the data-point. This operation is necessary to smooth the very strong biases that might cancel $\sigma_i$, standard deviation of the coupling matrix along line $i$, inducing some relevant quantities to diverge.\\The other set of memories under consideration is composed by configurations sampled from a 2-dimensional Ising model in the paramagnetic phase (i.e. with at different temperatures larger than the critical one). Each configuration is also presented as a $N = 784$ sized array with $\xi_i \in \{-1,+1\}$. Figures \ref{fig:mnist_a} and \ref{fig:ising_a} show examples of training configurations from these collections.

Both types of memories have internal correlations among their elements. MNIST data are very regular: the digit is written in the middle of the image and each digit is sufficiently symmetric, under the geometric point of view. On the other hand, Ising configurations have thermal correlations. Moreover, different memories can be mutually correlated across the set. In particular, MNIST digits are repeated across the set, and they might be generally similar with each other (e.g. a 'four' does not look very different from a 'one', as well as a 'eight' might look similar to a 'zero'). On the contrary, Ising configurations have low mutual correlations, because they are sampled in the paramagnetic phase, where there are no biases due to the non-zero magnetization. These aspects appear evident from the correlation matrices reported in figures \ref{fig:mnist_b}, \ref{fig:ising_b}.    

\subsection{Numerical analysis}
In order to characterize the quality of the noise contained into the training data, we generate Hebbian couplings $J^{H}_{ij}$ from the memories and perform a Monte Carlo routine at different temperatures $T$, to probe the energy landscape at different levels, as previously done for the set of random memories.
As we know from the theory in \cref{sec:appA2} the gradient of the Loss function of a SVM is given by 
\begin{equation}
\label{eq:deltaL_tot2}
\small
\delta\mathcal{L} \propto \lim_{m\rightarrow 1^-}\sum_{i,\mu}^{N,p}\frac{1}{2\sigma_i}\left[(m_{\mu}\chi_i^{1,\mu} + m_{1,\mu}\chi_i^{\mu}) - (m_{\mu}\xi_i^{\mu}\xi_i^{\mu_d} + M_{\mu}^{\mu_d}\chi_i^{\mu})\right] \exp{\left(-\frac{m^2\Delta_i^{\mu^2}}{2(1-m^2)}\right)}
\end{equation}
that can be rewritten in a more compact fashion as
\begin{equation}
    \delta \mathcal{L} = \delta \mathcal{L}_{N} + \delta \mathcal{L}_{U}.
\end{equation}
The quantity indicated as $M_{\mu}^{\mu_d}$ is an overlap between two memories, defined as $M_{\mu}^{\mu_d} = \frac{1}{N}\sum_{i=1}^N \xi_i^{\mu}\xi_i^{\mu_d}$, while $m_{1,\mu}$ is the overlap between the one-step evolution of the picked data-point and the $\mu$th memory, namely $m_{1,\mu} = \frac{1}{N}\sum_{j = 1}^N S_j^{1,\mu_d}\xi_j^{\mu}$ (see \cref{sec:appA2} for further details).
The main contribution to the gradient are then given by the two following quantities
\begin{equation}
    \label{eq:w12}
    \omega_i^{\mu} =\frac{1}{2\sigma_i}\left( m_{\mu}\chi_i^{1,\mu} + m_{1,\mu}\chi_i^{\mu}\right),
\end{equation}
and 
\begin{equation}
    \label{eq:w22}
    \Omega_i^{\mu} =\frac{1}{2\sigma_i}\left(m_{\mu}\xi_i^{\mu}\xi_i^{\mu_d} + M_{\mu}^{\mu_d}\chi_i^{\mu}\right).
\end{equation}
One is specifically interested in two features: the order of magnitude of $\omega_i^{\mu}$ and $\Omega_i^{\mu}$ in (\ref{eq:deltaL_tot}) and their values when $|\Delta_i^{\mu}| < 0^+$. In order for the optimal noise condition to apply, $\omega_i^{\mu}$ should be dominant with respect to $\Omega_i^{\mu}$ and negative in the relevant interval of values.
Hence we measure the Pearson coefficient between $\omega_i^{\mu}$, $\Omega_i^{\mu}$ and the stabilities $\Delta_i^{\mu}$ across different levels of the Hebbian landscape as well as the representative value for the same two quantities around $\Delta_i^{\mu} = 0$, i.e. $\omega_{emp}(0)$ and $\Omega_{emp}(0)$. We perform such measures for both the two sets of memories, across the Hebbian landscape.

\begin{figure*}[ht!]
\begin{tabularx}{\linewidth}{*{2}{X}}
     \centering
     \begin{subfigure}[b]{0.45\textwidth}
         \centering
         \includegraphics[width=\textwidth]{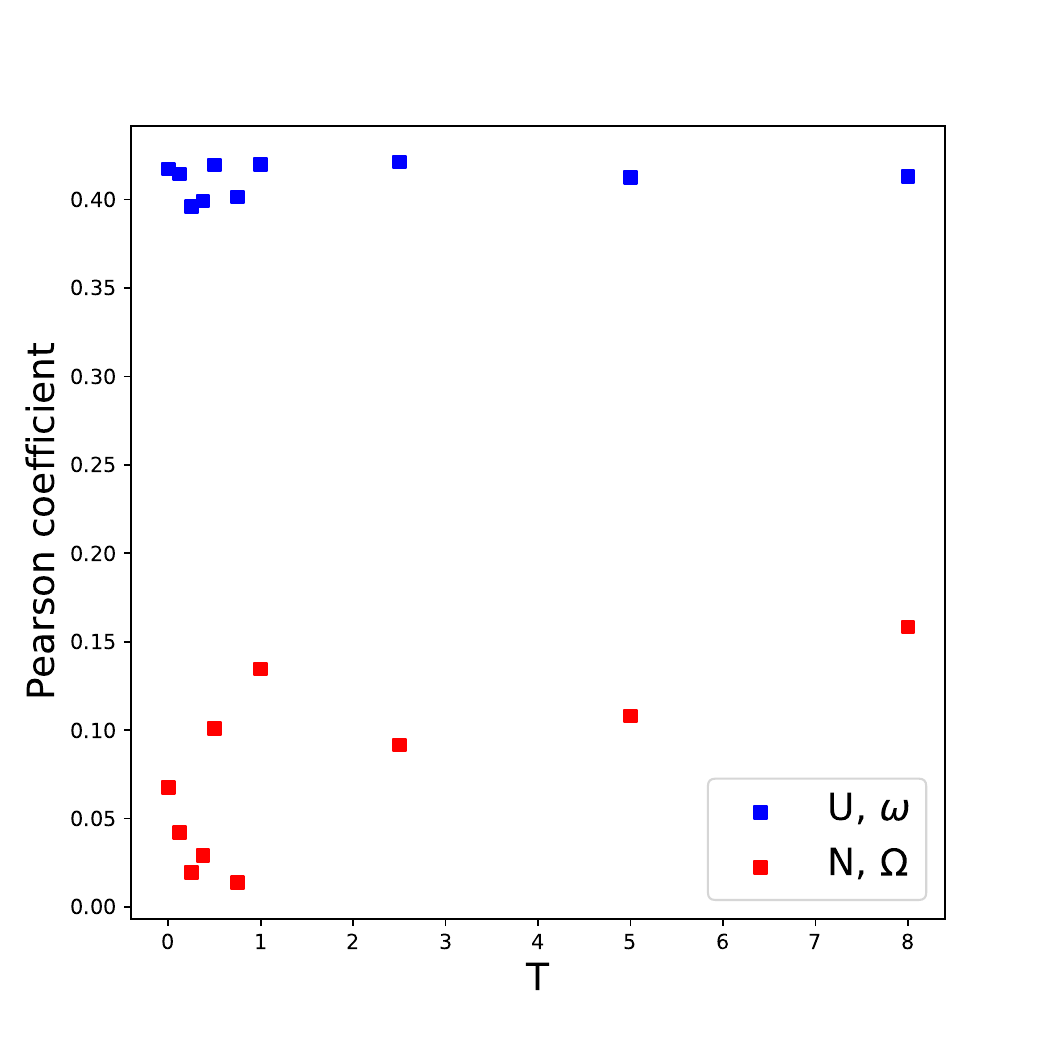}
         \caption{Pearson coefficient}
         \label{fig:mnist_analysis_a}
     \end{subfigure}
     \hfill
     \begin{subfigure}[b]{0.45\textwidth}
         \centering
         \includegraphics[width=0.9
        \textwidth]{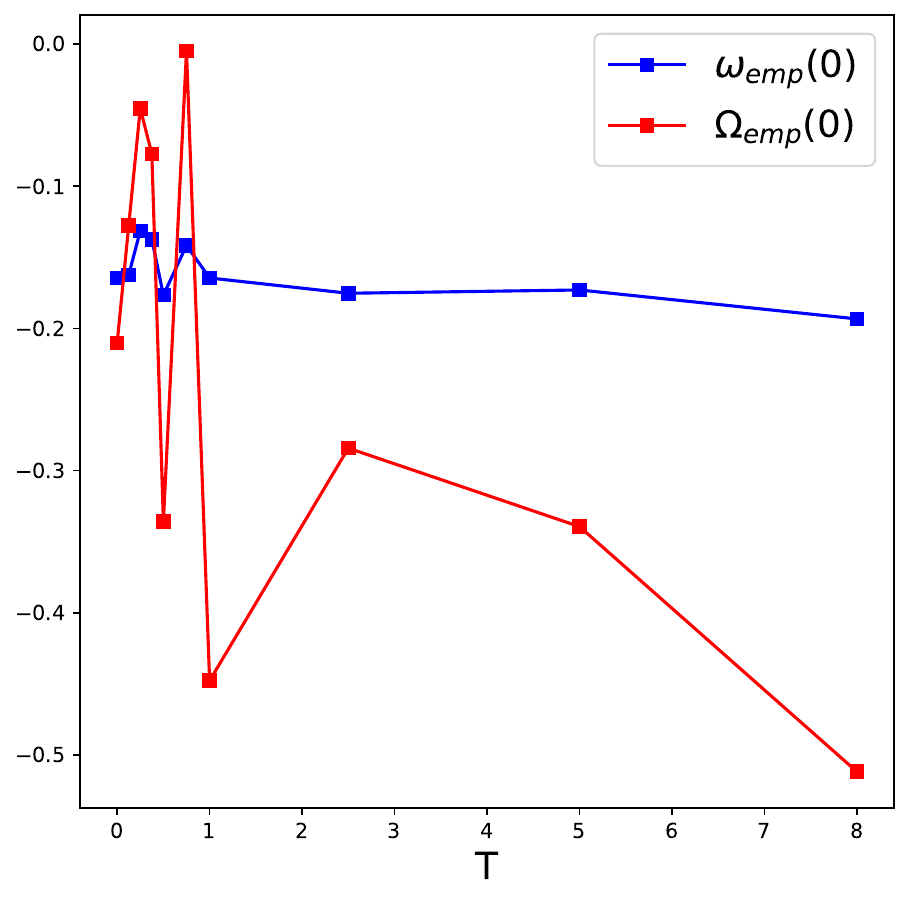}
         \caption{Quality of noise indicator}
         \label{fig:mnist_analysis_b}
     \end{subfigure}
     \hfill  
\end{tabularx} 
\caption{Numerical analysis of the noise contained in the configurations sampled from the Hebbian landscape for the MNIST data-set. The panel on the left reports a measure of the dependence of the relevant observables on the stabilities. The panel on the right reports an interpolation of the relevant observable for very small stabilities, indicating whether the optimal noise condition is well satisfied or not. Measures are averaged over $50$ landscapes. Choice of the parameters: $N = 784$, $\alpha = 0.5$.}
\end{figure*}

\begin{figure*}[ht!]
\begin{tabularx}{\linewidth}{*{2}{X}}
     \centering
     \begin{subfigure}[b]{0.45\textwidth}
         \centering
         \includegraphics[width=\textwidth]{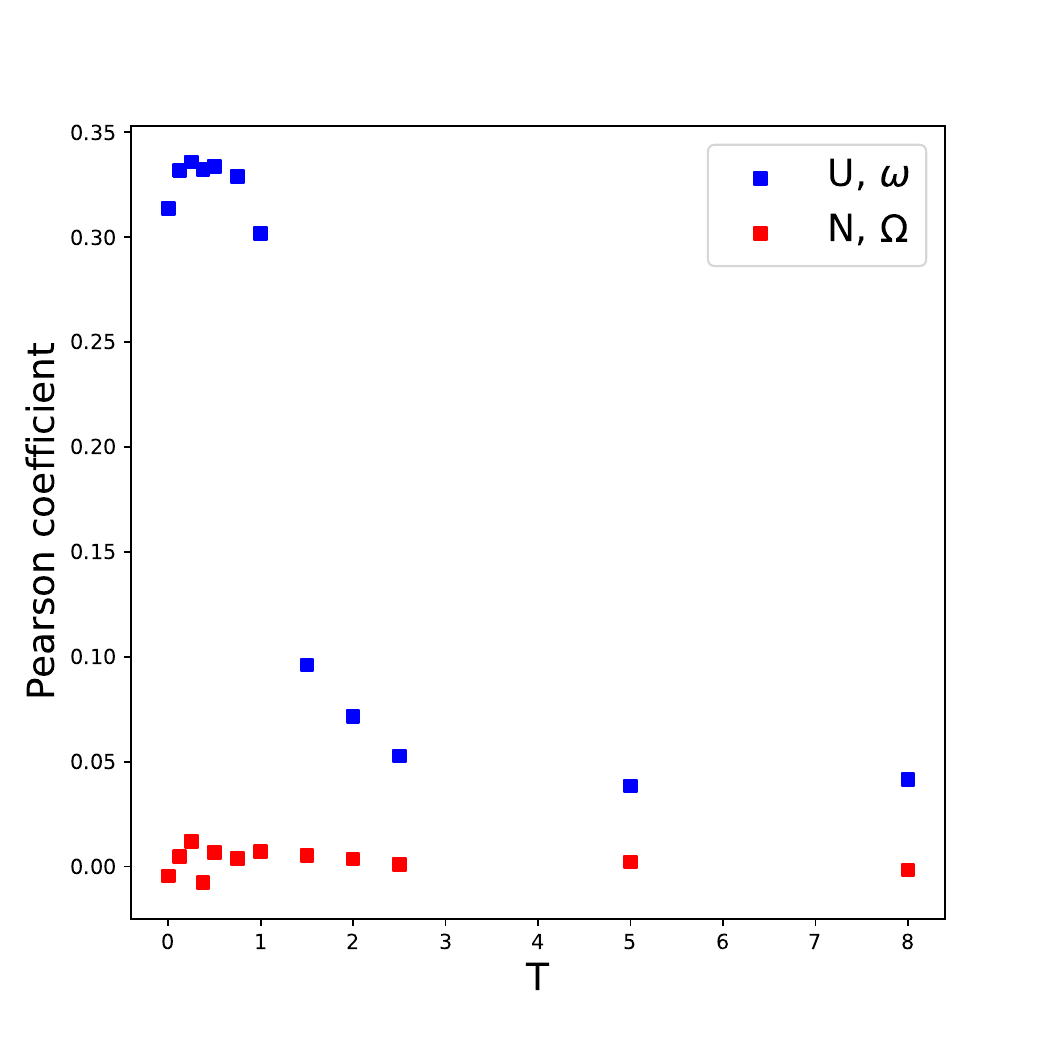}
         \caption{Pearson coefficient}
         \label{fig:ising_analysis_a}
     \end{subfigure}
     \hfill
     \begin{subfigure}[b]{0.45\textwidth}
         \centering
         \includegraphics[width=0.9
        \textwidth]{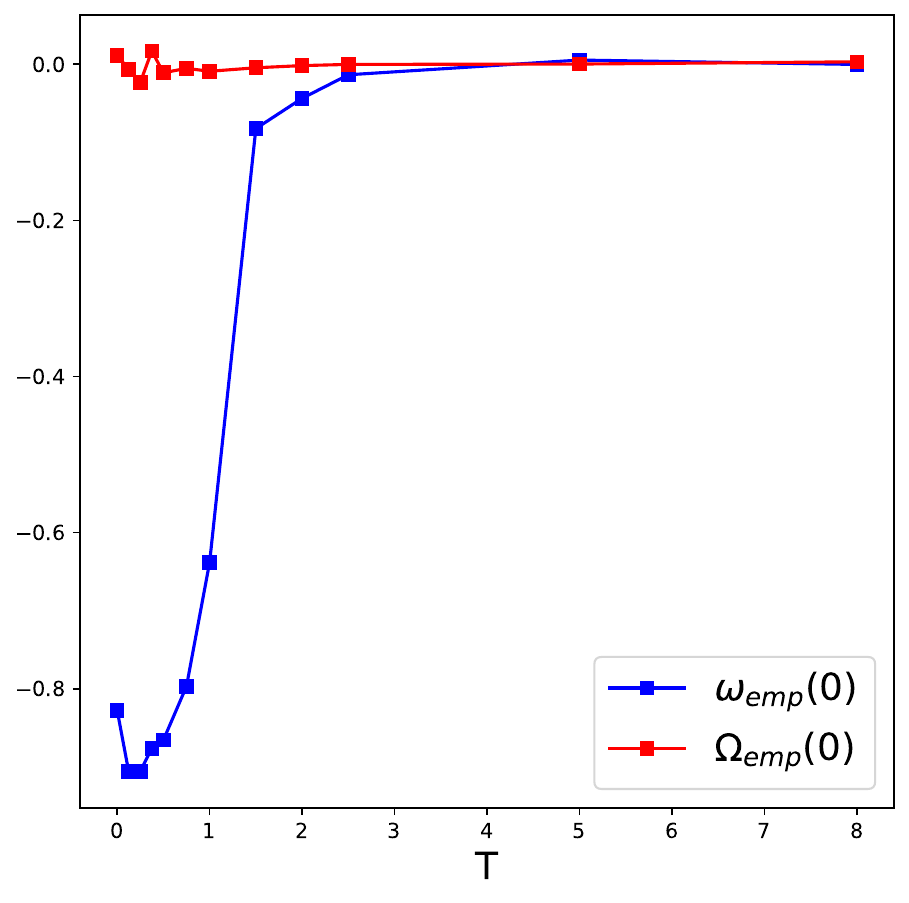}
         \caption{Quality of noise indicator}
         \label{fig:ising_analysis_b}
     \end{subfigure}
     \hfill  
\end{tabularx} 
\caption{Numerical analysis of the noise contained in the configurations sampled from the Hebbian landscape for the Ising data-set. The panel on the left reports a measure of the dependence of the relevant observables on the stabilities. The panel on the right reports an interpolation of the relevant observable for very small stabilities, indicating whether the optimal noise condition is well satisfied or not. Measures are averaged over $10$ landscapes. Choice of the parameters: $N = 784$, $\alpha = 0.5$.}
\end{figure*}
From the study of the disordered MNIST set shows we can conclude that such configurations would not lead to a good minimization of the SVM Loss at altitude in the Hebbian landscape. This is implied by two observations: the noise contribution to the gradient is not negligible due to a rather significant dependence on the stabilities (see \cref{fig:mnist_analysis_a}), and the values assumed by the same contribution relatively to small stabilities are large and negative (see \cref{fig:mnist_analysis_b}). \\On the other hand, Ising configurations are good training configurations at low temperatures $T$, showing a similar scenario to the one encountered with random configurations: the noise contribution does not depend on stabilities, suggesting it is going to cancel step by step in learning (see \cref{fig:ising_analysis_a}), and $\omega_{emp}(0)$ is negative and dominant with respect to $\Omega_{emp}(0)$ in the sensible range of $\Delta_i^{\mu}$ (see \cref{fig:ising_analysis_b}).\\The interpretation of the results is straight-forward. In contrast with the Unlearning contribution, the Noise contribution to the gradient of the Loss depends on the mutual correlations among memories in the set. Since digits in the MNIST set are similar with each other, such points result difficult to be learned by the TWN algorithm. On the contrary, Ising configuration are sufficiently separated with each other, implying the optimal noise condition to be satisfied for some regions of the landscape. We can also conclude that internal dependencies in the memories, which were present in both the MNIST and Ising data, are not relevant in the process, since they do not enter in both $\omega$ and $\Omega$.
\subsection*{Checkpoint}
In this section we have seen that:
\begin{itemize}
    \item The paramagnetic bi-dimensional Ising configurations are correctly memorized by the TWN algorithm with structured noise with $m_t = 0^+$  because, similarly to the random data scenario, the Hebbian landscape generated from these data contain good training configurations; the MNIST data, on the other hand, fail at reaching a good memory performance because the relative Hebbian landscape is poor of good training configurations. 
    \item One can approach a SVM that solidly stores a set of memories by means of the TWN algorithm as far as these memories are well separated from each other, i.e. they overlap weakly. Internal correlations inside the features of the memories are allowed as far as memories are mutually distant in the space of network configurations. 
\end{itemize}
\newpage
\section{\label{sec:cann}Unlearning on Continuous Attractor Neural Networks}
A Continuous Attractor Neural Network (CANN) represents a successful model that reproduces spatial associative memory \cite{battista_capacity-resolution_2020, rosay_statistical_2014, monasson_cross-talk_2013, monasson_crosstalk_2014, monasson_transitions_2015, cocco_statistical_2018}, i.e. the ability to retrieve real environments embedded in a Euclidean space. 
\\This time we consider a neural network of $N$ binary units $S_i\in\{0,1\}$. In analogy with the biology of the hippocampus, units are associated to \textit{place cells} i.e. special neurons associated to particular positions in space. We can generate such positions at random for each one of the $L$ environments that can be memorized by the neural network. We will employ the so called \textit{remapping} ansatz, i.e. 
$$ \vec{r}_i^{l} \in \left[-1/2, + 1/2\right]^{D}, \hspace{0.5cm}i = 1,..,N\hspace{0.5cm}l = 1,..,L.$$
Each place cell monitors a D-dimensional sphere of volume $w$, called \textit{place field}, also indicating the average fraction of the whole amount of place fields \textit{seen} by that particular neuron. The radius of such a sphere is $d_c$. The radius for $D = 1,2$ measures
\begin{equation}
    \label{eq:d_c}
    d_c = \frac{1}{2w}, \hspace{0.5cm}D = 1\hspace{2cm}d_c = \sqrt{\frac{w}{\pi}}, \hspace{0.5cm}D = 2.
\end{equation}
Imagining to pin one position in the real space $\vec{x}$ in one of the environments $l$, the relative neural activity configuration $\vec{S}$ will be given by the following rule
\begin{equation}
\centering
    \begin{cases}
        \label{eq:space_pattern1}
        S_i = 1 \hspace{0.5cm}\text{if}\hspace{0.5cm} |\vec{r_i}^l- \vec{x}| \leq d_c \\
        S_i = 0 \hspace{0.5cm} \text{otherwise}
    \end{cases}
\end{equation}
where $| \hspace{0.1cm}\cdot\hspace{0.1cm}|$ is the Euclidean norm operator. The spatial maps to be learnt by the network are assembled in the following manner: consider a number $L$ of physical environments in a dimension $D$; such environments have a cubic shape of unitary sides and periodic boundary conditions; then $p$ positions are generated in the environment, i.e. $$ \vec{R}^{l, \mu} \in \left[-1/2, + 1/2\right]^{D}, \hspace{0.5cm}l = 1,..,L \hspace{0.5cm}\mu = 1,..,p.$$ 
The real space maps to be stored can be transferred in the activity configuration space in the shape of memories by applying the usual rule \\
\begin{equation}
\centering
    \begin{cases}
        \label{eq:space_pattern2}
            \xi_i^{l,\mu} = 1 \hspace{0.5cm}\text{if}\hspace{0.5cm} |\vec{r_i}^{l} - \vec{R}^{l,\mu}| \leq d_c \\
        \xi_i^{l,\mu} = 0 \hspace{0.5cm} \text{otherwise}
    \end{cases}
\end{equation}
\subsection{Hebbian Learning}

Once there is a set of memories the Hebbian framework can be generalized to CANNs to build the couplings $J_{ij}$ among neurons as
\begin{equation}
    \label{eq:hebbs}
    J_{ij} = \frac{1}{p L}\sum_{l,\mu}^{L,p}(\xi_i^{l,\mu} - w)(\xi_j^{l,\mu}-w)\hspace{1cm} J_{ii} = 0\hspace{0.5cm} \forall i
\end{equation}
Also in this case we must define a particular dynamic rule to let the system evolve in time. The most reliable choice is the Glauber dynamics at temperature $T$ \cite{newman_monte_1999}. According to this rule we start from an initial configuration of the units and we update them asynchronously according to the following prescription   
\begin{equation}
    \label{eq:glauber_dynamics}
    P(S_i^{(t+1)} = x) = \frac{1}{1 + e^{-2\beta h_i^{(t)}\cdot(2x - 1)}},
\end{equation}
with $x \in\{0,1\}$, $\beta = 1/T$ and $h_i^{(t)} = \sum_{j=1}^N J_{ij}S_j^{(t)}$ being the \textit{local field}. The correspondent dynamic rule at $T = 0$ is then
\begin{equation}
    \label{eq:T0_dynamics}
    S_i^{(t+1)} = \frac{1}{2}\left[1 +  \text{sign}\left( h_i^{(t)}\right)\right]
\end{equation}
According to the rule ($\ref{eq:hebbs}$), for $\alpha$ smaller than a critical value and $T > 0$, the system explores the real space as a moving bump of activity, when Glauber dynamics is implemented \cite{monasson_crosstalk_2014, rosay_statistical_2014}. Notice that the average activity contained in the moving  bump is, by construction
\begin{equation}
    w = \frac{1}{N}\sum_{i = 1}^N S_i
\end{equation}
When $T = 0$, in the same region of the phase diagram, memories are closed to be fixed points of the dynamics. At larger values of a certain threshold $\alpha_c(T)$ the system falls in a glassy phase where maps are not correctly recalled \cite{monasson_cross-talk_2013, rosay_statistical_2014}. 

\subsection{Support Vector Machines Vs. Hebbian Rule at $T = 0$}
We are now interested in the numerical study of the CANN at $T = 0$. Let us pick a random position $\vec{x}$ in one of the environments and build an activity configuration from that, i.e. by doing
\begin{equation}
    \label{eq:initial_conf}
    \centering
    \begin{cases}
        S_i^{(0)} = 1 \hspace{0.5cm}\text{if}\hspace{0.5cm} |\vec{x} - \vec{r}_i^{l}| \leq d_c \\
        S_i^{(0)} = 0 \hspace{0.5cm} \text{otherwise}
    \end{cases}
\end{equation}
The \textit{spatial error} $\epsilon$ is defined as the average distance between the center of mass of the activity in space at the fixed point reached by implementation of rule ($\ref{eq:T0_dynamics}$) by initializing the network in $\vec{S}^{(0)}$ and the center of mass at the activity at initialization. The position of the center of mass of the bump in a given map $l$ at time $t$ of the dynamics is computed in the following way
\begin{equation}
    \label{eq:com}
    \vec{r}_{c.o.m.}^{(l)}(t) = \frac{\sum_{i = 1}^N S_i^{(t)} \vec{r}_i^{(l)}}{\sum_{i = 1}^N S_i^{(t)}}.
\end{equation}
Since both initial positions and maps are generated at random one expects that a good memory retrieves the stored positions in the environment by attracting the dynamics in their basins of attraction. In this case the maximal extension of the basins of attraction is expected to equal the average distance among the random stored positions, that is $\epsilon \sim p^{-\frac{1}{D}}$, in order for them to span the map uniformly.\\It has been showed by \cite{battista_capacity-resolution_2020} that this scenario is not reached by the more canonical CANNs by making use of a Hebbian-like criterion but rather by an optimized neural network such as a Support Vector Machine (SVM), where all the memories are perfectly retrieved by (\ref{eq:T0_dynamics}) and the stability of the fixed points is maximized (i.e. local fields in the memory configurations are maximal in absolute value). \\
\begin{figure}
\centering
\includegraphics[width=.65\textwidth]{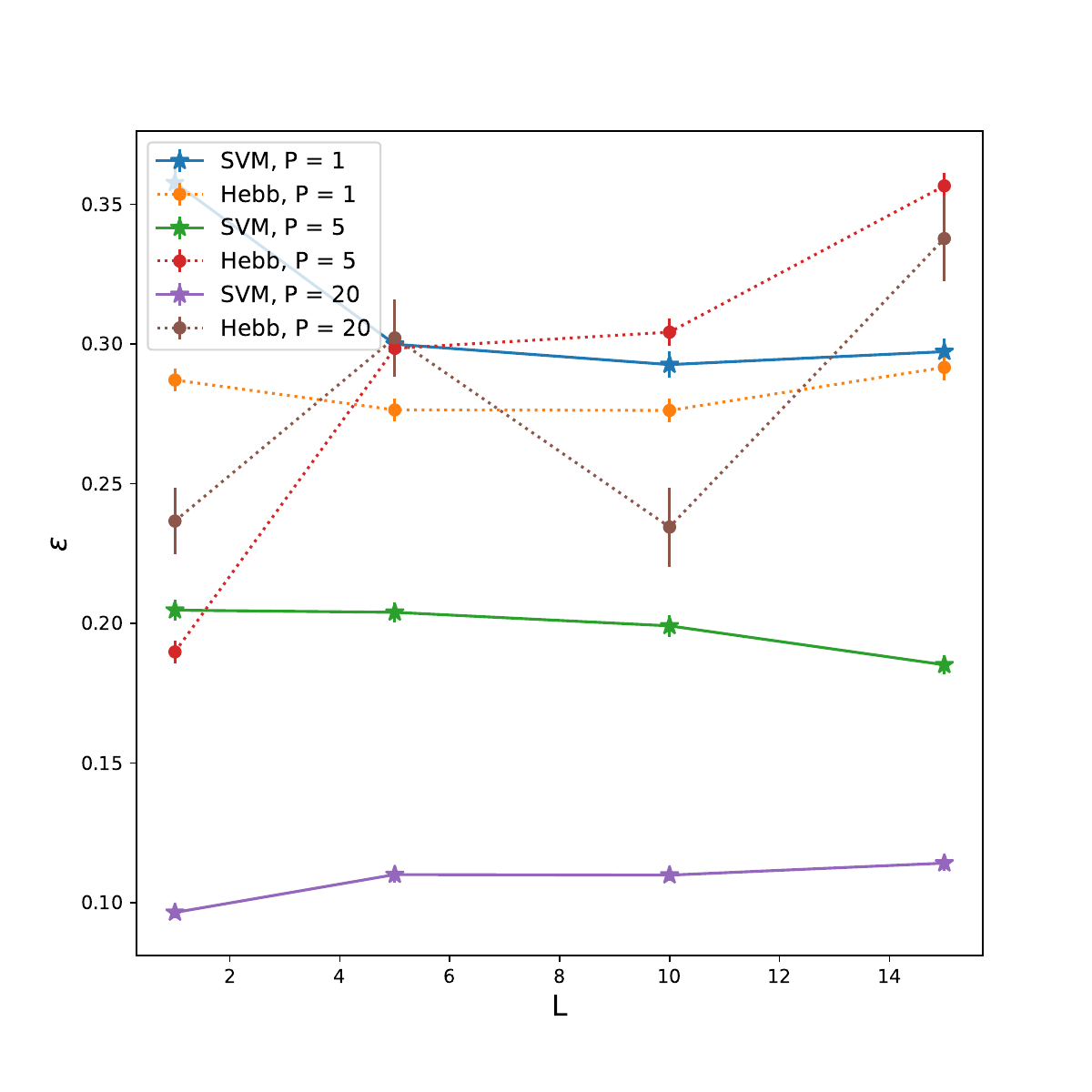}
    \caption{Spatial error $\epsilon$ measured as a function of $L$ and $p$ for both the SVM and Hebbian prescriptions. Measures have been averaged over $10$ samples for $p = 1,5$ and 1 sample for $p = 20$ for SVM, $10$ samples at all $p$ in the Hebbian case. Choice of the parameters: $N = 10^3$, $w = 0.3$, $D = 2$.}
\label{fig:eps}           
\end{figure}
These results are reproduced in \cref{fig:eps} for $D = 2$ and $w = 0.3$. Notice that $\epsilon(L,p)$ is an increasing quantity in $L$ that does not depend on $p$ in the Hebbian network, while the \textit{optimal packing} scenario is obtained in the SVM case.

\subsection{Hebbian Unlearning}
We now generalize the HU algorithm \cite{hopfield_unlearning_1983} for the case of a CANN.
$J^{(0)}$ is chosen to be the Hebbian connectivity matrix according to rule ($\ref{eq:hebbs}$). \\Then the following three steps are iterated:
    \begin{enumerate}
        \item Random shooting: a random initial configuration of the neurons is generated. 
        \item The initial configuration evolves until convergence into a stable state $\vec{S}^*$ by implementation of the dynamics (\ref{eq:T0_dynamics}).
        \item The connectivity matrix is updated as 
        \begin{equation}
            \label{eq:hebb_un}
            \delta J_{ij}^{(d)} =  - \frac{\lambda}{N}(S_i^*-w)( S_j^*-w) \hspace{1cm}J_{ii}^{(d)} = 0 \hspace{0.5cm}\forall i
        \end{equation}
        where $\lambda$ is a small learning rate and $d$ is the number of the algorithm iteration. 
    \end{enumerate}
\begin{figure*}
\begin{multicols}{3}
\centering
\subcaptionbox{L = 5, p = 1}{
    \includegraphics[width=\linewidth]{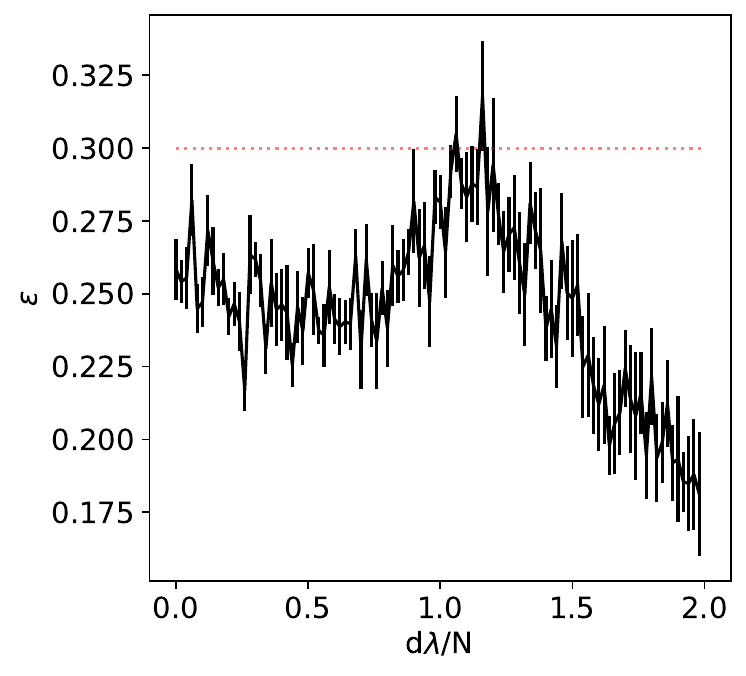}\par 
    \includegraphics[width=0.95\linewidth]{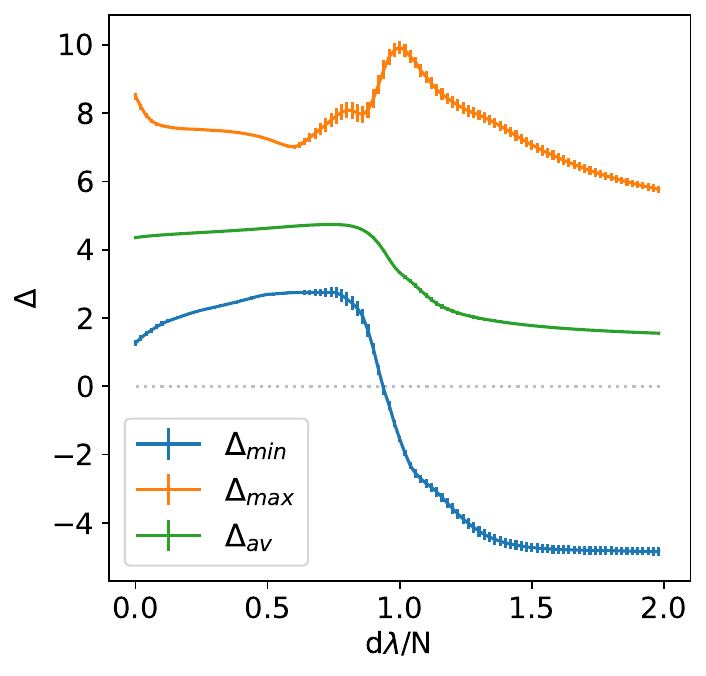}\par 
    \includegraphics[width=0.98\linewidth]{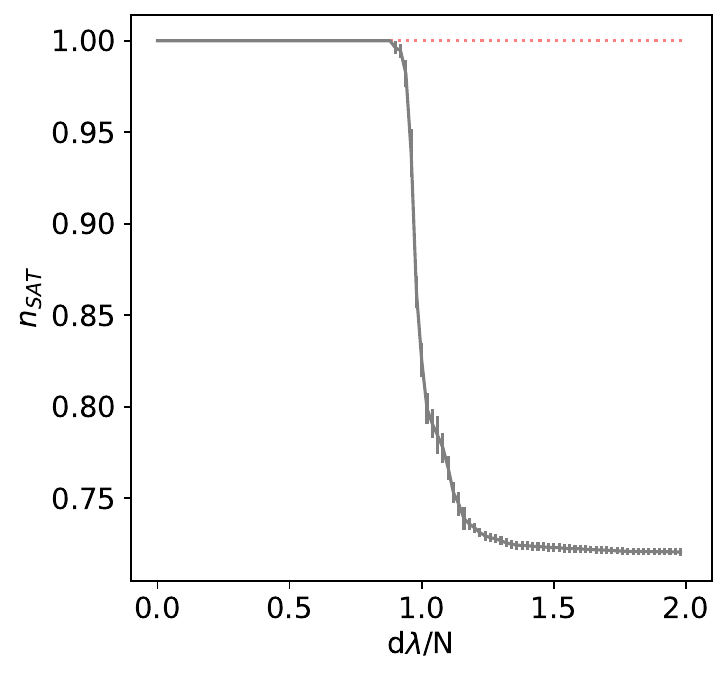} }\par
\end{multicols}
\begin{multicols}{3}
\subcaptionbox{L = 5, p = 5}{
    \includegraphics[width=\linewidth]{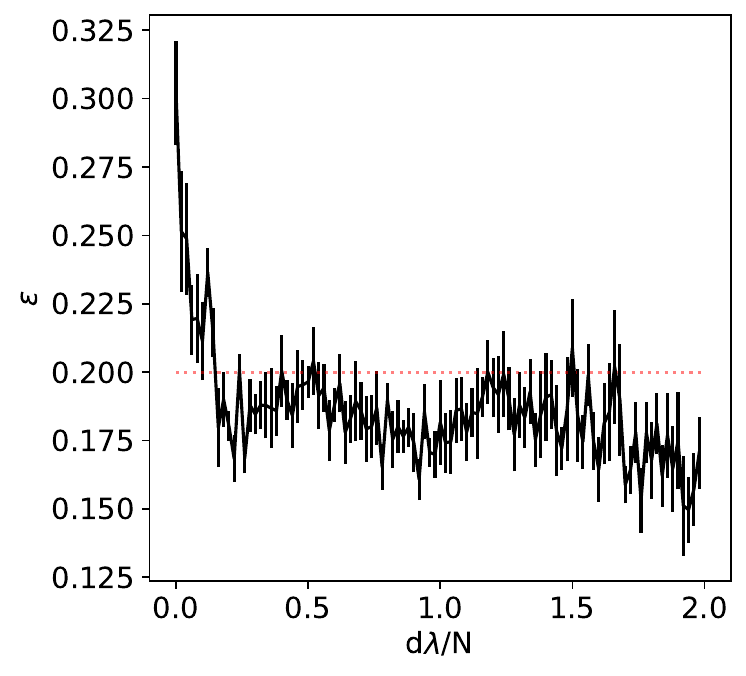}\par 
    \includegraphics[width=0.95\linewidth]{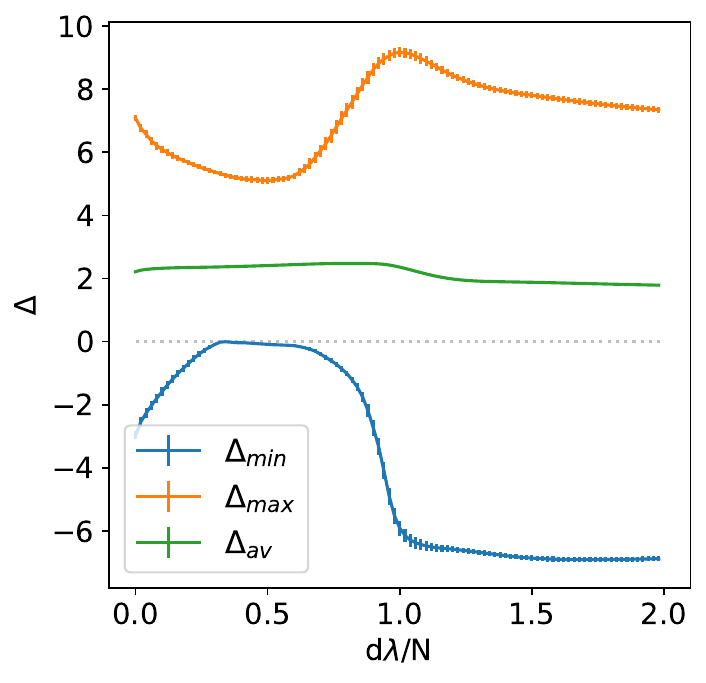}\par 
    \includegraphics[width=0.98\linewidth]{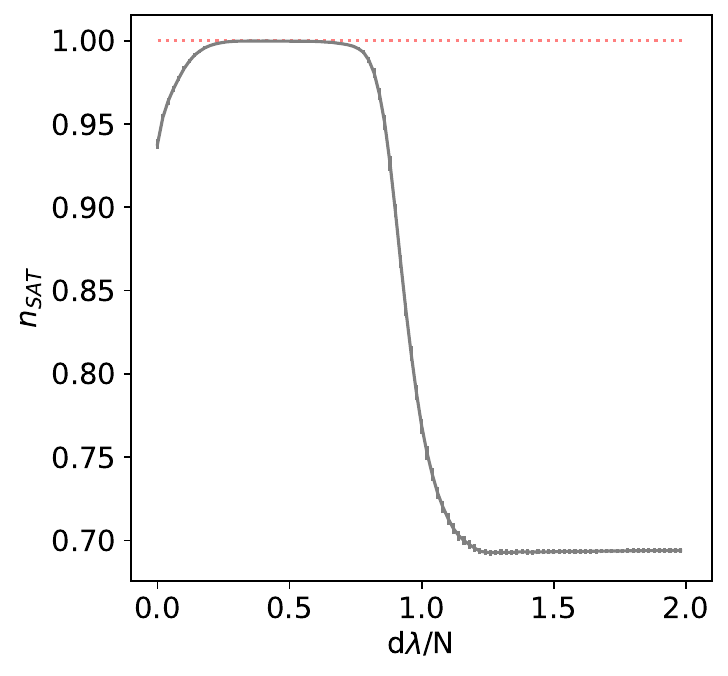}}\par 
\end{multicols}
\begin{multicols}{3}
\subcaptionbox{L = 5, p = 20}{
    \includegraphics[width=\linewidth]{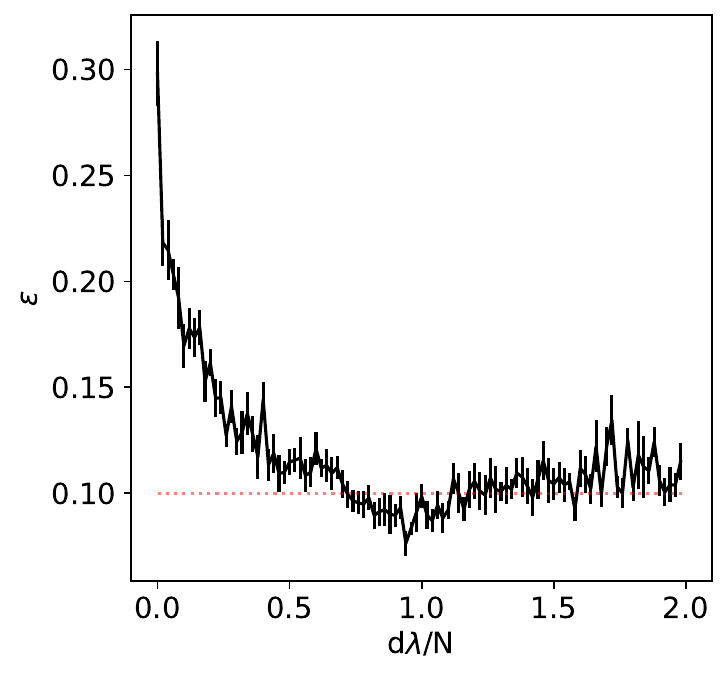}\par 
    \includegraphics[width=0.97\linewidth]{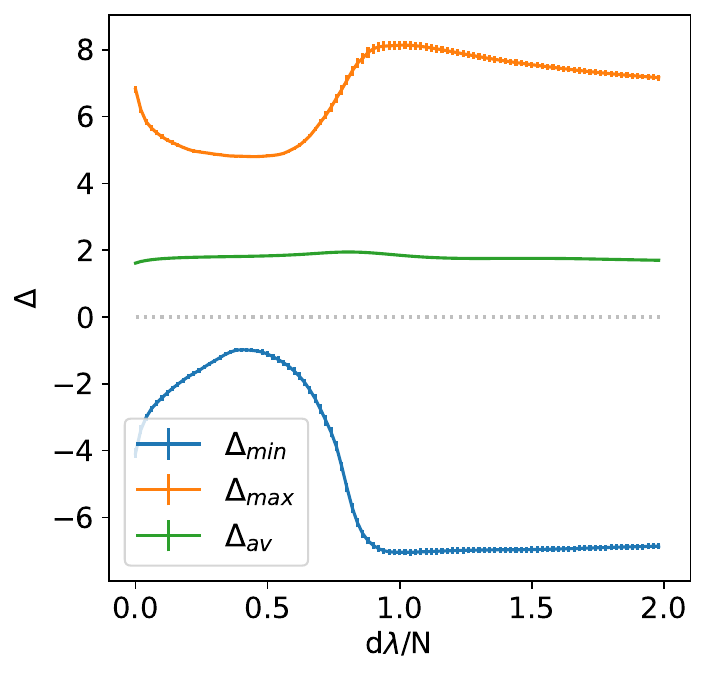}\par 
    \includegraphics[width=0.99\linewidth]{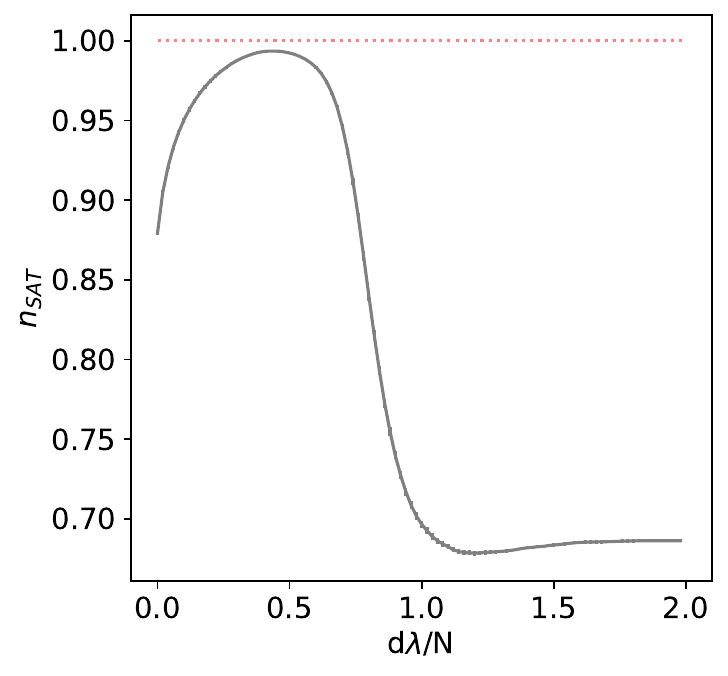}}\par 
\end{multicols}
\begin{multicols}{3}
\subcaptionbox{L = 5, p = 50}{
    \includegraphics[width=\linewidth]{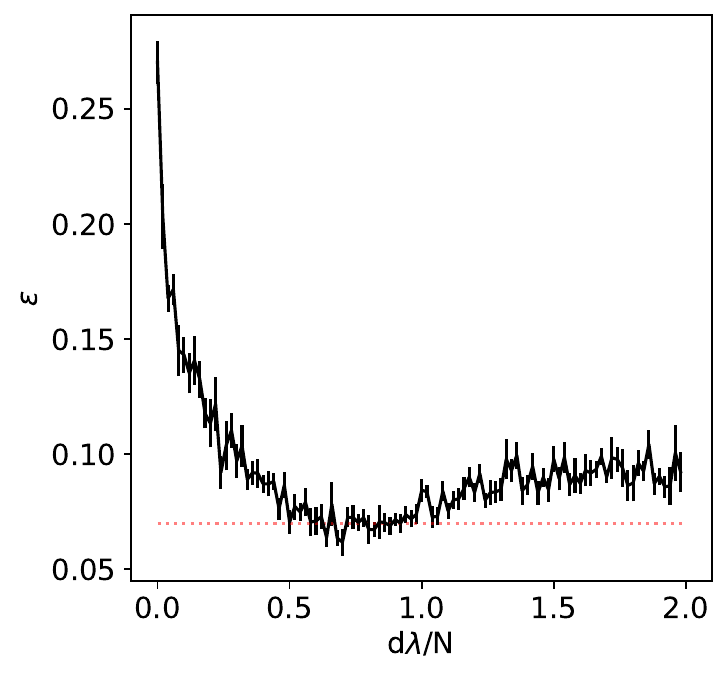}\par 
    \includegraphics[width=0.97\linewidth]{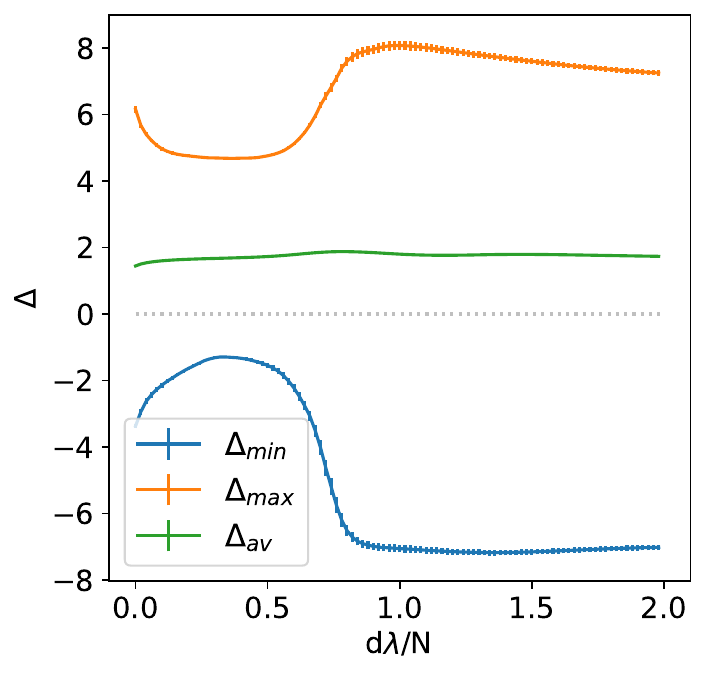}\par 
    \includegraphics[width=0.99\linewidth]{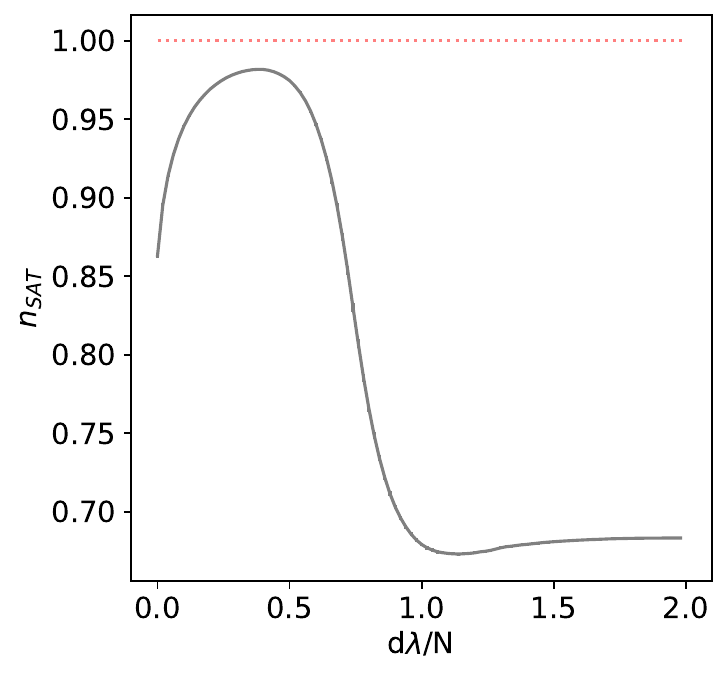}}\par 
\end{multicols}

\caption{Left: Spatial error $\epsilon$ as a function of the Unlearning steps. Center: average stabilities $\Delta$ as a function of the Unlearning steps. Right: fraction of satisfied constraints $n_{SAT}$ as a function of the Unlearning steps. Measures averaged over $10$ samples. Choice of the parameters: $D = 2$, $w = 0.3$, $\lambda = 10^{-2}$.}
\label{fig:unlearning_vs_time}
\end{figure*}

\begin{figure*}[ht!]
\begin{multicols}{3}
\centering
\subcaptionbox{Hebbian Learning}{
    \includegraphics[width=\linewidth]{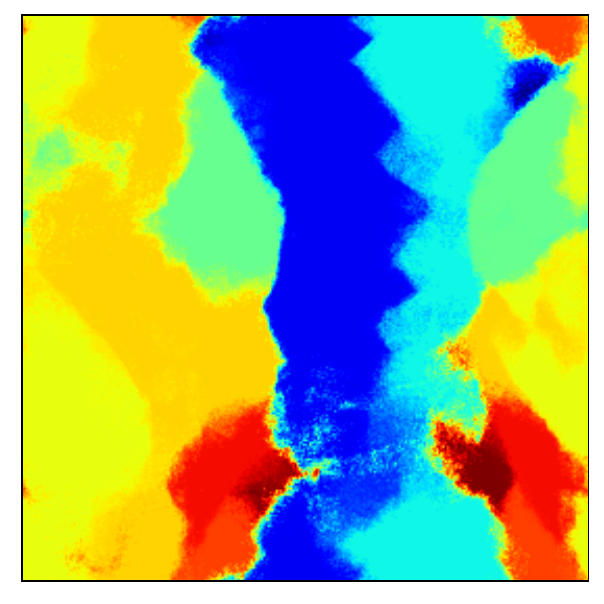}}\par
    \subcaptionbox{Hebbian Unlearning}{
    \includegraphics[width=\linewidth]{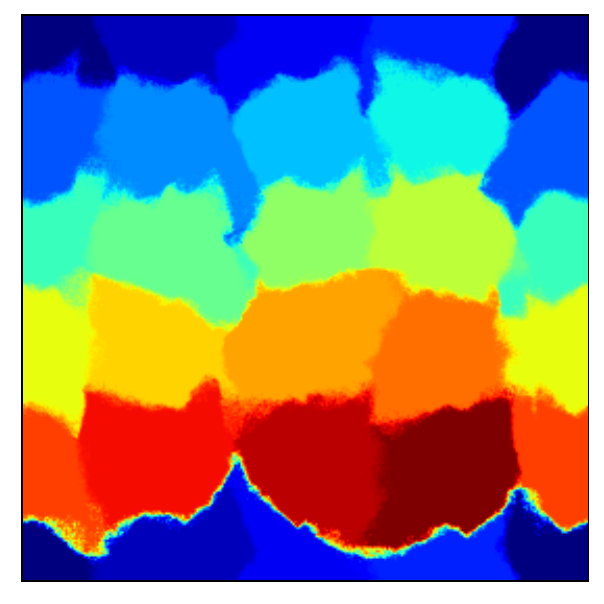} }\par
    \subcaptionbox{SVM Learning}{
    \includegraphics[width=\linewidth]{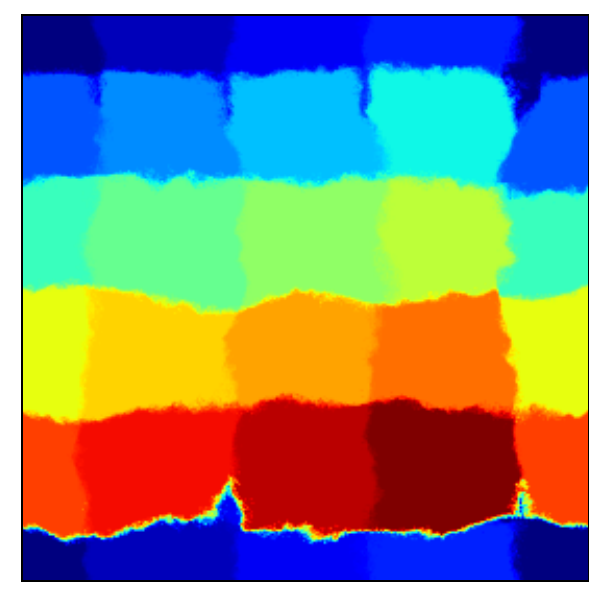}}\par
\end{multicols}
\caption{Representation of the basins of attraction in the real space after the three types of learning. Stored positions are disposed on a regular grid. Choice of the parameters: $D = 2$, $w = 0.3$, $L = 1$, $p = 20$, $N = 500$, $\lambda = 10^{-2}$.}
\label{fig:palettes}
\end{figure*}

We can now characterize the Unlearning network by measuring three relevant quantities, i.e.
\begin{itemize}
    \item The spatial error $\epsilon(L,p)$ as a function of the Unlearning steps. 
    \item The average stabilities of the memories defined as
    \begin{equation}
        \label{eq:stabs}
        \Delta_i^{l,\mu} = (2\xi_i^{l,\mu}-1)\sum_{j=1}^N \frac{J_{ij}}{\sqrt{\sum_k J_{i,k}^2}}\xi_j^{l,\mu}
    \end{equation}
    and we are specifically interested in $\Delta_{min} = \text{min}_{(i,l,\mu)}(\Delta_i^{l,\mu})$.
    %, $\Delta_{max} = \text{max}_{(l,\mu)}(\Delta_i^{l,\mu})$ and $\Delta_{av} = \frac{1}{LpN}\sum_{i,l,\mu}\Delta_i^{l,\mu}$.
    \item The fraction of satisfied constraints in the relative percepton problem $n_{SAT}$: a constraint labelled by $(i,l,\mu)$ is SAT when $\Delta_i^{l,\mu} \geq 0$, UNSAT otherwise.
\end{itemize}
\begin{figure*}[ht!]
\begin{tabularx}{\linewidth}{*{2}{X}}
     \centering
     \begin{subfigure}[b]{0.45\textwidth}
         \centering
         \includegraphics[width=\textwidth]{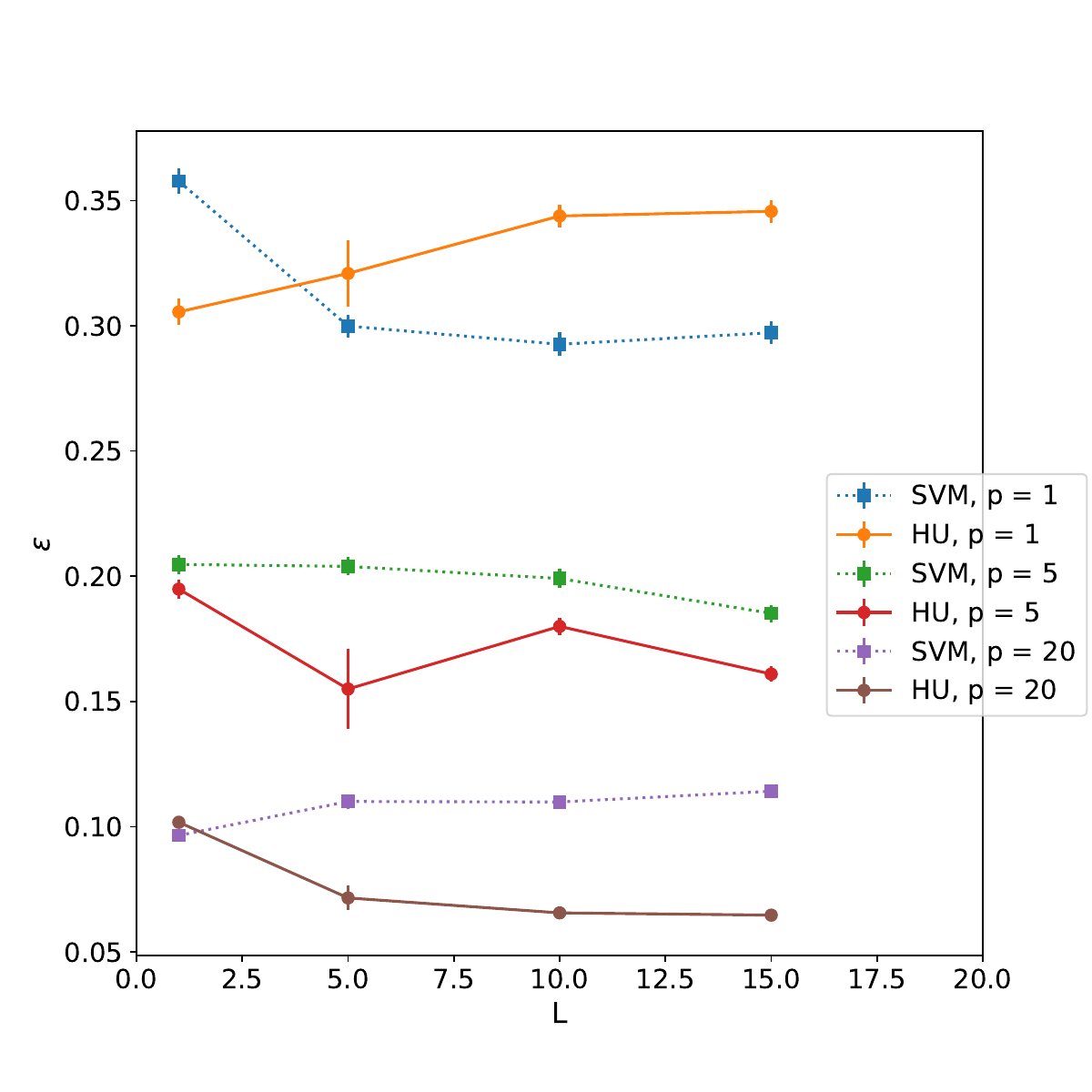}
         \caption{}
         \label{fig:eps_unlearning_a}
     \end{subfigure}
     \hfill
     \begin{subfigure}[b]{0.45\textwidth}
         \centering
         \includegraphics[width=
        \textwidth]{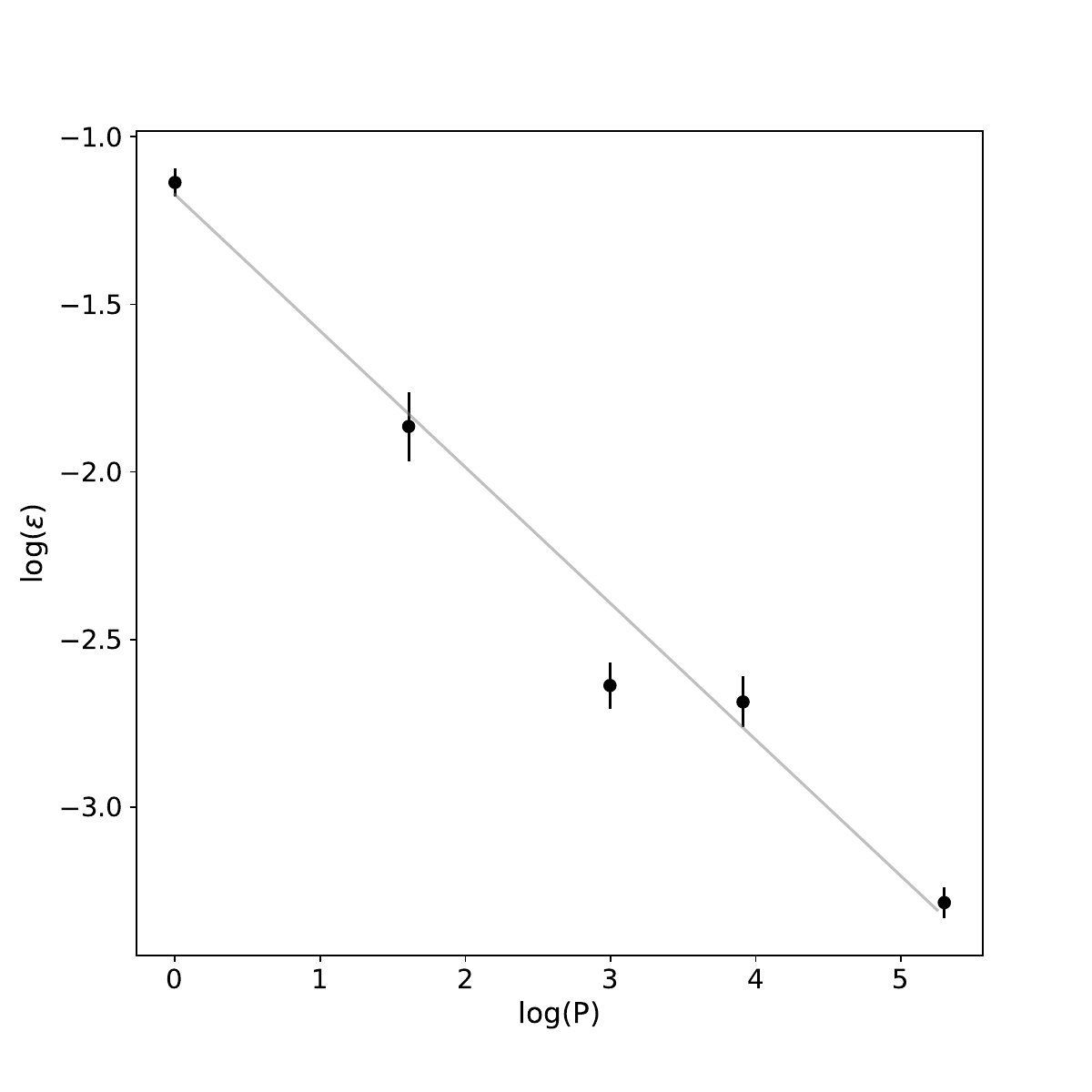}
         \caption{}
         \label{fig:eps_unlearning_b}
     \end{subfigure}
     \hfill  
\end{tabularx} 
\caption{(a): Spatial error $\epsilon$ measured as a function of $L$ and $p$ for both the SVM and after Hebbian Unlearning (HU). Measures have been averaged over $10$ samples for $p = 1,5$ and 1 sample for $p = 20$ for SVM, with $N = 10^3$, $10$ samples at all $p$ in the Unlearning case, with $N = 500$. (b): log-log plot of $\epsilon$ versus $p$ at $L = 5$ for HU, best fit line in grey. Choice of the parameters: $w = 0.3$, $D = 2$, $\lambda = 10^{-2}$.}
\end{figure*}
The behaviour of these three quantities is represented in \cref{fig:unlearning_vs_time} for $L = 5$ and $p = 1,5,20,50$. As one can see, the minimum value of the spatial error is reached in correspondence of the peak of $\Delta_{min}$ and $n_{SAT}$. However, perfect-retrieval is reached by the system only for small values of $p$, signaling the presence of a critical capacity for the HU procedure, as it was valid for random memories studied in \cite{benedetti_supervised_2022, van_hemmen_increasing_1990}.\\At this point we can experimentally extrapolate the number of iterations at which $\epsilon$ reaches its minimum value and we measure such value at different values of $L$ to compare it with the results from the SVM. It should be noticed that the spatial error $\epsilon$ does not depend on $L$ and $N$ but only on $p$, as found for SVMs in \cite{battista_capacity-resolution_2020}: this result is displayed in \cref{fig:eps_unlearning_a}. Numerical values are also comparable between the SVM and the Unlearning network. Figure \ref{fig:eps_unlearning_b} reports the expected behaviour of $\epsilon(L,p)$. The opposite of the coefficient of the best fit line is found to be consistent in a $3\sigma$ confidence interval with the real dimension of space $D$. In fact, we have $$D_{exp}^{-1} = 0.41 \pm 0.03$$Another experiment is performed in order to show the effectiveness of HU in approaching the CANN to the optimal SVM scenario. 
A single 2-dimensional environment with $p = 20$ points disposed on a regular grid and learned by the network according to the Hebbian learning SVM and HU. Then the map is scanned in space and neurons are initialized according to the usual criterion (see \cref{eq:space_pattern2}): $T = 0$ dynamics is iterated until convergence and the node of the grid that is closest to the final position of the center of mass of the neural activity labels the starting point. 
Figure \ref{fig:palettes} reports the results. Same colours are associated to same basins of attraction of the stored points on the grid. Notice that Hebbian learning (\cref{fig:palettes}a) completely destroys the regularity of the grid, while HU and SVM (\cref{fig:palettes}b, \cref{fig:palettes}c) center basins of attraction, that are squared in shape with each side $\sim (1/p)^{\frac{1}{2}}$, around the grid points, maximizing their size. SVM performs this job optimally, as it could be predicted from \cite{battista_capacity-resolution_2020} but HU seems to approach that limit very well.

\subsection{Diffusion dynamics at $T > 0$}
\begin{figure*}[ht!]
\begin{multicols}{3}
\centering
    \includegraphics[width=0.9\linewidth]{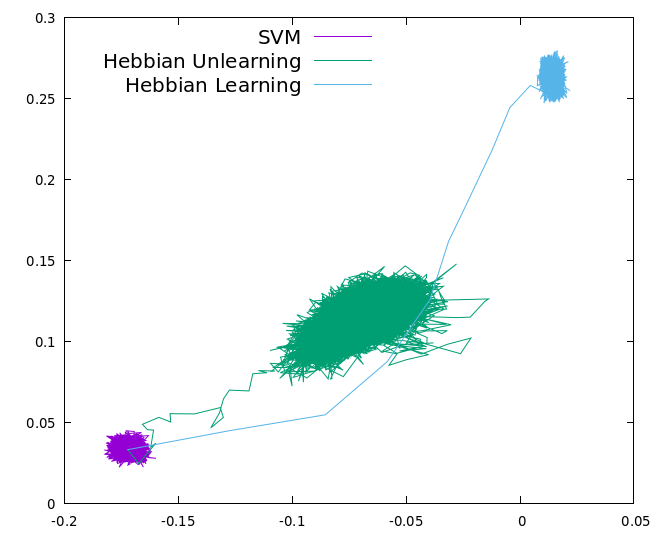}\par
    \includegraphics[width=\linewidth]{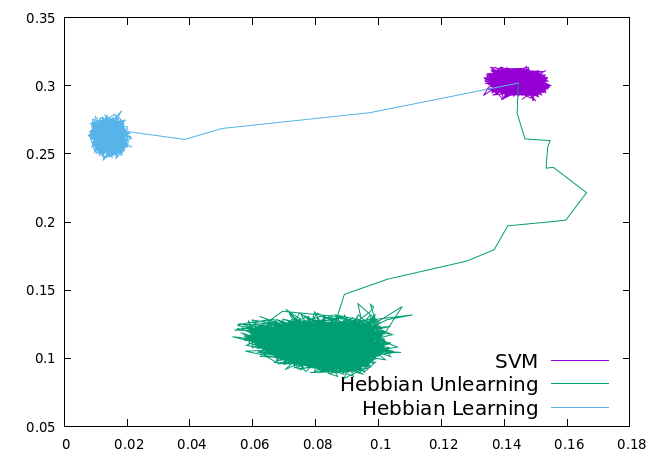}\par
    \includegraphics[width=\linewidth]{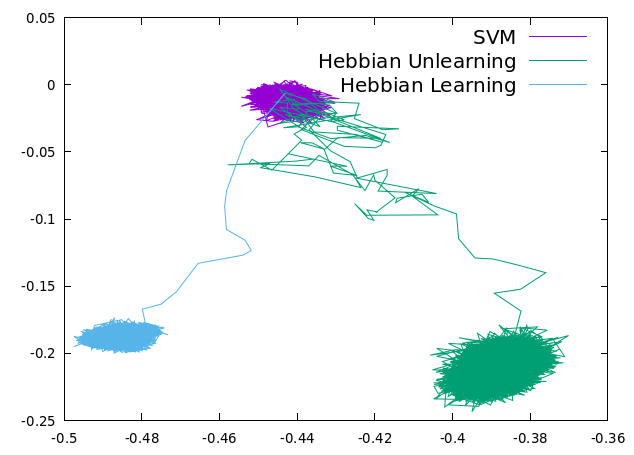}\par
\end{multicols}
\caption{Diffusive trajectories of the moving bump in the landscape resulting from the Hebbian Learning, Hebbian Unlearning and SVM. Each time step is chosen to be one sweep of the network, i.e. $N$ spins being evaluted by the the MCMC algorithm, and simulations have been run over $5\cdot10^4$ sweeps in total. Choice of the parameters: $D = 2$, $w = 0.3$, $\lambda = 10^{-2}$,  $N = 500$, $L = 1$, $p = 20$, $T = 1.0$.}
\label{fig:diff_p20}
\end{figure*}

\begin{figure*}[ht!]
\begin{multicols}{1}
\centering
    \subcaptionbox{}{
    \includegraphics[width=\linewidth]{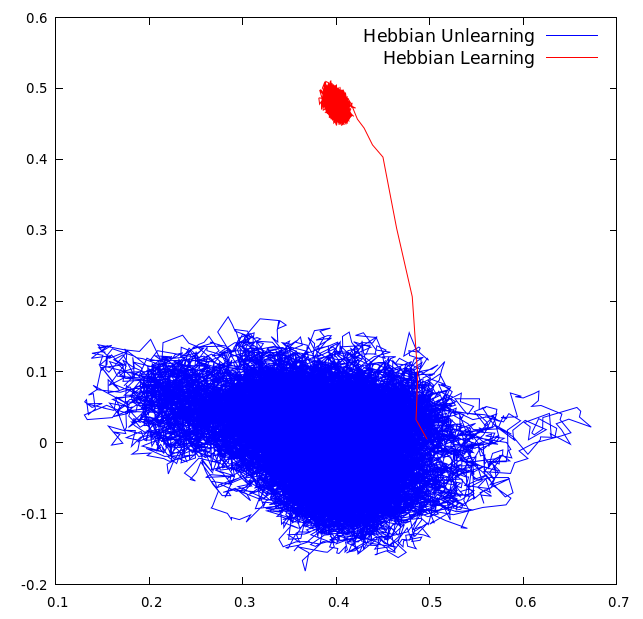}
    }\par
   
    \includegraphics[width=\linewidth]{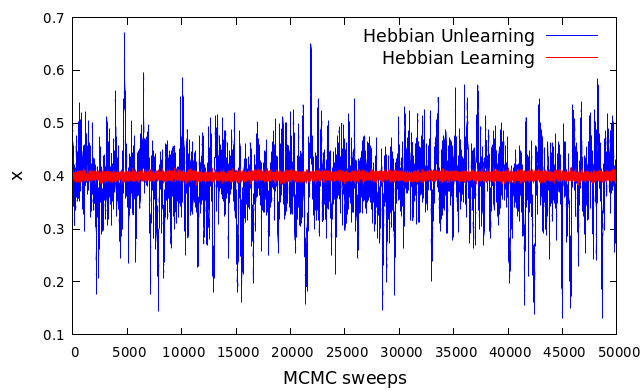}
     \subcaptionbox{}{
    \includegraphics[width=\linewidth]{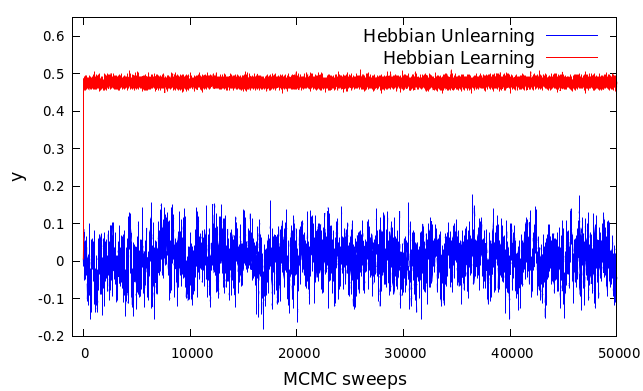}
    }
\end{multicols}
\caption{Hebbian Unlearning (blue) and Hebbian Learning (red). Each time step is chosen to be one sweep of the network, i.e. $N$ spins being evaluated by the the MCMC algorithm, and simulations have been run over $5\cdot10^4$ sweeps in total. Choice of the parameters: $D = 2$, $w = 0.3$, $\lambda = 10^{-2}$,  $N = 500$, $L = 1$, $p = 200$, $T = 1$.}
\label{fig:diff_p200}
\end{figure*}

We now switch on the temperature $T > 0$ and study the dynamics of the activity bump in the space map as done in \cite{monasson_cross-talk_2013, rosay_statistical_2014}. Since we are interested in the diffusion in space of the bump only one environment will be stored in the network ($\alpha = 0$ case) in order to impede the activity to change map \cite{monasson_transitions_2015}. 
Figure \ref{fig:diff_p20} represents diffusion when $p$ is low, i.e. $p = 20$, for all the three considered learning rules: the position of the center of mass of the neuronal activity is plotted as a function of the time steps. Before choosing the temperature it has been checked whether the network is in the \textit{clump} phase \cite{monasson_cross-talk_2013, monasson_crosstalk_2014}, i.e. neuronal activity is well clustered on the map. Dynamics is initialized in one given stored position on the map. Disorder appears to be very strong as the system is stuck for a very long time in one basin of attraction separated by the others by high barriers in free energy. The SVM explores a region that is well centered around the stored position, that is reasonable since all memories are stable fixed points of the $T = 0$ dynamics. On the other hand, when Hebbian Learning and HU are implemented the clump descends a bit in the Lyapunov function before finding a metastable minimum to probe.   
Another experiment is performed by increasing the number of stored positions to $p = 200$.
Just HU and Hebbian Learning are evaluated this time, because of the high computational cost needed to find the connectivity matrix of the SVM. 
\\Figure \ref{fig:diff_p200} displays the results for one realization of the network and one common initial condition. We notice that diffusion of the Hebbian network is still ineffective since it gets easily stuck in one metastable state. On the other hand diffusion of the system after the performance of HU is more efficient since it explores a wider portion of space. The 2-dimensional track of the center of mass is represented in \cref{fig:diff_p200}a  while the motion in the two separated dimensions are reported in \cref{fig:diff_p200}b.

%Figure (\ref{fig:diff_p200_2}) reports the diffusion of the center of mass in another case in which Unlearning was performed. In particular the whole motion is reported in (\ref{fig:diff_p200_2}, a) while a zoomed version of the figure is in (\ref{fig:diff_p200_2}, b). The system seems to diffuse even more effectively along the $y$-axis as it is also noticeable by the projection in (\ref{fig:diff_p200_2}, c).

\begin{figure}[ht!]
\centering
\includegraphics[width=.65\textwidth]{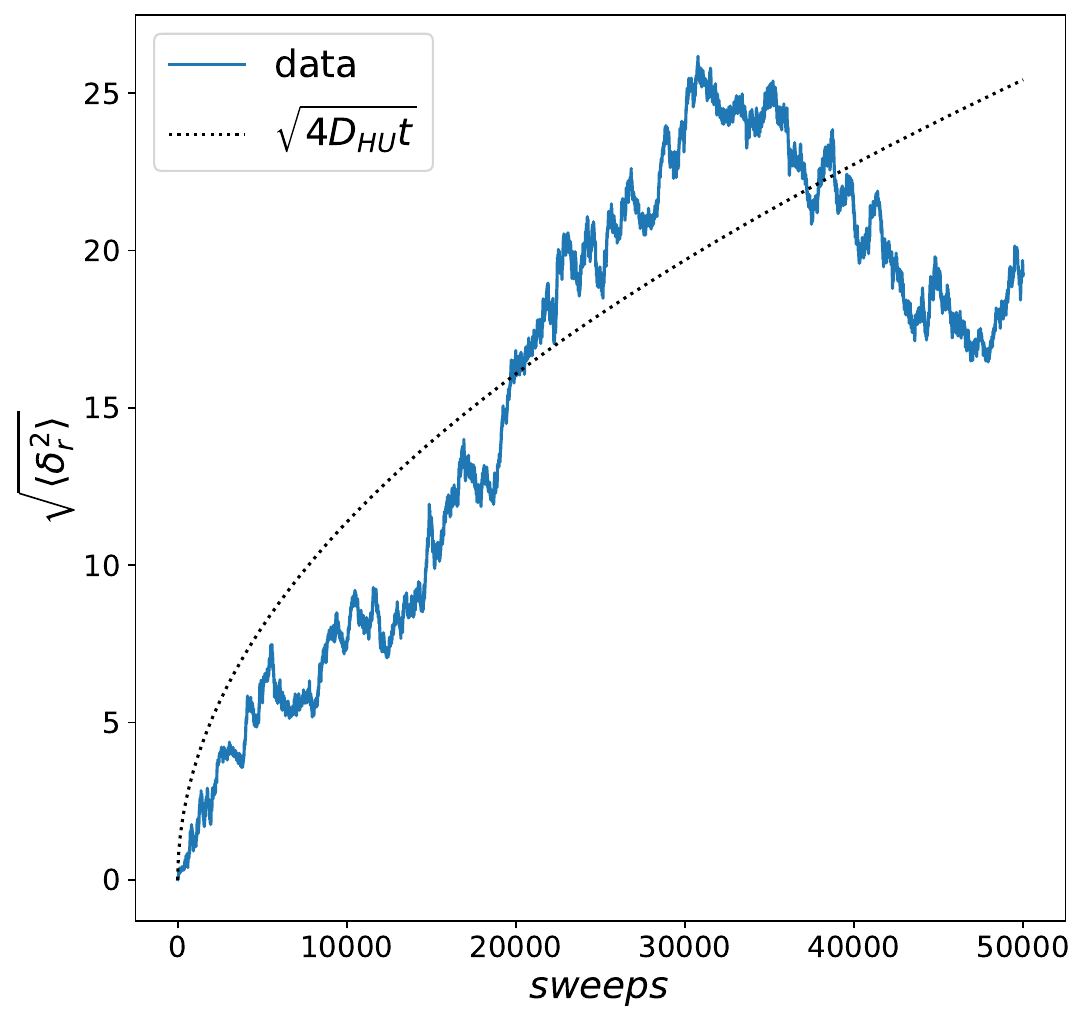}
    \caption{Mean square displacement of the bump after the optimal amount of Unlearning. The diffusion coefficient $\mathcal{D}_{HU}$ has been fitted from the data. Data have been averaged over $10$ realizations. Choice of the parameters: $D = 2$, $w = 0.3$, $\lambda = 10^{-2}$,  $N = 500$, $L = 1$, $p = 200$, $T = 1$.}
\label{fig:brown_diff}           
\end{figure}

A diffusion coefficient $\mathcal{D}$ can be measured by simulations as done in \cite{monasson_crosstalk_2014}. In particular, for $D = 2$, one has

\begin{equation}
    \label{eq:coeff_diff}
    \mathcal{D} = \frac{1}{4 N_{\text{sweeps}}}\sum_{t=1}^{N_{\text{sweeps}}-1} \left( \delta x_t^2 + \delta y_t^2\right)
\end{equation}
where $\delta x_t = x(t+1) - x(t)$ and $\delta y_t = y(t+1) - y(t)$. \\
We thus analyse the case of $N = 500$, $D = 2$, $w = 0.3$, $L = 1$, $p = 200$ at a temperature $T = 1$ for both the learning procedures. Ten samples of the system are evaluated. We obtain
\begin{equation}
    \mathcal{D}_{\text{Hebbs}} = (7.0\pm 2.0) \cdot 10^{-5}, \hspace{1cm}\mathcal{D}_{\text{HU}} = (3.23\pm 0.51) \cdot 10^{-3},
\end{equation}
where errors are the standard deviation of the mean over the samples. To prove that we can effectively refer to the motion as a pure diffusive process, the mean square displacement of the bump after HU is reported in fig. \ref{fig:brown_diff} as a function of time: notice the good agreement of data with the Einstein relation for Brownian motion.
One concludes that diffusion with Unlearning is $\sim10^2$ times more effective than in the standard Hebbian scenario, given this choice of the parameters. The supplemental material in \cite{battista_capacity-resolution_2020} suggests that this is the case for SVM too. In conclusion, HU reproduces, yet again, the optimal conditions for diffusion in one single environment.  
\subsection*{Checkpoint}
In this section we have seen that:
\begin{itemize}
    \item HU approaches the performance of a SVM even when implemented on spatially correlated memories. This result is consistent with the analysis advanced in \cref{sec:choice_of_training_conf}. 
    \item When spatial maps are learned through SVM, or nearly equivalently HU, a Monte Carlo samples the real Euclidean space very effectively, without getting trapped in a limited region.  
\end{itemize}
\newpage

\section{\label{sec:samp}Sampling the optimal noise}
%Selecting training data as fixed points of the dynamics amounts to imposing specific internal dependencies among the noise units $\vec{\chi}$, which are no more i.i.d. random variables, as it was in \cite{gardner_training_1989}. We refer to such dependencies an \textit{structure} of noise. 
Selecting training data that satisfy equation (\ref{eq:deltaL_lwsn3}) amounts to imposing specific internal dependencies among the noise units $\vec{\chi}$, which are no more i.i.d. random variables, as it was in \cite{gardner_training_1989}. We refer to such dependencies an \textit{structure} of noise, and this is what apparently characterizes particular states in the Hebbian landscape of attractors, such as stable fixed points and surrounding saddle points. 
\\Insights from the previous sections on the structure of well performing training data can be used to sample good training configurations according to other strategies. Namely, one can use a supervised Monte Carlo routine that searches for maximally noisy configurations (i.e. $m_t = 0^+$) satisfying condition (\ref{eq:deltaL_lwsn3}). The coupling matrix is initialized according to Hebb's rule (\ref{eq:hop}), and updated recursively according to either TWN (see \cref{eq:lwn}) or HU (see \cref{eq:unl_rule}). We first introduce the sampling algorithm and then report some numerical results regarding both the TWN and HU routines. 
We sample maximally noisy training configurations $m_t = 0^+$ such that  
    $E(\vec{\chi}|m, J)<0$, where
    \begin{equation}
        \label{eq:E_MC}
        \small
        E(\vec{\chi}|m, J) :=  \frac{m}{\sqrt{2\pi(1-m^2)}}\sum_{i,\mu}^{N,p}\omega_i^{\mu}\exp{\left(-\frac{m^2\Delta_i^{\mu^2}}{2(1-m^2)}\right)},
    \end{equation}
%where $\Tilde{\omega}_i^{\mu}$ depends on $\chi_i$ through \cref{eq:omegatilde}.
that is a function of the noisy variables $\chi_i^{\mu}$, conditioned on a reference overlap $m$ and the couplings $J$.
%The definition of $\Tilde{\omega}_i^{\mu}$ depends on the chosen update rule for the couplings. In the case of the training-with-noise algorithm, we use \cref{eq:omegatilde} whereas, in the case of the Unlearning algorithm, we use $\omega_i^{\mu} = m_{\mu}\chi_i^{\mu}$. 
Sampling is done through the following procedure:
\begin{enumerate}
    
    \item The network is initialized in a random configuration and the asynchronous dynamics in \cref{eq:dynamics} is run until convergence on a fixed point. The final state $\vec{S}^{\mu_d}$ must have an overlap $m_t$ in the interval $(0,1/\sqrt{N})$ with one memory $\mu_d$, otherwise the procedure is repeated. 
    
    \item A $T = 0$ temperature dynamics in the landscape of $E(\vec{\chi}|m, J)$ is performed until $E(\vec{\chi}|m, J)<0$.  We use a Kawasaki kind of dynamics over the noisy variables $\vec{\chi}$ to make sure that $m_t$ maintains the prescribed value.

\end{enumerate}
Since $E(\vec{\chi}|m, J)$ is proportional to $\delta \mathcal{L}_U$ (see \cref{eq:deltaL_lwsn}), the procedure will lead to a reduction in the Loss in \cref{eq:loss_ws}. In this setting, the perfect-retrieval and robustness properties can be tuned by the parameter $m$, while the training configurations always have a fixed value of $m_t=0^+$. In particular, to require a performance that is most similar to the one of a SVM, we will set $m \rightarrow 1^-$. \\The sampling procedure starts from fixed points because we know, from the previous sections, that they are close to be the most effective configurations. Interestingly, the described procedure results significantly more effective than a standard minimization of $\mathcal{L}(m = 1^-, J)$: in the latter case training stops when $\mathcal{L} = -1$, while the former technique apparently pushes the stabilities further in the positive values. 
%Nevertheless, the goal of this section is not to propose a winning, or more computationally efficient training routine, but rather to search for the most virtuous training data points in the context of maximal training noise, and reveal their identity.  
\subsection{\label{sec:5b} Algorithm performance}
The sampling procedure results in a  better performance for both TWN and the HU update rules.
\begin{figure*}[ht!]
\begin{tabularx}{\linewidth}{*{3}{X}}
     %\centering
     \begin{subfigure}[b]{\linewidth}
         \centering
         \includegraphics[width=\textwidth]{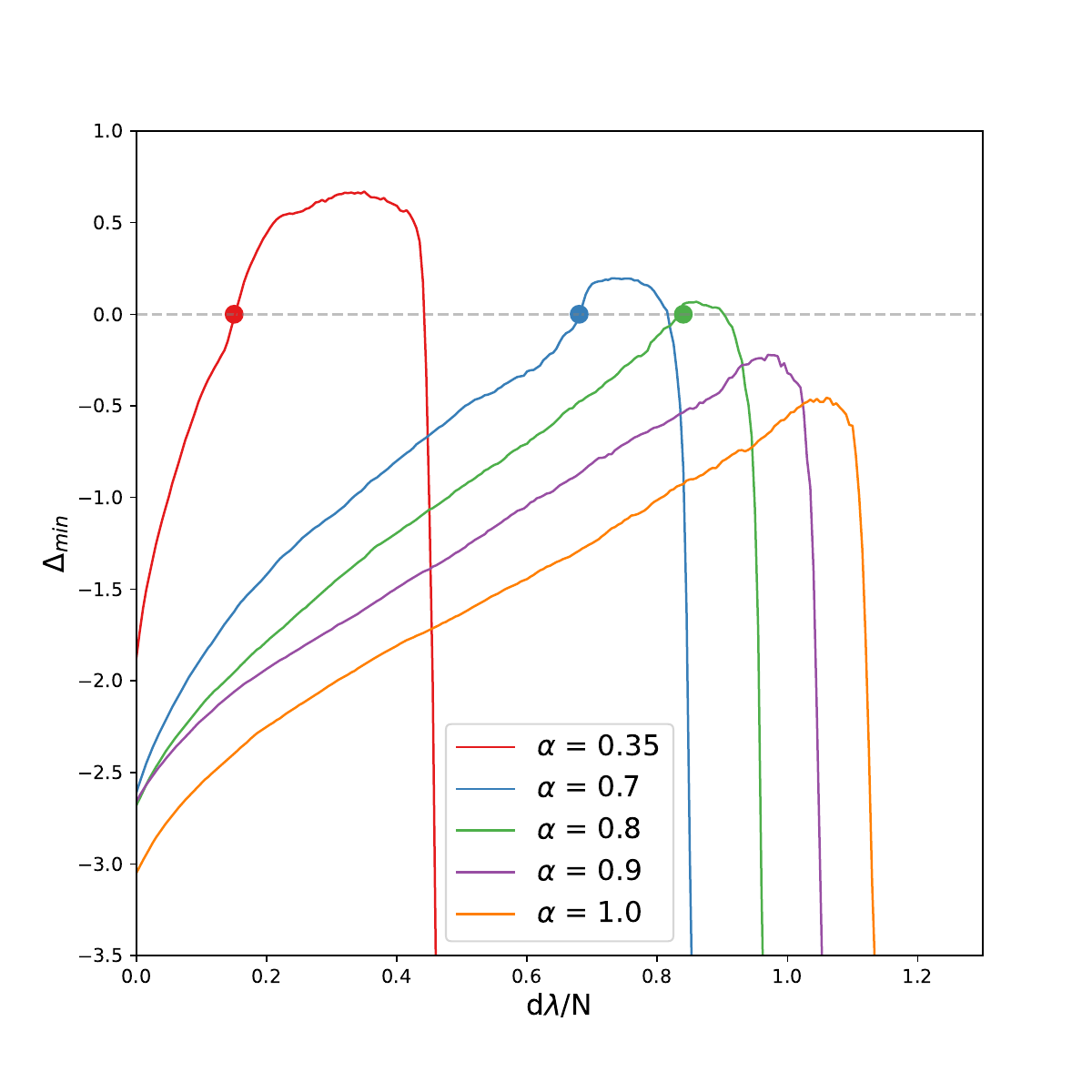}
         \caption{}
     \end{subfigure}
     &
     \hfill
     \begin{subfigure}[b]{\linewidth}
         \centering
         \includegraphics[width=\textwidth]{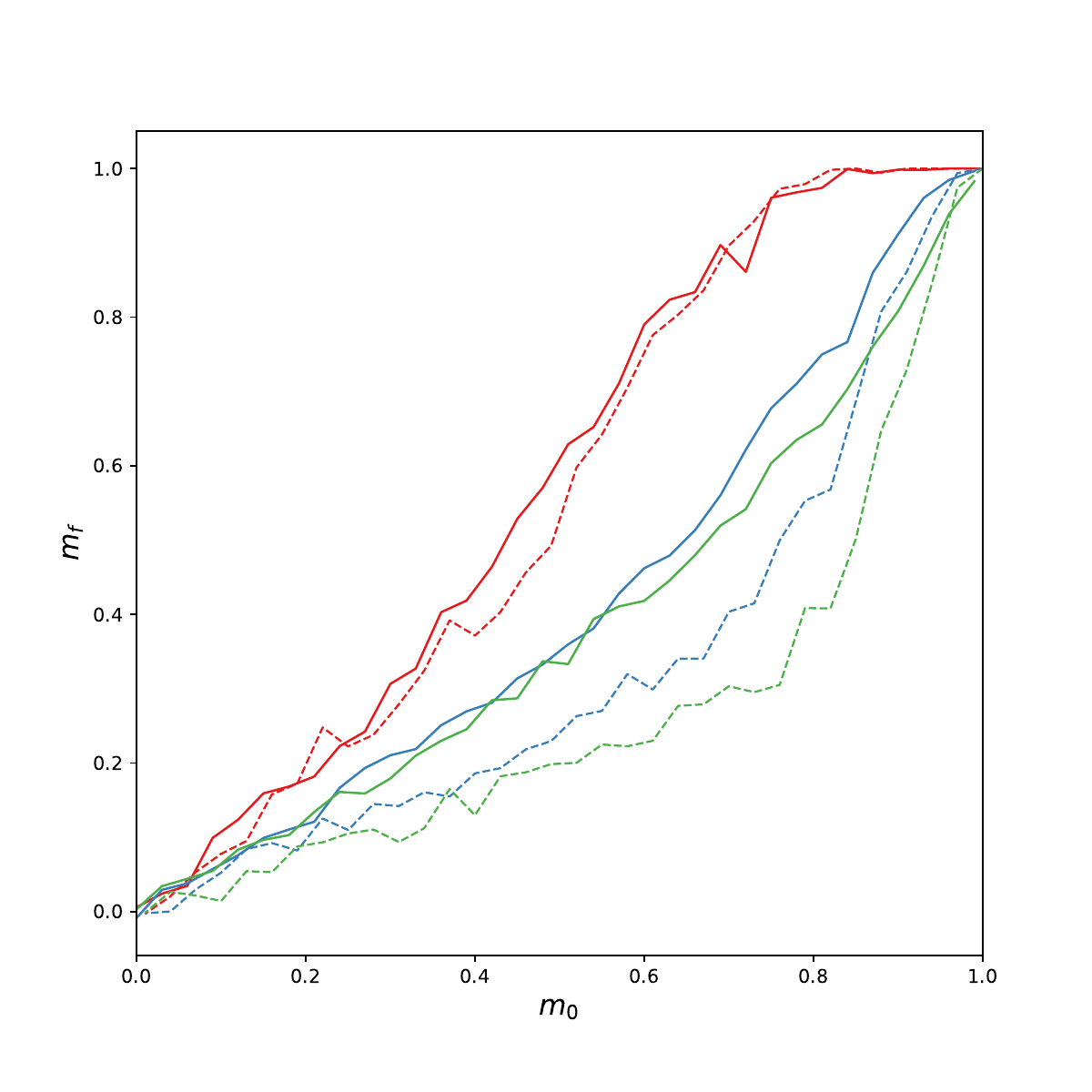}
         \caption{}
     \end{subfigure}
     &
     \hfill
     \begin{subfigure}[b]{\linewidth}
         \centering
         \includegraphics[width=\textwidth]{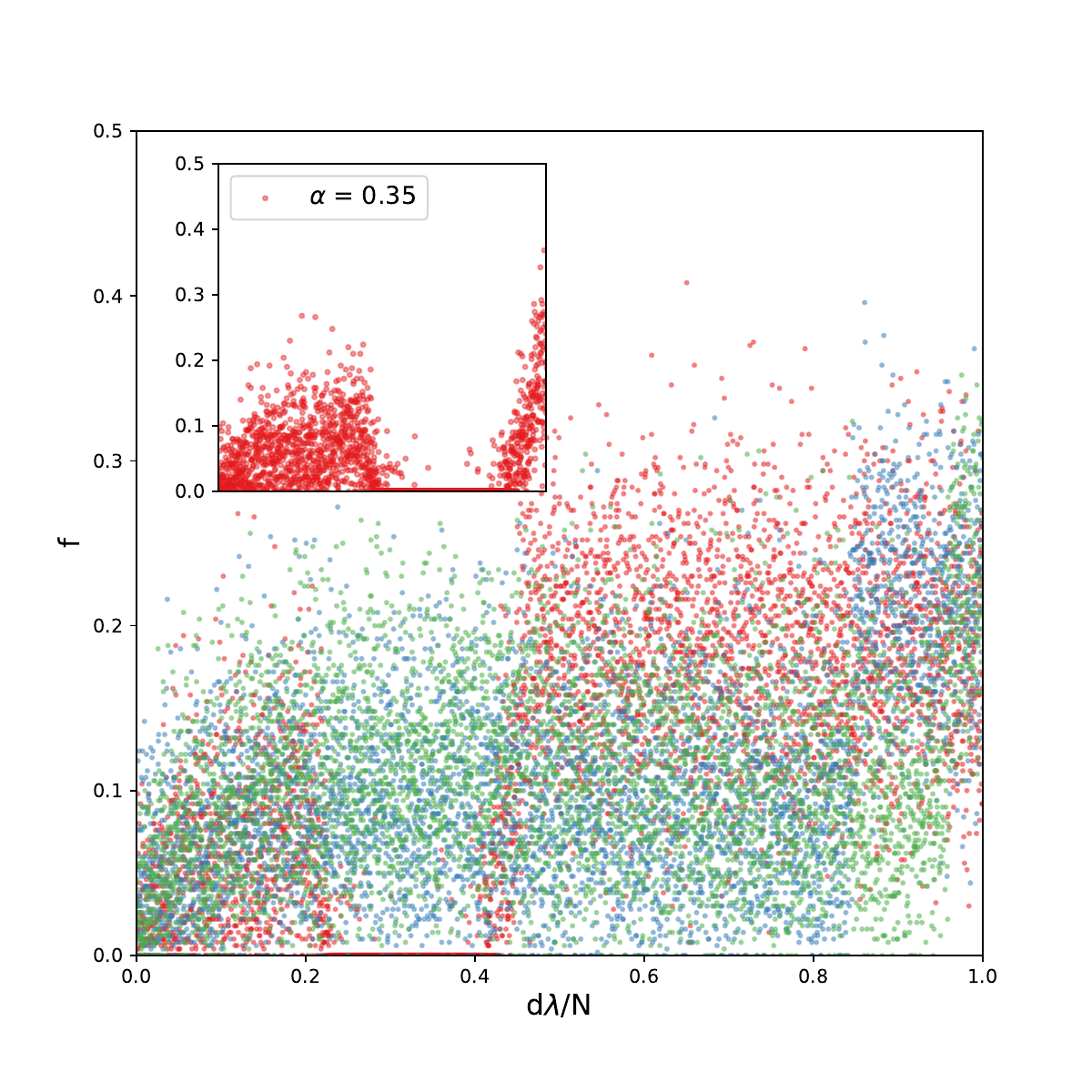}
         \caption{}
     \end{subfigure}
\end{tabularx}
\caption{Performance of the TWN algorithm taught with noisy configurations sampled according to section \ref{sec:samp} regarding four progressing values of the load $\alpha$. (a) Minimum stability as a function of the algorithm time: \textit{full} line is the HU with sampling, \textit{dotted} line is the traditional HU. (b) Retrieval map $m_f(m_0)$, relatively to the \textit{circles} in panel (a) for $\alpha \in [0.35, 0.7, 0.8]$: \textit{full} line is the HU with sampling, \textit{dashed} line is a SVM trained with no symmetry constraints with the same control parameters. (c) Saddle index $f$ as a function of the algorithm steps for $\alpha \in [0.35, 0.7, 0.8]$. The sub-panel zooms over $\alpha = 0.35$ alone. All data points have been averaged over 20 samples in (a),(b) and 5 samples in (c). Errors are neglected for clarity of the image. The choice of the parameters: $N = 100$, $\lambda = 10^{-3}$, $m = 0.9999$.}
\label{fig:lwn_sampled}       
\end{figure*}
\begin{figure*}[ht!]
\begin{tabularx}{\linewidth}{*{3}{X}}
     %\centering
     \begin{subfigure}[b]{\linewidth}
         \centering
         \includegraphics[width=\textwidth]{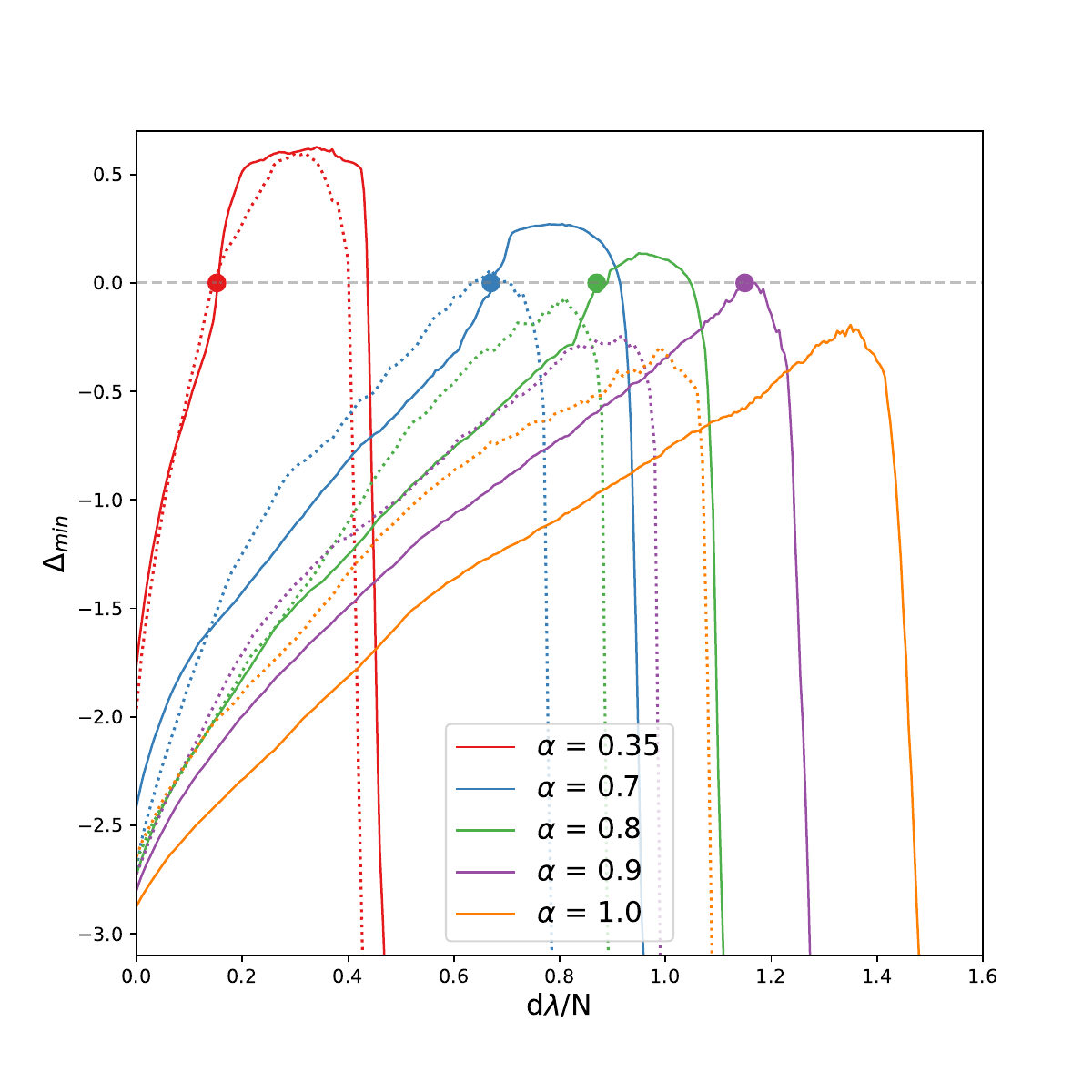}
         \caption{}
     \end{subfigure}
     &
     \hfill
     \begin{subfigure}[b]{\linewidth}
         \centering
         \includegraphics[width=\textwidth]{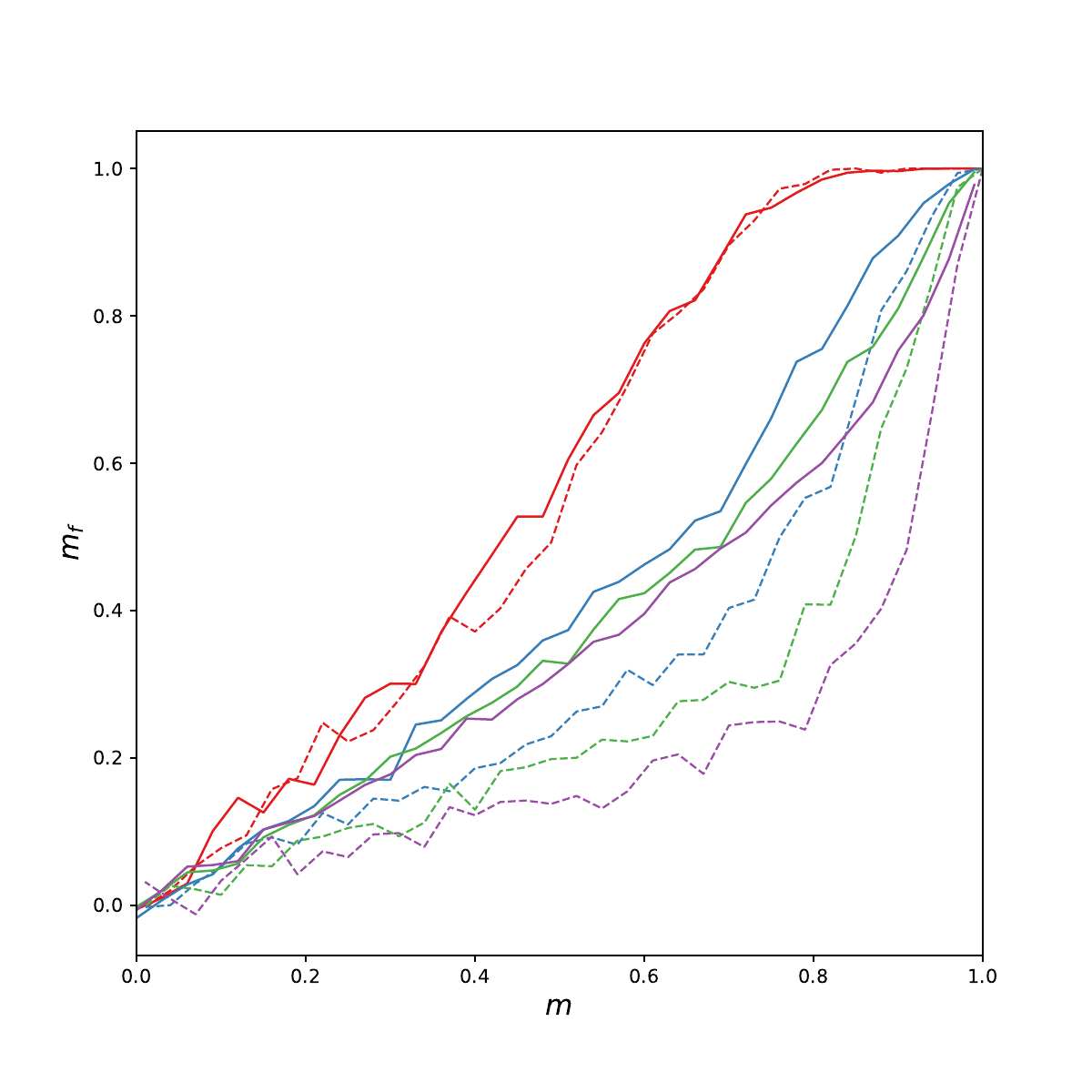}
         \caption{}
     \end{subfigure}
     &
     \hfill
     \begin{subfigure}[b]{\linewidth}
         \centering
         \includegraphics[width=\textwidth]{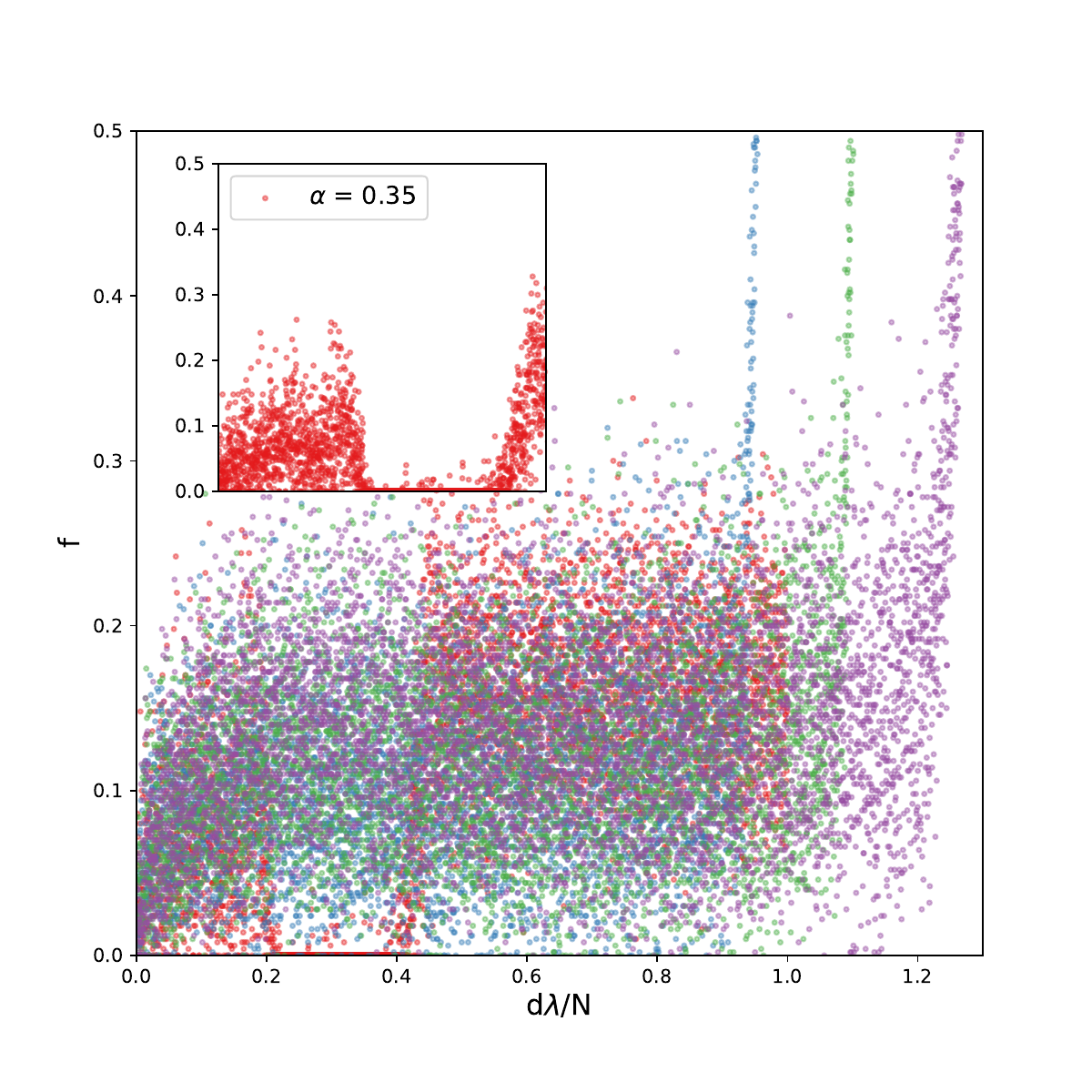}
         \caption{}
     \end{subfigure}
\end{tabularx}
\caption{Performance of the HU algorithm trained with noisy configurations sampled according to section \ref{sec:samp} regarding four progressing values of the load $\alpha$. (a) Minimum stability as a function of the algorithm time: \textit{full} line is HU with sampling, \textit{dotted} line is the traditional HU. (b) Retrieval overlap $m_f(m_0)$, relatively to the \textit{circles} in panel (a) for $\alpha \in [0.35, 0.7, 0.8, 0.9]$: \textit{full} line is HU with sampling, \textit{dashed} line is a SVM trained with no symmetry constraints with the same control parameters. (c) Saddle index $f$ as a function of the algorithm steps for $\alpha \in [0.35, 0.7, 0.8, 0.9]$. The sub-panel zooms over $\alpha = 0.35$ alone. All data points have been averaged over 20 samples in (a),(b) and 5 samples in (c). Errors are neglected for clarity of the image. The choice of the parameters: $N = 100$, $\lambda = 10^{-3}$, $m = 0.9999$.}
\label{fig:unl_sampled}        
\end{figure*}
Results for TWN are reported in \cref{fig:lwn_sampled}. Panel (a) shows that perfect-retrieval is reached up to $\alpha \simeq 0.8$ for a network of size $N = 100$. Panel (b) shows the retrieval map $m_f(m_0)$, for different values of $\alpha$ and for the lowest number of algorithm iterations leading to perfect-retrieval. The high values of $m_f$ for $m_0 \sim 1$ indicates that the network, while achieving perfect-retrieval, maintains good robustness properties. Results for a SVM with the same control parameters are also shown, for comparison. The curves obtained from TWN are higher than ones relative to the SVM, signaling a better recalling performance, even though quantifying this effect will require a more detailed study of finite size effects. 
Fig. \ref{fig:unl_sampled} shows analogous plots for the HU update rule. Again, the network compares favorably with traditional HU in terms of memory capacity, that is increased up to $\alpha \simeq 0.9$ (see panel (a)) with respect to the maximum capacity of $\alpha \simeq 0.7$ obtained with fixed points only. Panel (b) represents an indication for the sizes of the basins of attraction at the lowest number of algorithm iterations leading to perfect-retrieval. Yet again basins appear to be larger than the ones from a SVM trained with the same control parameters.

Panel (c) in \cref{fig:lwn_sampled} and \cref{fig:unl_sampled} report the behavior of the saddle index $f$ of the sampled configurations as a function of the number of steps of the algorithm, relatively to TWN and HU respectively. Most of the sampled training configurations are low index saddles. When $\alpha$ is low, there exists an interval in the training process when stabilities are all far above zero. During this interval, the exponential factors in \cref{eq:E_MC} are all extremely small, resulting in negligible improvement of the Loss function regardless of the training configuration chosen by the sampler. Correspondingly, the sampler contents itself with fixed points of the dynamics, resulting in the gap in the saddle bands visible in the insets of \cref{fig:lwn_sampled} and \cref{fig:unl_sampled} (c) for $\alpha=0.35$. For higher values of $\alpha$, i.e. when the training task is more challenging, the algorithm never succeeds to increase $\Delta_{min}$ to the point of making sampling trivial, and there is no gap in the saddle band.
\subsection*{Checkpoint}
In this section we have seen that:
\begin{itemize}
    \item Instead of sampling specific states in the energy landscape of the model, i.e. saddles of a given index or local minima, one can directly sample the optimal noisy training configurations, i.e. the best configurations satisfying \cref{eq:deltaL_lwsn3} for each step of the training algorithm. 
    \item The network obtained by sampling directly the best noisy training data outperforms both the TWN algorithm and HU in terms of maximum retrieval capacity and robustness.  
\end{itemize}
\newpage

\section{Summary \& Conclusions}

In this chapter we have seen that:
\begin{itemize}
    \item The training-with-noise algorithm \cite{gardner_training_1989} converges to a network that is fully described by the statistical mechanics developed by Wong and Sherrington in \cite{wong_optimally_1990, wong_neural_1993}. For any value of $\alpha$, when the training overlap is $m_t = 0^+$ the algorithm trains a Hebbian network; when $m_t = 1^-$ the same procedure trains a Support Vector Machine. 
    \item The symmetric version of the training-with-noise algorithm can modified to learn configurations having $m_t = 0^+$ and satisfying the condition (\ref{eq:deltaL_lwsn3}): while Gardner's standard algorithm would have trained a Hebbian network, now the resulting network approaches a Support Vector Machine. 
    \item The initial conditions for the algorithm are important. When the model is initialized in a Hebbian network, stable fixed points and low saddles in the energy landscape having $m_t = 0^+$ with respect to the memories abundantly satisfy the optimal noise condition in \cref{eq:deltaL_lwsn3}, favouring the training of a Support Vector Machine. Since these states are very far from the memories the procedure can go fully unsupervised (by means of a relaxation dynamics or a Monte Carlo sampling at low temperatures), increasing in speed and biological plausibility. 
    \item The traditional Hebbian Unlearning algorithm is contained into the training-with-noise algorithm, when $m_t = 0^+$ and the sampled states are stable fixed points of the dynamics. Since such states are good noisy configurations, according to the criterion in \cref{eq:deltaL_lwsn3}, the Unlearning routine must train a network that resembles a Support Vector Machine. This statement is supported by detailed numerical evidence.
    \item Memories containing an internal structure in the features (e.g. memories that can be mapped into positions in a Euclidean space, hence having strong spatial correlations), yet being well separated in the configuration space, can be learned by the modified training-with-noise algorithm and Unlearning. Memories that have both an internal structure and mutual correlations among each other fail to be learned by the same methods. 
\end{itemize}

For the purpose of statistical mechanics, our focus in this study is on disordered systems, specifically in the context of associative memory, which is a prominent area of research. 
The work by Abbott and Kepler \cite{abbott_universality_1989} investigates the learning of memory collections as a flux in Gardner's space of interactions \cite{gardner_space_1988}. 
This flux exhibits three distinct fixed points, or universality classes, representing \textit{unique} (or \textit{saturated}) solutions to the perfect-retrieval problem: the Hebbian matrix \cite{hebb_organization_1949, hopfield_neural_1982, amit_modeling_nodate}, the Pseudo-inverse matrix \cite{personnaz, dotsenko_statistical_1991}, and the Support Vector Machine (SVM) \cite{battista_capacity-resolution_2020, marvin_minsky_seymour_papert_perceptrons_1969}. 
While we understand the retrieval properties of Hebbian networks, which do not achieve proper perfect-retrieval, it is important to note that Pseudo-inverse matrices achieve perfect-retrieval up to $\alpha$ being unity. However, they also represent overfitting systems, as memories have null basins of attraction when $\alpha > 1/2$. 
In this context, the training-with-noise algorithm serves as an interesting interpolation between two of these fixed points: the Hebbian point (corresponding to $m_t = 0^+$) and the SVM (corresponding to $m_t = 1^-$). By considering the noise injection of the training-with-noise algorithm as a regularization in the perceptron learning process, we introduce another degree of freedom, which is the \textit{noise structure}. 
When the noise is i.i.d., an SVM can be trained with $m_t \rightarrow 1^-$. However, if the training data satisfy the condition for optimal noise, we only require $m_t = 0^+$ to approach the SVM fixed point, at least for $\alpha$ values that are not too high. 
Our work demonstrates that Hebbian Unlearning represents this type of flux in the space of interactions, challenging the prevalent belief in the statistical mechanics community that Unlearning leads the connectivity matrix to resemble a Pseudo-inverse matrix rather than a perceptron, particularly the most stable one. This supports the idea that Unlearning serves as an effective and unsupervised (hence more biologically relatable) training procedure for associative memory models.

Moving on to the potential implications of our work in biology and artificial intelligence, empirical observations indicate that the mammal brain is a highly efficient learner. It requires only a few examples to memorize concepts and generalize them \cite{kahnt_dopamine_2016, johnston_abstract_2023}. Some brain regions seem to function similarly to perceptrons \cite{brunel_optimal_2004, brunel_is_2016}, even though natural learning is unlikely to be supervised. Neuroscientists increasingly emphasize the significance of sleep for memory consolidation \cite{creery_electrophysiological_2022, maingret_hippocampo-cortical_2016}, and intriguing connections emerge between unsupervised training algorithms in machines and the necessity of specific synaptic plasticity processes occurring separately from daily experiences \cite{crick_function_1983, girardeau_brain_2021, hinton_forward-forward_2022, hinton_unsupervised_1999, tadros, audio}. Hoel \cite{erik_hoel_overfitted_2021} speculates on the importance of dream-sleep to avoid overfitting in the brain by injecting noise through hallucinoid contents, thereby enhancing robustness. Based on these works, a local Hebbian-like action on synapses can ensure decorrelation of stored memories and prevent confusion. Contrary to past literature repeatedly stating Hebbian Unlearning as a form of \textit{reverse} learning, our research reveals that it is responsible for learning coherent noisy versions of memories, helping minimize overfitting.

It is essential to emphasize that the theoretical neuroscientists \cite{foldiak_forming_1990, barlow_adaptation_1989, zhang_anti-hebbian_1997} simultaneously, yet independently, proposed the importance of an anti-Hebbian rule in learning. 
This rule emerged as a useful mechanism for decorrelating information in the system and maximizing the quality of the retrieval process. Scientists also conjectured the potential homeostatic role of such inverse processes, observed indirectly in real networks of neurons, indicating a tendency for the total synaptic volume to be conserved in the brain \cite{turrigiano_homeostatic_1999, turrigiano_homeostatic_2012}. 
The fact that this inverse rule spontaneously arises from noisy perceptron learning (see \cref{eq:lp_update}), encourages the possibility of observing it in future neurophysiology experiments.

In light of these findings and the work presented in this article, one could conceive natural learning as a \textit{two-phase} process, making use of a single synaptic modification rule. In the first \textit{online} phase, external stimuli are processed by the network using the standard training-with-noise algorithm. These stimuli can be imagined as maximally noisy versions of unknown \textit{archetypes} embedded in the environment. Consequently, the training shapes a pure Hebbian landscape of attraction outside the retrieval regime. In the second \textit{offline} phase, the early-formed network samples structured noisy neural configurations, weakly correlated with the archetypes, from the landscape of attractors. These states could be lower saddles or stable fixed points of the neural dynamics. Subsequently, when these neural configurations undergo the same kind of training-with-noise algorithm, memory is consolidated by centering the unconscious archetypes in the middle of large basins of attraction.

In summary, our work makes progress on three fronts. Firstly, it advances a new type of regularization for associative memory models that strongly relies on the topology of the landscape of attractors in spin-glass-like neural networks. Future developments of these studies might focus on characterizing the structure of noise more rigorously, possibly computing a proper density distribution for the effective weights in \cref{eq:deltaL_lwsn3}. \\Secondly, it sheds light on the specific structure of noise that is optimal for learning in neural networks, which may contribute to develop a more refined theory underlying the empirical techniques of noise injection used in training deep networks \cite{drop, shorten_survey_2019, tomasini_how_2023, bonnasse-gahot_categorical_2022}. Finally, it establishes a connection between unsupervised learning processes, which are more biologically relevant, and the supervised ones that underpin most modern neural network theory. This, in turn, encourages a deeper investigation into the noise and its structure present in the stimuli used by the brain to effectively shape associative memory.

\chapter{\label{sec:chap3} The inferential power of Unlearning}

We are now going to deal with Boltzmann Machines (BMs), a class of neural networks introduced in \cref{sec:BML}. BMs accomplish a generative task, i.e. the typical neural states are meant to be indistinguishable from the data that were used to train the network. One can thus sample new data from the probability distribution of the model. For what concerns this kind of systems it is useful to define the concepts of \textit{overfitting} and \textit{generalization}.
\\In the context of BMs, three probability distributions play a role: the inferred model $P_{mod}$, the empirical distribution of the training data $P_{data}$ and $P_{true}$, i.e. the real hidden distribution from which data are sampled. 
We say that the inferred model is \textit{overfitting} (the data) when $P_{mod}$ resembles $P_{data}$ more than $P_{true}$, i.e. the model fits the training data better than it does with unseen examples generated from the same distribution. This translates into an over-specialization of the model, namely it becomes too focused on the details of the specific training-set, instead of understanding the broader structure highlighted by the data. 
Conversely, when the performance of the model does not change significantly when it is trained with a new unseen set of data generated from the same source, we will say that it \textit{generalizes} well. 
\\There are many aspects that impede neural networks to generalize well. For instance, BMs always tends to overfit the training data, by construction. Though, if we assume the data to satisfy the law of large number, and the size of the training-set is sufficiently large, one can approach the true hidden distribution. In principle, if all the possible data are used for training, all the realizable testing data will be included in the statistics. However, when $N$ is large enough, the number of available data might not be enough to reach a good degree of generalization. 
A general idea from statistics wants overfitting to be related to an over-parametrization of the model: there are too many degrees of freedom that the model can employ to reproduce the training data in detail. This argument holds for BMs, that fit the first two moments of a probability distribution by minimizing a Loss is a simple way. Nevertheless, this idea seems not to be valid in deep neural networks, which instead benefit from over-parametrization \cite{overpara, baldassi_overpara} . In this case the gradient of the Loss must be performed by computing quantities over the several layers: this makes the process complex and barely controllable. 
%In the context of shallow models, such as BMs, generalization is favoured by a reduction of the parameters.
Since real data are not generated by a Boltzmann distribution, the real statistics of the data cannot be fully known, i.e. the loss of the BM learning problem cannot be fully minimized. Nevertheless one can employ regularization methods to minimize the distance between the the model and the real statistics of the data while, at the meantime, selecting specific models in this region of comparable loss. This provides different inferred systems with different properties, depending on additional requirements imposed on the inferential problem. For instance, some of these models can be sparse or even contain null parameters (e.g. $L_0$, $L_1$-norm and information based regularization schemes \cite{barrat-charlaix_sparse_2021}), others are fully connected (e.g. $L_2$-norm method).  
Past literature \cite{tkacik_thermodynamics_2015, mora_are_2011, mastromatteo_criticality_2011} highlighted that critical models, i.e. inferred systems being highly susceptible to small changes in the parameters, can be attractive for BM learning. In fact, if we consider training as an homogeneous sampling of the models that minimize $\mathcal{L}$, critical models have a large basin of attraction and are thus sampled most often~\cite{mastromatteo_criticality_2011}.
Criticality can be problematic for data generation: it implies a long correlation time in the Monte Carlo dynamics, slowing down the sampling process; it makes the model susceptible under rescaling of the parameters, reducing its predictivity in real data applications. Hence, avoiding criticality might help increasing generalization.
To summarize, we define the generalization performance of the model according to two properties: \textit{accuracy}, i.e. the capability of the model to be as similar as possible to the distribution that generated the data; \textit{robustness}, i.e. the tendency of the model not to change its typical configurations when slightly changing (e.g., rescaling by a common temperature factor) the parameters. 
The goal of this chapter is to introduce a new type of regularization in order to find a better compromise between these two requirements. 
This new tool, that we named Unlearning regularization, appears to be more effective than other techniques in approaching a higher generalization performance of the neural network. 
The performance of the Unlearning regularization is evaluated from artificial data generated by both a Curie-Weiss (CW) and a Sherrington-Kirkpatrick (SK) model with a small number of neurons. In this way the original parameters are known, the fixed points of the learning equations are reached precisely and the useful quantities can be computed exactly. \\A particular limit of this type of regularization coincides with a thermal version of the traditional Hebbian Unlearning (HU) algorithm studied in \cref{sec:noise_inject}. We hence evaluate this particular case of study and conclude that HU is able to infer the original data distribution with a very good degree of generalization. The performance shows an optimum in time that scales with the control parameters, as it was for the associative memory task. We conclude that HU can be interpreted as a two-steps BM learning procedure. Finally, our last contribution is a study of the under-sampling regime of BMs in the case of random data, which provides for a new linking trait between generative models and the best performing associative memory models.  \\ \\
The structure of this chapter is the following. Some extensively employed regularization techniques (i.e. $L1$ and $L2$) are firstly introduced in \cref{sec:regs}. Then the new Unlearning regularization is defined in \cref{sec:u_reg} and its generalization performance is analyzed for two different data sources: a CW model (\cref{sec:CWdata}) and a SK model (\ref{sec:SKdata}). The study of a particular limit of the regularization technique leading to Unlearning follows in \ref{sec:Ulimit}, with an explanation for its inferential behaviour. Eventually, in \cref{sec:overfit}, we identify a common trait between SVMs and BMs in their under-sampling regime, i.e. when the number of training data is small. 
\newpage
\section{\label{sec:regs} Regularization in Boltzmann Machines}
$L_p$ regularizations are certainly the most famous regularization methods, also due to their intuitive interpretation. This approach penalises high values of the parameters by adding a $L_p$ norm of the same variables in the expression of the Loss function. As a consequence, the sparsity of the graph is incentivized: the redundant parameters are weakened with respect to the relevant ones. The expression of the Loss of a BM with $L_1$ and $L_2$ regularizations is given by the following equations
\begin{equation}
    \label{eq:L1}
    \mathcal{L}_{1} = \mathcal{L}^{BM}(J, \vec{h}) + \gamma_1\sum_{i,j>i}|J_{ij}| + \gamma_1\sum_{i}|h_i|, 
\end{equation}

\begin{equation}
    \label{eq:L2}
    \mathcal{L}_2 = \mathcal{L}^{BM}(J, \vec{h}) + \frac{\gamma_2}{2}\sum_{i,j>i}J_{ij}^2 + \frac{\gamma_2}{2}\sum_{i}h_i^2, 
\end{equation}
where $\mathcal{L}^{BM}$ is defined in \cref{eq:kull} and $\gamma_1,\gamma_2$ two small regularization rates. This translates into new updating rules for the parameters, i.e. 
\begin{equation}
    \label{eq:L1up}
    \delta J_{ij}^{(1)} = \delta J_{ij}^{BM} - \lambda\gamma_1\text{sign}(J_{ij})\hspace{1cm}\delta h_i^{(1)} = \delta h_{i}^{BM} - \lambda\gamma_1\text{sign}(h_i), 
\end{equation}
\begin{equation}
    \label{eq:L2up}
    \delta J_{ij}^{(2)} = \delta J_{ij}^{BM} - \lambda\gamma_2 J_{ij}\hspace{1cm}\delta h_i^{(2)} = \delta h_{i}^{BM} - \lambda\gamma_2 h_i. 
\end{equation}
%Another type of regularization is described in \cite{barrat-charlaix_sparse_2021} and it consists on a information-based decimation method for the couplings $J$: equations \ref{eq:Jup} and \ref{eq:hup} are iterated until convergence; for each coupling $J_{ij}$ a virtual joint distribution of the model is computed as if the coupling were absent from the graph; a small percentage of the couplings that show the smallest symmetric Kullback-Leibler divergence is erased from the graph without being updated anymore in the training equations; the procedure is re-iterated until a limit connectivity is reached in the graph. This method appeared to be more effective at reducing overfitting than other types of regularizations.\\
We are now interested in a criterion that describes the generalization performance of a BM. 
As mentioned before, we divide generalization into two features: accuracy, i.e. the similarity of $P_{mod}$ to $P_{true}$, in terms of a measure in the space of the probability distributions; robustness, i.e. capability of the model to change slightly under variations of the parameters.\\
The first aspect can be quantified by the Kullback-Leibler divergence $D_{KL}(true | mod)$ between the inferred model and the original one, defined as
\begin{equation}
    \label{eq:kl_div}
    \small
    D_{KL}(true|mod) = \sum_{\vec{S}}P_{true}(\vec{S})\log\left(\frac{P_{true}(\vec{S})}{P_{mod}(\vec{S})}\right).
\end{equation}
This quantity is not symmetric under the exchange of the two distributions, which can be inconvenient to define a distance. We can then adopt a symmetric version of the divergence as
\begin{equation}
    \label{eq:skl_div}
    \small
    sD_{KL}(true,mod) = D_{KL}(true|mod) + D_{KL}(mod|true).
\end{equation}
Even if numerical results do not show a significant difference between these two quantities, we will mainly adopt $sD_{KL}$. \\
Regarding the second trait, it is known that Ising-like models inferred via BM learning are close to be critical \cite{tkacik_thermodynamics_2015, mora_are_2011, mastromatteo_criticality_2011}. Criticality is associated with a susceptibility of the system to small variations of the temperature and can be measured through the \textit{specific heat} \cite{landau_statistical_1980}. This observable is defined as  
\begin{equation}
    \label{eq:cv}
    C_v(\beta) = -\beta\frac{\partial S_{\beta}}{\partial \beta} = \beta^2\left(\langle E^2 \rangle_{\beta} - \langle E \rangle_{\beta}^2 \right),
\end{equation}
where $S_{\beta}$ is the entropy of the model at a given inverse temperature $\beta$. Notice that \cref{eq:cv} defines a susceptibility with respect to the energy $E$, which also uniquely determines the model with its parameters. 
The fact that inferred neural networks typically display a peak in $C_v$ around the value of $\beta$ used for training implies the approaching of a critical behaviour at that same temperature (the finite size of the system impedes $C_v$ to properly diverge and to show a real criticality). 
Therefore, since all parameters naturally scale with $\beta$ in \cref{eq:boltzpdf}, $C_v$ is a measure of the sensibility of the model to a small perturbation of the parameters: high values reached by $C_v$ suggest strong variations of the inferred statistics to small variations of the parameters. From now on we will consider the quantity $C_v(\beta)/\beta$ for $\beta \neq 1$ or, equivalently, $C_v(1)$ since they properly represent the variation of a thermodynamic quantity, i.e. the entropy of the model, with respect to the inverse temperature $\beta$.
\subsection*{Checkpoint}
In this section we have seen that:
\begin{itemize}
    \item $L_p$ regularizations are used in BMs to weaken the redundant parameters with respect to the relevant ones. 
    \item The generalization capability of a BM is declined in two different properties: the similarity between $P_{mod}$ and $P_{true}$; the robustness of $P_{mod}$ under perturbations of the parameters. 
    \item The distance between the inferred model and the true distribution of data can be measured through the Kullback-Leibler divergence; the robustness of the model under a perturbation of the parameters is signaled by a flattening of the rescaled specific heat $C_v(\beta)/\beta$, or equivalently $C_v(1)$ when the model is inferred at $\beta = 1$. 
\end{itemize}
\newpage

\section{\label{sec:u_reg} Unlearning regularization}
We now propose a new type of regularization that has, as a goal, to impose the maximum robustness under rescaling of the parameters, i.e. 
\begin{equation}
    J \longrightarrow a J\hspace{2cm}\vec{h} \longrightarrow a \vec{h} \hspace{1.5cm}\text{with}\hspace{0.5cm}a \geq 0.
\end{equation}
We can interpret this operation as a redefinition of the inverse temperature $\beta$ in the model. 
We will evaluate the performance of this method on different types of data-sets, then we will compare it to $L_p$ regularizations showing a gain in the generalization capability of the network. 
\\\\For the model to be robust under a redefinition of the parameters, we can add a regularization term to the traditional BM loss function that shifts the peak of the specific heat (i.e. the critical temperature) away from $\beta = 1$, where the data are generated. 
Hence, let us define the following loss function
\begin{equation}
    \label{eq:obj}
    \small
    \mathcal{L}(J,h | a) = D_{KL}(data|1) + \left(\frac{a-1}{a}\right) D_{KL}(data|a),
\end{equation}
with
\begin{equation}
    \label{eq:crossent}
    D_{KL}(data|\beta) = \sum_{\vec{S}}P_{data}(\vec{S})\log{\left(\frac{P_{data}(\vec{S})}{P_{\beta}(\vec{S})}\right)}. 
\end{equation}
Notice that the chosen distance between the two distributions is asymmetric, but the reason for this choice will be evident in a few lines.   
The gradient descent equations for this Loss function become
\begin{equation}
    \label{eq:Juprob}
    \delta J_{ij} = \lambda\left[ a\left(\langle S_i S_j \rangle_{data} - \langle S_i S_j \rangle_{a}\right) - (\langle S_i S_j\rangle_1 - \langle S_i S_j \rangle_a )\right],
\end{equation}
\begin{equation}
    \label{eq:huprob}
    \delta h_{i} = \lambda\left[ a\left(\langle S_i \rangle_{data} - \langle S_i \rangle_{a}\right) - (\langle S_i \rangle_1 - \langle S_i\rangle_a )\right].
\end{equation}
When $a = 1$ equations (\ref{eq:Juprob}), (\ref{eq:huprob}) coincide with the original BM learning. In the limit $a \rightarrow 0$ one has $\langle S_i S_j \rangle_{a = 0} \rightarrow 0$ when $N\gg 1$ and \cref{eq:Juprob} tends to a thermal version of HU (see \cref{eq:unl_rule}) performed at $\beta = 1$, i.e.
$$\delta J_{ij} = -\lambda \langle S_i S_j \rangle_1, $$
which is similar to previous attempts from the associative memory literature \cite{nokura_paramagnetic_1996}. On the other hand, when $a \rightarrow \infty$, and the learning rate is redefined as $\lambda = \mathcal{O}(a^{-1})$, the algorithm becomes
$$\delta J_{ij} = \lambda\left( \langle S_i S_j \rangle_{data} - \langle S_i S_j \rangle_{\infty} \right) $$
which also resembles the Unlearning procedure. As a difference with the standard routine, there is a Hebbian \textit{input} term $\langle S_i S_j \rangle_{data}$ which impedes the couplings to vanish at convergence.\\
At this point, the choice of the expression for $\mathcal{L}$ should appear more clear: we search for an algorithm that interpolates between the BM learning algorithm and HU. In this way HU emerges as a particular limit of a regularization method on BMs.  
\\\\For the rest of our study we will deal with a small sized network, i.e. $N \sim 18 \div 20$, which will allow to reach the fixed points of equations (\ref{eq:Juprob}) and (\ref{eq:huprob}) precisely, whether such points are admitted, with no errors due to the finite sampling. We can then can compute the observables $C_v/\beta$ or $sD_{KL}$ more easily, in order to determine the generalization performance of the system. For simplicity of notation we will rename $$ \langle S_i S_j \rangle_{data} =c_{ij}^d \hspace{1.5cm}\langle S_i S_j \rangle_{\beta} = c_{ij}^{\beta},$$ where we will mainly deal with $\beta = a$ and $\beta = 1$. The inverse temperature of the generating model is $\beta = \beta_d$.
The generating models for the data will be of two kinds: a mean field fully connected Ising network (i.e. Curie-Weiss \cite{zamponi_mean_2014}) and a fully disordered network (i.e. Sherrington-Kirkpatrick \cite{sherrington_solvable_1975}) which are both critical in the thermodynamic limit, i.e. they admit a divergence of the specific heat. The divergence appears as a peak in the finite size case. Solving the BM learning algorithm equations would lead to the exact parameters of the generating model, that approaches a criticality at finite $N$. Nevertheless, solving equations (\ref{eq:Juprob}) and (\ref{eq:huprob}) leads to a non-trivial network with its own generalization capability.
\subsection{\label{sec:CWdata} Data generated from a Curie-Weiss model}
Let us consider the case of inferring a Curie-Weiss model (CW), that is defined by the following energy function
\begin{equation}
E_{CW}[\vec{S} | J] = -\sum_{i,j>i}S_i J_{ij}S_j \hspace{2cm} J_{ij} = \frac{J}{N} \hspace{1cm} \forall i,j.
\end{equation}
This model can be treated fully analytically in the finite size case, and the solution procedure for the main quantities is here reported.
We will make use of the algorithm presented in \cref{eq:Juprob} and \cref{eq:huprob}. In this case correlations can be obtained exactly even in the finite $N$ case. Since fields are zero, by construction of the model, we are interested in the evolution equation for the couplings, i.e.
\begin{equation}
    \label{eq:JuprobCW}
    \Dot{J} = \lambda\left[ a\left(c_d - c_a(t)\right) - (c_1(t) - c_a(t) )\right],
\end{equation}
where we used the fact that both couplings and correlation functions assume the same values $\forall i,j$. Notice that all elements of the matrices are identical, because each node receives the same fields from its neighbours.  
The fixed point equation for the correlation functions from \cref{eq:JuprobCW} is
\begin{equation}
    \label{eq:CWfixed}
    c_a^* = \frac{c_1^*}{1-a} - \frac{a}{1-a}c_d,
\end{equation}
where the star indicates the value of the correlations at convergence. 
Moreover we know that 
\begin{equation}
\label{eq:CWpart}
Z_{\beta J} = \sum_m \binom{N}{\frac{N}{2}(1+m)}\exp{\left(\frac{\beta J}{2}(N m^2-1)\right)},
%\int_{-1}^{+1}dm\cdot \binom{N}{\frac{N}{2}(1+m)}\exp{\left(\beta(N m^2-1)J(t)\right)}
\end{equation}
where $m \in [-1, -1 + \frac{2}{N}, .., 1-\frac{2}{N}, +1]$ and we used
\begin{equation}
E[\vec{S}] = - J\left(N\left(\frac{1}{N}\sum_i S_i\right)^2 - 1\right) = - J\left(N m^2 - 1\right).
\end{equation}
Then one has that the second moment of the magnetizations $\langle m^2 \rangle_{\beta}$ with respect to the Gibbs-Boltzmann measure can be written as
\begin{equation}
    \label{eq:msqav}
    I_N(\beta J) =  \frac{\sum_m m^2 \binom{N}{\frac{N}{2}(1+m)} \exp{\frac{\beta}{2} J(N m^2-1)}}{\sum_m\binom{N}{\frac{N}{2}(1+m)} \exp{\frac{\beta}{2} J(N m^2-1)}} = \frac{1 + (N-1)c(\beta J)}{N}.
\end{equation}
Thus
\begin{equation}
    \label{eq:c_beta}
    c(\beta J) = \frac{N I_N(\beta J)-1}{N-1}
\end{equation}
The fourth moment of the magnetization
$\langle m^4 \rangle_{\beta}$ can be written as
\begin{equation}
    \label{eq:msqsqav}
    I^{(4)}_N(\beta J) =  \frac{\sum_m m^4 \binom{N}{\frac{N}{2}(1+m)} \exp{\frac{\beta}{2} J(N m^2-1)}}{\sum_m\binom{N}{\frac{N}{2}(1+m)} \exp{\frac{\beta}{2} J(N m^2-1)}} = \frac{1 + (N^3-1)k(\beta J)}{N^3},
\end{equation}
where $k(\beta J)$ is the fourth-order correlation among spins.
Eq. (\ref{eq:CWfixed}) becomes
\begin{equation}
    \label{eq:CWfixed2}
    c(aJ) = \frac{c(J)}{1-a} - \frac{a}{1-a}c(\beta_d J),
\end{equation}
or equivalently
\begin{equation}
    \label{eq:CWfixed3}
    I_N(aJ) = \frac{I_N(J)}{1-a} - \frac{a}{1-a}I_N(\beta_d J),
\end{equation}
%One can then find the value of the coupling $J(a)$ by substituting \cref{eq:c_beta} in equation \cref{eq:CWfixed} for both $\beta = a$ and $1$ with $c_d$ computed from \cref{eq:c_beta} with $\beta = \beta_d$ and $J = 1$.
that has be to solved for $J$.\\ 
The specific heat of the inferred model can be computed as
\begin{equation}
    \label{eq:Cv_CW_analytic}
    \small
    C_v(\beta) = \left(\frac{\beta N J(a) }{2}\right)^2\left(I^{(4)}_N(\beta J(a)) - {I^{(2)}_N}^2(\beta J(a)) \right).
\end{equation}
\begin{figure}[ht!]
\begin{multicols}{2}
\centering
\subcaptionbox{}{
\includegraphics[width=0.95\linewidth]{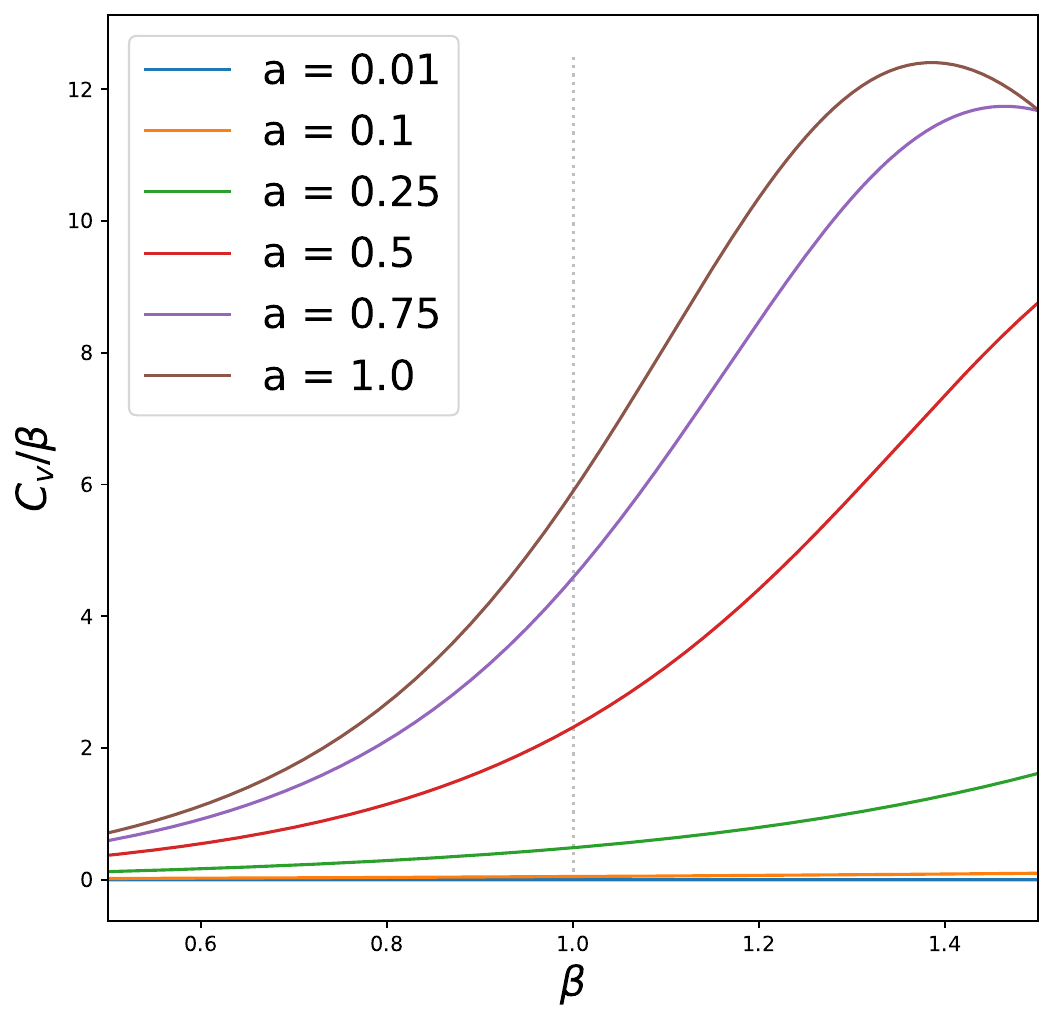}}\par
\subcaptionbox{}{
\includegraphics[width=\linewidth]{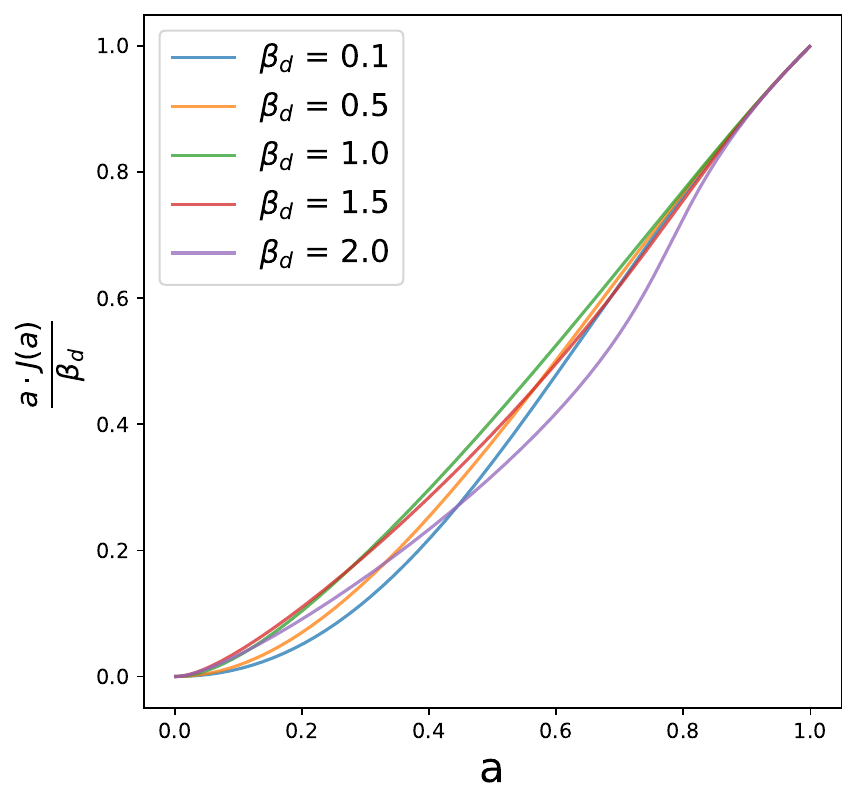}}\par
\end{multicols}
\caption{Inferred couplings and specific heat for a small network trained with a CW model with $N = 20$ spins. (a) The specific heat is reported in a window centered around $\beta = 1$ at different values of the parameter $a$. (b) The inferred couplings multiplied by $a/\beta_d$ as a function of $a$ at different values of $\beta_d$: the linear regime displays the robustness the parameters under rescaling.}
\label{fig:CW}
\end{figure}
\\Equation (\ref{eq:CWfixed3}) is solved for a network of $N = 20$ neurons and results are reported in \cref{fig:CW}.
As showed in \cref{fig:CW}a the specific heat has a peak slightly after $\beta = 1$ that flattens progressively when $a$ is decreased. Note that, for each value of $a$ the couplings $J = J(a)$ are obtained from the calculations, but quantities are computed from the Gibbs-Boltzmann distribution with the rescaled parameters $\beta\cdot J(a)$. Moreover, in \cref{fig:CW}b $J(a)$ is computed and the quantity $aJ(a)/\beta_d$ is plotted as a function of $a$ for different choices of $\beta_d$. As one can notice, the lines approach a linear regime before reaching $J(1)/\beta_d = 1$. This behaviour signals that $J(a)$ is nearly constant in $a$, hence the inferred model changes little passing from $\beta = 1$ to $\beta = a$, as required by the regularization. 

\subsection{\label{sec:SKdata} Data generated from a Sherrington-Kirkpatrick model}
\begin{figure}[ht!]
\centering
\includegraphics[width=0.5\linewidth]{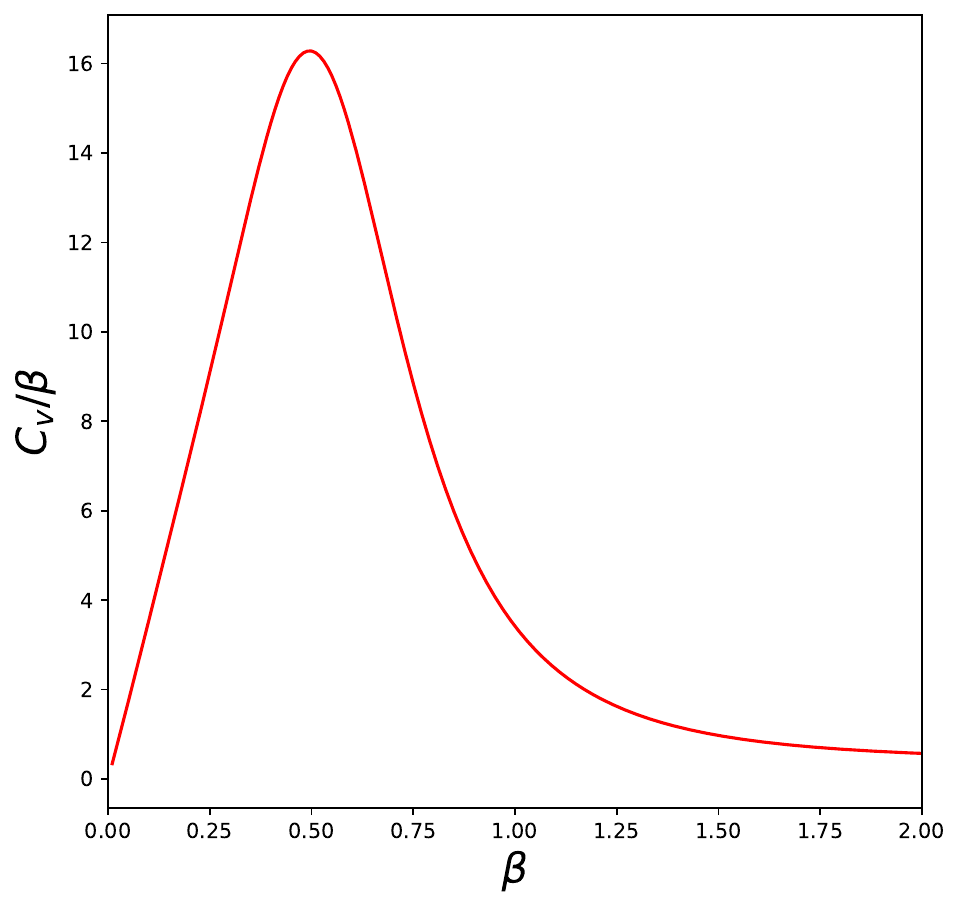}
\caption{Specific heat $C_v/\beta$ as a function of the inverse temperature $\beta$ for the single SK spin-glass used in our study. The network contains $N = 18$ neurons.}
\label{fig:cvSK}
\end{figure}

\begin{figure}[ht!]
\begin{multicols}{2}
\centering
\subcaptionbox{}{
\includegraphics[width=\linewidth]{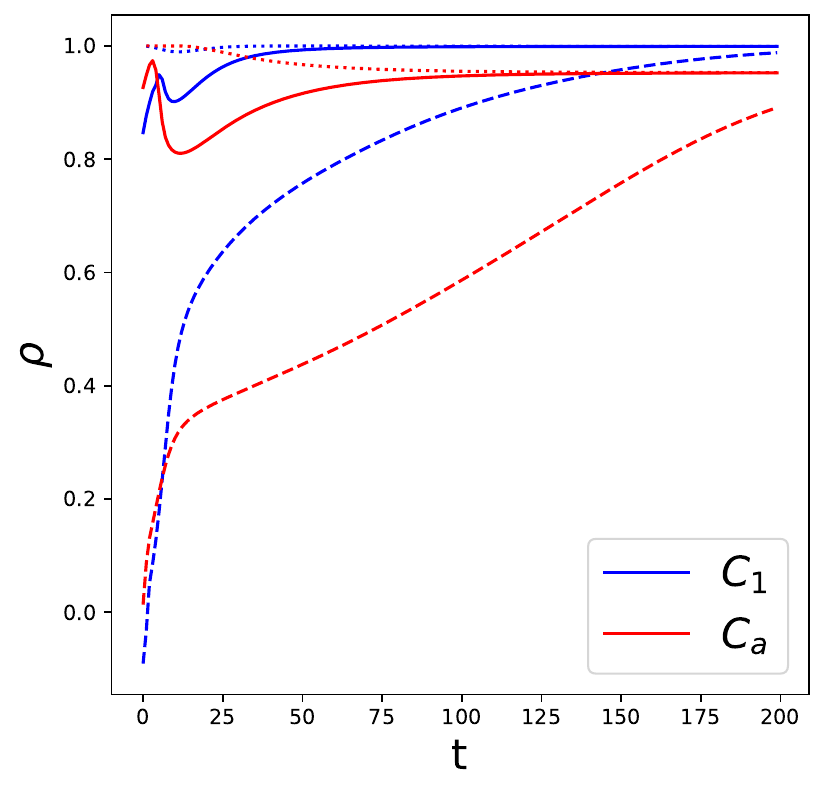}}\par
\subcaptionbox{}{
\includegraphics[width=\linewidth]{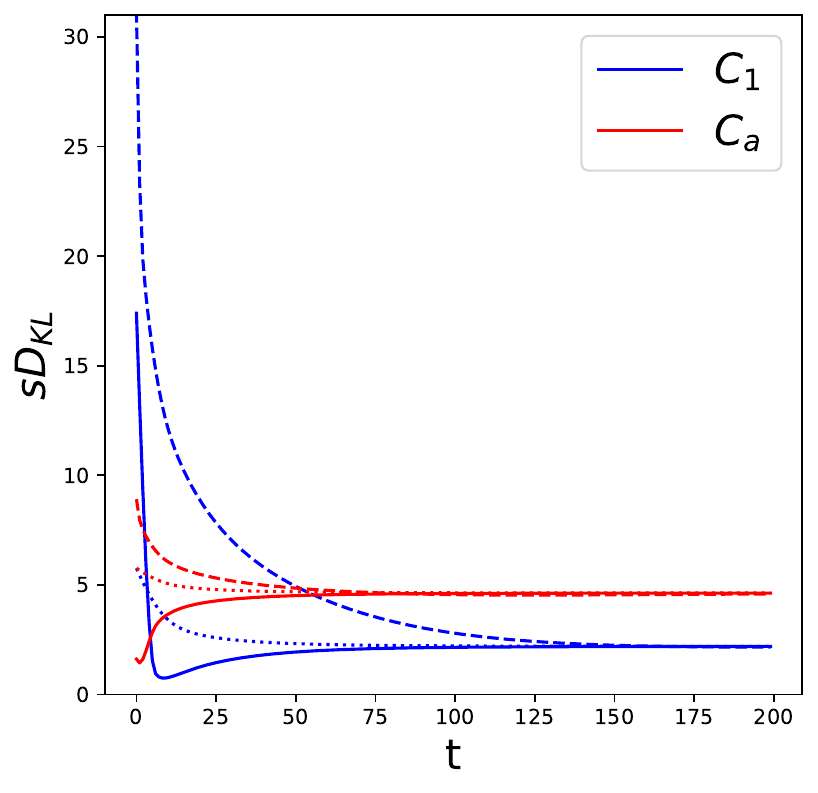}}\par
\end{multicols}
\caption{(a) The Pearson coefficient $\rho$ between the correlation matrix $c_{\beta}$ with $\beta = a$ and $\beta = 1$ and the data correlation matrix $c_d$. (b) The symmetric Kullback-Leibler divergence between the Gibbs-Boltzmann distribution at $\beta = a$ and $\beta = 1$ and the one that has generated the data, i.e. with $\beta_d = 0.4$. The \textit{full} line reports the case of Hebbian initialization of the couplings; the \textit{dashed} line corresponds to a random initialization; the   \textit{dotted} line represents the initialization to $J^{(0)} = 0$. Choice of the parameters: $N = 18$, $a = 0.3$, $\lambda = 0.06$.}
\label{fig:SK_inits}
\end{figure}

Let us consider the case of inferring a Sherrington-Kirkpatrick model (SK), i.e.
\begin{equation}
\label{eq:HSK}
E_{SK}[\vec{S} | J] = -\sum_{i, j>i}S_i J_{ij}S_j \hspace{2cm} J_{ij} \sim \mathcal{N}(0,N^{-1/2})
\end{equation}
For clarity of the results, we will use only one realisation of the parameters for the SK, having a peak in the specific heat represented in \cref{fig:cvSK}.
A network of $N = 18$ neurons is trained with a SK model at $\beta_d = 0.4$ with the new BM learning algorithm. The parameters are initialised as $J^{(0)} = c_d$ and $\vec{h}^{(0)} = 0 $. 
\begin{figure}[ht!]
\begin{multicols}{2}
\centering
\subcaptionbox{}{
\includegraphics[width=\linewidth]{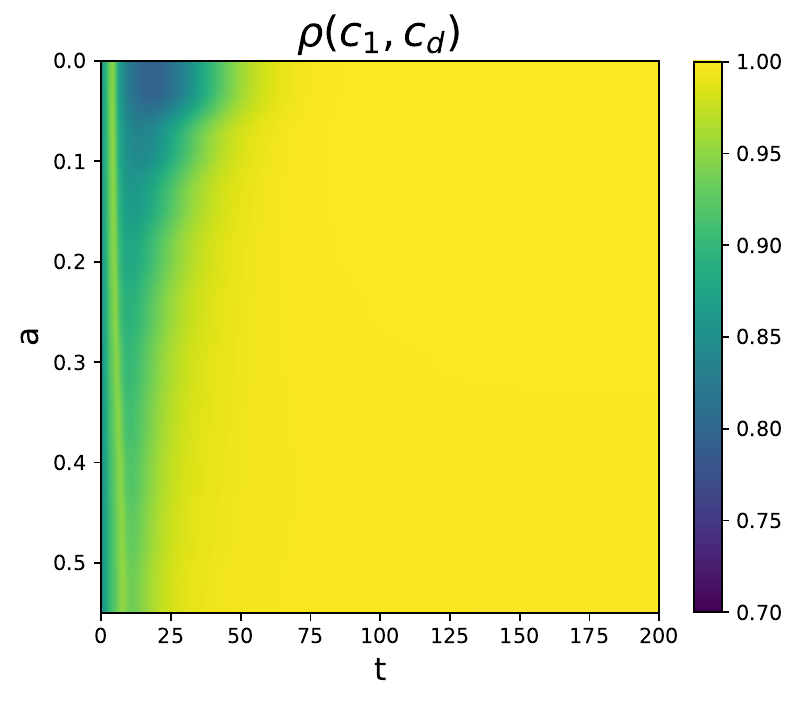}}\par
\subcaptionbox{}{
\includegraphics[width=\linewidth]{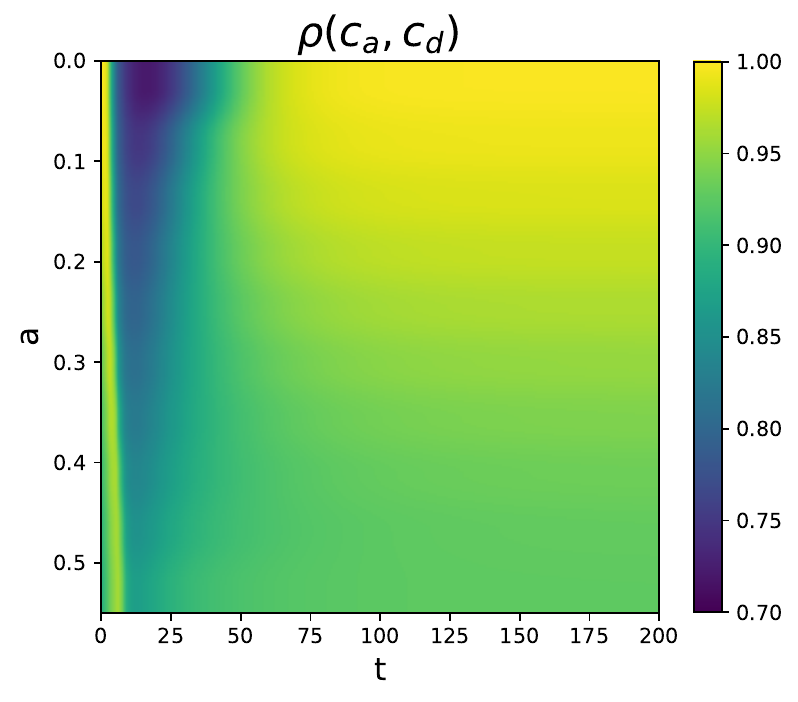}}\par
\subcaptionbox{}{
\includegraphics[width=0.98\linewidth]{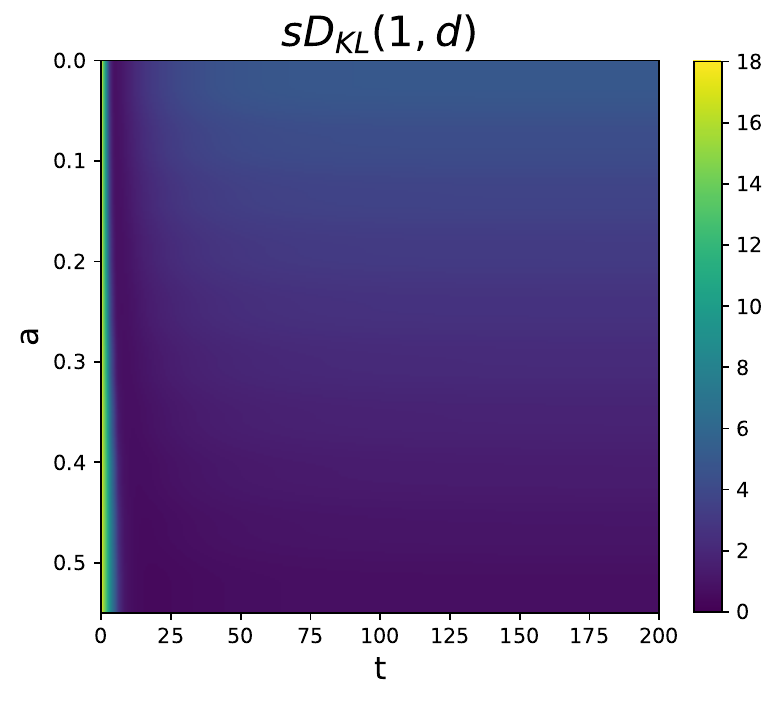}}\par
\subcaptionbox{}{
\includegraphics[width=0.98\linewidth]{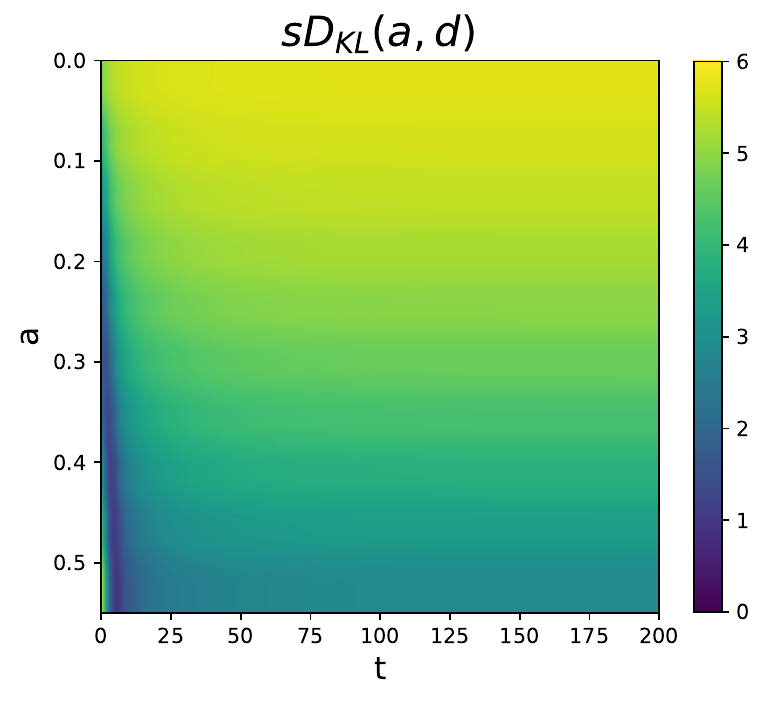}}\par
\end{multicols}
\caption{Relevant observables for a small network trained from a SK model as functions of the parameter $a$ and the algorithm time $t$. The generating model for the data has a inverse temperature $\beta_d = 0.4$. Parameters are initialized in the standard way, i.e. $J^{(0)} = c_d$ and $\vec{h}(0) = 0$. $\rho(A,B)$ is the Pearson coefficient between the elements of the matrices $A$ and $B$, while $sD_{KL}(\beta,d)$ is the symmetrized Kullback-Leibler divergence betwen the Gibbs-Boltzmann distribution at $\beta$ and the original one that generated the data. Choice of the parameters: $N = 18$, $\lambda = 0.06$.}
\label{fig:SK}
\end{figure}
Figure \ref{fig:SK} plots the Pearson coefficient $\rho$ for the 2-point correlation matrices and the $sD_{KL}$ between the inferred model with $\beta = a$ and $\beta = 1$ and the original $SK$, both as functions of the algorithm steps and the parameter $a$. As one can conclude from \cref{fig:SK}a and \cref{fig:SK}b, the quality of the correlation between $c_d$ and $c_{\beta}$ is better for $\beta = 1$ with respect to $\beta = a$. 
This is probably due to the fact that the gradient deriving from $D_{KL}(data|a)$ in \cref{eq:Juprob} is weighted by the factor $(a-1)/a$, that in our experiment is smaller than unity. The same trend is observed with $sD_{KL}$ in \cref{fig:SK}c and \cref{fig:SK}d: the model with $\beta = 1$ approaches the original system way more closely than the one with $\beta = a$ at any value of $a$. 
Another important aspect is that both $\rho$ and $sD_{KL}$ show a transient regime in the first few training steps. In this regime $\rho$ oscillates while the $sD_{KL}$ reaches its minimum value, pushing the model to its best generalizing configuration of the parameters. This holds for both the models with $\beta = a$ and $\beta = 1$. Fig. \ref{fig:SK_inits} compares the standard Hebbian initialization of the couplings with other kinds of initializations, when $a = 0.3$: $J_{ij}^{(0)} = 0$ and $J_{ij} \sim \mathcal{N}(0,N^{-1/2})$ $\forall i,j$. The study is repeated for both $\beta = 1$ and $\beta = a$, showing a general better performance for $\beta = 1$, as was observed from the previous colour plots. Since the transient regime does not appear in the other two initializations, we conclude that the initialization of couplings plays a fundamental role in the training of a BM. 

%We can conclude that the Pearson coefficients between the two-point correlations $c_a$, $c_1$ are maximised and $\text{max}_{\beta\in[0,1]}(C_v)$ is minimized, with $\text{argmax}_{\beta\in[0,1]}(C_v) = 1$, after a certain number of time steps $t \sim \mathcal{O}(N)$ and for low $a$. In this phase the inferred model is both robust under rescaling of $J_{ij}$ and well generalising. On the boundary of this region the training procedure is even capable of finding couplings that are very similar to the original model, while flattening the specific heat. Such results are general for different choices of $\beta_d$ and initial conditions for the parameters.  

\begin{figure}[ht!]
\centering
\includegraphics[width=0.5\linewidth]{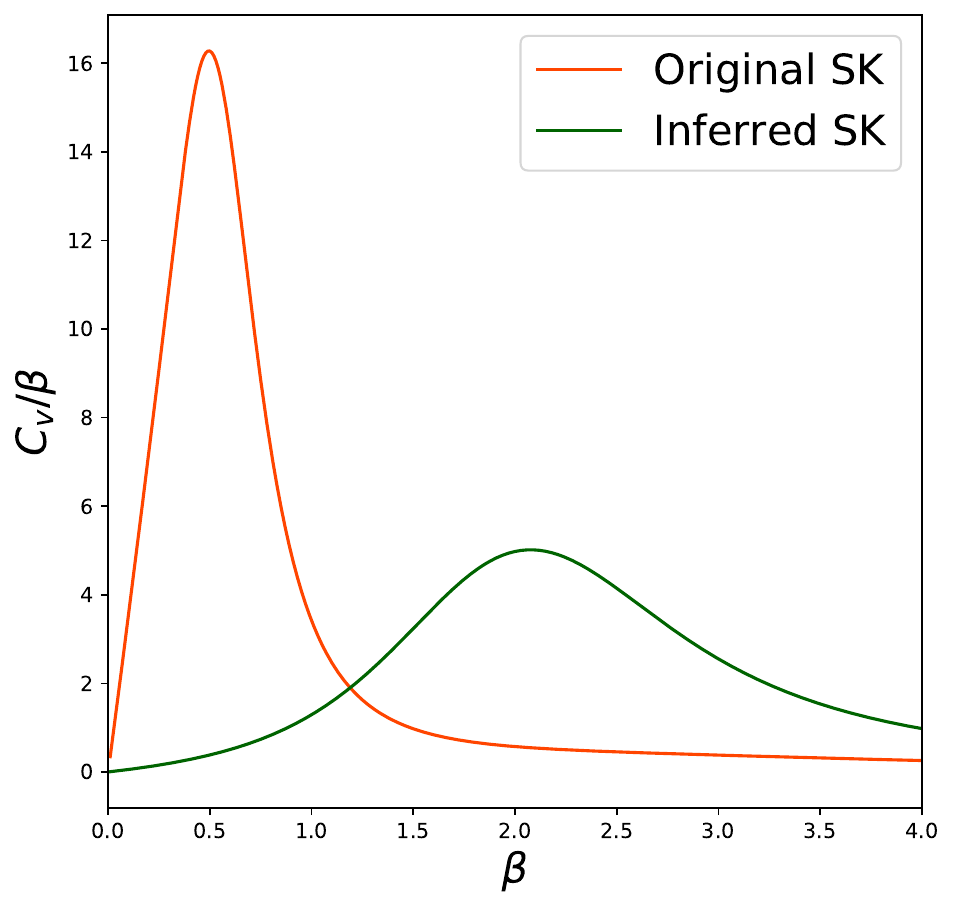}
\caption{Specific heat $C_v/\beta$ as a function of the inverse temperature $\beta$ for the original SK spin-glass and the inferred one via the Unlearning regularization. Choice of the parameters: $N = 18$, $a = 0.3$, $\beta_d = 0.4$, $\lambda = 0.06$.}
\label{fig:cvSK_compare}
\end{figure}
In conclusion, we compare the specific heat $C_v/\beta$ of the original model with the curve obtained through the Unlearning regularization, at convergence of the update equation (\ref{eq:Jup}), always for $a = 0.3$. We want to stress that the experiments permits to compute the matrix $J$ associated to the particular choice of $a$, but the specific heat at each $\beta$ is relative to a Gibbs-Boltzmann distribution with rescaled parameters $\beta J$. The results displayed in \cref{fig:cvSK_compare} show that the inferred system has a lower peak, that is also shifted towards higher values of $\beta$, as we had observed in the CW case. 
\\We conclude that the regularization has both reproduced the statistics of the original spin system at $\beta = \beta_d$ and reached a higher robustness under variation of the parameters, constructing a good generative model.

\subsection{\label{sec:comp_regs} Comparing different regularization techniques}
It is now necessary to compare the generalization performance of the Unlearning method with other kinds of regularization. We will consider, in particular the $L_1$ and $L_2$ types described in \cref{sec:regs}. Data will be generated from a SK model with $\beta_d = 0.4$, as in \cref{sec:SKdata}.
\begin{figure}[ht!]
\begin{multicols}{3}
\centering
\subcaptionbox{}{
\includegraphics[width=\linewidth]{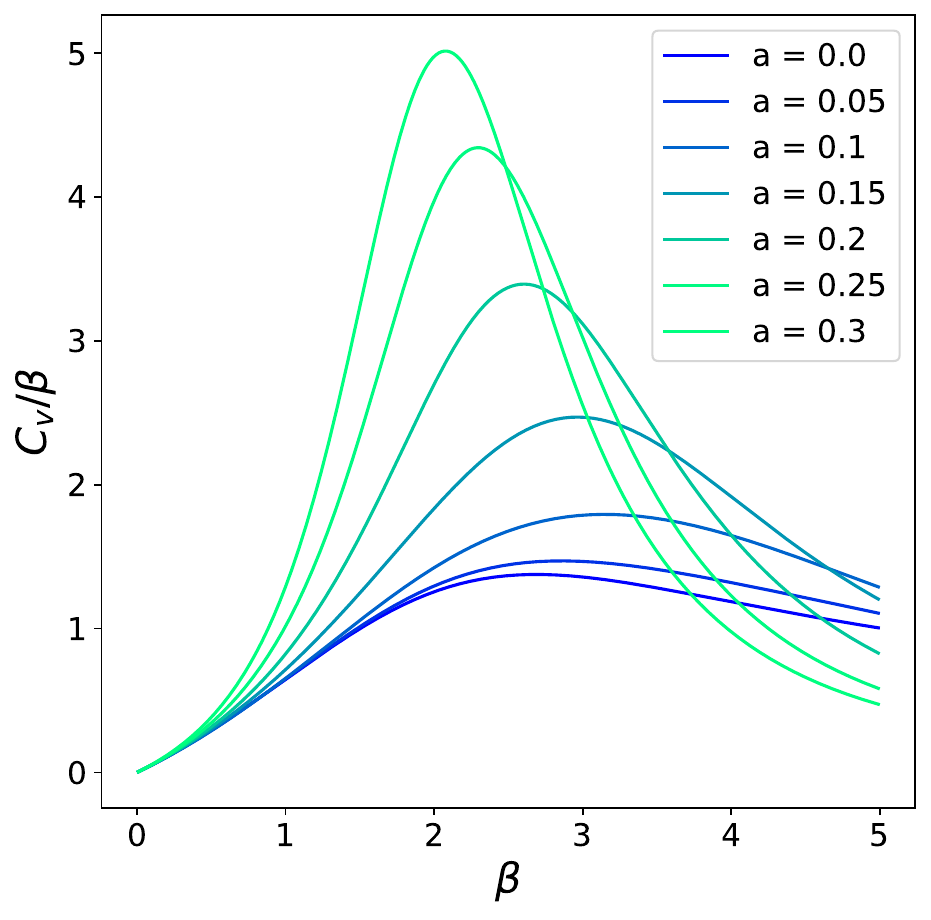}}\par
\subcaptionbox{}{
\includegraphics[width=\linewidth]{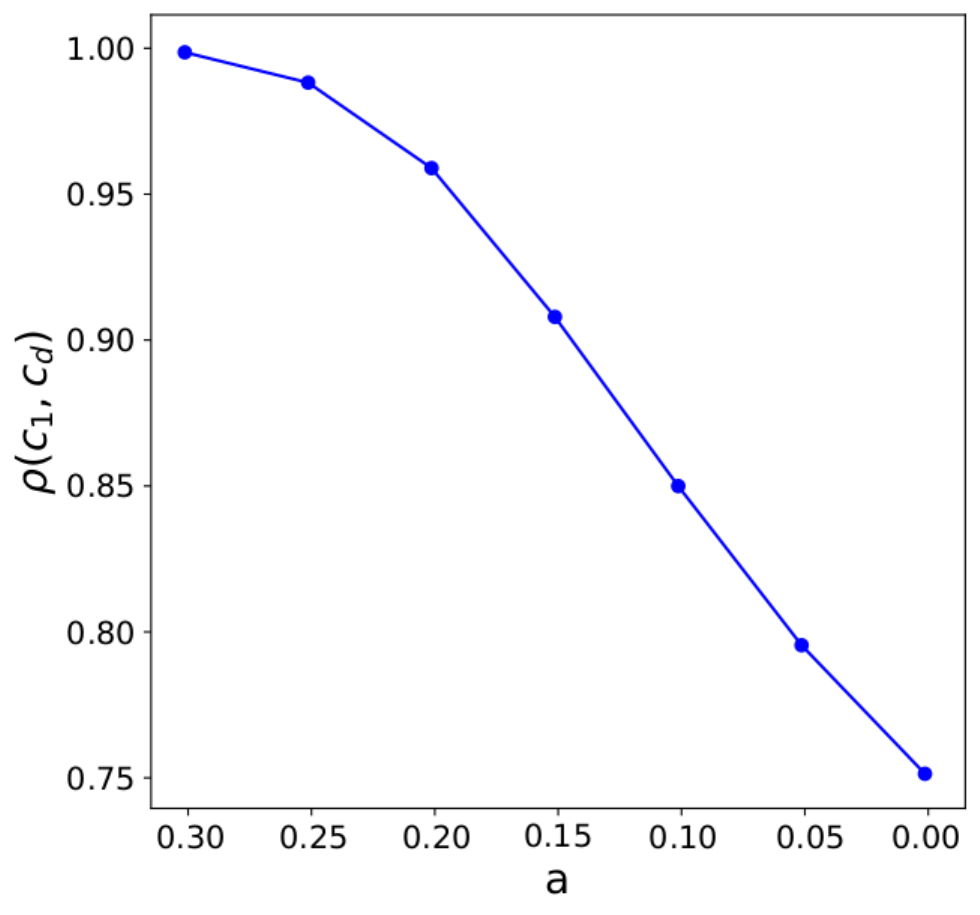}}\par
\subcaptionbox{}{
\includegraphics[width=\linewidth]{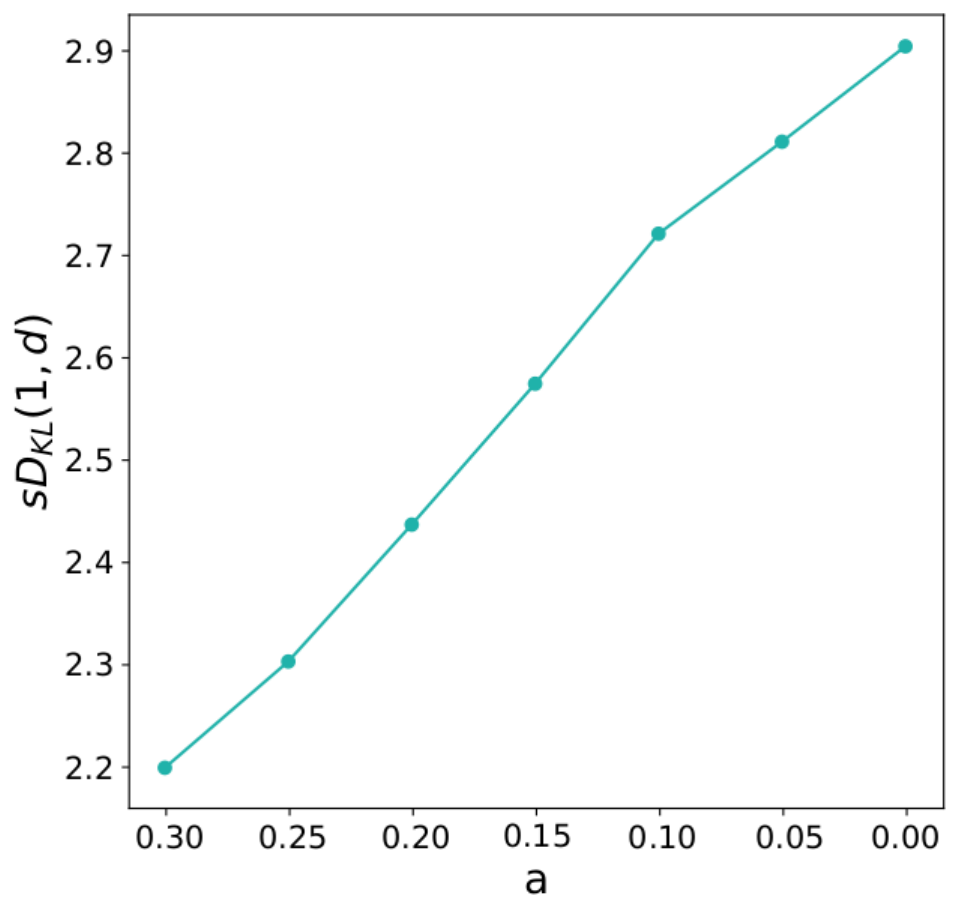}}\par
\end{multicols}
\caption{Results relative to a Boltzmann-Machine trained with the Unlearning regularization. (a) Specific heat $C_v/\beta$ of the model as a function of the inverse temperature $\beta$. (b) Pearson coefficient between the correlation matrix of the model at $\beta = 1$ and the data correlation matrix $c_d$. (c) Symmetric Kullback-Leibler divergence between the model at $\beta = 1$ and the generating model. Choice of the parameters: $N = 18$, $\lambda = 0.06$.}
\label{fig:SK_regs_U}
\end{figure}
\begin{figure}[ht!]
\begin{multicols}{3}
\centering
\subcaptionbox{}{
\includegraphics[width=\linewidth]{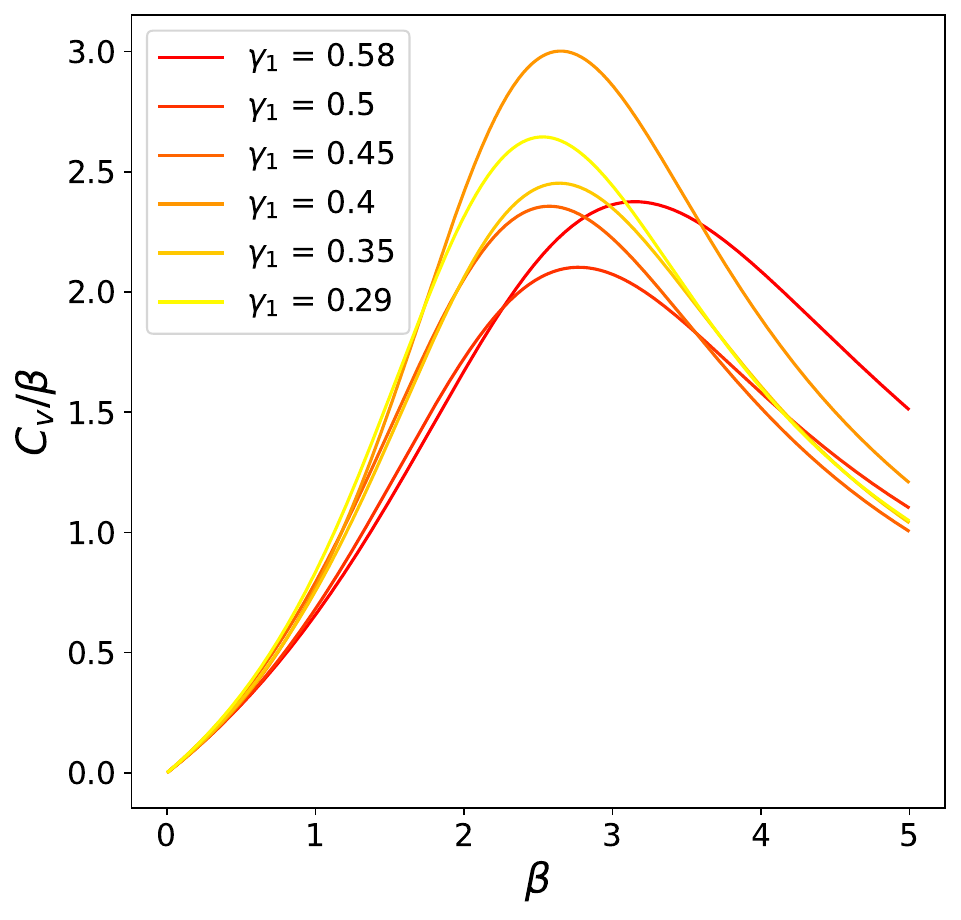}}\par
\subcaptionbox{}{
\includegraphics[width=\linewidth]{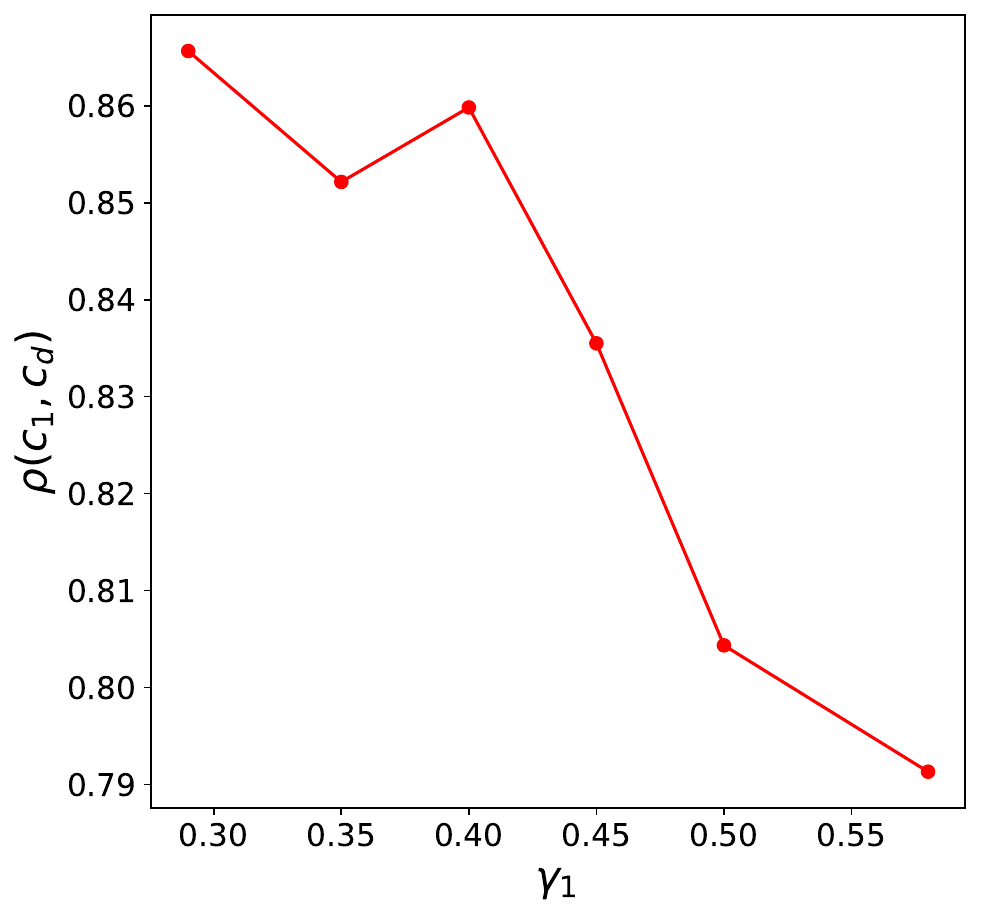}}\par
\subcaptionbox{}{
\includegraphics[width=\linewidth]{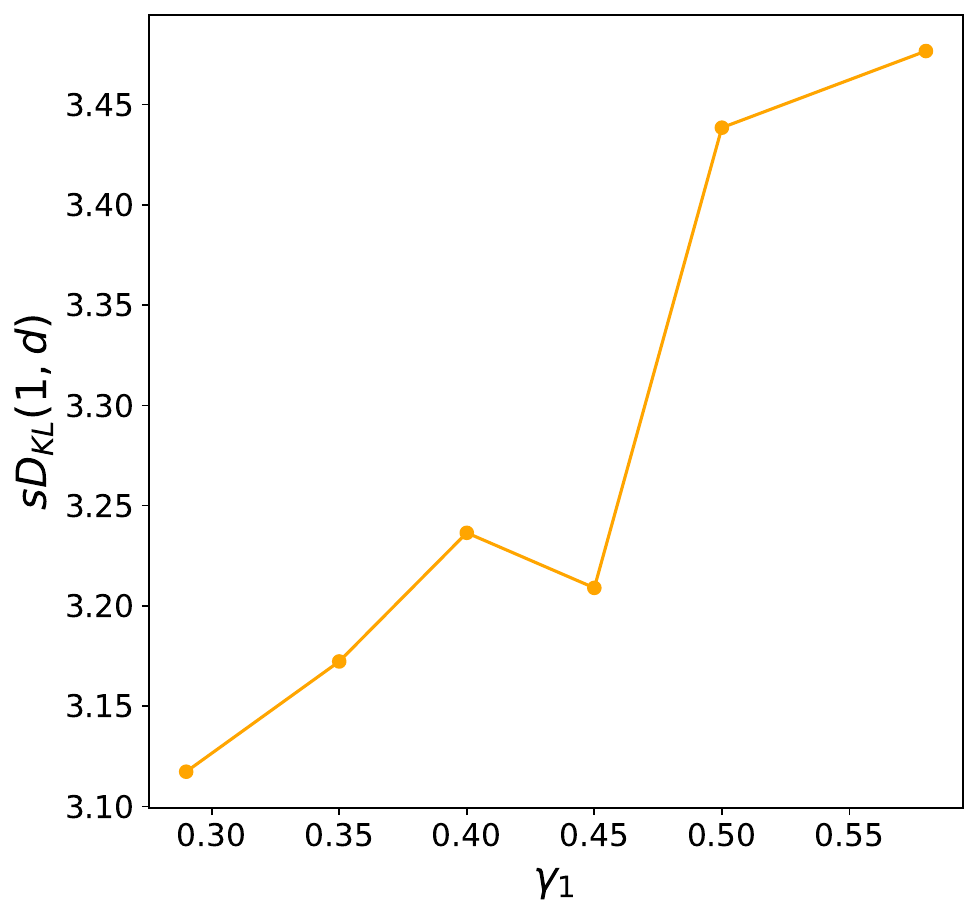}}\par
\end{multicols}
\caption{Results relative to a Boltzmann-Machine trained with the $L_1$ regularization. (a) Specific heat $C_v/\beta$ of the model as a function of the inverse temperature $\beta$. (b) Pearson coefficient between the correlation matrix of the model at $\beta = 1$ and the data correlation matrix $c_d$. (c) Symmetric Kullback-Leibler divergence between the model at $\beta = 1$ and the generating model. Choice of the parameters: $N = 18$, $\lambda = 0.06$. }
\label{fig:SK_regs_L1}
\end{figure}
\begin{figure}[ht!]
\begin{multicols}{3}
\centering
\subcaptionbox{}{
\includegraphics[width=\linewidth]{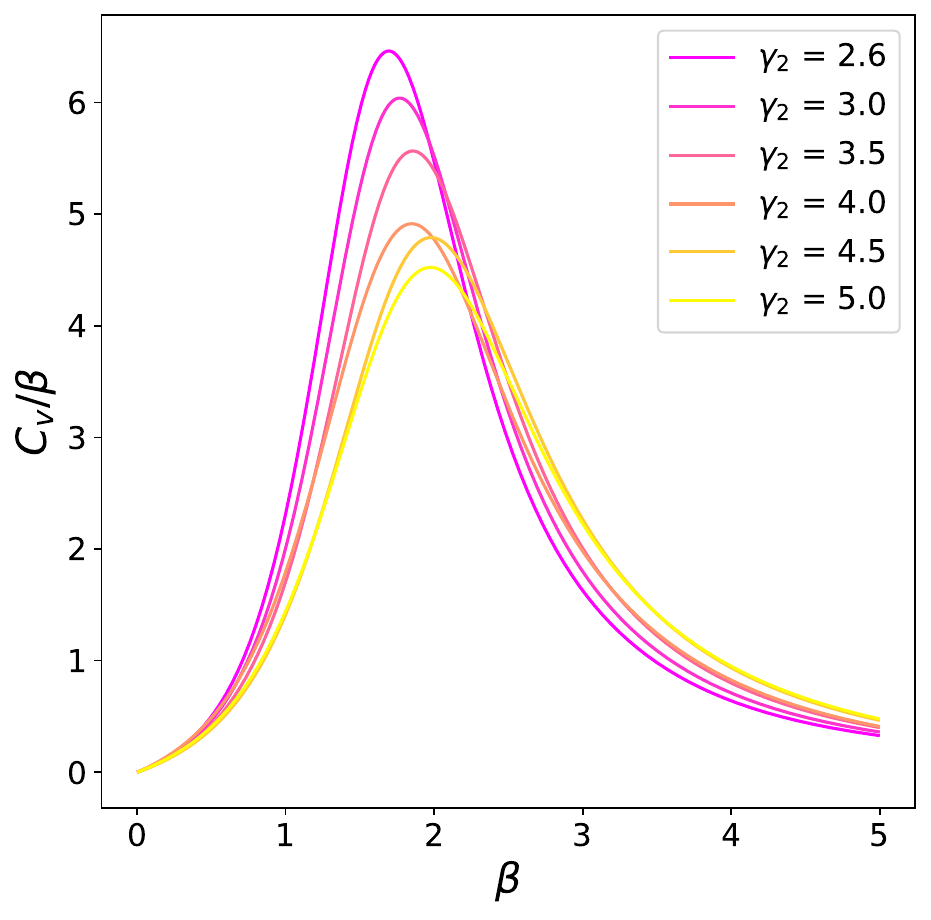}}\par
\subcaptionbox{}{
\includegraphics[width=\linewidth]{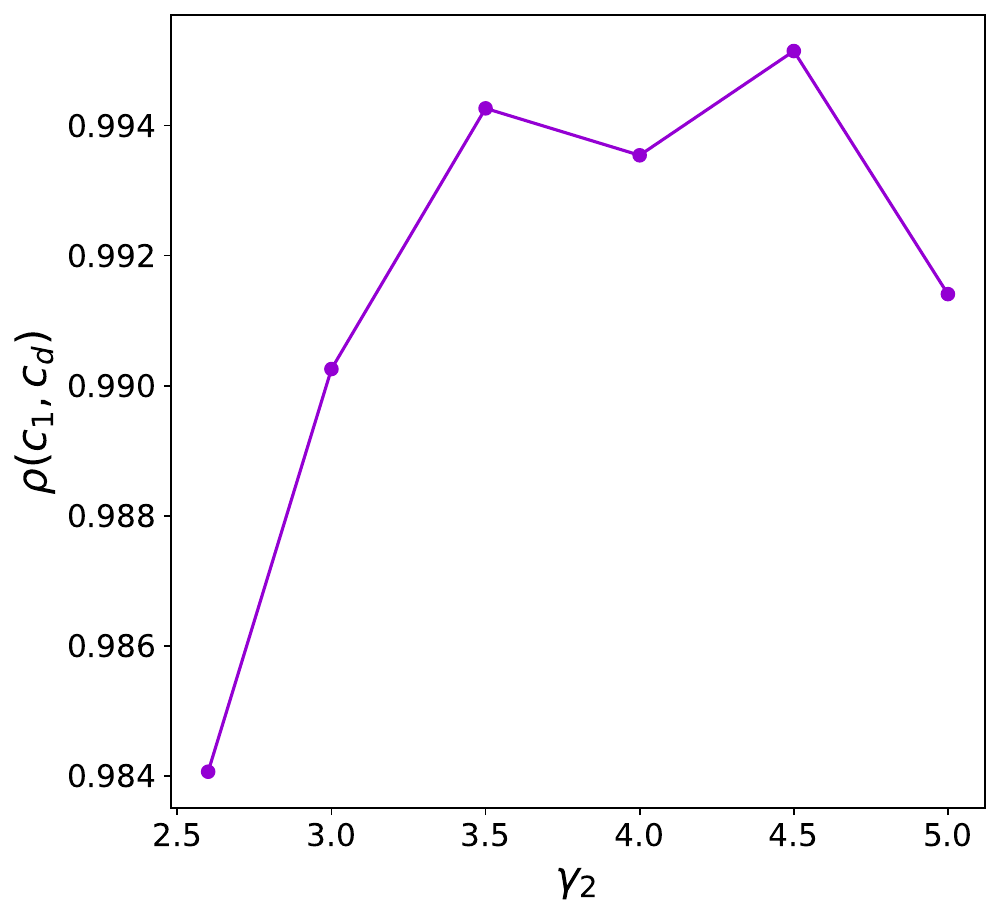}}\par
\subcaptionbox{}{
\includegraphics[width=\linewidth]{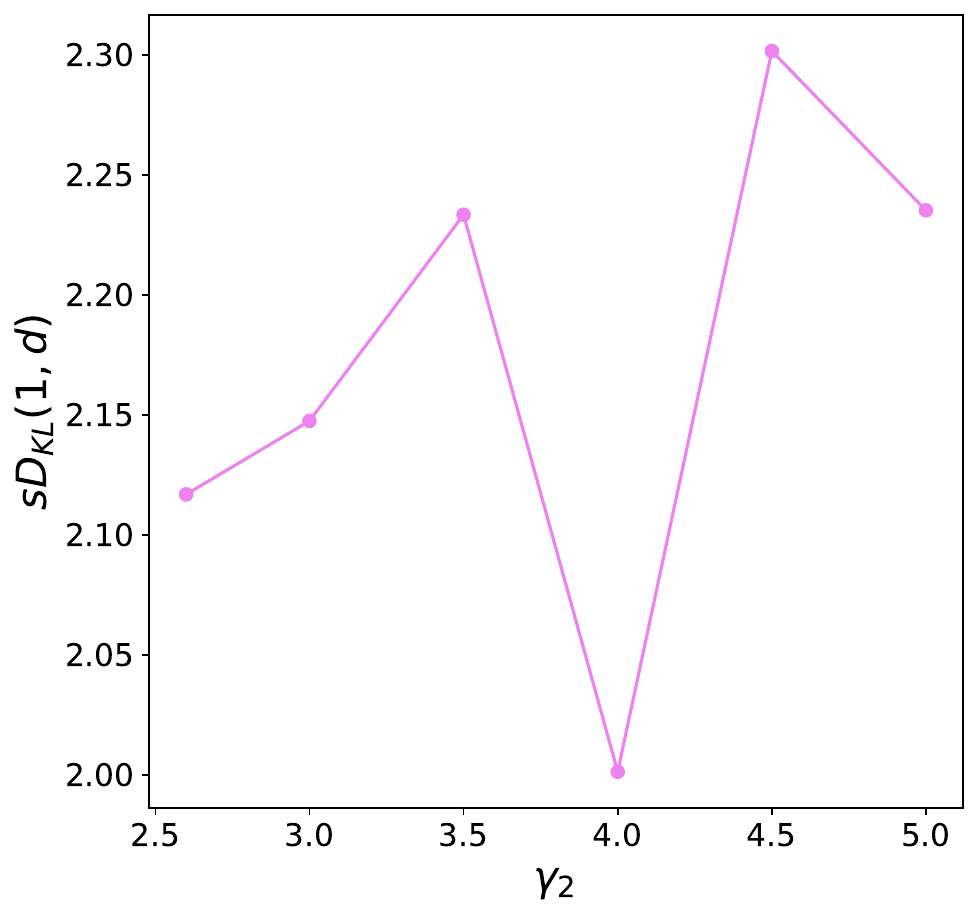}}\par
\end{multicols}
\caption{Results relative to a Boltzmann-Machine trained with the $L_2$ regularization. (a) Specific heat $C_v/\beta$ of the model as a function of the inverse temperature $\beta$. (b) Pearson coefficient between the correlation matrix of the model at $\beta = 1$ and the data correlation matrix $c_d$. (c) Symmetric Kullback-Leibler divergence between the model at $\beta = 1$ and the generating model. Choice of the parameters: $N = 18$, $\lambda = 0.06$. }
\label{fig:SK_regs_L2}
\end{figure}

\begin{figure}[ht!]
\centering
\includegraphics[width=0.65\linewidth]{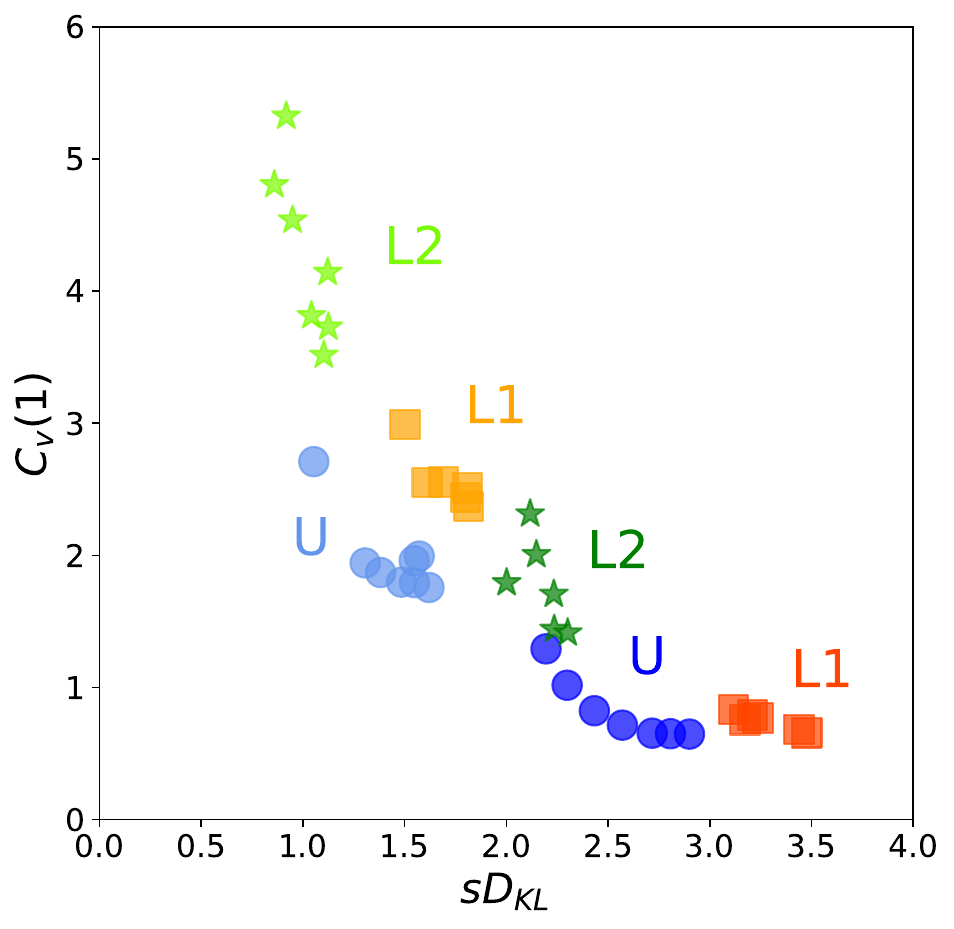}
\caption{Graphical representation of the comparison among the three regularization techniques: U, $L_1$ and $L_2$. The performance of the network is benchmarked through measuring the specific heat at $\beta = 1$ versus the symmetric Kullback-Leibler divergence between the original model and the inferred one at $\beta = 1$. The point $(0,0)$ represents the best generalization performance of the generative model. The resulting networks from the three regularizations have the same standard deviation $\sigma_J$ of the couplings, specifically darker colours are relative to $\sigma_J \simeq 0.17$ while lighter colours are associated to $\sigma_J \simeq 0.26$. For each cluster of symbols, i.e. one regularization at a given $\sigma_J$, only one point is related to the fixed point of the BM learning equations: the rest of the points are obtained by early-stopping the algorithm at different values of $a$ (U), $\gamma_1$ ($L_1$) and $\gamma_2$ ($L_2$) when $\sigma_J$ matches the prescribed value.}
\label{fig:SK_regs_plot}
\end{figure}
To confront the networks obtained by the different methods we require the standard deviations of the coupling matrices to be comparable among each other. The network is initialized in the Hebbian fashion, i.e. $J^{(0)} = c_d$. The value of $a = 0.3$ is thus chosen and the couplings are optimized through \cref{eq:Juprob} until convergence. 
The standard deviation of the couplings is measured obtaining $\sigma_J \simeq 0.17$. Consequently the BM learning is repeated by adding the $L1$ and $L2$ regularizations. The rates $\gamma_1$ and $\gamma_2$ are chosen in order to obtain the same standard deviation $\sigma_J$ at convergence of the algorithm. 
At this point the specific heat $C_v/\beta$, the Pearson $\rho(c_1,c_d)$ and the symmetric divergence $sD_{KL}(1,d)$ are measured across the three different techniques. 
The same procedure is reiterated at different values of $a,\gamma_1,\gamma_2$ that lead to the same value of $\sigma_J$: the only difference with the original experiment is that now \cref{eq:Juprob} is not iterated until convergence to the fixed point, but it is instead early-stopped as soon as $\sigma_J \simeq 0.17$ is reached by the system.
\\ The results of the experiment are reported in figures \ref{fig:SK_regs_U}, \ref{fig:SK_regs_L1} and \ref{fig:SK_regs_L2} for the three methods.
The general trend shows that one cannot obtain a minimum specific heat (let us here consider $C_v(1)$ as the most representative observable to measure) without a Loss in the distance $sD_{KL}$ between the inferred and the original model. Regarding the Pearson coefficient $\rho$, its behaviour can slightly change across the regularizations, but it is in general anticorrelated with $sD_{KL}$. Stronger fluctuations in the measures of $\rho$ and $sD_{KL}$ for the $L_1$ and $L_2$ regularizations are due to a the choice of $\lambda$ that needs to be improved for further experiments.\\  
The same results are resumed by the scatter plot in \cref{fig:SK_regs_plot}. The generalization performance of each run is represented as a point with coordinates $(sD_{KL}(1,d), C_v(1))$: each regularization has a different symbol and colour, darker colours represent the previous experiment with $\sigma_J \simeq 0.17$ while lighter ones refer to the same experimental procedure with $\sigma_J \simeq 0.26$. 
In both the experiments we can evidently see the Unlearning method to be closest to the origin of the plane, i.e. the best, yet unreachable, generalization power. The $L_2$ method generally reaches a lower $sD_{KL}$ while keeping the specific heat high. 
Conversely, the $L_1$ technique flattens the specific heat significantly, while gaining in distance between the original and the inferred models. 
Unlearning, on the other hand, places itself in between the two methods, appearing as the best choice to increase the way the network predicts the original model.   
\subsection*{Checkpoint}
In this section we have seen that:
\begin{itemize}
    \item The Unlearning regularization method requires the inferred model to be more robust under variation of the temperature (i.e. a total rescaling of the parameters) and at the same time interpolates between two different training algorithms: the standard BM learning (i.e. $a = 1$) and HU (i.e. $a = 0$). 
    \item Both the cases of data sampled from a Curie-Weiss and a Sherrington-Kirkpatrick model show a good generalization performance obtained through the Unlearning regularization method. 
    \item The Unlearning regularization outperforms $L_1$ and $L_2$ methods in generalization. 
\end{itemize}
\newpage

\section{\label{sec:Ulimit} The Hebbian Unlearning limit}
We will now study the specific limit of the Unlearning regularization leading to an algorithm which strongly resembles the HU routine. The inferential performance of the learning procedure is examined by measuring the distance between the inferred model and the original one at different training steps. Results show, for the first time, that HU can be employed as an inferential tool, due to the beneficial effect of the Hebbian initialization of the couplings. We advance an explanation for its performance that is supported by further numerical evidence.  \\\\
In the limit $a \rightarrow 0$ the updating rules (\ref{eq:Juprob}) and (\ref{eq:huprob}) become
\begin{equation}
    \label{eq:Juprob1}
    \delta J_{ij} = -\lambda \langle S_i S_j \rangle_1,
\end{equation}
\begin{equation}
    \label{eq:huprob1}
    \delta h_{i} = -\lambda \langle S_i \rangle_1.
\end{equation}
Eq. (\ref{eq:Juprob1}) strongly resembles the traditional HU algorithm in \cref{eq:unl_rule}. By contrast with the original rule, the new regularization samples configurations at $\beta = 1$ instead of stable fixed points of the neural dynamics, and makes use of a thermal average, rather than summing each contribution at each time step. 
We will keep dealing with small networks so that, given the initial conditions for the parameters, the Loss function of the problem can be minimized exactly. 
As a numerical experiment, we train a BM with $N = 18$ neurons to learn a SK model at $\beta_d = 0.4$. The Unlearning regularization is performed with $a = 0$ and the usual initial conditions for the parameters, i.e. $J^{(0)} = c_d$ and $h_i^{(0)} = 0$ $\forall i$. 
\begin{figure}[ht!]
\begin{multicols}{3}
\centering
\subcaptionbox{}{
\includegraphics[width=\linewidth]{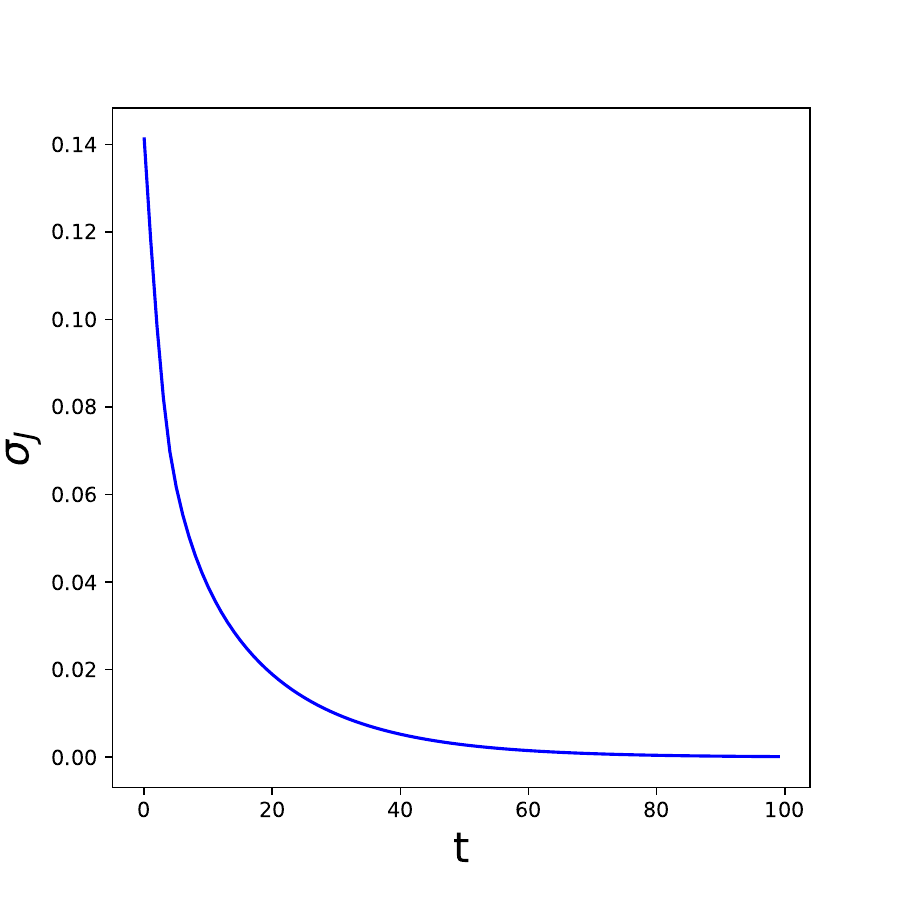}}\par
\subcaptionbox{}{
\includegraphics[width=\linewidth]{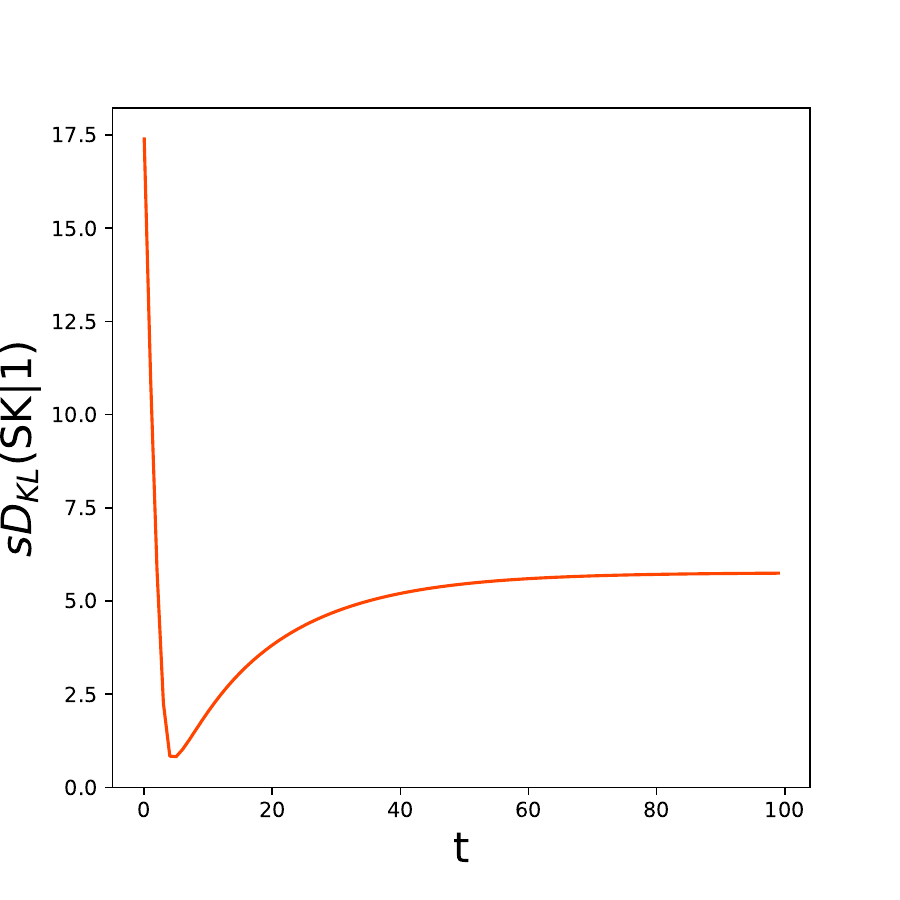}}\par
\subcaptionbox{}{
\includegraphics[width=\linewidth]{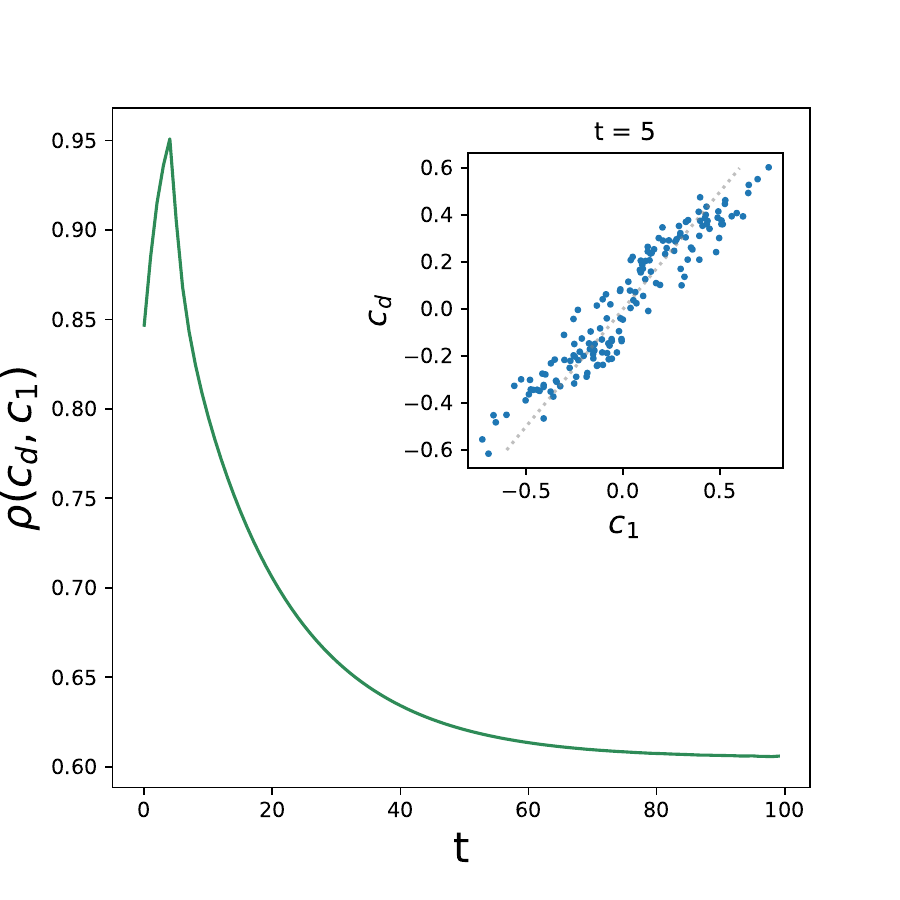}}\par
\end{multicols}
\caption{Three relevant observables to benchmark the inferential performance of the Unlearning regularization with $a = 0$: the standard deviation of the couplings $\sigma_J$ (a), the Kullback-Leibler divergence between the original SK model and the inferred one at $\beta = 1$ (b), the Pearson coefficient between the data correlation matrix $c_d$ and $c_1$, i.e. the correlation matrix for the model at $\beta = 1$ (c). Choice of the parameters: $N = 18$, $\lambda = 0.06$, $\beta_d = 0.4$.}
\label{fig:UNLa0_1}
\end{figure}

Even if training is performed over both the couplings and the fields, according to the rules (\ref{eq:Juprob1}) and (\ref{eq:huprob1}), the fields $\vec{h}$ do not contribute to the energy of the original model, hence we will focus exclusively on the evolution of the interactions $J$.   
Fig. \ref{fig:UNLa0_1}a displays the standard deviation of the couplings: as we can see the general trend shows an exponential decay of the interactions, as one typically observes in the HU algorithm \cite{van_hemmen_increasing_1990}. The decay to zero is implied by the fact that the fixed point of the gradient descent equations for the Boltzmann Machine must be found when $c_1 \equiv 0$. 
Fig. \ref{fig:UNLa0_1}b depicts the symmetric Kullback-Leibler divergence between the Gibbs-Boltzmann distribution of the original SK model, from which data were sampled, and the inferred model with $\beta = 1$. The $sD_{KL}$ is high at the beginning, it decreases until reaching a global minimum signaling an optimum in the inferential performance, then it increases again and stabilizes on a plateau. The minimum is sufficiently low to indicate a good statistical consistency between the model and data. 
The Pearson coefficient between the correlation matrix of the data $c_d$ and the one of the inferred model at unitary temperature $c_1$ is measured step-by-step in the learning process and reported in \cref{fig:UNLa0_1}c: as one can notice, there is a global maximum near the position of the global minimum of the $sD_{KL}$ that we previously identified. 
A sub-plot displays the good agreement between the two matrices, the inferred $c_1$ and $c_d$, at the position of the peak.
\\This analysis suggests that the Unlearning regularization with $a = 0$ displays two working regimes: one transient regime where the model at $\beta = 1$ shows a good statistical agreement with the generating model; another regime where $c_1 \rightarrow 0$, and the total performance deteriorates. 
The first transient regime is the most important one, because it suggests that HU can be read as inferential tool, aside of its associative memory use. We know from \cref{sec:BML} that in BM learning data are shown to the model each time-step of the algorithm and this imposes the moment matching.
The co-existence of a positive Hebbian term and a negative Unlearning one in the traditional BM learning has already been pointed out by different works in the literature \cite{hinton_unsupervised_1999, hinton_forward-forward_2022}. In fact, the variation of each pair of couplings $J_{ij}$ is given by
\begin{equation}
    \label{eq:gradJ_new}
    \delta J_{ij} = \delta^H J_{ij} + \delta^U J_{ij}, 
\end{equation}
where we neglected the dependence on time, and
\begin{equation}
    \label{eq:deltaJ_H}
    \delta^H J_{ij} = \langle S_i S_j \rangle_{data}, 
\end{equation}
\begin{equation}
    \label{eq:deltaJ_U}
    \delta^U J_{ij} = -\langle S_i S_j \rangle_{mod}.
\end{equation}
We can interpret each step in the minimization of $\mathcal{L}$ as the result of two contributions, one constant quantity deriving from the data, and another one deriving from the evolving model. 
While a standard BM can start wherever in the space of the parameters and ends up at the fixed point of \cref{eq:gradJ_new}, in the HU case the update is performed by using only the negative unlearning contribution. Nevertheless, the data have been seen by the model, specifically through the Hebbian choice of the initial conditions. This observation suggests that HU approaches the minimum of the loss function in two temporally separated steps: by first descending along the \textit{data} direction and then, progressively, along the \textit{model} one. As a consequence, we can assume that such a two-steps minimization equals the standard BM learning only while
\begin{equation}
    \label{eq:cond_2stepsBML}
    |\langle S_i S_j \rangle_{mod}| \gg |\langle S_i S_j \rangle_{data}|
\end{equation}
for most of the pairs $i,j$. \\To test whether this condition holds while training a BM regularized with $a = 0$, and initialized with $J^{(0)} = c_d$, we measure the standard deviation of ratio between the elements  of $c_1$ and $c_d$ as a function of time. 
\begin{figure}[ht!]
\begin{multicols}{2}
\centering
\subcaptionbox{}{
\includegraphics[width=\linewidth]{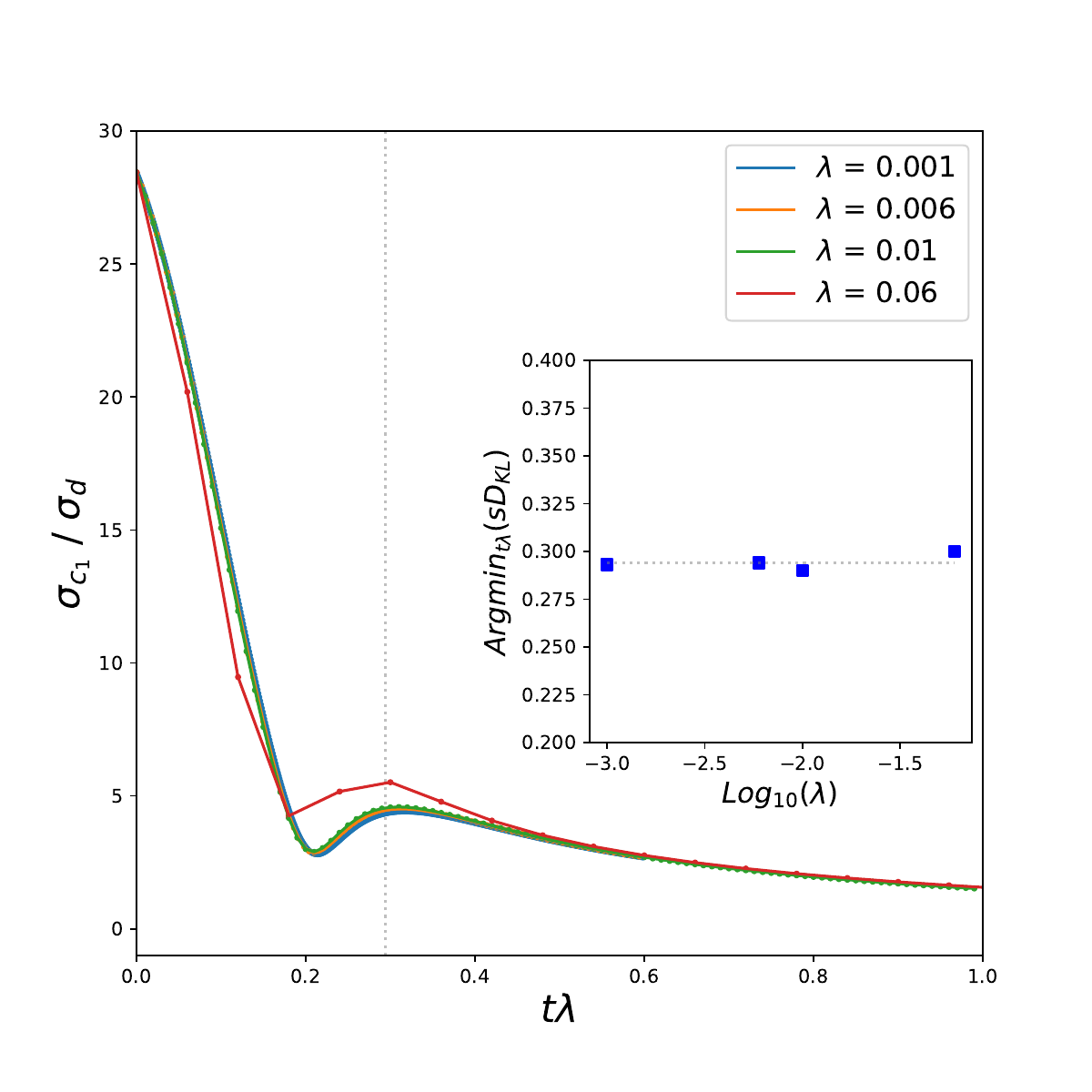}}\par
\subcaptionbox{}{
\includegraphics[width=\linewidth]{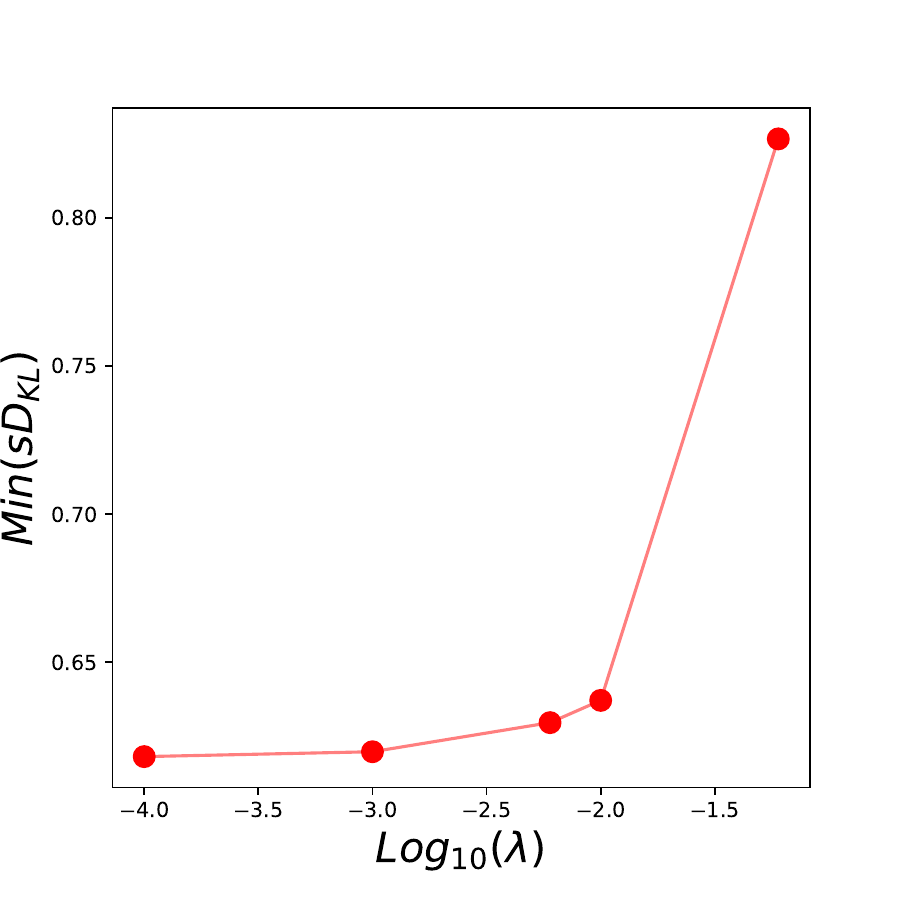}}\par
\end{multicols}
\caption{(a) Standard deviation of the ratio between the elements of $c_1$ and $c_d$ as a function of the normalized time $t\lambda$; the subplot reports the values of $t\lambda$ associated to global minima of the symmetric Kullback-Leibler divergence ($sD_{KL}$) between the generating model and the inferred one for different choices of $\lambda$; both the vertical and horizontal gray dotted lines report the mean of the points in subplot. (b) Global minimum of the symmetric Kullback-Leibler divergence between the generating model and the inferred one as a function of $\lambda$ in log-scale. Choice of the parameters: $N = 18$, $a = 0$, $\beta_d = 0.4$.}
\label{fig:UNLa0_ratio}
\end{figure}
Results are presented in fig. \ref{fig:UNLa0_ratio}a, where the standard deviation of the matrix obtained from the ratio of the elements of $c_1$ over $c_d$ is plotted as a function of $t \lambda$ for different choices of $\lambda$. The curves collapse well from $\lambda$ smaller than $\mathcal{O}(10^{-2})$. 
The system starts with $c_1$ generally being much larger than $c_d$, which satisfies the condition (\ref{eq:cond_2stepsBML}). Around $t \sim 0.2 \lambda^{-1}$ the curves reach a local minimum, then increases until a local maximum at $t \sim 0.29 \lambda^{-1}$, to start a slow decay right after. Both these two stationary points are located where $c_1$ and $c_d$ share the same order of magnitude, i.e. where the condition for the two-steps minimization of the Loss ceases to hold. 
Specifically, the second stationary point is associated to the minimum of the $sD_{KL}$ between the original SK and the model, as reported in the subplot contained in \cref{fig:UNLa0_ratio}a. Moreover, \cref{fig:UNLa0_ratio}b reports the values assumed by $sD_{KL}$ at its global minimum for various choices of $\lambda$. As one can notice, the generalization of the model increases when $\lambda$ is small: we can imagine that the smaller is $\lambda$ the stronger is the effect of the initial overshooting at minimizing $\mathcal{L}_{data}$. \\We have thus showed that, when the network is initialized in the Hebbian fashion, the dominant contribution to the BM learning is the Unlearning one, and this is the reason for the HU algorithm to be a good inferential device.   
\subsection*{Checkpoint}
In this section we have seen that:
\begin{itemize}
    \item The HU algorithm can be implemented as an inferential tool, i.e. to learn the joint probability distribution of the data, instead of storing a small subset of data-points. 
    \item The quality of the inferential performance is due to the initial conditions for the parameters. When couplings are initialized in the Hebbian way a BM can significantly reduce the Loss function, and thus obtain a good neural network in two separate steps, and this two-steps procedure is the HU algorithm. 
    \item Similarly to the associative memory scenario, the best inferential performance of HU is reached after an optimal amount of iterations of the training routine. 
\end{itemize}
\newpage

\section{\label{sec:overfit} The analogy between a SVM and a BM in the under-sampling regime.}
\begin{figure}[ht!]
\begin{multicols}{3}
\centering
\subcaptionbox{}{
\includegraphics[width=\linewidth]{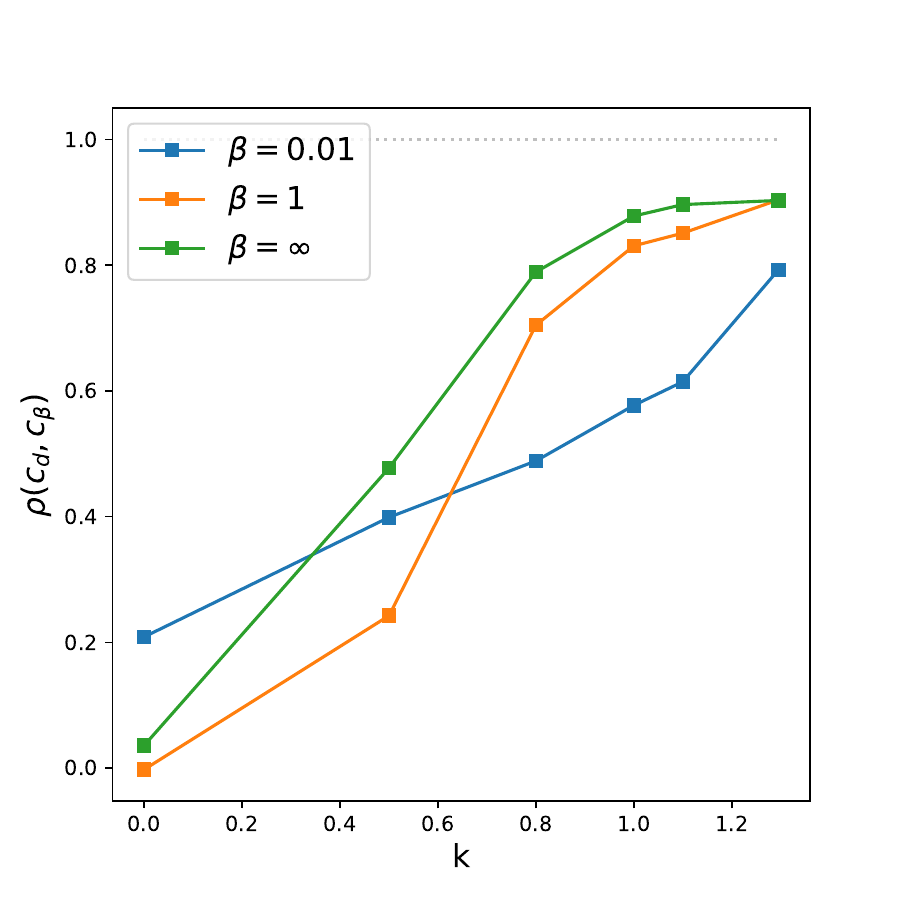}}\par
\subcaptionbox{}{
\includegraphics[width=\linewidth]{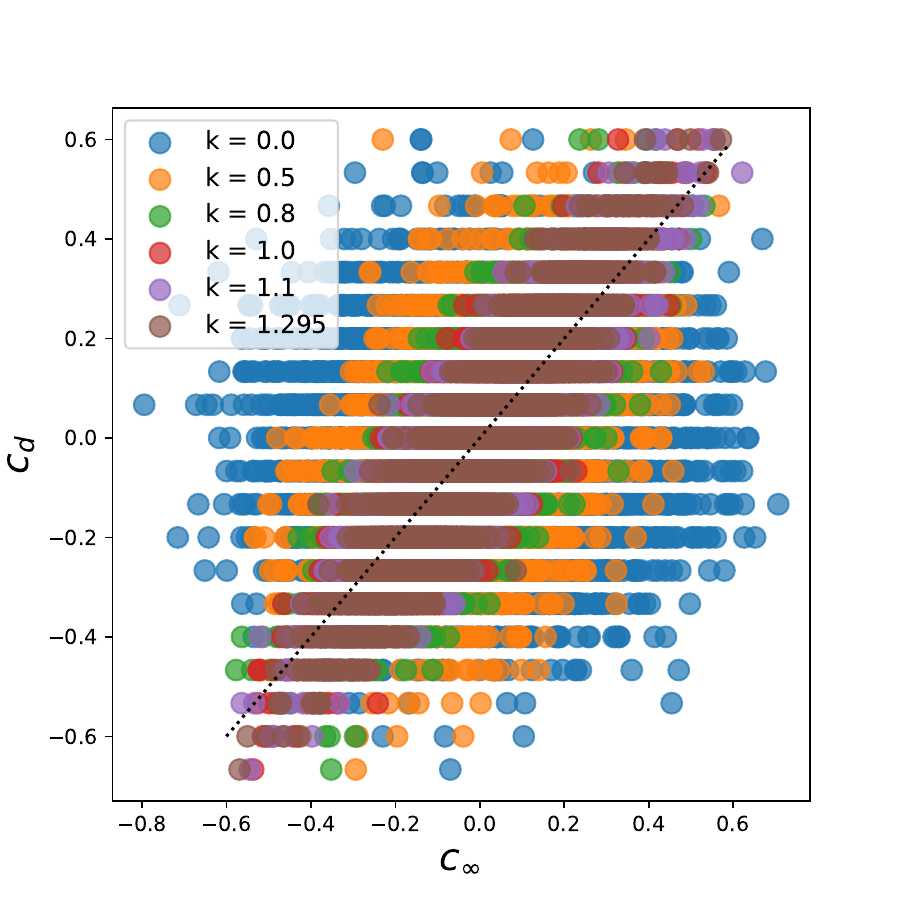}}\par
\subcaptionbox{}{
\includegraphics[width=\linewidth]{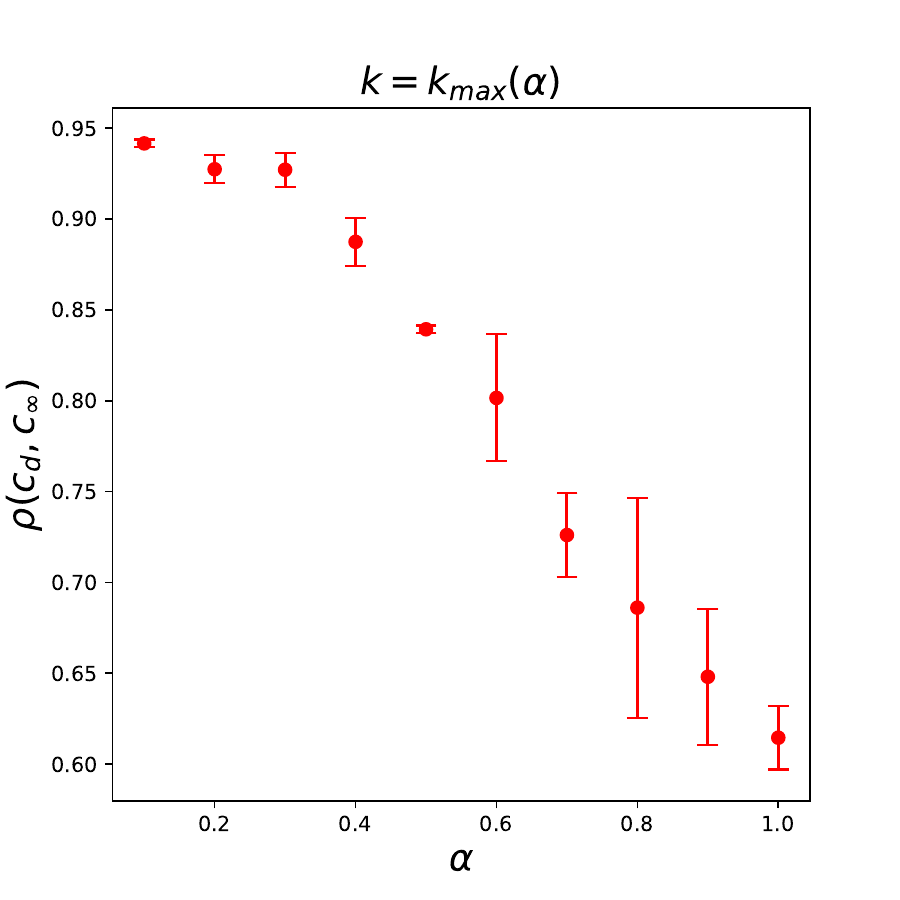}}\par
\end{multicols}
\caption{Numerical measures relative to a symmetric linear perceptron trained through (\ref{eq:lp_sym_update}). The maximum margin is estimated empirically for each realization of the memories and random initial conditions of $J$. (a) Pearson coefficient between $c_d$ and $c_{\beta}$ as a function of the margin $k$, $\alpha = 0.3$. The plots for different values of $\beta$ are displayed. (b) Correlation matrix $c_d$ plotted as a function of $c_{\infty}$, $\alpha = 0.3$. The black dotted line reports the line $c_d = c_{\beta}$. All points are relative to one realization of the network. (c) Pearson coefficient between the elements of $c_d$ and $c_{\beta}$ in a SVM (i.e. $k = k_{max}(\alpha)$) as a function of $\alpha$. Points are averaged over five realizations of both memories and random initial conditions for $J$, the errors are the standard deviations of the measures. Choice of the parameters: $N = 100$, $\lambda = 0.01$. }
\label{fig:over_rho}
\end{figure}

The aim of this section is to show that symmetric SVMs tend to satisfy the moment matching condition $c_d = c_{\beta}$  when $\beta$ is sufficiently large (i.e. $\beta > 1$ in our case of study) and $\alpha$ is small, where we recall $\alpha$ to control the number of stored memories $p = \alpha N$. Since the moment matching condition must hold in BMs, the emergence of this property in SVMs must underline an important similarity between the memory performance of maximally stable linear perceptrons and networks trained through BM learning. The analysis of the landscape of attractors of a SVM can be a hard problem to tackle \cite{kepler_domains_1988, gyorgyi_techniques_2001}. However, the discovery of this new signature of the model can shed light on new dynamic properties of this type of networks. A physical interpretation of this observation, also given in the course of this section, relates this property of SVMs to the depth reached by the memories in the energy landscape, a topological trait that is, in turn, linked to the notable size of the basins of attraction. \\Eventually, by reason of the detailed comparison between SVMs and the HU algorithm advanced in \cref{sec:noise_inject}, we will conclude that the main learning algorithm examined in this work, i.e. HU, SVMs and BMs, all tend to train the same types of neural networks when the number of memories is small. This result is the very conclusive point of this thesis and it will be reached at the end of this section. 
%\\An example of a property of SVMs that was empirically discovered and then employed for calculations is the one found by Kepler and Abbott in \cite{kepler_domains_1988, gyorgyi_techniques_2001}, according to which, configurations belonging to the basins of attraction of a SVM must satisfy
%\begin{equation}
%    \label{eq:ka_cond}
%    m^{\mu}(1) > \frac{1}{2}\left(1 + m^{\mu}(0) \right)
%\end{equation}
%where $m^{\mu}(0)$ is the overlap of the configuration with the $\mu$ memory and $m^{\mu}(1)$ is the same overlap after one step of the \textit{synchronous} retrieval dynamics in \cref{eq:dynamics}. This condition has been used to effectively obtain a closed expression of the typical radius of the basins of attraction of the memories in this type of networks.
\\\\As usual in the associative memory framework we consider a number $p = \alpha N$ of $N$-dimensional memories that are generated as Rademacher variables $\{\vec{\xi}^{\mu}\}_{\mu=1}^p$ with
\label{eq:rade2}
$$P(\xi^{\mu}_i= +1) = P(\xi^{\mu}_i= -1) = 1/2\hspace{0.5cm}\forall i,\mu.$$
These memories are thus memorized by a symmetric linear perceptron with tunable margin $k$, by means of the algorithm (\ref{eq:lp_sym_update}). Let us define
\begin{equation}
    \label{eq:cd_svm}
    c_{ij}^d = \frac{1}{p}\sum_{\mu = 1}^p\xi_i^{\mu}\xi_j^{\mu},
\end{equation}
that is the Hebbian matrix of the memories. We also introduce
\begin{equation}
    \label{eq:cbeta}
    c_{ij}^{\beta} = \langle S_i S_j \rangle_{\beta},
\end{equation}
that is the thermal 2-point correlation matrix of the system, where $\langle \hspace{0.1cm}\cdot \hspace{0.1cm}\rangle_{\beta}$ is the probability measure according a Gibbs-Boltzmann distribution of the states at a temperature $\beta^{-1}$ (see \cref{eq:boltzpdf}). This quantity can be practically estimated through a Monte Carlo routine.
\\We now provide for some numerical evidence that SVMs approach a good matching between $c_d$ and $c_{\beta}$ for high values of $\beta$ and lower values of $\alpha$.
We first evaluate the correlation between $c_d$ and $c_{\beta}$ in the linear perceptron trained at different values of $k$ with the same set of memories. For this purpose, we measure the Pearson coefficient between the elements of the two matrices (see \cref{eq:rho}). 
Fig. \ref{fig:over_rho}a shows an increasing trend of $\rho$ with respect to $k$. The maximum value is reached when $k = k_{max}(\alpha)$. 
Moreover, the correlation is higher for smaller temperatures. The same behaviour is displayed in \cref{fig:over_rho}b for $\beta = 1$: $c_d$ is not related to $c_1$ at all when $k = 0$, but the two quantities correlate very well when we tune the stability to its maximal value $k_{max}$. We have also measured how the quality of the moment matching approached when $k = k_{max}(\alpha)$ changes with $\alpha$. Results are reported in \cref{fig:over_rho}c and display a deterioration of this property when $\alpha$ progressively increases. As a conclusion, we expect the moment matching condition to be approached for sufficiently low temperatures and lower values of $\alpha$.         
\\We also propose another equivalent experimental study, that compares the empirical spectrum of the eigenvalues of the matrices $c_d$ and $c_1$ for a symmetric SVM with the theoretical expected distribution. The network is chosen to have $N = 500$ and $\alpha = 0.3$. Since $c_d$ is a Wishart matrix the density function of its eigenvalues if given by the Marchenko-Pastur distribution
\begin{equation}
    \label{eq:mp_pdf}
    MP(e| \alpha) = \left(1-\alpha\right)\theta(1-\alpha)\delta(e) + \frac{1}{2\pi}\frac{\sqrt{(e_{+} -e )(e - e_{-})}}{\alpha^{-1}e},
\end{equation}
with 
\begin{equation}
    e_{\pm} = (1 - \alpha^{-1/2})^2,
\end{equation}
and $\theta(x)$ being the Heaviside function.
\begin{figure}[ht!]
\centering
\includegraphics[width=0.65\linewidth]{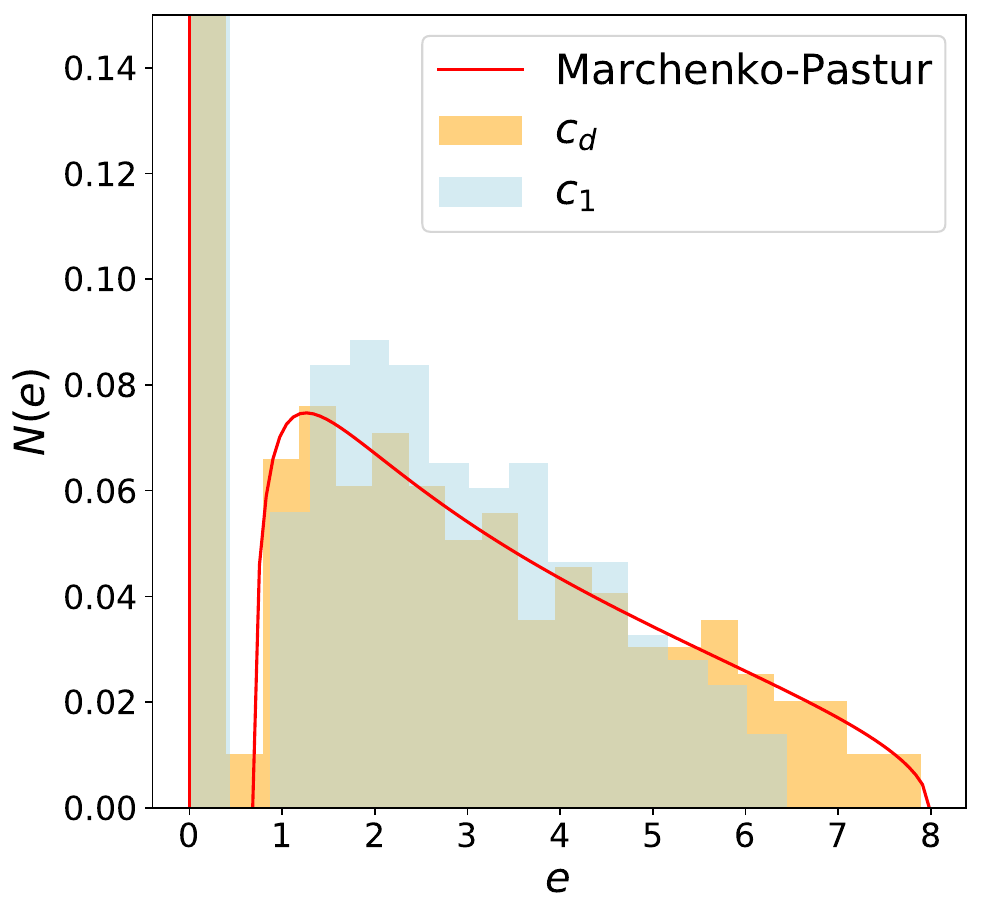}
\caption{Comparison between the theoretical Marchenko-Pastur distribution of the eigenvalues of a Wishart matrix with a parameter $\alpha = 0.3$, and the empirical histogram of the eigenvalues of $c_1$ and $c_d$ for a SVM with $N = 500$.}
\label{fig:over_eigen}
\end{figure}
This suggests that, for an associative memory model that stores a number of memories in a very robust way, i.e. resembling a SVM, one can compute the thermal correlations at $\beta \geq 1$ and measure the spectrum of the eigenvalues of the covariance matrix. 
Whether this fits a MP distribution with a parameter $\alpha^*$, then one can infer the number of dominant basins of attraction. 
Of course this is valid only if such robust memories, that are hidden in the landscape of attractors, are treatable as random binary vectors, i.e. they do not contain a particular internal structure or mutual dependencies. 
\\A physical interpretation for the moment matching condition, and in particular why $c_d \simeq c_{\beta}$ in the SVM is now proposed. The thermal correlation $c_{\beta}$ can rewritten explicetely as
\begin{equation}
    \label{eq:cbeta_exp}
    c_{ij}^{\beta} = \frac{\sum_{\vec{S}}S_i S_j \exp{\left(-\beta E[\vec{S}]\right)}}{\sum_{\vec{S}}\exp{\left(-\beta E[\vec{S}]\right)}},
\end{equation}
that one can rewrite in terms of the typical states at that temperature
\begin{equation}
    \label{eq:cbeta_exp2}
    c_{ij}^{\beta} = \frac{\sum_{\alpha}S_i^{\alpha}S_j^{\alpha}\exp{\left(-\beta E_{\alpha} \right)}}{\sum_{\alpha}\exp{\left(-\beta E_{\alpha} \right)}}.
\end{equation}
When $\beta \rightarrow \infty$ the exponential terms in \cref{eq:cbeta_exp2} are peaked over the ground state of the energy of the model $E_{GS} < 0$. Such state can be single or degenerate with multiplicity $p$.
Hence, the index $\alpha$ is substituted with $\mu = 1,..,p$, the degenerate states are named $\vec{\xi}^{\mu}$ obtaining
\begin{equation}
    \label{eq:cinf_exp}
    c_{ij}^{\infty} = \lim_{\beta \rightarrow \infty}\frac{\sum_{\mu = 1}^p \xi_i^{\mu}\xi_j^{\mu}\exp{\left(-\beta E_{GS} \right)}}{p\exp{\left(-\beta E_{GS} \right)}} = \frac{1}{p}\sum_{\mu = 1}^p \xi_i^{\mu}\xi_j^{\mu},
\end{equation}
that corresponds to 
\begin{equation}
    \label{eq:cinf_exp2}
    c_{\infty} = c_d,
\end{equation}
if $\{\vec{\xi}^{\mu}\}_{\mu}^p$ are also the binary memories. 
As a result, approaching the condition (\ref{eq:cinf_exp2}) implies that, when $k \rightarrow k_{max}(\alpha)$, memories become close to be degenerate ground states of the energy landscape. 
This condition is rare in recurrent neural networks. In a Hebbian network, for instance, \cref{eq:cinf_exp2} holds exclusively for $\alpha = 0$ in the thermodynamic limit, and $\alpha \sim 0$ when $N$ is finite. 
In the Pseudo-inverse model \cite{personnaz, dotsenko_statistical_1991}, that is represented by the following connectivity matrix
\begin{equation}
    \label{eq:pseudoinverse}
    J_{ij} = \frac{1}{N}\sum_{\mu,\nu}\xi_i^{\mu}C^{-1}_{\mu,\nu}\xi_j^{\nu}\hspace{1cm}C_{\mu,\nu} = \sum_{k=1}^N\xi_{k}^{\mu}\xi_k^{\nu},
\end{equation}
perfect-retrieval is admitted up to $\alpha = 1$, with memories lying at a fixed energy $E = -N$ independently of $\alpha$; moreover, the fact that the basins of attraction are absent when $\alpha > 1/2$ implies that memories cannot be ground states of the landscape, since one single step of the dynamics started in the proximity of each memory would decrease the energy. 
A similar behaviour would be observed in a symmetric linear perceptron with $k = 0$, where memories lack of basins of attraction.
\\The relation between the size of the basins of attraction and the depth of memories in the energy landscape is still an open question in the spin-glass community. Previous studies \cite{benedetti_supervised_2022, forrest_content-addressability_1988} relate the typical size of the basins of attraction to the margin $k$ of a linear perceptron.  
Hence, the moment matching condition reached at certain values of $\alpha$ by SVMs must be connected to the size of the basins of attraction.
\\ \\As we have seen from the previous sections, BMs satisfy a moment matching condition. As a consequence, one can imagine to train a BM in a under-sampling regime, i.e. when the number of data-points is small and $P_{data}$ must differ from $P_{true}$ (see the discussion in \cref{sec:data_gen}). To be more specific, we are considering the typical associative memory scenario, where data-points are also binary memories $\{\vec{\xi}^{\mu}\}_{\mu = 1}^p$ being randomly generated. \\Since symmetric SVMs approach the moment matching condition for certain values of $\alpha$, we should expect that the same network can be found by training a BM with the same control parameters. 
\begin{figure}[ht!]
\centering
\includegraphics[width=0.65\linewidth]{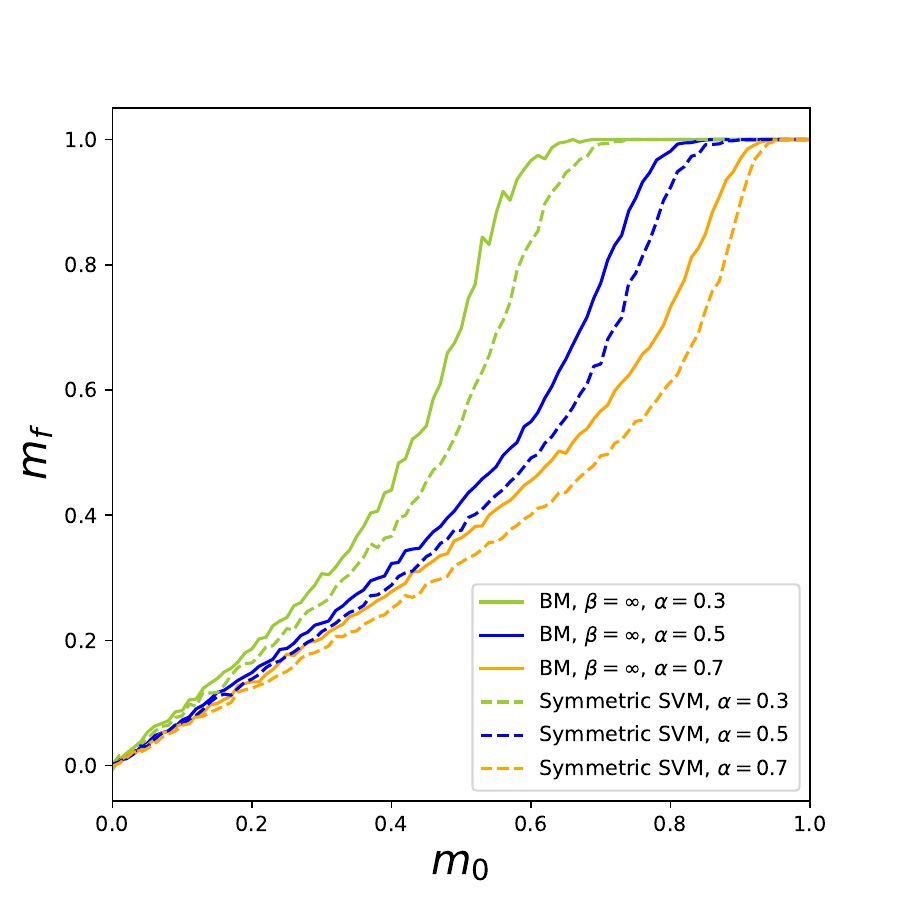}
\caption{Retrieval map $m_f(m_0)$ for a symmetric SVM and BM with $\beta = \infty$ at different values of $\alpha$. BMs are reported with a \textit{full} line while SVMs are represented by the \textit{dashed} line. Measures are averaged over $5$ samples. Errors are neglected because small. Choice of the parameters: $N = 500$, $\lambda = 1$.}
\label{fig:ret_map_svm_bm}
\end{figure}
As a numerical experiment, we trained a BM with $\beta = \infty$ and a SVM with the same set of memories, for different values of $\alpha$. Even if the couplings were initialized completely at random at each training for each model, both the BMs and the SVM reached a mutual overlap of the final $J$ being superior to $90\%$. Figure \ref{fig:ret_map_svm_bm} shows the retrieval map (see \cref{eq:m}) for the three final models with $N = 500$ neurons at $\alpha = \{0.3,0.5,0.7\}$. As one can notice, they all reach perfect-retrieval with comparable sizes of the basins of attraction. To be more precise, BMs display basins that are slightly larger than SVMs, yet more detailed numerical studies should be performed to exclude eventual finite size effects. The memory performance of BMs with $\beta = 1$ appears to be experimentally consistent with the one at $\beta = \infty$ for each choice of $\alpha$.\\ 
%As we know, BM learning naturally converges to the moment matching in \cref{eq:cinf_exp2}. However, a zero temperature Monte Carlo would simply converge to the first stable fixed point in the landscape that, generally speaking, is not representative for the ground state of the energy. Nevertheless, for $\alpha < \alpha_c \simeq 0.6$, fixed points found in the training phase appear to be sufficiently close to the ground state to satisfy the real condition, as evaluated by the real Gibbs-Boltzmann measure. 
%Even if previous works in literature \cite{kojima_capacity_1995} suggested that a BM trained at $\beta = 1$ or $\beta = \infty$ reaches perfect-retrieval up to $\alpha < \alpha_c \simeq 0.6$, new evidence, probably.
Hence, numerical evidence suggests that BMs being trained on a small number of random memories can reproduce the SVM performance pretty well, up to consistent values of $\alpha$. 
Since we proved in \cref{sec:noise_inject} that HU tend to the SVM, and thus to the same moment matching condition, for $\alpha < \alpha_c$, one can conclude that the three learning rules in question, i.e. HU, SVM and BM, must approach the same spot in the space of the couplings.   
%The point of contact between SVMs and BMs in the under-sampling regime, when trained with Rademacher data, must be the HU algorithm, that appears, from the previous analysis, to behave equally to other two algorithms. 
%This aspect is supported by the numerical study in \cref{sec:Ulimit}, where it was showed that a two-steps BM (namely the thermal version of the HU algorithm) and standard BM reached similar performances at different time scales. 
\subsection*{Checkpoint}
In this section we have seen that:
\begin{itemize}
    \item SVMs tend to match the 2-points correlation matrix measured at temperature $\beta$ (i.e. $\langle S_i S_j \rangle_{\beta}$, with $\langle \cdot \rangle_{\beta}$ being the Gibbs-Boltzmann measure) and the Hebbian matrix of the memories. Specifically, this condition, named moment matching, holds for high values of $\beta$ and low values of $\alpha$. 
    \item The moment matching condition when $\beta \rightarrow \infty$, with the 2-points correlations being measured according by the Gibbs-Boltzmann distribution of the model, means that memories are degenerate ground states of the energy. This condition is peculiar of SVMs, with respect of other neural network models. 
    \item Since moment matching is intrisically satisfied by BMs, we conclude that, given the same choice of the control parameters, BM learning and SVMs train similar recurrent neural networks. Moreover, because \cref{sec:noise_inject} has proved that HU approaches a SVM when $\alpha < \alpha_c^{HU}$, we conclude that, given the same choice of the control parameters, BM learning, SVM and HU train very similar recurrent neural networks.
\end{itemize}
\newpage

\section{Summary \& Conclusions}

In this chapter we have seen that:
\begin{itemize}
    \item In the context of Boltzmann Machines, and specifically for what concerns the generalization performance of the inferred network, the standard $L_1$ and $L_2$ regularization techniques are outperformed by the Unlearning regularization.  
    \item A particular case of the Unlearning regularization reproduces a thermally averaged Hebbian Unlearning rule, that shows good inferential capabilities when initialized in the standard Hebbian fashion. Hebbian Unlearning can be interpreted as a two-steps Boltzmann-Machine learning.
    \item In the framework of associative memory models, that learn how to retrieve an extensive number of randomly generated memory vectors, there is a strong analogy between Support Vector Machines, Boltzmann Machines and the Unlearning algorithm. Such analogy lies into a moment matching condition which is common to all these three learning routines. This property, and so the similarity among the models, is stronger for smaller values of $\alpha$, i.e. $\alpha < 0.6$. 
\end{itemize}

The goal of this chapter is twofold: 
\begin{enumerate}
    \item to show the effectiveness of Unlearning as a form of regularization in Boltzmann-Machines;
    \item to establish a connection between three learning rules: Support Vector Machines, Boltzmann Machines, Unlearning. 
\end{enumerate}
For what concerns the first goal, we have used small networks that allowed to compute all quantities in closed form. Precisely, the ground-truth distribution belonged to two models, such as the Curie-Weiss and then Sherrington-Kirkpatrick, known to approach criticality even for finite $N$. 
As a result, we could not only test the proximity of the inferred model to the original one, but also the susceptibility of the network under variations of the parameters, by comparing the original true trend of the specific heat with respect to the inverse temperature, with the trend obtained after training. 
We could conclude that the two main components that define the generalization of the model, i.e. its consistency with the ground-truth distribution and its stability under rescaling of the parameters, are enhanced by the Unlearning regularization. \\
Moreover we analyzed the particular limit of the regularization that gives as learning rules the ones expressed in equations (\ref{eq:Juprob1}) and (\ref{eq:huprob1}). 
This procedure, which provides that a thermal 2-point correlation is computed at $\beta = 1$, strongly resembles the traditional Hebbian Unlearning routine: previous works in literature \cite{nokura_paramagnetic_1996} support the fact that the associative memory performance of a \textit{thermal} or \textit{paramagnetic} Unlearning does not deviate much from the original rule (see \cref{eq:unl_rule}). 
As a new contribution to the Unlearning literature we tested its inferential power, i.e. the capability of reproducing the original model by evaluating the entire set of data (which remains feasible due to the reduced dimension of the network). 
So far the Unlearning algorithm was exclusively tested in an associative memory framework, where data are actually memories, that are sub-dominant in number with respect to the entire set of possibile configurations. 
Results show a good generalization performance of the algorithm. The explanation for this success relies on the choice of the initialization of the parameters. An initial Hebbian network (where the Hebbian matrix is constructed over the entire set of data), by contrast with other types of initialization, implies that the contribution to the gradient descent of the Loss (see \cref{eq:newBMLloss}) deriving from the data (i.e. the positive \textit{Hebbian} contribution in the Boltzmann Machine learning) is much smaller than the contribution deriving from the cross-entropy of the model given the data (i.e. the negative \textit{Unlearning} contribution in the Boltzmann Machine learning). 
This condition holds until an optimal amount of iterations that can be estimated from the numerical simulations.
We conclude that Boltzmann Machine learning can be performed in two steps: an initial abrupt descent over the Loss (i.e. \textit{overshooting}) along the direction of the data and then a gradual adjustment along the direction of the model with increments being updated at each step of the algorithm. 
The effectiveness of a two-step training for a Boltzmann Machine encourages a biological interpretation of this kind of training, as many members in the artificial intelligence and computational neuroscience community seem convinced that optimization of synapses in the brain must occur through the alternation of the daily online experience and offline sleep \cite{sleep, girardeau_brain_2021}.
This point, together with the importance of the Hebbian initialization of the network, appears to be consistent with our previous conclusions regarding Unlearning from the point of view of the associative memory modeling, treated in \cref{sec:noise_inject}. \\
At last, we have showed a new property of Support Vector Machines. These types of models satisfy the same moment matching condition that Boltzmann Machines meet. 
This implies that using the Boltzmann Machine learning algorithm or training a Support Vector Machine should lead to similar outcomes in terms of an associative memory model at least when data are randomly generated, they are few in number, and training is performed at sufficiently low temeperature. 
Moment matching for high $\beta$ and low $\alpha$ is related to stable and deep memories, with the depth directly associated to large basins of attraction. This aspect also gives interesting insights regarding the under-sampling regime of a Boltzmann Machine, where the model is overfitting the data while surrounding them with wide basins of attraction. 
An interesting approach to the overfitting regime of generative recurrent neural network models has already been tackled in \cite{agliari_emergence_2022} for a controlled set-up where data were learned in a Hebbian fashion: in this case overfitting did not imply perfect-retrieval and robustness as it actually does for a proper Boltzmann Machine. 
\\Eventually, since we have evidence that Boltzmann Machine learning in the under-sampling regime can be effectively trained in two steps and thus reproduced by the Hebbian Unlearning algorithm at the same temperature, the similarity between Support Vector machines and Boltzmann Machines implies that these two methods, along with Unlearning, belong to the same kind of training procedures. This is the most important contribution of this thesis work. 

\chapter{Discussion \& Future perspectives}
We will now discuss the future perspectives for this research, dedicated to the study of the Unlearning rule across different learning frameworks.\\\\
Concerning the study of the optimal noise structure to train linear perceptrons, contained in \cref{sec:noise_inject}, the future will certainly see a simplification of the model possibly aimed at obtaining an analytical phase diagram for the network performance. One possible way would be the generalization of the calculations contained in \cref{app:rep_twn} with some structured data that can be treated rigorously. 
One might start by trying different types of structure and see how the phase diagram changes. Two possible candidates are memories containing combinatorial disorder \cite{mezard_mean-field_2017} or memories embedded in a manifold \cite{goldt_modeling_2020, negri_storage_2023}. 
These types of data-set would even satisfy the requirement for the functioning of the training-with-noise algorithm in presence of structured noise: the memories can contain internal correlations among their features, but they should be well separated from each other in the space of configurations. 
This aspect has been clearly pointed out in \cref{sec:corr}.\\Another important point stressed by the analysis in \cref{sec:noise_inject} is the importance of the Hebbian initialization of the network, due to its topological properties. 
The disposition of the glassy states and the surrounding saddles in a Hebbian network out of the retrieval regime appears non-trivial and extremely important in the perspective of finding beneficial noisy configurations to employ in the training. 
Specifically, the distribution of the observables $\omega_i^{\mu}$ with respect to the stabilities $\Delta_i^{\mu}$ (see \cref{sec:choice_of_training_conf}) was crucial to determine whether the class of configurations under consideration decreased the Loss of a maximally stable perceptron problem or not. 
The physical interpretation of this quantity emerges more clearly in the case of fixed points, where $\omega_i^{\mu} = \xi_i^{\mu}S_i(\vec{\xi}^{\mu}\cdot\vec{S})/N$: while a normal projection of the memories over the glassy states would show an isotropic scenario in the landscape, this new scalar product shows a peculiar anisotropy that is responsible for the functioning of the Unlearning algorithm and its generalizations. Understanding the reason for this particular topological property would be fundamental to close the Unlearning problem. \\ \\

In \cref{sec:chap3} we have showed that the \textit{negative learning} principle behind Unlearning exceeds the associative memory application and it can be applied to inferential problems, where the goal is to use a model to reproduce the real distribution of a data-set. 
Even though this idea was already suggested by past works by Hinton \cite{hinton_unsupervised_1999, hinton_forward-forward_2022}, nobody in literature had isolated and singularly evaluated the Unlearning contribution to Boltzmann Machine learning. A follow-up of our study will be testing the Unlearning regularization in the case of real data-sets, and comparing it to the usual regularization techniques. This would imply to deal with the empirical sampling errors and the intrinsic noise contained into data. 
\\Moreover, the moment-matching property of Support Vector Machines needs further investigations. From the point of view of statistical mechanics, calculations are hard to perform when symmetry is constrained: Gardner solved the issue by limiting the problem to a diluted network \cite{gardner_phase_1989}. 
Even though it has been showed that this regime of connectivity affects the perfect-retrieval behaviour of the Support Vector Machine \cite{benedetti_supervised_2022}, things in practice do not qualitatively change. 
One should be now interested in computing the distribution of the energy for the memories as a function of the load parameter $\alpha$, and compare it with the one of the ground state. 
In light of our results, we would expect the memories to be close to the ground state for small values of $\alpha$ and then to be progressively lifted up when $\alpha$ increases. For what concerns the inferential problem instead, and the reason why Boltzmann Machines in the under-sampling regime can resemble maximally stable perceptrons, the attention should be focused on the topological properties of the Hebbian energy landscape.
Since this type of model can be seen as an interpolation (controlled by $\alpha$) between the a Curie-Weiss and a Sherrington-Kirkpatrick model, it might be that fixed points sampled by the dynamics at $\beta = \infty$ are still representative of the ground states of the energy. 
%Specifically, it should be important to examine the dynamic properties of the landscape across the typical stable fixed points out of the retrieval phase. 
%The value of the load $\alpha \simeq 0.6$ might signal a topological change in the landscape that we conjecture here: when $\alpha < 0.6$ the fixed points sampled by the dynamics are close to the ground state of the energy, which is thus accessible, by contrast with other spin-glass models (e.g. the SK model); when $\alpha > 0.6$ the same configurations do not reproduce the Gibbs-Boltzmann statistics at $\beta = \infty$ anymore. 
As a consequence Hebbian landscapes, especially at lower values of $\alpha$ can be employed to train a Support Vector Machine in a faster unsupervised manner, by means of a Boltzmann Machine algorithm or, equivalently, the Hebbian Unlearning routine. \\ \\
Ideally, this research line would lead to the construction of a model that is able to perform both the associative memory and the generative task at the same time: one single training algorithm assembling a network that both retrieves a number of \textit{important} concepts, and generates new examples that are consistent with the stimuli received from the environment. This idea is close to what specialists in statistical learning call \textit{benign overfitting} \cite{bartlett_benign_2020, frei_benign_2023}: the model can both remember the data and correctly fit new unseen examples.  
As a further speculation, we might conjecture that, by implementing algorithms that reach deeper states in the Hebbian energy landscape (e.g. \cite{s_optimization_1983, el_alaoui_sampling_2022}) at higher values of $\alpha$, we might use such states to train a Boltzmann Machine. At this point, if moment matching is satisfied (i.e. \cref{eq:cinf_exp2}) we might reach a condition of benign overfitting: a robust associative memory coexisting with an inferential device. 
This objective might be gained through the choice of continuous neural representations, as it is done for the rate models used in theoretical neuroscience \cite{gerstner_neuronal_2014, ventura_memory_2020}, that would make the dynamics smoother in the landscape. On the other hand, the maximum storage limit of the symmetric Support Vector Machine, i.e. $\alpha_c \simeq 1.3$ as obtained by \cite{gardner_phase_1989} for diluted symmetric perceptrons, would be overcome by imposing sparser representation of the memories \cite{gardner_maximum_1987}.\\\\
Finally, another aim of this thesis is to incentive a stronger experimental effort for the investigation of anti-Hebbian synaptic mechanisms in the brain. We believe that, in light of the analysis exposed in this work, and consistently with the intuition of a series of scientists from past to recent times \cite{foldiak_forming_1990, barlow_adaptation_1989, crick_function_1983, sleep, girardeau_brain_2021, KK}, this counter-intuitive and apparently not biological mechanism could, instead, be at the basis of the formation and the consolidation of memory in the brain.  

\appendix

\chapter{Analytical expression of the one-step retrieval map \label{app:m1}}
As done in \cite{kepler_domains_1988, gardner_optimal_1989}, one can define the following quantity
\begin{equation}
\label{eq:field}
    h_i^{\mu} = \frac{\xi_i^{\mu}}{\sqrt{N}}\sum_{j = 1}^NJ_{ij}S_j,
\end{equation}
that we will call $\textit{cross-stability}$ of the configuration $\vec{S}$. Configurations $\vec{S}$ have an overlap $m_0$ with the memory $\vec{\xi}^{\mu}$ and they are generated by the probability distribution in \cref{eq:p_s}.
As a consequence, $h_i^{\mu}$ is a random variable distributed according to 
\begin{equation}
    P(h_i^{\mu}|m_0, J) = \frac{\sum_{\vec{S}}\left[\delta\left(m_0 - \frac{1}{N}\sum_{k = 1}^N S_k\xi_k^{\mu}\right)\delta\left( h_i^{\mu} - \frac{\xi_i^{\mu}}{\sqrt{N}}\sum_{k\neq i}J_{ik}S_k\right) \right]}{\sum_{\vec{S}}\left[\delta\left(m_0 - \frac{1}{N}\sum_{k = 1}^N S_k\xi_k^{\mu}\right) \right]},
\end{equation}
which can computed by using the Fourier transform of the Dirac-delta function and summing over $\{-1,+1\}^N$. 
%We neglected the memory index $\mu$ without any Loss of generality. 
This leads to the following probability density function 
\begin{equation}
\label{eq:p_hi}
    P(h_i^{\mu}|m_0, J) = \frac{1}{\sqrt{2\pi(1-m_0^2)}}\exp{\left(-\frac{(h_i^{\mu} - m_0\Delta_i^{\mu}(J))^2}{2(1-m_0^2)}\right)},
\end{equation}
where the neuron index $i$ was also dropped for simplicity, and the stabilities depend on the realization of the couplings. The one-step retrieval map averaged over all the possible $\vec{S}$ configurations is a function of $m_0$ and $J$, and it is given by
\begin{equation}
\small
    m_1(m_0, J) = \frac{1}{\alpha N^2}\sum_{i,\mu}^{N,p}\int dh_i^{\mu}P(h_i^{\mu}|m_0,J)\text{sign}\left(h_i^{\mu}\right) = \frac{1}{\alpha N^2}\sum_{i,\mu}^ {N,p}\text{erf}\left(\frac{m_0\Delta_i^{\mu}}{\sqrt{2(1-m_0^2)}} \right),
\end{equation}
%\begin{equation}
%    m_1(m) = \frac{1}{\alpha N^2}\sum_{i,\mu}^{N,p}\text{sign}\left(h_i^{\mu}\right) \xrightarrow{N\rightarrow \infty}\int_{-\infty}^{+\infty}d\Delta\rho(\Delta)\langle \text{sign}\left(h\right) \rangle_{h,\Delta},
%\end{equation}
%where $\langle \hspace{0.1cm}\cdot\hspace{0.1cm}\rangle_{h,\Delta}$ is the expected value over the distribution of the cross-stabilities in \cref{eq:p_hi} and $\rho(\Delta)$ is the probability density function of the stabilities. We assumed the stabilities of the network to be self-averaging quantities. The resulting quantity is
and this result is fully general and it is relative to one realization of the network.
If we assume that stabilities are self-averaging quantities and they satisfy the probability distribution function $\rho(\Delta)$, then we can compute the one-step overlap as averaged over many realizations of the network, i.e. for many training processes. One obtains
\begin{equation}
\small
\label{eq:m1_Ninf}
m_1(m_0) = \lim_{N\rightarrow\infty}\frac{1}{\alpha N^2}\sum_{i,\mu}^ {N,p}\text{erf}\left(\frac{m_0\Delta_i^{\mu}}{\sqrt{2(1-m_0^2)}} \right) = \int_{-\infty}^{+\infty}d\Delta \rho(\Delta)\text{erf}\left( \frac{m_0\Delta}{\sqrt{2(1-m_0^2)}}\right),
\end{equation} 
which is the quantity defined in \cref{eq:onestep_ret_map}. 
This observable can provide with some rough intuition about the shape of the basins of attraction in any fully connected neural network, where no exact calculation exists to estimate their typical radius. It will also be crucial for the definition of a Loss function for the \textit{training-with-noise} algorithm \cite{gardner_training_1989} presented in section \ref{sec:twn}. 
\chapter{Gradient of the Loss function of the Training-with-noise algorithm\label{app:grad_twn}}
\section{Training with noise}
\label{sec:appA1}
At each step of the algorithm a memory label $\mu_d$ is sampled at random and the update (\ref{eq:lwn}) is performed over the couplings. The new value of the $\mathcal{L}$ function (\ref{eq:loss_ws}) is
%\begin{widetext}
\begin{equation}
    \label{eq:new_loss}
    \mathcal{L}^{'} = -\frac{1}{\alpha N^2}\sum_{i,\mu }^{N,p}\text{erf}\left(\frac{m\Delta_i^{\mu}}{\sqrt{2(1-m^2)}} + \frac{\lambda m}{N\sigma_i\sqrt{2N(1-m^2)}}\epsilon_i^{\mu_d}\xi_i^{\mu}\xi_i^{\mu_d}\sum_{j \neq i}S_j^{\mu_d}\xi_j^{\mu}\right).
\end{equation}
Since $\delta\sigma_i \propto \frac{\lambda}{N}\frac{J_i}{\sigma_i} + \mathcal{O}\left(\frac{\lambda}{N} \right)^3$ and the mean $J_i$ of the couplings along line $i$ equals zero by initialization and it is naturally maintained null during the algorithm, we have considered 
$\sigma_i^\prime\simeq \sigma_i$. 
Then $\mathcal{L}'$ can be rewritten as
\begin{multline}
\label{eq:new_loss2}
    \mathcal{L}^{'} = -\frac{1}{\alpha N^2}\sum_{i,\mu\neq \mu_d}^{N, p}\text{erf}\left(\frac{m\Delta_i^{\mu}}{\sqrt{2(1-m^2)}} + \mathcal{O}\left( \frac{1}{N}\right)\right) -\\-\frac{1}{\alpha N^2}\sum_{i}^{N}\text{erf}\left(\frac{m\Delta_i^{\mu_d}}{\sqrt{2(1-m^2)}} + \frac{\lambda m\cdot m_t}{\sigma_i\sqrt{2N(1-m^2)}}\epsilon_i^{\mu_d} + \mathcal{O}\left(\frac{1}{N}\right)\right),
\end{multline}
%\end{widetext}
where we have used that $\frac{1}{N}\sum_{j\neq i}\xi_j^{\mu}S_j^{\mu_d} = \mathcal{O}(N^{-1/2})$ when $\mu \neq \mu_d$ and $m_t = \mathcal{O}(1)$. We thus expand the error-function at the first order in $\mathcal{O}(N^{-1/2})$ obtaining the variations to $\mathcal{L}$ in equation (\ref{eq:delta_loss}). 
\section{\label{sec:appA2}Training with structured noise}
\begin{figure*}[ht!]
\begin{tabularx}{\linewidth}{*{2}{X}}
     %\centering
     \begin{subfigure}[b]{\linewidth}
         \centering
         \includegraphics[width=\textwidth]{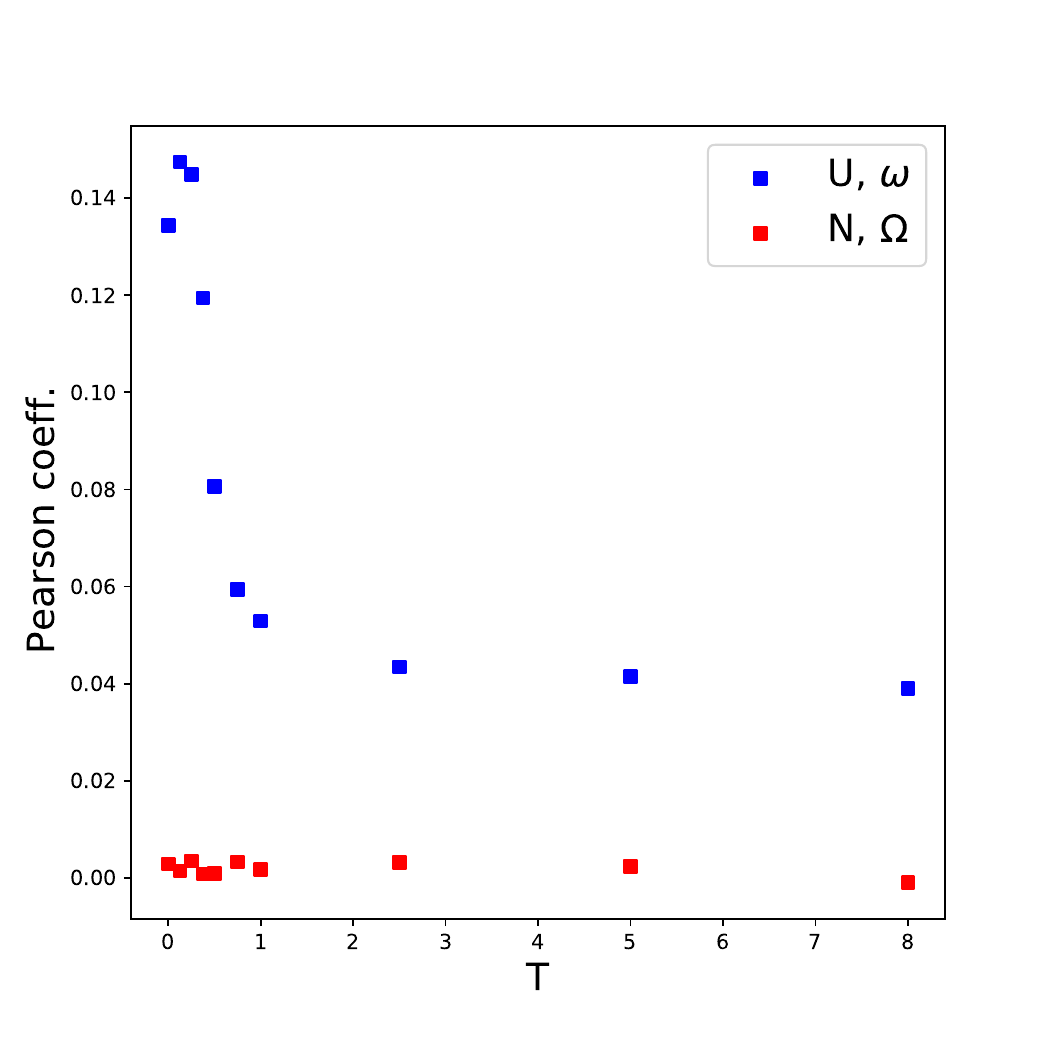}
         \caption{}
         \label{fig:ptest_a}
     \end{subfigure}
     &
     \hfill
     \begin{subfigure}[b]{\linewidth}
         \centering
         \includegraphics[width=\textwidth]{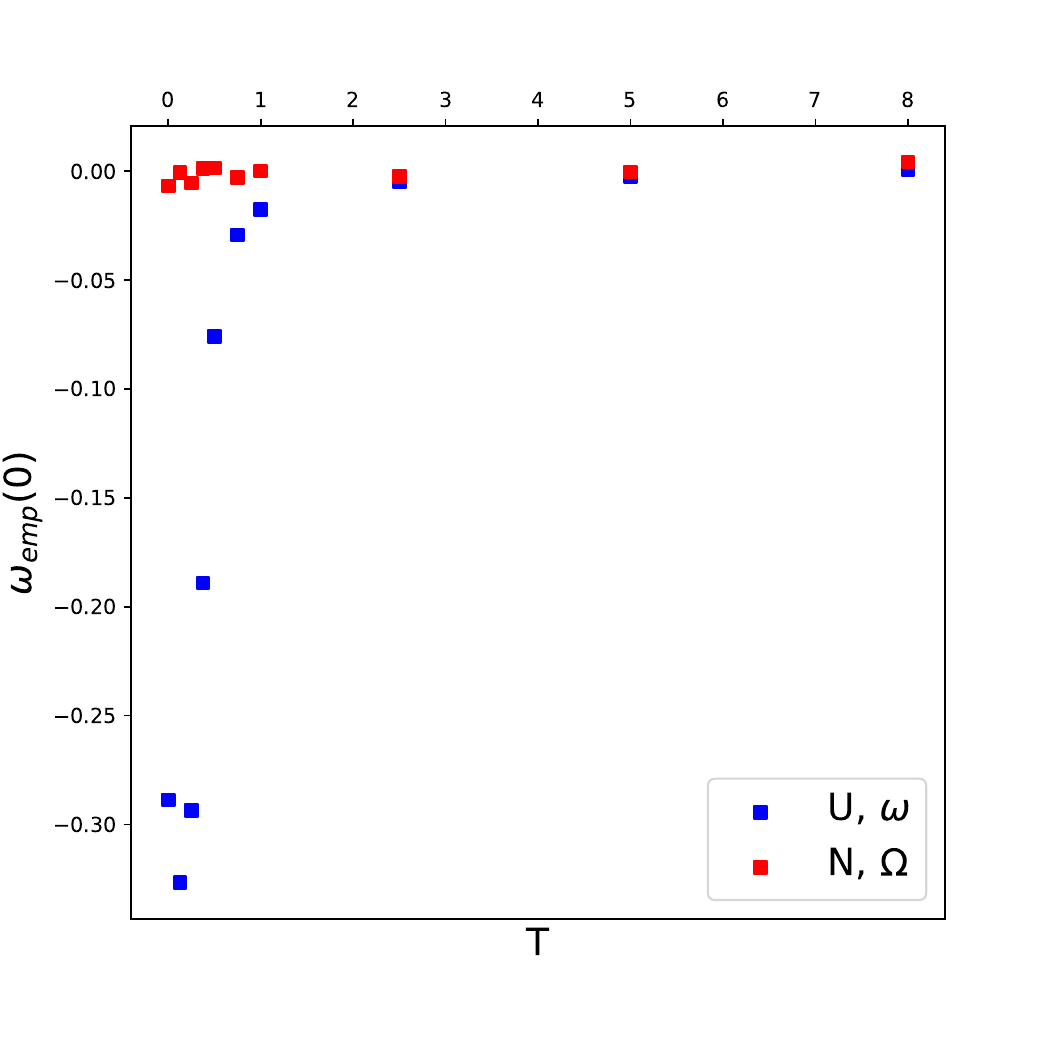}
         \caption{}
         \label{fig:ptest_b}
     \end{subfigure}
\end{tabularx}
\caption{(a) Pearson coefficient between $\omega_i^{\mu}$, $\Omega_i^{\mu}$ and the stabilities $\Delta_i^{\mu}$, (b) $\omega_{emp}(0)$ for the case of a Hebbian network at different temperatures $T$. Configurations at a given temperature $T$ have been sampled by a Monte Carlo of the Kawasaki kind, in order to choose only maximally noisy states ($m_t = 0^+$). Points are collected from $15$ samples of the network. Choice of the parameters: $N = 500$, $\alpha = 0.5$.}
\end{figure*}

The new value of $\mathcal{L}$ is derived by using equation (\ref{eq:lwn2}) to evaluate the variation of stabilities
%\begin{widetext}
\begin{multline}
    \label{eq:new_lossApp}
    \mathcal{L}^{'} = -\frac{1}{\alpha N^2}\sum_{i,\mu }^{N,p}\text{erf}\Bigg(\frac{m\Delta_i^{\mu}}{\sqrt{2(1-m^2)}} + \frac{\lambda m \xi_i^{\mu}\xi_i^{\mu_d}}{2N\sigma_i\sqrt{2N(1-m^2)}}\sum_{j=1}^N S_j^{\mu_d}\xi_j^{\mu} + \\+\frac{\lambda m \xi_i^{\mu} S_i^{\mu_d}}{2N\sigma_i\sqrt{2N(1-m^2)}}\sum_{j=1}^N \xi_j^{\mu_d}\xi_j^{\mu} - \\ -\frac{\lambda m \xi_i^{\mu}S_i^{1,\mu_d}}{2N\sigma_i\sqrt{2N(1-m^2)}}\sum_{j=1}^N S_j^{\mu_d}\xi_j^{\mu} - \frac{\lambda m \xi_i^{\mu}S_i^{\mu_d}}{2N\sigma_i\sqrt{2N(1-m^2)}}\sum_{j=1}^N S_j^{1,\mu_d}\xi_j^{\mu}\Bigg),
\end{multline}
where $\sigma_i^{'} \simeq \sigma_i$ as in the previous paragraph. We now recall $\chi_i^{\mu} = \xi_i^{\mu}S_i^{\mu_d}$,  $\chi_i^{1,\mu} = \xi_i^{\mu}S_i^{1,\mu_d}$, $m_{\mu} = \frac{1}{N}\sum_{j = 1}^N S_j^{\mu_d}\xi_j^{\mu}$ and $m_{1,\mu} = \frac{1}{N}\sum_{j = 1}^N S_j^{1,\mu_d}\xi_j^{\mu}$ and expand the error-function at the first order in $\mathcal{O}(N^{-1/2})$ obtaining
\begin{multline}
\label{eq:deltaL_tot}
\delta\mathcal{L} = \frac{m\lambda}{\sqrt{2\pi\alpha^2 N^5 (1-m^2)}}\sum_{i,\mu}^{N,p}\frac{1}{2\sigma_i}\Large[(m_{\mu}\chi_i^{1,\mu} + m_{1,\mu}\chi_i^{\mu}) - \\- (m_{\mu}\xi_i^{\mu}\xi_i^{\mu_d} + M_{\mu}^{\mu_d}\chi_i^{\mu})\Large] \exp{\left(-\frac{m^2\Delta_i^{\mu^2}}{2(1-m^2)}\right)}
\end{multline}
%\end{widetext}
where $M_{\mu}^{\mu_d} = \frac{1}{N}\sum_{i=1}^N \xi_i^{\mu}\xi_i^{\mu_d}$. Equation (\ref{eq:deltaL_tot}) can be decomposed in
\begin{equation}
    \delta \mathcal{L} = \delta \mathcal{L}_{N} + \delta \mathcal{L}_{U}
\end{equation}
where $\delta \mathcal{L}_{U}$ contains the weight 
\begin{equation}
    \label{eq:w1}
    \omega_i^{\mu} =\frac{1}{2\sigma_i}\left( m_{\mu}\chi_i^{1,\mu} + m_{1,\mu}\chi_i^{\mu}\right),
\end{equation}
while $\delta \mathcal{L}_{N}$ contains 
\begin{equation}
    \label{eq:w2}
    \Omega_i^{\mu} =\frac{1}{2\sigma_i}\left(m_{\mu}\xi_i^{\mu}\xi_i^{\mu_d} + M_{\mu}^{\mu_d}\chi_i^{\mu}\right).
\end{equation}
We study the two contributions numerically, on a Hebbian network, i.e. with no learning going on, for the case of $m_t = 0^+$. The Pearson coefficient is measured between the vector of the stabilities $\Delta_i^{\mu}$ and the weights $\omega_i^{\mu}$ as well as with  $\Omega_i^{\mu}$ separately. This quantity should underline an eventual reciprocal dependence between $\omega_i^{\mu}, \Omega_i^{\mu}$ and $\Delta_i^{\mu}$. The test is repeated over states sampled by a Monte Carlo at different temperatures $T$.
Results are reported in \cref{fig:ptest_a}
where it is evident that $\Omega_i^\mu$ does not have any correlation with $\Delta_i^\mu$, while the dependence of $\omega_i^\mu$ on the stabilities is evident. Moreover, we measured the indicator $\omega_{emp}(0)$, signaling the typical values of the weights $\omega_i^{\mu}$ and $\Omega_i^{\mu}$ when $\Delta_i^{\mu} \sim 0$ (see section \ref{sec:trademark_virtuous} for further details), as reported in \cref{fig:ptest_b}. The plot clearly shows that $\Omega_i^{\mu}$ is small and generally fluctuating around zero. These aspects hold during the training procedure also, as it can be observed by performing the same measure at different step of the TWN procedure over states with $m_t = 0^+$. 
We will thus refer to $\delta \mathcal{L}_U$ as the relevant contribution to the variation of the function $\mathcal{L}$.

\chapter{Replica symmetric analysis of the Training-with-noise algorithm \label{app:rep_twn}}
In this Appendix we resume the replica calculations in the replica-symmetric ansatz \cite{mezard_spin_1986} performed by \cite{wong_optimally_1990, wong_neural_1993} showing their agreement with the numerical results obtained through Gardner's TWN algorithm \cite{gardner_training_1989}.  
\section{Analytics}
\begin{figure*}[ht!]

\centering

    \includegraphics[width=0.6\linewidth]{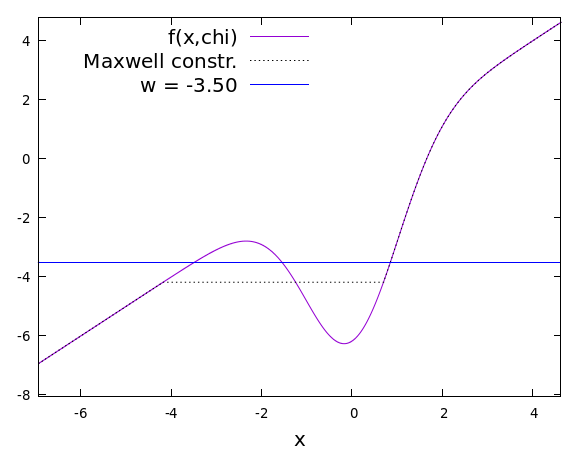}
  
%    \subcaptionbox{m = 0}{
%    \includegraphics[width=\linewidth]{delta_histo_m0.pdf}}\par

\caption{Graphic representation of the saddle point equation (\ref{eq:sad1}). The choice of the parameters is: $\alpha = 0.3$, $m = 0.7$, $\chi = 7.89$.}
\label{fig:sad1}
\end{figure*}

The partition function of the model is given by
\begin{equation}
\label{eq:Z}
    Z = \int\prod_{i,j}dJ_{ij}\delta\left(\sum_{j\neq i}J_{ij}^2 - N\right)\exp{\left(\beta\sum_i\sum_{\mu}\text{erf}\left(\frac{m\xi_i^{\mu}\sum_jJ_{ij}\xi_j^{\mu}}{\sqrt{2N(1-m^2)}}\right)\right)},
\end{equation}
while the distribution of the stabilities is
\begin{equation}
    \small
    \label{eq:rho1}
    \rho(\Delta) = \overline{\frac{1}{Z}\int \prod_{j}dJ_{ij}\delta\left(\sum_{j}J_{ij}^2 - N\right)\exp{\left(\beta\sum_{\mu}\text{erf}\left(\frac{m\xi_i^{\mu}\sum_jJ_{ij}\xi_j^{\mu}}{\sqrt{2N(1-m^2)}}\right)\right)}\delta\left( \xi_i^1\sum_j\frac{J_{ij}}{\sqrt{N}}\xi_j^1-\Delta\right)},
\end{equation}
where we have neglected the factorization over $i$, since we treat the optimization process as independent along the lines of the $J_{ij}$ matrix. Index $i$ will be nevertheless indicated for the sake of completeness.\\  
Replicas can be used to evaluate the normalisation, i.e.
\begin{equation}
    1/Z = \lim_{n\rightarrow 0}Z^{n-1}
\end{equation}
Consequently, equation (\ref{eq:rho1}) can be rewritten as
\begin{equation}
    \small
    \label{eq:rho2}
    \rho(\Delta) = \lim_{n\rightarrow 0}\overline{\int \prod_a^n\prod_{j}^NdJ_{ij}^a\delta\left(\sum_{j}(J_{ij}^a)^2 - N\right)\exp{\left(\beta\sum_{a,\mu}\text{erf}\left(\frac{m\xi_i^{\mu}\sum_jJ_{ij}^a\xi_j^{\mu}}{\sqrt{2N(1-m^2)}}\right)\right)}\delta\left( \xi_i^1\sum_j\frac{J_{ij}^1}{\sqrt{N}}\xi_j^1-\Delta\right)}
\end{equation}
that becomes
\begin{multline}
  \label{eq:rho3}
    \rho(\Delta) = \lim_{n\rightarrow 0}\int \prod_a^n\prod_{j}^NdJ_{ij}^a\delta\left(\sum_{j}(J_{ij}^a)^2 - N\right) \prod_{\mu,a}\int_{-\infty}^{+\infty}dx_{\mu}^{a}\exp{\left(\beta\sum_{\mu,a}\text{erf}\left(\frac{m\cdot x_i^{\mu}}{\sqrt{2(1-m^2)}}\right)\right)}\cdot \\ \cdot \overline{\prod_{\mu,a}\delta\left(\xi_i^{\mu}\sum_j\frac{J_{ij}^a}{\sqrt{N}}\xi_j^{\mu} - x_{\mu}^{a}\right)\delta\left( \xi_i^1\sum_j\frac{J_{ij}^1}{\sqrt{N}}\xi_j^1-\Delta\right)}
\end{multline}
by rewriting the Dirac delta in its integral representation we get
\begin{multline}
    \small
    \label{eq:rho4}
    \rho(\Delta) = \lim_{n\rightarrow 0}\int \prod_a^n\prod_{j}^NdJ_{ij}^a\delta\left(\sum_{j}(J_{ij}^a)^2 - N\right)\cdot \\ \cdot \prod_{\mu,a}\int_{-\infty}^{+\infty}dx_{\mu}^a\int_{-\infty}^{+\infty}\frac{dy_{\mu}^a}{2\pi}\int_{-\infty}^{+\infty}dz\exp{\left[\beta\sum_{\mu,a}\text{erf}\left(\frac{m\cdot x_i^{\mu}}{\sqrt{2(1-m^2)}}\right)-\sum_{\mu,a}x_{\mu}^a y_{\mu}^a - iz\Delta\right]}\cdot \\ \cdot \overline{\prod_{\mu\neq 1}\prod_j\exp{\left(i\sum_a\frac{J_{ij}^a}{\sqrt{N}}\xi_i^{\mu}\xi_j^{\mu}y_{\mu}^a\right)}}\cdot\overline{\prod_j\exp{\left(+i\frac{y_{1}^a}{\sqrt{N}}\sum_a J_{ij}^a\xi_i^{1}\xi_j^{1}+ i\frac{z}{\sqrt{N}}J_{ij}^1\xi_i^1\xi_j^1\right)}}
\end{multline}
We will now compute the averages over the disorder, i.e. the random memories. The first one, relative to $\mu \neq 1$ variables, becomes
\begin{equation}
    \label{eq:dis_A}
    \overline{\prod_{\mu\neq 1}\prod_j\exp{\left(i\sum_a\frac{J_{ij}^a}{\sqrt{N}}\xi_i^{\mu}\xi_j^{\mu}y_{\mu}^a\right)}} = \exp{\left(-\frac{1}{2}\sum_{\mu\neq 1}\sum_{a,b}Q_{ab}y_{\mu}^ay_{\mu}^b\right)}
\end{equation}
where we have Taylor expanded at the second order around the $\mathcal{O}(1/\sqrt{N})$ term and we have introduced the overlap matrix $Q_{ab}$ defined as
\begin{equation}
    \label{eq:Q}
    Q_{ab} = \frac{1}{N}\sum_{j = 1}^N J_{ij}^a J_{ij}^b
\end{equation}
The second averaged expression becomes
\begin{equation}
    \label{eq:dis_B}
\small
    \overline{\prod_j\exp{\left(+i\frac{y_{1}^a}{\sqrt{N}}\sum_a J_{ij}^a\xi_i^{1}\xi_j^{1}+ i\frac{z}{\sqrt{N}}J_{ij}^1\xi_i^1\xi_j^1\right)}} =
\end{equation}
$$
    = \overline{\prod_j\exp{\left( i\sum_a\frac{J_{ij}^a}{\sqrt{N}}\xi_i^1\xi_j^1 y_1^a + i\frac{J_{ij}^1}{\sqrt{N}}(y_1^1 + z)\xi_i^1\xi_j^1\right)}} =$$
$$
    =\prod_j\cos{\left(\sum_{a\neq 1}\frac{J_{ij}^a}{\sqrt{N}} y_1^a + \frac{J_{ij}^a}{\sqrt{N}}(y_1^1+z)\right)}= $$ $$= \exp{\left(-\frac{1}{2}\sum_{a,b\neq 1}Q_{ab}y_1^a y_1^b - \sum_{a \neq 1}Q_{a1}y_1^a(y_1^1 + z) - \frac{1}{2}(y_1^1 + z)^2\right)} = $$
where, once again, a Taylor expansion at the second order in $\mathcal{O}(1/\sqrt{N})$ has been performed. We now make the following change of variables
\begin{equation}
    \label{eq:c_o_v}
    \Tilde{y}_1^1 = y_1^1 + z\hspace{0.5cm}d\Tilde{y}_1^1 = dy_1^1
\end{equation}
Hence equation (\ref{eq:dis_B}) becomes
\begin{equation}
    \label{eq:dis_B2}
    = \exp{\left(-\frac{1}{2}\sum_{ab}Q_{ab}y_1^a y_1^b\right)}
\end{equation}
where, instead of $y_1^1$ one considers the new one given by (\ref{eq:c_o_v}).\\ The averaged expression for the distribution of the stabilities is now given by
\begin{multline}
    \label{eq:rho5}
    \small
    \rho(\Delta) = \lim_{n\rightarrow0}\int DQ\det{Q}^{\frac{N}{2}}[ \prod_a\int_{-\infty }^{+\infty}\frac{dx^a}{2\pi}dy^a\exp( \beta\sum_a\text{erf}\left(\frac{mx^a}{\sqrt{2(1-m^2)}}\right)-i\sum_ax^ay^a-\\-\frac{1}{2}\sum_{a,b}Q_{ab}y^ay^b)]^{\alpha N - 1}\cdot\prod_{a\neq 1}\int_{-\infty}^{+\infty}\frac{dx^a}{2\pi}dy^a dx^1 dy^1d\Tilde{y}^1\exp(\beta \sum_a \text{erf}\left( \frac{m x^a}{\sqrt{2(1-m^2)}}\right)- \\ -i\sum_ax^ay^a-i(\Tilde{y}^1 - y^1)\Delta - \frac{1}{2}\sum_{a,b}Q_{ab}y^ay^b) = 
\end{multline}
with $$ DQ = dQ\prod_a\delta\left(Q_{aa} - 1\right)$$
We now assume the replica symmetry of $Q_{ab}$, i.e.
\begin{equation}
    \label{eq:rs}
    Q_{ab} = \delta_{ab} + (1 - \delta_{ab})q
\end{equation}
Hence the computation follows
\begin{multline}
    \label{eq:rho6}
    = \lim_{n\rightarrow 0}\int Dq\exp{(Ng_n(q))}\int D_q w\left[ \int\frac{dx}{2\pi}\int dy \exp{(-\beta\mathcal{L}_m(x) - i(x-w)y - \frac{(1-q)}{2}y^2} \right]^{n-1}\cdot \\ \cdot\int\frac{dx}{2\pi}dyd\Tilde{y}\exp{\left(-\beta\mathcal{L}_m(x)-i(x-\Delta)y - i(\Delta - w)\Tilde{y}-\frac{(1-q)}{2}\Tilde{y}^2\right)}
\end{multline}
where 
\begin{equation}
    D_q w  = \frac{1}{\sqrt{2\pi q}}e^{-\frac{w^2}{2q}}
\end{equation}
and $g_n(q)$ is the \textit{Cramer} function, given by
\begin{multline}
    \label{eq:ldf}
    \small
    g_n(q) = \frac{1}{2}\log(1 + (n-1)q) + \frac{n-1}{2}\log{(1-q)} + \\ +\alpha\log{\left[ \int D_q w\left( \int \frac{dx}{\sqrt{2\pi(1-q)}}\exp{(-\beta\mathcal{L}_m(x)-\frac{(x-w)^2}{2(1-q)}})\right)^n\right]}
\end{multline}
The limit for $n\rightarrow 0$ of $g_n(q)$ gives
\begin{multline}
    \label{eq:ldf2}
    \lim_{n\rightarrow 0}g_n(q) = n\cdot g_{\beta,m}(q) = \\ =\small n\left[\frac{1}{2}\log{(1-q)} + \frac{q}{2(1-q)} + \alpha\int D_q w\log{\left[ \int_{-\infty}^{+\infty}\frac{dx}{\sqrt{2\pi(1-q)}}\exp{\left(-\beta\mathcal{L}_m(x) - \frac{(x-w)^2}{2(1-q)} \right)}\right]}\right]
\end{multline}
Consequently, the distribution of the stabilities becomes
\begin{equation}
    \label{eq:rho_final}
    \rho(\Delta) = \int_{-\infty}^{+\infty} D_{q^*} w \frac{\exp{\left( -\frac{(\Delta - w)^2}{2(1-q^*)} + \beta\text{erf}\left(\frac{m\Delta}{\sqrt{2(1-m^2)}}\right)\right)}}{\left[ \int_{-\infty}^{+\infty}dx\exp{\left(\beta\text{erf}\left(\frac{m\Delta}{\sqrt{2(1-m^2)}}\right) - \frac{(x-w)^2}{2(1-q^*)}\right)} \right]}
\end{equation}
where $q^*$ is the particular value of the overlap s.t. $$\frac{dg_{\beta,m}}{dq}|_{q  = q^*} = 0$$
Following the same procedure one is able to compute the replica symmetric free-energy of the model, being
\begin{equation}
    \label{eq:f}
    f_{RS,\beta}(m,\alpha) = -\frac{1}{N\beta }\overline{\log{(Z_\beta)}} = -\lim_{n\rightarrow 0}\frac{1}{nN\beta}(\overline{Z^n}-1) = -\frac{g_{\beta,m}(q^*)}{\beta}
\end{equation}
%\subsection{$T\rightarrow 0$ and $q\rightarrow 1$ limit}
We are now interested in the $\beta \rightarrow \infty$ limit with $q = 1$. Such a limit reproduces the ground truth of the problem, i.e. the very minimum of the Loss function. \\One can thus couple $q$ with the annealing temperature $T = 1/\beta$ by introducing an interaction susceptibility $\chi$, such that
\begin{equation}
\label{eq:chi}
    \chi = \frac{1}{T}\left[\langle J^2 \rangle_T - \langle J \rangle_T ^2\right] = \frac{1}{T}(1 - q) 
\end{equation}
where $\langle \hspace{0.1cm}\cdot\hspace{0.1cm}\rangle_T$ is the thermal average and the last passage is justified by the spherical constraint over the couplings. One can now substitute the recurrent term $(1-q)$ in both the $g$ function and the $\rho$ with $\chi T$. The first one becomes
\begin{multline}
    \label{eq:g_new}
    \small
    g_{T,m}(\chi) = \frac{1}{2}\log{(\chi T)} + \frac{1 - \chi T}{2\chi T} + \\ +\alpha\int D_{1 - \chi T} w \log{\left[  \int_{-\infty}^{+\infty}\frac{dx}{\sqrt{2\pi\chi T}}\exp{\left(-\frac{1}{T}\left(-\text{erf}{\left(\frac{mx}{\sqrt{1-m^2}}\right)} + \frac{(x-w)^2}{2\chi}\right)\right)}\right]}
\end{multline}
One can now substitute $g$ into the expression for the free energy given by (\ref{eq:f}) and perform the $T \rightarrow 0$ limit which simultaneously pushes $q \rightarrow 1$ according to (\ref{eq:chi}), since $\chi \geq 0$ by definition. We get
\begin{equation}
\label{eq:new_f}
    f_m(\alpha, \chi) = -\frac{1}{2\chi} - \alpha\int Dw\left[\text{erf}{\left(\frac{mx^*}{\sqrt{1-m^2}}\right)} - \frac{(x^*-w)^2}{2\chi}\right]
\end{equation}
where $x^*$ is obtained by a saddle point equation and $\chi$ should be found by partial derivation of the free energy. By identifying the exponent in the gaussian integral of the energetic contribution to the free energy with
\begin{equation}
    \label{eq:e}
    e_m(\chi, x) = -\text{erf}{\left(\frac{mx}{\sqrt{1-m^2}}\right)} + \frac{(x-w)^2}{2\chi}
\end{equation}
we have
\begin{equation}
\label{eq:sad1}
    \frac{\partial e_m(\chi,x)}{\partial x} = 0 \Longrightarrow w = x - \frac{\sqrt{2}m\chi}{\sqrt{\pi(1 - m^2)}}\exp{\left(-\frac{m^2x^2}{2(1-m^2)}\right)}
\end{equation}
from which one gets $x^*$ as a function of $w$ and $\chi$. On the other hand, one must differentiate the entire free energy in $\chi$ to obtain $\chi^*$. Hence we get
\begin{equation}
\label{eq:sad2}
    \frac{\partial f_m(\chi, x^*)}{\partial \chi} = 0 \Longrightarrow \int Dw\left(x^*(w,\chi) - w\right)^2 = \alpha^{-1}
\end{equation}
Equation (\ref{eq:sad1}) leads to the graphic intersection problem depicted in figure (\ref{fig:sad1}). Given $w$ and $\chi$, one finds $x^*$. Then one uses such a $x^*$ to solve equation (\ref{eq:sad2}), which permits to find $\chi^*$. Once one has $\chi$ and $x^*(w,\chi)$, the free energy in (\ref{eq:new_f}) can be computed. Nevertheless, the support of the function in $x$ in figure (\ref{fig:sad1}) is not monotonic, and so not invertible, which makes the solution of the second saddle equation (\ref{eq:sad2}) not feasible. Hence one can exploit the Maxwell construction of the thermodynamics, as represented by the dashed curve in figure (\ref{fig:sad1}), while solving the problem. 
It can be verified easily that the new expression for the density function of $\Delta_i^{\mu}$ when $T\rightarrow 0$ and $q \rightarrow 1$ is given by
\begin{multline}
    \label{eq:new_rho}
    \small
    \rho(\Delta) = \int Dw \delta\left( \Delta - x^*(w,\chi^*)\right) =\\ = \frac{1}{\sqrt{2\pi}}\left(1  + \sqrt{\frac{2}{\pi}}\frac{m^3 \chi}{(1 - m^2)^{3/2}}\Delta\exp\left( -\frac{m^2\Delta}{2(1-m^2)}\right)\right)\exp{-\frac{w(\Delta)^2}{2}}
\end{multline}
where we have made use of the invertibility of $w(x,\chi)$ and also the following property of the Dirac delta
\begin{equation}
    \label{eq:dirac}
    \delta\left( \Delta - x^*(w,\chi)\right) = \delta(F(w)) = \delta(w - w(x^*,\chi^*))\frac{dw(x^*,\chi^*)}{dx^*}|_{x^* = \Delta}
\end{equation}

\section{Numerics}
\begin{figure*}[ht!]

\centering

    \includegraphics[width=0.6\linewidth]{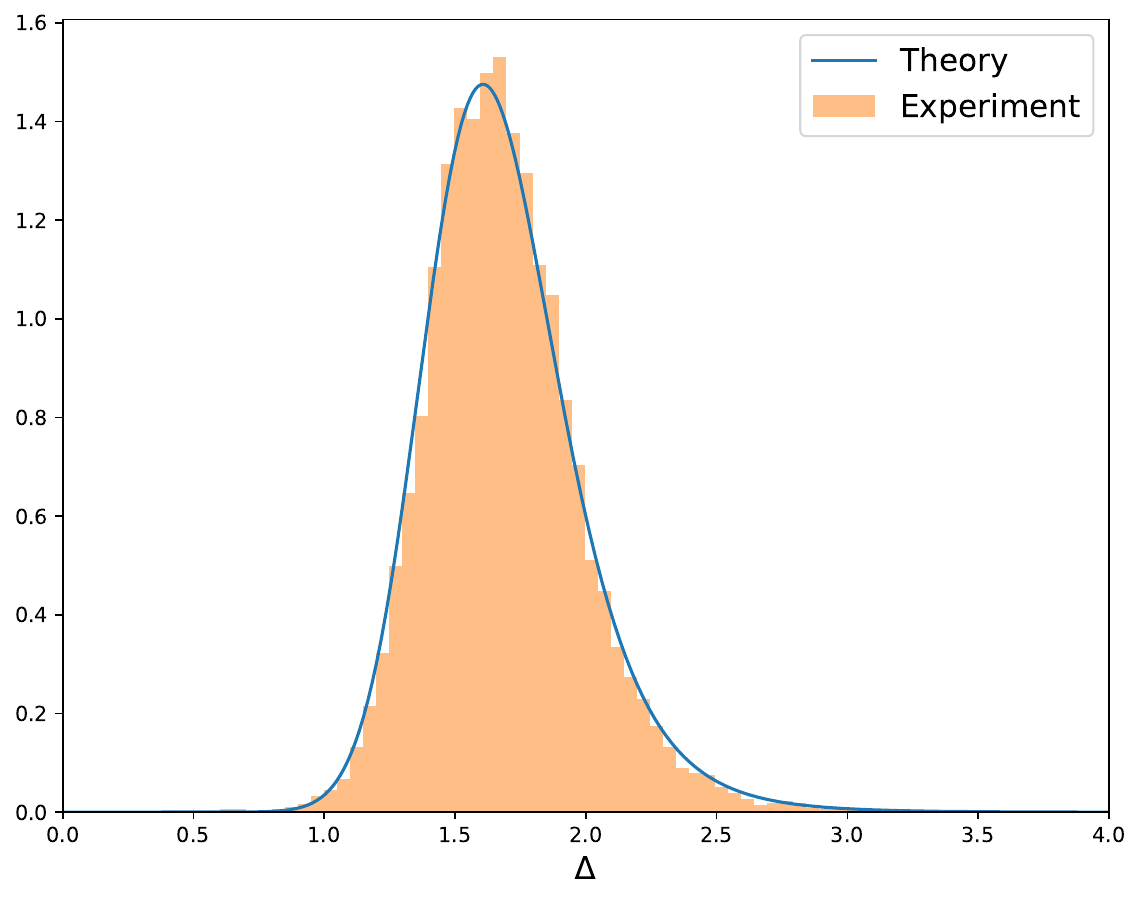}
  
%    \subcaptionbox{m = 0}{
%    \includegraphics[width=\linewidth]{delta_histo_m0.pdf}}\par

\caption{Normalised histogram of the stabilities measured from the experiment compared with Wong and Sherrington's prevision. The choice of the parameters is: $N = 100$, $\alpha = 0.3$, $m = 0.7$, $\lambda = 10^{-2}$.}
\label{fig:pdf}
\end{figure*}

\begin{figure*}[ht!]
\begin{tabularx}{\linewidth}{*{2}{X}}
     %\centering
     \begin{subfigure}[b]{\linewidth}
         \centering
         \includegraphics[width=\textwidth]{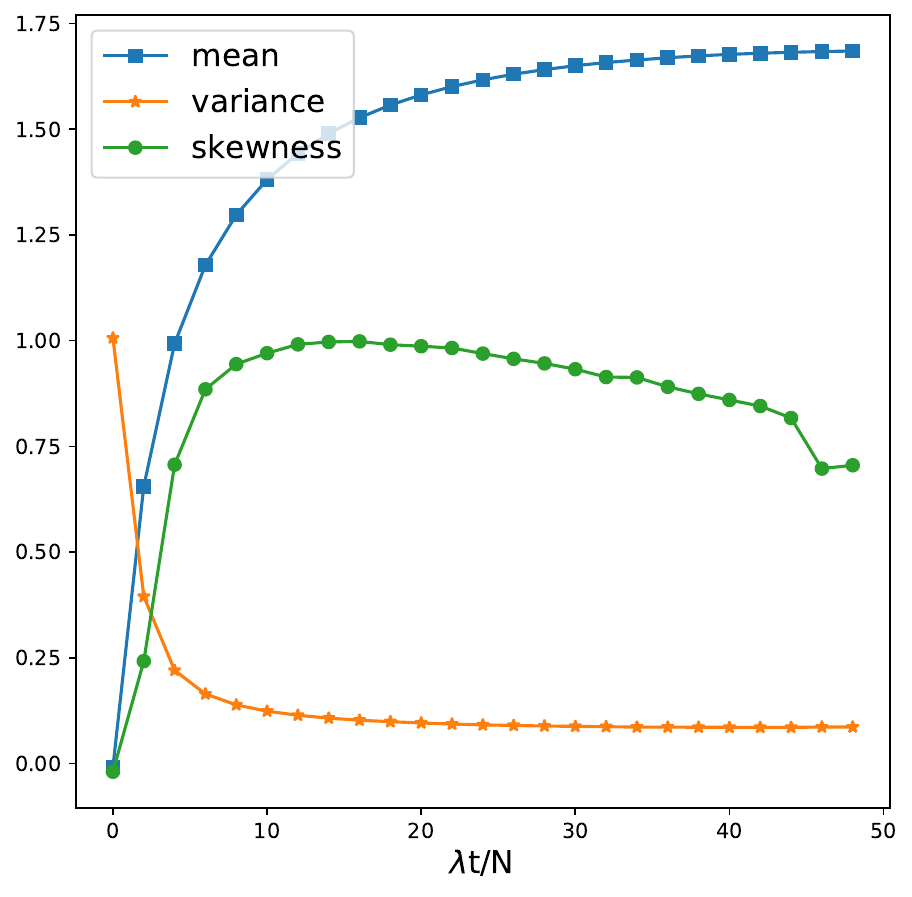}
         \caption{}
         \label{fig:trends_a}
     \end{subfigure}
     &
     \hfill
     \begin{subfigure}[b]{\linewidth}
         \centering
         \includegraphics[width=\textwidth]{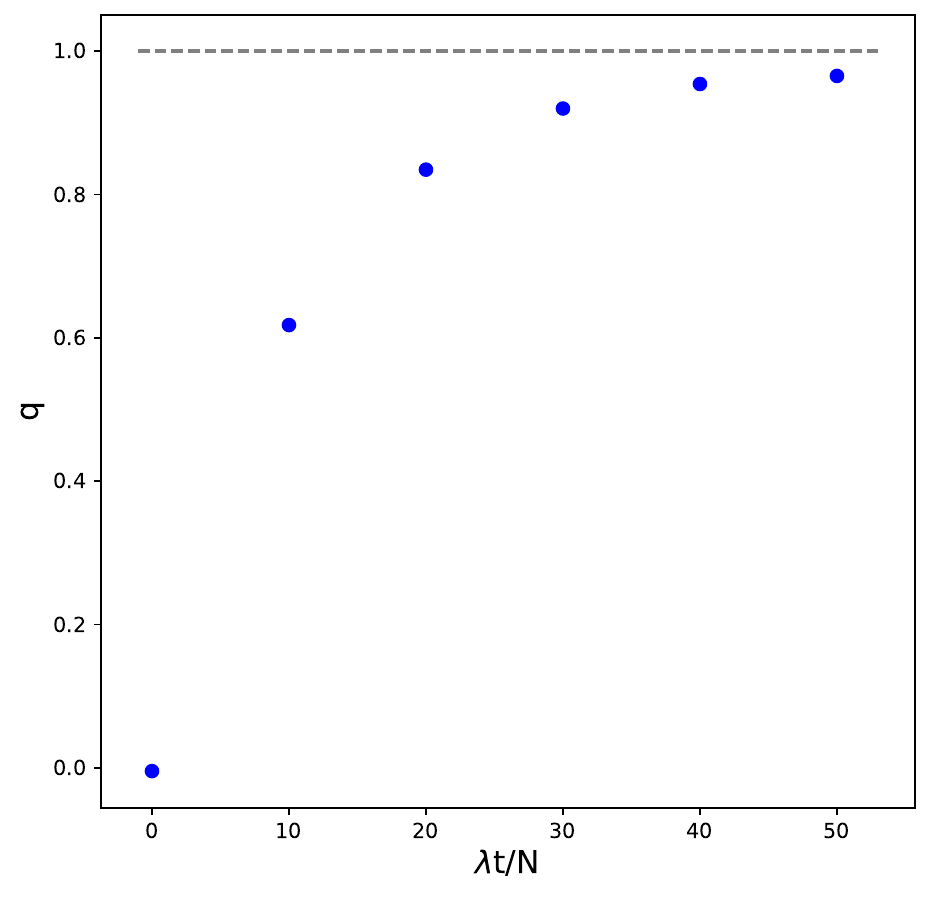}
         \caption{}
         \label{fig:trends_b}
     \end{subfigure}
\end{tabularx}
\caption{Moments of the empirical distribution of the stabilities (a) and overlap $q$ (b) as a function of the normalised time $\lambda t/N$ while implementing the TWN algorithm. Errorbars are not indicated because smaller than the symbols. The choice of the parameters is: $N = 100$, $\alpha = 0.3$, $m = 0.7$, $\lambda = 10^{-2}$.}
\end{figure*}

The numerical procedure has consisted of a random initialization of the couplings $J_{ij}$ followed by the implementation of the rule described in section \ref{sec:twn}. The values of the parameters are $N = 100$, $\alpha = 0.3$, $m = 0.7$, $\lambda = 10^{-2}$ and $t_{max} = 5\cdot 10^{5}$. Data have been collected over five repetitions of the experiments. The histogram of stabilities (\ref{eq:stab}) is empirically measured and compared to equation (\ref{eq:new_rho}), as reported in figure (\ref{fig:pdf}). There is a good accordance between theory and experiment. 
Fig. \ref{fig:trends_a} shows the evolution of the mean, variance and skewness of the $pdf$ of the stabilities as a function of the algorithm steps. Fig. \ref{fig:trends_b}, instead, depicts the evolution of the overlap $q$ measured between couples of connectivity matrices obtained from the same realisation of the memories, as a function of the algorithm steps: $q$ correctly converges to unity while approaching the global minimum of the Loss.  

\chapter{Measure of the basins of attraction in a recurrent neural network \label{app:gen_twn}}
\begin{figure}[ht!]
\centering
\includegraphics[width=0.7\linewidth]{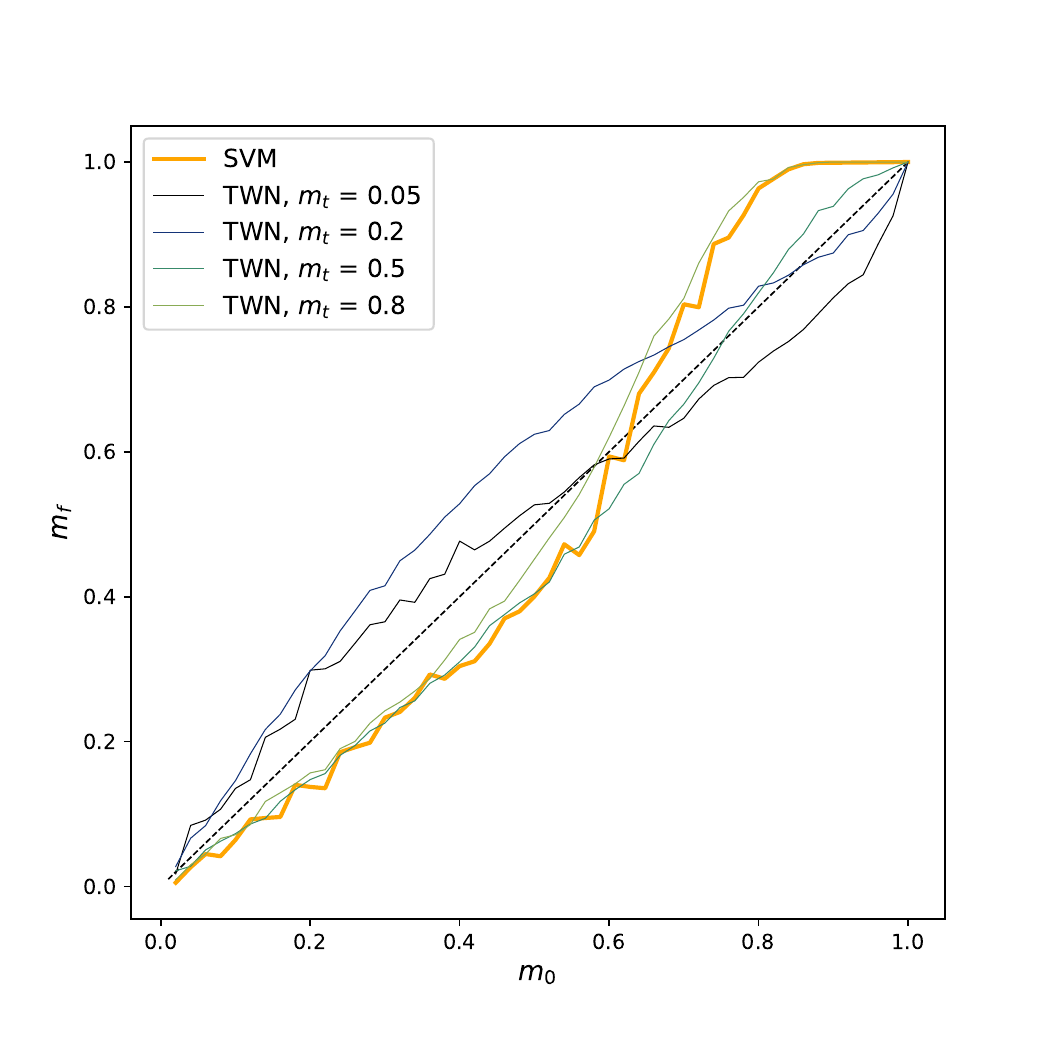}
\caption{Retrieval map $m_f(m_0)$ in the case of networks trained through SVM and TWN algorithms. Curves are shown as a function of $m_t$ and compared with the bisector, indicated with a \textit{dashed black} line. Points are averaged over $10$ samples. Choice of the parameters: $N = 200$, $\alpha = 0.45$.}
\label{fig:mfmi_evo}
\end{figure}

We report here a general experimental procedure to measure the average size of the basins of attraction of a fully connected neural network of finite size $N$ and a given choice of the control parameters.\\ The network is firstly trained according to an algorithm of our choice. Once the couplings have been found, the asynchronous version of dynamics (\ref{eq:dynamics}) is initialized in one of the memories. The dynamics is run until convergence onto the attractor associated to the basin of belonging of the memory. Now the retrieval map $m_f(m_0)$ is measured with respect to that particular attractor and the procedure is repeated over different memories and realizations of the network. The average radius of the basin of attraction is then measured as the value of $1-m_0$ where $m_f(m_0)$ equals a reference value. In our case such value is $m_f = 0.98$. 
\\We have applied this procedure on networks trained either as SVMs and with the TWN algorithm. In the former case a convex algorithm contained in the $cvxpy$ Python domain \cite{diamond_cvxpy_nodate} is implemented to train the network. To be more specific, $N$ independent machines are trained to correctly classify $p = \alpha N$ binary memories of the kind of $\vec{\xi}^{\mu} \in \{-1,+1\}^N$ having as labels $\xi_i^{\mu}$ with $i \in [1,..,N]$.\\Regarding the dynamics, the stability of fixed points is in general implied by some properties of the couplings, mainly their degree of symmetry. For the case of the TWN algorithm we start from a random symmetric matrix, as done in \cite{kepler_domains_1988}: the update of the couplings will only perturb the initial symmetry yet allowing the measures to be still consistent with the theory. On the other hand, numerics show that SVMs are sufficiently symmetric to let the asynchronous dynamics converge. The comparison between the retrieval maps obtained for the two algorithms with $\alpha = 0.45$ and $N = 200$ is reported in \cref{fig:mfmi_evo}. The curve relative to the SVM is fixed, while the one associated with the TWN is changing with respect to the training overlap $m_t$.\\Even if the SVM always reaches perfect-retrieval of the memories for $\alpha < 2$ \cite{gardner_space_1988}, a network trained with noise might show two different behaviors: one associated to \textit{retrieval} where each memory is close to an attractor, and one related to \textit{non-retrieval} where the memory is far from its attractor and the basins of attraction might contain orthogonal configurations with respect to the central attractor. In particular network models where couplings are assembled according to particular rules (e.g. \cite{hopfield_neural_1982}, \cite{dotsenko_statistical_1991}) the transition between these two regimes can be computed analytically. In the case of TWN this cannot be done. It is then important to find an empirical criterion to divide the two behaviors as a function of $(\alpha, m_t)$. \\ Let us assume that when $N \gg 1$ the retrieval map $m_f(m_0)$ develops a plateau starting from $m_0 = 1$ and ending in some limit value $m_c < 1$ such that $m_f = 1$ along all this interval. Hence one can associate the formation of such a plateau with the existence of a cohesive basin of attraction, where close configurations in hamming distance to the attractor converge to the attractor. One then wants to measure the value of $m_t$ at which such property of the basin disappears. As a possible estimate, it is convenient to consider the $m_t^*$ such that
\begin{equation}
\label{eq:change_conv}
    \frac{dm_f}{dm_0}(m_t^*)\Big|_{m_0 = 1} =  1.
\end{equation}
The numerical extrapolation of the overlap in \cref{fig:mf1} at different values of $N$ shows a good agreement between the approximate separation between the two regions showed in \cref{fig:mf1} and the line estimated by condition (\ref{eq:change_conv}).

\backmatter
\cleardoublepage
\phantomsection % Give this command only if hyperref is loaded
\chapter{Ringraziamenti}
Questo capitolo della tesi è in realtà il più importante. \\\\
La Fisica è meravigliosa, ne sono innamorato: anche se ho imparato a concepirla come un lavoro, sarà sempre la mia più grande passione. Però nulla di nulla avrebbe senso, nemmeno la Fisica, senza le persone che rendono il percorso così ricco, e significativo per me. Nella speranza che anche io sia stato significativo per questi miei compagni di viaggio, è giunto il momento di esprimere, per iscritto, l’immensa gratitudine che nutro nei loro confronti. \\\\ 
Ringrazio innanzitutto la mia famiglia. Mamma e Papà per avermi dato tutto, e Maria, perché qualunque cosa accada mi prendi sempre per mano. Grazie ai Nonni, per il loro amore e la loro complicità, grazie alle Zie, a Zio, ma soprattutto a Francesco e Giorgia, due cugini fraterni. \\\\
Ricordo coloro che mi hanno allevato nella Scienza. Ringrazio prima di tutto Francesco e Giancarlo, per essere stati così umani e comprensivi, oltre che due mentori. Grazie inoltre a Simona, Rémi, Enzo e Marco, da cui ho imparato tantissimo. \\\\  
 Grazie ai numerosi specialisti che si sono presi cura della mia salute fisica e mentale in questi anni. Sono molto riconoscente a Barbara, e a tutta la squadra di fisioterapisti tra Roma e Parigi. La fisioterapia e la psicoterapia, oltre ad essere simili, sono materie molto più complicate della Fisica. Vorrei infine ricordare la flotta di Kugoo Kirin S3, audaci monopattini da combattimento, specialmente Kugoo n.1 e Kugoo n.2, caduti sotto le mani degli scassinatori: mi avete sempre riportato a casa. \\\\
Il passaggio dagli studi al mestiere di ricercatore è stato accompagnato da un mio importante, seppur lento e graduale, mutamento interiore, che ha spostato l’attenzione dalla ricerca di me in me stesso alla ricerca di me negli altri. Lasciate dunque che mi rivolga a quelle persone per me così fondamentali, alle quali vorrei dedicare questo manoscritto: i miei Amici.\\
Voglio ringraziare i miei compagni fraterni: Cami, la mia migliore amica, e poi Carlo, Fede, Jambo, Filo, Gianmarco, Marika, Manu, Ale, da quel turbinio inarrestabile che chiameremo Malatesta Nightloop. Grazie a Ianno ed il Leo, sempre disponibili per un supplì, Luca e Lucia per la loro creatività, e l’inarrestabile Ago. 
Grazie alle anime della Garbatella, Morlu, Tere, Francesca, Vanessa, per le serate passate a Biffi o sul tetto.\\ 
Grazie a quei merenderos de La Sapienza, i.e. Simone, Anto, Stefano, Giovanni, Caja e i Chimerini, Mattia, Edoardo e Giorgio dell'IIT.\\
Ed ora ricordo chi mi ha accompagnato nei due anni passati a Parigi, una città alla quale sarò eternamente legato. Ringrazio Gianloca che nun se ferma n’attimo, Gianmarco, Antonio e Maria, Matte, Giulia e Marghe, Alberto e Federico, Sam, Niccolò, Leah, Aram: vi ricorderò stesi e ammiccanti sulla sponda della Senna altezza Tino Rossi, con camembert e vino rosso.       
Grazie ai miei colleghi ed amici all’ENS, in particolare a Felix, Giampaolo, Silvia, Simone, Laura e Ada, Flavio, Jaron, Joseph, Jeanne, Sabrina, Jules, i ragazzi del gruppo Cocco-Monasson e Walzckak-Mora, e poi la Gardner Squad, ovvero gli irreplicabili Mauro e Fabian.\\
Arrivo adesso ad un luogo speciale, così bello non perché una pagota in mezzo all’Île-de-France, ma per via delle splendide persone che lo abitano: la Maison Du Japon. Ripenso ai bei momenti spesi con Juan, Matheus, Martina, Mohak, Lore, Gianluca, Jaz e Paola, Igor, Anouk, Taka, Leire, Francesco, alle serate passate nella cucina del secondo piano oppure in giardino a cantare. Ricordo con affetto Shashwath e Kai del terzo piano, e poi Paula, Nara e Monica, e i coniugi Nishiumi, che ci hanno custoditi nel loro nido. E infine ringrazio Ester, una stella: con te ho imparato qualcosa che va ben oltre le pagine di una tesi di dottorato. \\\\

Vi porto sempre con me.\\ 

Enrico. 
\nocite{*}
\bibliographystyle{unsrt}
\bibliography{biblio}
%\addcontentsline{toc}{chapter}{\bibname}
% Here put the code for the bibliography. You can use BibTeX or
% the BibLaTeX package or the simple environment thebibliography.

\end{document}